\documentclass[aps,prb,showpacs,twocolumn,superscriptaddress]{revtex4-2}
\usepackage{bm,color,amsmath,amssymb,mathrsfs,latexsym,graphicx,psfrag,mathtools}
\usepackage{color}
\usepackage[dvipsnames]{xcolor}
\usepackage[hidelinks,colorlinks,linkcolor=blue,
citecolor=blue,urlcolor=blue]{hyperref}
\usepackage{dsfont}
\usepackage{xspace}

\usepackage{empheq}

\newcommand{\bsl}[1]{\boldsymbol{#1}}

\newcommand{\bra}[1]{\langle #1|}
\newcommand{\ket}[1]{|#1 \rangle}
\newcommand{\braket}[2]{\left\langle #1 | #2  \right\rangle}

\newcommand{\ii}{\mathrm{i}}

\newcommand{\Tr}{\mathop{\mathrm{Tr}}}
\renewcommand{\Re}{\mathop{\mathrm{Re}}}
\renewcommand{\Im}{\mathop{\mathrm{Im}}}

\newcommand{\eqnref}[1]{Eq.\,\eqref{#1}}

\newcommand{\appref}[1]{supplementary information\,\ref{#1}}
\newcommand{\refcite}[1]{Ref.\,\cite{#1}}

\newcommand{\mat}[1]{\left(\begin{matrix}#1\end{matrix}\right)}

\newcommand{\eq}[1]{\begin{equation} #1 \end{equation}}

\newcommand{\eqa}[1]{\begin{align}\begin{split} #1 \end{split}\end{align}}

\let\oldAA\AA
\renewcommand{\AA}{\text{\normalfont\oldAA}}

\newcommand{\ie}{{\emph{i.e.}}}
\newcommand{\eg}{{\emph{e.g.}}}

\newcommand{\G}{\mathcal{G}}
\newcommand{\E}{\mathcal{E}}

\renewcommand{\H}{\mathcal{H}}

\newcommand{\V}{\mathcal{V}}

\newcommand{\mgb}{\text{MgB$_2$}}
\newcommand{\mcomega}{\left\langle \omega^2 \right\rangle}
\newcommand{\BZ}{\text{1BZ}}

\newcommand{\bd}{\bsl{d}\xspace}
\newcommand{\rr}{\bsl{r}\xspace}
\newcommand{\kk}{\bsl{k}\xspace}

\newcommand{\qq}{\bsl{q}\xspace}
\newcommand{\MM}{\bsl{M}\xspace}

\newcommand{\GG}{\bsl{G}\xspace}

\newcommand{\gmm}    {g_{ij}}
\newcommand{\gm}     {$\gmm$\xspace}
\newcommand{\dsm}    {\left[D_{\rm s}\right]_{ij}}
\newcommand{\ds}     {$\dsm$\xspace}
\newcommand{\ceq}[1] {(\ref{#1})}
\usepackage{comment}

\def\ba#1\ea{\begin{align}#1\end{align}}
\def\bg#1\eg{\begin{gather}#1\end{gather}}
\def\bpm{\begin{pmatrix}}
	\def\epm{\end{pmatrix}}
\newcommand{\nn}{\nonumber}

\renewcommand{\b}[1]{{\boldsymbol #1}}

\newcommand{\bk}{{\b k}}

\renewcommand{\v}[1]{{\bf #1}}

\newcommand{\der}{\partial}

\newcommand{\vep}{\varepsilon}

\newcommand{\brk}[2]{\langle #1 | #2 \rangle}

\newcommand{\eqn}[1]{Eq.~\eqref{#1}}

\newcommand{\proj}[1]{\left| #1 \right>\!\left< #1 \right|}
\newcommand{\ev}[1]{\left< #1 \right>} 
\def\bk{\bsl{k}}
\def\br{\bsl{r}}
\def\bR{\bsl{R}}
\def\Q{\mathcal{Q}}
\def\D{\mathcal{D}}
\def\S{\mathcal{S}}
\def\E{\mathcal{E}}

\newcommand{\kkb}{\bsl{k}}

\usepackage{cleveref}
\crefname{appendix}{App.\,}{Apps.\,}
\crefname{equation}{Eq.\,}{Eqs.\,}
\crefname{figure}{Fig.\,}{Figs.\,}
\crefname{table}{Tab.\,}{Tabs.\,}
\crefname{section}{Sec.\,}{Secs.\,}
\creflabelformat{appendix}{[#2#1#3]}
 
\begin{document}
\title{Quantum Geometry in Quantum Materials}

\author{Jiabin Yu}
\affiliation{Department of Physics, University of Florida, Gainesville, FL, USA}
\affiliation{Department of Physics, Princeton University, Princeton, New Jersey 08544, USA}

\author{B. Andrei Bernevig}
\thanks{These authors are listed alphabetically.}
\affiliation{Department of Physics, Princeton University, Princeton, New Jersey 08544, USA}
\affiliation{Donostia International Physics Center, P. Manuel de Lardizabal 4, 20018 Donostia-San Sebastian, Spain}
\affiliation{IKERBASQUE, Basque Foundation for Science, Bilbao, Spain}

\author{Raquel Queiroz}
\thanks{These authors are listed alphabetically.}
\affiliation{Department of Physics, Columbia University, New York, NY, USA}

\author{Enrico Rossi}
\thanks{These authors are listed alphabetically.}
\affiliation{Department of Physics, William \& Mary, Williamsburg, Virginia 23187, USA}

\author{P\"aivi T\"orm\"a}
\thanks{These authors are listed alphabetically.}
\affiliation{Department of Applied Physics, Aalto University School of Science, FI-00076 Aalto, Finland}

\author{Bohm-Jung Yang}
\thanks{These authors are listed alphabetically.}
\affiliation{Department of Physics and Astronomy, Seoul National University, Seoul 08826, Korea}
\affiliation{Center for Theoretical Physics (CTP), Seoul National University, Seoul 08826, Korea}
\affiliation{Institute of Applied Physics, Seoul National University, Seoul 08826, Korea}

\begin{abstract}
Quantum geometry, characterized by the quantum geometric tensor, is pivotal in diverse physical phenomena in quantum materials. In condensed matter systems, quantum geometry refers to the geoemtric properties of Bloch states in the Brillouin zone. This pedagogical review provides an accessible introduction to the concept of quantum geometry, emphasizing its extensive implications across multiple domains. Specifically, we discuss the role of quantum geometry in optical responses, Landau levels, and fractional Chern insulators, as well as its influence on superfluid weight, spin stiffness, exciton condensates, electron-phonon coupling, etc. By integrating these topics, we underscore the pervasive significance of quantum geometry in understanding emergent behaviors in quantum materials. Finally, we present an outlook on open questions and potential future directions, highlighting the need for continued exploration in this rapidly developing field.
\end{abstract}

\maketitle

\section{Introduction}

Quantum materials can be loosely defined as materials for which quantum mechanical
effects manifest on a macroscopic scale. Two classes of quantum materials are paradigmatic:
superconductors, and quantum Hall systems.
For superconductors, electron-electron interaction is the key ingredient that leads to a macroscopic manifestation of quantum mechanics:
such interaction causes the electrons to form phase-coherent Cooper pairs and this
results in the Meissner effect and the dissipationless transport of charge current.
For a two-dimensional (2D) electron gas in the integer quantum Hall regime, the perfect quantization of the Hall conductivity can be understood without explicitly taking any effects of electron-electron interactions into account.
The integer quantum Hall effect (QHE) can be attributed
to the unique topology of the free-electrons' ground state~\cite{thouless1982b}.
Such topology is encoded by the Chern number, $C$, given by the integral over the Brillouin zone
of the Berry curvature that measures the change of the eigenstate's phase as the momentum $\bk$ is varied.
The Berry curvature is part of the {\em quantum geometry} of a material.
The QHE is the archetypical demonstration that quantum geometry is one of the key quantities
that make a material a {\em quantum material}.
As we will discuss in the remainder of the review, the Berry curvature turns out to be the anti-symmetric 
part of a tensor $Q$, the quantum geometric tensor (QGT)~\cite{provost1980}. 
In recent years it has become apparent that the symmetric part of this tensor, the quantum metric, $g$,
also plays a key role in making a material, quantum. 
In a loose sense, the quantum metric appears to be the key quantity to understand the properties
of materials in which both interactions and quantum geometry lead to macroscopic manifestations
of quantum mechanics.

Quantum geometry is the geometric structure that naturally arises in the space of quantum states
when such states depend on continuous parameters.
One classic example of quantum geometry is the geometric phase of a quantum state under adiabatic evolution, in which case the continuous parameter is time.
Within condensed matter physics, the continuous parameters are the components of the crystal momentum $\kk$, and quantum geometry refers to the geometric properties of the Bloch states, more precisely, the periodic part of the Bloch states $\ket{u_{\bsl{k}}}$.
In this context, quantum geometry is also called band geometry, which includes long-known concepts such as the spread of the possible Wannier basis and the parallel transport of the electronic states.  

The quantum geometric tensor (also called the Fubini-Study metric~\cite{Fubini1904,Study1905}) has components:
\eq{
Q_{ij}(\bsl{k}) = \bra{\partial_{k_i} u_{\bsl{k}} }[{\mathds{1}} - \ket{u_{\bsl{k}}}\bra{u_{\bsl{k}}}] \ket{\partial_{k_j} u_{\bsl{k}}}\ ,
\label{eq:qgt}}
where $k_i$ is the $i$th component of the Bloch momentum $\bsl{k}$. 
For simplicity in writing \eqnref{eq:qgt} we have considered the case of a well-isolated band.
The antisymmetric part of $Q_{ij}(\bsl{k})$ is $\ii  B_{ij}(\bsl{k})= [ Q_{ij}(\bsl{k})-Q_{ji}(\bsl{k})]/2 $ is related to the well known Berry curvature~\cite{Berry1984,Simon1983,Aharonov1987} $F_{ij}(\bsl{k})$ as $B_{ij}(\bsl{k})=-F_{ij}(\bsl{k})/2$,
and the symmetric part $g_{ij}(\bsl{k})= [ Q_{ij}(\bsl{k})+Q_{ji}(\bsl{k})]/2$ is the quantum metric 
$g$ given that corresponds to the metric for infinitesimal distances 
of the Hilbert-Schmidt quantum distance $d_\mathrm{HS}(\bsl{k},\bsl{k}^\prime)\equiv\sqrt{1-\left|\langle u_{\bsl{k}} | u_{\bsl{k}^\prime} \rangle \right|^2}$: $ds^2 = \sum_{i,j} g_{ij}(\bsl{k}) dk_i dk_j$.

In two dimensions (2D) the integral over the Brillouin Zone (BZ) 
of $B_{xy}(\bsl{k})/\pi$ for the states of an occupied band is quantized and equal to the Chern number $C$.
Conversely, the integral of $g_{ij}(\bsl{k})$ over the BZ is in general not quantized.
However, in 2D, the positive semidefinite nature of $Q$
(combined with inequality between trace and determinant) implies  the following inequalities~\cite{roy2014} 
\begin{align}
  {\rm Tr}g(\bsl{k})  \geq
  2\sqrt{{\rm det}g(\bsl{k}) } \geq 2|B_{xy}(\bsl{k})|
  \label{eq.det.in}
\end{align}
We can introduce the tensor 
$M\equiv (1/\pi)\int_{BZ} d^d\bsl{k}\ Q(\kkb)$.
Because $M$ is a sum 
of positive semidefinite tensors, it is itself positive semidefinite, and
so ${\rm det}M\geq 0$. In 2D this leads to the inequality
${\rm det(Re}(M))\geq {\rm det(Im}(M))$, that can be seen as the
integral equivalent of \eqnref{eq.det.in}, and can be written as
\eq{
 {\rm det}\left[\frac{1}{\pi}\int d^2k\ g(\bsl{k}) \right] \geq
 {\rm det}\left[\frac{1}{\pi}\int d^2k\ B(\bsl{k}) \right] = C^2.
 \label{eq.ineq}
}
\eqnref{eq.ineq} is a classic example of topology bounding quantum geometry from below~\cite{10.1063/1.530758}.
The generalization of \eqnref{eq.ineq} leads to the lower bound of quantum geometry due to the Euler number~\cite{Xie2020TopologyBoundSCTBG,Yu2022EOCPTBG,BJY2024EulerBoundQG,Slager2024EulerOptical} (the generalization is most natural in the Chern gauge for the Euler bands), and the lower bound has also been derived for obstructed atomic limits~\cite{Herzog-Arbeitman2021} and chiral winding number~\cite{PhysRevB.94.245149}.
Recently, the lower bound of quantum geometry has also been derived~\cite{Yu2024Z2boundQG} for the time-reversal protected $Z_2$ topology~\cite{Kane2005Z2,Zhang2006QSH,Kane2005QSH,Bernevig2006BHZ}. 
These topological bounds allow us to put a lower bound to the geometric contribution
to quantities such as the superfluid weight, as discussed in Sec.~\ref{SandS}. 

The quantization in 2D of the integral of $B_{ij}(\kk)/(2\pi)$ over the BZ,
and its direct relation to the off-diagonal conductivity $\sigma_{xy}$~\cite{thouless1982b}
made the study of the physical consequences of the anti-symmetric part of 
$Q(\bsl{k})$ one of the most active areas of research in condensed matter physics for the past twenty years.  It has
led to several discoveries, such as topological insulators (TIs) and superconductors~\cite{hasan2010a,qi2011a,chiu2016}, 
Weyl and Dirac semimetals (SMs)~\cite{Vafek2014,Wehling2014,Burkov2018}, and, more recently,
higher order topological materials~\cite{Benalcazar2017a,schindler2018a,Song2017,khalaf2018a,Ezawa2018,Trifunovic2019}. 
Conversely, the study of the symmetric part of $Q(\bsl{k})$ has received much less attention largely due to the fact
until recently $g(\bsl{k})$ had only been shown to contribute to quantities that are challenging to measure experimentally,
like the Hall viscosity~\cite{Avron1995,Read2009,haldane2009b,Read2011,Hoyos2012,bradlyn2012b,Haldane2015,Shapourian2015,holder2019},
and the `Drude weight'', $D$, of the electrical conductivity of 
{\em clean systems at zero temperature}~\cite{Resta2011b,gao2015,resta2018,Marrazzo2019,Bellomia2020}.
Theoretical and experimental developments in the last few years have profoundly changed the situation.
First, it was shown that $g$ is related to nonlinear responses~\cite{gao2014,morimoto2016b,nagaosa2017a,kolodrubetz2017,ahn2020b,ahn2022a,wang2021a,ma2023,Liu_2024QGreview,avdoshkin2024,sala2024}.
It was further pointed out that $g$ contributes to
the superfluid weight (same as superfluid stiffness) \ds 
of a superconductor~\cite{peotta2015,Julku2016,liang2017,Hu2019,xie2020,julku2020,rossi2021,Torma2022ReviewQuantumGeometry,torma2023,tian2023},
and that such contribution is significant when the bandwidth of the bands
crossing the Fermi energy is smaller than the superconducting gap.
Both nonlinear responses and \ds
could in principle be measured in realistic experimental conditions.
In addition, the realization of magic-angle twisted bilayer graphene (TBG)~\cite{Cao2018c,Cao2018a,Yankowitz2019,Chen2019,lu2019a,Choi2019,sharpe2019,Polshyn2019,Codecido2019,Liu2020a,Shen2020,Chen2020,Cao2020,andrei2020} and other twisted materials \cite{Zhang2023MoTe2, Dong2023CFLtMoTe2, Young2024MagtMoTe2} introduced  experimentally accessible systems with extremely flat bands exhibiting superconductivity and Fractional Chern Insulator/Ferromagnetism
for which the quantum metric contribution to the superconducting \ds or to the magnon stiffness can be large.
These developments have motivated a huge interest in understanding the role of \gm in quantum materials.

We now discuss more in detail how the quantum metric affects the properties of condensed matter systems.
However, it is worth emphasizing that, besides condensed matter physics, 
the quantum metric plays a role in many other areas of physics, such as metrology, via the closely 
related concept of quantum Fisher information~\cite{helstrom1976}, non-equilibrium dynamics~\cite{kolodrubetz2017},
and quantum information science~\cite{pezze2009,paris2009b}.

\subsection{Simple two-bands model}
To gain some intuition about  quantum geometry, it is useful to consider a
simple two-band model described by the following Hamiltonian $h(\theta,\varphi)=\bd(\theta,\varphi)\cdot\bm{\sigma}$
where $\bm{\sigma}=(\sigma_x,\sigma_y,\sigma_z)$ is the vector formed by the $2\times 2$ Pauli matrices and 
\eq{
  \bd = (\sin(J_\theta\theta)\cos(J_\varphi\varphi), \sin(J_\theta\theta)\sin(J_\varphi\varphi), \cos(J_\theta\theta)).
 \label{eq.d.two.bands}
}
with $J_\theta$, $J_\varphi$, two integers. The energy eigenvalues are $\epsilon_{\pm}=\pm|\bd(\theta,\varphi)|=\pm 1$. 
The eigenvalues $\epsilon_{\pm}$ do not depend on the variables $\theta$ and $\varphi$ that parametrize $h$
and therefore describe two flat bands. The variables $\theta$ and $\varphi$ only affect the energy eigenstates: 
$v_-=(\sin(J_\theta\theta/2)e^{-\ii J_\varphi\varphi}, -\cos(J_\theta\theta/2))^T$,  
$v_+=(\cos(J_\theta\theta/2)e^{-\ii J_\varphi\varphi},  \sin(J_\theta\theta/2))^T$.

\begin{figure}[!!!t]
 \begin{center}
  \centering
  \includegraphics[width=0.49\textwidth]{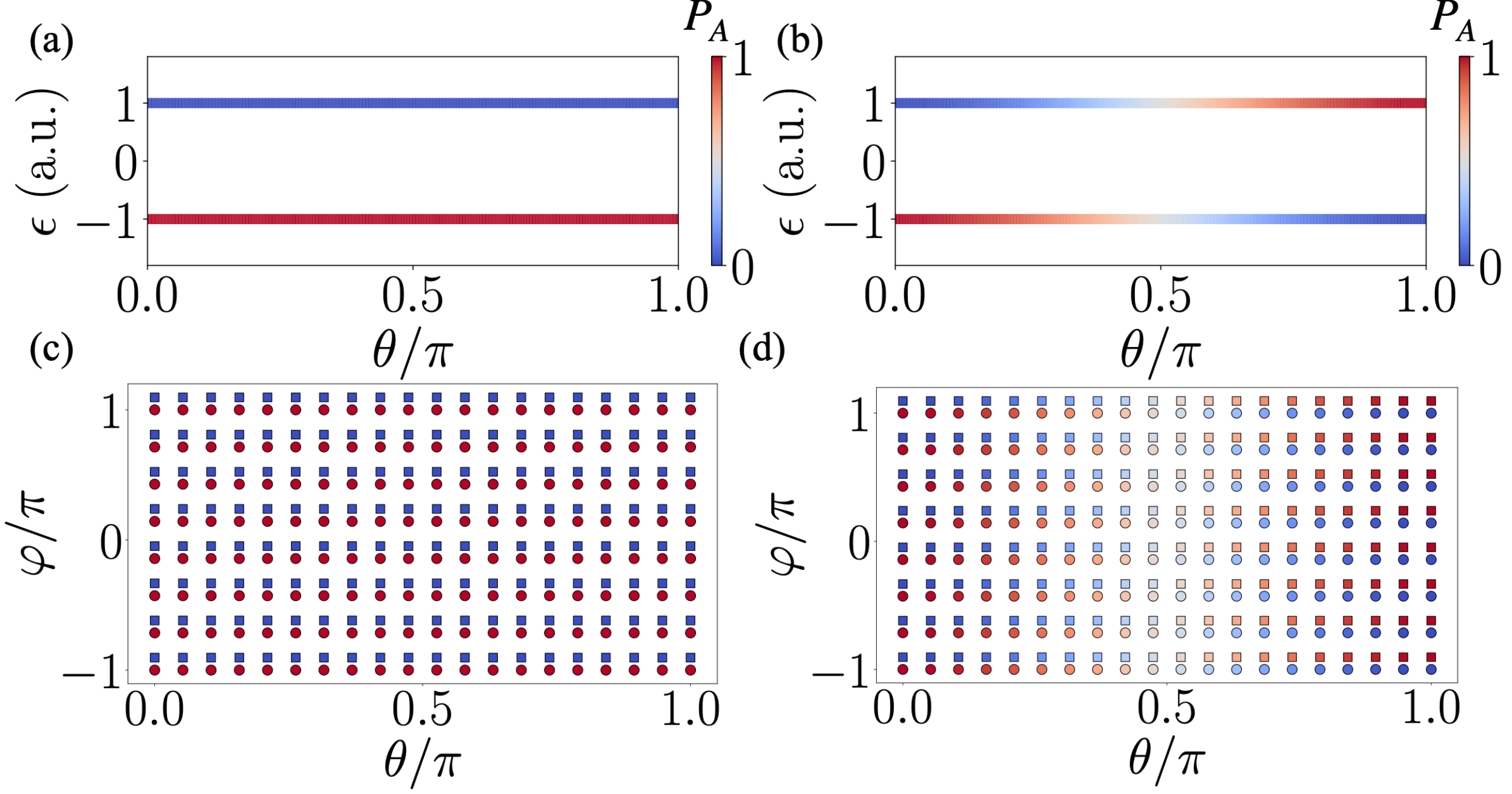}
  \caption{
           Schematic spectra and states hybridization for a simple two-level model in which the two-fold degree of freedom is mapped to a sublattice, $\{A,B\}$.
           (a), (b) Energy eigenvalues as a function of $\theta$
           for the case when $J_\theta=J_\phi=0$
           and $J_\theta=J_\phi=1$, respectively. The colors show
           the occupation probability of sublattice $A$, $P_A$.
           (c), (d) Representation of the lowest energy eigenstate
           distribution, as function of $\theta$ and $\varphi$,
           between sublattice $A$, represented by circles, and sublattice $B$, represented by squares, and the color showing the probability occupation of each sublattice, for $J_\theta=J_\phi=0$ and
           $J_\theta=J_\phi=1$, respectively. For the probabilities, $P_A$, $P_B$, on each sublattice
           we use the same color scheme used for $P_A$ in panels (a) and (b).
          } 
  \label{fig.model}
 \end{center}
\end{figure}

In the limit $J_\theta=J_\phi=0$ also the eigenstates do not depend on $\theta$ and $\varphi$ a
and therefore the QGT is identically zero;
the Hamiltonian describes a system with no quantum geometry.
If we associate the degree of freedom described by the Pauli matrices to a sublattice
degree of freedom, this case can be visualized as the situation in which in
each energy eigenstate the electrons are completely localized on one of the two ``sublattices'' (entries of the spinor wavefunction),
as shown schematically in Fig.~\ref{fig.model}~(a,c).

When $J_\theta=J_\phi=1$ the eigenstates depend on $\theta$ and $\varphi$ and therefore
the two bands possess a non-trivial quantum geometry. 
Using the expression above for $v_-$, and the definition~\eqnref{eq:qgt} of the QGT (with $k_i$ ($i=1,2$) running over the labels $(\varphi,\theta)$) for the lowest band we find
\eq{
 Q_{ij}(\varphi,\theta) = 
 \begin{pmatrix}
 \frac{\sin^2\theta}{4} & \ii \frac{\sin\theta}{4}\\
 -\ii \frac{\sin\theta}{4} & \frac{1}{4}
\end{pmatrix}
}
As expected the anti-symmetric part of the QGT is equal to $-1/2$ the Berry curvature~\cite{bernevig2013}.
It is straightforward to verify that the 
inequalities in \cref{eq.det.in} are satisfied.
\cref{fig.model}~(b,d) illustrate the dependence of the eigenstates on 
the variables that parametrize the Hamiltonian: the dependence on $\theta$ and $\varphi$ 
of the hybridization of the two degrees of freedom is responsible for 
a nonzero QGT. In the case of a sublattice, this can be visualized as a dependence 
on $\theta$ of the relative weight of the electron wave function between the sublattices, Fig.~\ref{fig.model}~(d).

For the case in which $H$ describes electrons with a two-fold degree of freedom 
(sublattice, spin, or a generic orbital degree of freedom) a nonzero 
anti-symmetric part of $Q_{ij}$ can result in a contribution to off-diagonal (transverse) transport coefficients.
For instance, for $J_\theta=J_\phi=1$ our simple model exhibits a Chern number $C=1$~\cite{bernevig2013} that can be associated with a non-zero and quantized Hall conductivity,
and therefore the presence of delocalized electronic states.
Similarly, the real part of $Q_{ij}(\bsl{k})$ can result in a contribution to {\em diagonal} (longitudinal) transport coefficients. 
For instance, in this example, in the metallic regime, when the band is not completely filled,
one would expect the Drude weight $D$ to be zero given that the band is completely flat 
and $D\sim n/m^*$, with $n$ the electron's density and $m^*$ the electron's effective mass, i.e.,
the curvature of the bands. 
However, the fact that $g_{ij}(\bsl{k})$ is nonzero results in a nonzero 
Drude weight~\cite{gao2015,resta2018}, current noise~\cite{neupert2013}, and a quantum geometric dipole~\cite{fertig2024}, signaling that even if the band is completely flat, the system
can respond, in the ideal case, to an external d.c. electric field. This is an indication that some states in this band are not completely on-site localized.

It is interesting to consider the limit $J_\varphi=0$, $J_\theta=1$.
In this case, $h$ is parametrized by only one variable, $\theta$, and therefore
the Berry curvature, and so the anti-symmetric part of $Q_{ij}(\bsl{k})$, are identically zero.
Nevertheless, the quantum metric is still not zero: $g_{\theta\theta}=1/4$.
This simple example is an extreme case of the important fact that the quantum metric
can be nonzero even if the Berry curvature is zero. As \cref{eq.det.in} shows, the quantum metric is only bounded from below by the Berry curvature.

\subsection{Localization tensor}

The QGT tensor is not limited to the single-band case---it can be defined for an isolated set of any number of bands.
Of common interest is the set of occupied bands of a band insulator, for which the QGT reads
\eq{
\left[ Q_{ij}(\bsl{k}) \right]_{mn} = \bra{\partial_{k_i} u_{m\bsl{k}} }[{\mathds{1}} - P_{\bsl{k}}] \ket{\partial_{k_j} u_{n\bsl{k}}}\ ,
\label{eq:qgt_multiband}
}
where $P_{\bsl{k}} = \sum_{n\in\text{occ.}}\ket{u_{n\bsl{k}}}\bra{u_{n\bsl{k}}}$.
There is a very physical connection between the QGT in \cref{eq:qgt_multiband} and the localization of the electronic wavefunctions.
This connection is imprinted in the linear response of materials when subjected to applied electric or magnetic fields; it follows naturally from the fact that the QGT defined in \cref{eq:qgt_multiband}, when integrated over the Brillouin zone and traced over all occupied bands, 
\eq{
\mathcal{Q}_{ij}=\sum_{n\in\text{occ.}} \int_{\text{BZ}}\frac{d^d k}{(2\pi)^d}\left[Q_{ij}(\bk) \right]_{nn}\ ,
}
with $d$ the spatial dimension. $\Q_{ij}$
can be recast as the ground state dipole-dipole correlator
\begin{equation}
    \mathcal{Q}_{ij}=\langle r_i(1-P)r_j\rangle \label{eq:groundstateqgt} =  {1\over \V}\Tr[ P r_i(1-P)r_j P]
\end{equation}
where $P=\sum_{n\in\rm occ.}\sum_{\bsl{k}\in\text{BZ}}\proj{\psi_{n\bk}}$ is the projector in the occupied subspace.
Here $\langle\br\ket{\psi_{n,\bsl{k}}} = e^{\ii \bsl{k}\cdot \bsl{r}} \langle\br\ket{u_{n,\bsl{k}}}$ is the Bloch state for the $n$th band, and henceforth, we choose the unit system in which
\eq{
\hbar = 1\ ,
}
unless specified otherwise
The position matrix elements are defined for Bloch states\cite{Blount1962} 
\begin{equation}
    \langle \psi_{n\bk}  | {r}_i | \psi_{m\bk'} \rangle = \delta_{\bk,\bk'} \left[A_i(\bsl{k})\right]_{mn}+ i\delta_{mn}\partial_{k_i} \delta_{\bk,\bk'} \ ,
    \label{eq:blount-matrix-element}
\end{equation}
where $\delta_{\bk,\bk'}= (2\pi)^d  \delta(\bk - \bk') / \mathcal{V}$, $\mathcal{V}$ the total volume, and $\left[A_i(\bsl{k})\right]_{nm}=i\langle u_{n\bsl{k}} | \partial_{k_i} | u_{m\bsl{k}} \rangle$. 
The projector $1-P$ guarantees that Eq.\eqref{eq:groundstateqgt} is gauge independent, by removing the diagonal contributions of the position operator. The integrated geometric tensor $\mathcal{Q}_{ij}$, whose symmetric part and anti-symmetric parts contain the integrated quantum metric and the Chern number of the ground state, can be interpreted as a localization tensor originating in the uncertainty of the position operator in the ground state \cite{Resta.Resta.2011}. First discussed by Kohn \cite{Kohn.Kohn.1964}, a divergent $\Q_{ij}$ would correspond to an infinite sensitivity of the ground state to a shift in momentum or a twist in boundary conditions. Among other definitions based on a spectral gap, this definition based on spatial delocalization of the wavefunction is one of the most suited ways to discriminate between a metal and an insulator. Its geometrical interpretation was then put forward by Resta \cite{Sorella.Resta.1999}, and Souza, Wilkens, and Martin~\cite{Martin.Souza.1999}.

The localization tensor has deep consequences in both the constraints on the basis functions that can be used to describe a given band subspace. 
To understand the spatial extent of electronic bands, it is useful to adopt Wannier states, localized in real space to describe Bloch bands \cite{MarzariRMP2012}. They are defined as
\[|w_{n\bR}\rangle=\frac{1}{\sqrt{N}}\sum_{\bsl{k}\in\mathrm{BZ}} e^{-i \bk \cdot \bR+\varphi(\bk)}\left|\psi_{n\bk}\right\rangle\]
where $N$ is the number of lattice sites, and $\varphi_n(\bk)$ is a momentum-dependent phase redundancy, which can be tuned to optimize the localization of the Wannier states $\ket{w_{n\bR}}$. This localization is characterized by the localization functional 
$\Omega=\sum_n\left[\langle w_{n\bf 0}| r^2|w_{n\bf 0}\rangle-\langle w_{n\bf 0}| \bsl{r}|w_{n\bf 0}\rangle^2\right]$ \cite{Vanderbilt.Marzari.1997}. While not gauge independent, $\Omega$ can be separated into a gauge independent part that coincides with the trace over spatial indices of the integrated quantum metric, often referred to as $\Omega_I=(\mathcal{V}/N)\Re(\Tr\Q)$, and a gauge dependent part $\tilde\Omega$ (see Appendix[\ref{sec:metric}]). The gauge-dependent part $\tilde\Omega$ diverges in the absence of exponentially localized Wannier functions. This happens for metals, or in the presence of a nonzero Chern number\cite{Marzari.Brouder.2007}. An extended review of these results is presented in Appendix[\ref{sec:metric}].

\section{Quantum Geometry and Correlated States}

\subsection{Superconductivity and superfluidity} \label{SandS}

Superconductivity is fundamentally influenced by the spread of the Wannier states and hence by quantum geometry: the superfluid weight (superfluid stiffness) $D_{\rm s}$ has a contribution that arises from quantum geometry, $D_{\rm s,geom}$, in addition to the conventional one $D_{\rm s,conv}$ given by band dispersion~\cite{peotta2015}:
\begin{equation}
D_s = D_{\rm s,conv} + D_{\rm s,geom} \ .
\end{equation}
The superfluid weight relates the 
DC, long-wavelength supercurrent $\bsl{J}$ to
the static, long-wavelength, transverse 
vector potential $\bsl{\mathcal{A}}$:
\begin{equation}
   J_i = - \sum_{j} \left[ D_{\rm s} \right]_{ij} \mathcal{A}_{j}
   \label{eq.london}
\end{equation}
and needs to be positive-definite for supercurrent to exist. Moreover, in the simplest picture, large $\mathcal{D}_{\rm s}=\Tr[D_{\rm s}]/d$ means a large critical current of superconductivity, $j_{\rm c} \propto \mathcal{D}_{\rm s}/\xi$, where $\xi $ is the coherence length of the superconductor. In two dimensions, $D_{\rm s}$ also determines the critical temperature of superconductivity since the Berezinskii-Kosterlitz-Thouless (BKT) temperature $T_{\rm BKT}$ depends on $\mathcal{D}_{\rm s}$. 
In 3D the penetration depth $\lambda$ is directly proportional to $1/\sqrt{\mathcal{D}_{\rm s}}$. A finite value of 
$\lambda$ is crucial for the Meissner effect. 
Given that $\mathcal{D}_{\rm s}$ is the proportionality constant entering the London equation~\eqnref{eq.london}, the equation responsible for the Meissner effect, and its relation to $T_{\rm BKT}$ in 2D and to $\lambda$ in 3D, it is a fundamental defining quantity of superconductivity. Interestingly, $D_{\rm s}$ has an intrinsic connection to quantum geometry.  

The quantum geometric contribution of superconductivity becomes dramatic in a flat Bloch band. The conventional contribution of superfluid weight $D_{\rm s,conv}$ is inversely proportional to the effective mass of the band and vanishes in a flat band. Superconductivity in a flat band is thus completely based on quantum geometric effects. Such effects arise since $D_s$ is defined via the current-current correlator~\cite{Scalapino1993}, and the current operator of a multiband system has two parts ($m$,$n$ are band indices and $i=x,y,z$): 
\begin{equation}
\bra{u_{m\bsl{k}}}J_i\ket{u_{n\bsl{k}}} = \delta_{mn} \partial_{k_i}\epsilon_{n\bsl{k}} + (\epsilon_{m\bsl{k}} - \epsilon_{n\bsl{k}}) \bra{\partial_{k_i} u_{m\bsl{k}}} u_{n\bsl{k}} \rangle,  \label{simplecurrentequation}
\end{equation}
Here $\bsl{k}$ is the momentum and $\epsilon_{n\bsl{k}}$ gives the dispersion for the $n$th band. The last term, which contains a derivative of the Bloch function, connects $D_{\rm s}$ to the quantum geometric quantities defined in the introduction.

Full formulas of $D_{\rm s} = D_{\rm s,conv} + D_{\rm s,geom}$ are available in the literature~\cite{peotta2015,liang2017,Huhtinen2022FlatBandSCQuantumMetric}. 
The result is the following in the limit of $N_{\rm f}$ completely flat degenerate bands, isolated from other bands by large gaps compared to the attractive interaction energy scale $|U|$, assuming zero temperature and time-reversal symmetry, and under so-called uniform pairing condition where pairing is the same in all the flat band orbitals:
\begin{align}
  [D_{s}]_{ij}  &= \frac{4e^2 N_{\rm f}}{(2\pi)^{d-1} N_{\rm orb}}|U|f(1-f)\mathcal{M}^{\rm min}_{ij}, \label{eq.Ds_qm}\\
  \mathcal{M}^{\rm min}_{ij} &= \frac{1}{2\pi} \left[\int {\rm d}^d\bsl{k} \, g_{ij}(\bsl{k})\right]_{\rm min} .
    \label{eq:DsIsolatedBand}
\end{align}
Here, $f\in[0,1]$ is the filling fraction of the isolated flat band, $N_{\rm orb}$ is the number of orbitals where the flat band states have a nonzero amplitude, $-e$ is the electron charge, $d$ is the space dimension, and $g_{ij}(\bsl{k})$ is the quantum metric defined in the introduction. The label ``min'' refers to the integrated quantum metric whose trace is minimal under variation of the orbital positions while keeping all other parameters, e.g.~hoppings, the same. (Equivalently, it is minimal under the change of Fourier transformation convention of the atomic basis.) This result is in striking contrast with the simple Bardeen-Cooper-Schrieffer (BCS) formula for a single band, $\mathcal{D}_{\rm s} = e^2 n_{\rm s}/m^*$, where $n_{\rm s}$ is the density of Cooper pairs (superfluid density) and $m^*$ the effective mass. The result \eqref{eq.Ds_qm}-\eqref{eq:DsIsolatedBand} was essentially derived in Ref.~\cite{peotta2015}, but in Ref.~\cite{Huhtinen2022FlatBandSCQuantumMetric} it was noted that the $\mathcal{M}_{ij}$ of the original work~\cite{peotta2015} has to be replaced by $\mathcal{M}^{\rm min}_{ij}$ because the quantum metric is a basis dependent quantity~\cite{simon2020} while $D_{\rm s}$ and $\mathcal{M}^{\rm min}_{ij}$ are basis-independent~\cite{Huhtinen2022FlatBandSCQuantumMetric,Tam2024GeomIndependence}. These results are derived within multiband mean-field theory, but the general idea has been confirmed by exact,
perturbative and beyond-mean-field numerical calculations~\cite{Julku2016,tovmasyan2016,liang2017,Mondaini2018,Hofmann2020_SC,Peri2021TBGFragileAndSC,Herzog2022ManyBodySCFlatBand} of some carefully chosen attractive interacting flat band models (see the reviews~\cite{Rossi2021review,Torma2022ReviewQuantumGeometry,Peotta2023review} for more examples). In Ref.~\cite{bouzerar2024hidden,penttila2024flatband}, it has been shown that many of the analytical results presented in~\cite{peotta2015,Huhtinen2022FlatBandSCQuantumMetric} can be extended to several cases of non-uniform pairing and the results remain essentially similar. The effect of closing the gap between the flat band and other bands has been studied as well, see ~\cite{jiang2024superfluidweightcrossovercritical} and references therein. 

How should one physically understand the role of the quantum metric in flat-band superconductivity? One way to gain intuition is to consider the two-body problem, the Cooper problem~\cite{Cooper1956}, in a flat band. 
In this case
there is a massive degeneracy which, however, is lifted by the interaction between the two particles: the bound pair becomes dispersive, with an inverse effective mass given by the quantum metric~\cite{Torma2018SelectiveQuantumMetric}! Similar to the  Fermi surface Cooper problem, the two-body problem in the flat band gives essentially the same answer as the mean-field approach. Further insight into why quantum geometry may be critical for pair mobility is provided by its connection to the localization of Wannier functions, as discussed in the introduction~\cite{marzari2011}. Indeed, by projecting the interacting multiband model to a flat band~\cite{tovmasyan2016} one can show that interactions induce pair hopping that is linearly proportional to the interaction $U$ –- and overlap integrals of Wannier functions at neighboring sites. In 2D this relates nicely to the lower bound of superconductivity derived in Ref.~\cite{peotta2015}:
$\mathcal{D}_{\rm s} \geq |C|$ (in appropriate units), where $C$ is the spin Chern number of a time-reversal symmetric system; 
as Wannier functions cannot be exponentially localized in a topological band~\cite{Brouder2007Wannier}, their overlaps guarantee interaction-induced motion and eventually superconductivity.
The role of Wannier functions in superconductivity offers routes for deriving upper bounds too. For example, the optical spectral weight of a superconductor and superfluid weight were considered in~\cite{Verma2021FlatBandSC,doi:10.1073/pnas.2217816120}, where quantum geometric quantities appeared as key quantities.  For further information and discussion of the large literature on this topic, we refer to existing review articles~\cite{Rossi2021review,Torma2022ReviewQuantumGeometry,Peotta2023review}. 
Here we would like to mention only a few interesting developments published after these review articles.

In Ref.~\cite{Iskin2023GL,Chen2024Ginzburg-Landau,hu2024anomalouscoherencelengthsuperconductors}, a Ginsburg-Landau theory was developed for multiband systems, with quantum geometry in focus. According to these works, in the isolated flat band limit and with uniform pairing, the coherence length of the superconductor is determined by the minimal quantum metric. For non-isolated flat bands, the coherence length can be smaller than the quantum geometry length~\cite{iskin2024coherencelengthquantumgeometry}. Furthermore, for strong interactions the Cooper pair size and the coherence length may be distinct, resembling the BEC-end of the BEC-BCS crossover~\cite{iskin2024pairsizequantumgeometry}. In Ref.~\cite{thumin2024correlation} a definition of the coherence length based on the exponential decay 
length of the anomalous Green’s function was used, leading to a result that differs from the minimal quantum metric. 
The decay of the pair correlation function or the anomalous Green’s function can be non-trivial in flat bands. For example, they may completely 
vanish beyond a few lattice sites, instead of exhibiting 
a continuous decay~\cite{virtanen2024,thumin2024correlation}. This happens in flat bands that 
host compactly localized Wannier functions (such as obstructed atomic limits~\cite{yuanfengOAI}).

The apparent discrepancy between the coherence length results can be explained through the subtlety of the definition of this concept in flat bands. The mean-field anomalous Green’s function and the pair correlation function may decay rapidly in length scales different from the minimal quantum metric, however, when one includes \textit{fluctuations} of the order parameter and calculates the spread of the pair correlation function, a coherence length given by the minimal quantum metric is obtained. Fluctuations are included in the Ginzburg-Landau formalism like in the calculation of the superfluid weight (stiffness), so, naturally, dependence on quantum geometry emerges from both. Ref.~\cite{virtanen2024} studied a superconductor-normal-superconductor (SNS) Josephson junction where the normal part is a flat band system longer than the coherence length. It was found that supercurrent over the junction was only possible by contributions from nearby dispersive bands or by interaction-mediated transport.

One might worry that disorder would kill flat band superconductivity. However, 
$D_s$ for a flat band, s-wave,  superconductor with non-trivial quantum metric
appears as robust against non-magnetic disorder as $D_s$ for a superconductor
with dispersive bands and trivial quantum metric~\cite{Lau2022Disorder}.
Interestingly, the dispersive band superfluid weight acquires a geometric contribution in the presence of disorder that 
at low disorder strengths compensates the suppression of the conventional contribution; 
this is intuitive as disorder hinders conventional ballistic transport given by the band dispersion. 
Quantum geometry has been shown to be relevant for correlations in disordered systems also in other contexts than superconductivity~\cite{Marsal2024,romeral2024}.
Another salient feature of flat band superconductors is that quasiparticles seem to be localized~\cite{Pyykkonen2023Non-Eq}. Finally, it is important to keep in mind that although the quantum geometry of the band guarantees superconductivity to be possible, sometimes another competing order, e.g., a charge density wave or phase separation, can win~\cite{Hofmann2023SC_CWD} even with attractive interactions. Quantum geometry can also lead to pair-density wave order instead of superconductivity, signified by a negative superfluid weight~\cite{Jiang2023PWD,Chen2023PWD}. 
Quantum geometry may also affect the Kohn-Luttinger mechanism of superconductivity because the form factor in the polarization function responsible for screening depends on the geometric properties of the wave functions~\cite{shavit2024,jahin2024}; remarkable enhancements of the critical temperature were found in these works for certain model systems.

\subsection{Spin-wave stiffness}
\label{sec.spin-wave}

The close analogy between superconductivity and the XY 
model~\cite{roddick1995,emery1995} 
suggests
the connections shown in the previous section between quantum geometry and the properties of superconductors
should be relevant for ferromagnetic states in which the ground state is characterized by an order
parameter $\MM$ that breaks a continuous spin, or pseudospin, symmetry.
In the continuum limit, this can be seen by considering the effective Ginzburg-Landau action 
of a ferromagnet
\eq{
 S = S_0(\MM) + \beta\frac{1}{2}\int d\rr \mathcal{D}_s^{(s)} [|\nabla M_x|^2 + |\nabla M_y|^2 + |\nabla M_z|^2]
 \label{eq.action.fm}
}
where $\MM$ is the magnetization and $S_0$ is the part of the action that does not depend on the gradient of $\MM$.
In Eq.~(\ref{eq.action.fm}), we have assumed the spin stiffness tensor to be diagonal and isotropic: $[D^{(s)}_s]_{ij} = \mathcal{D}^{(s)}_s\delta_{ij}$.

Similar to superconductivity, the spin-stiffness $[D^{(s)}_s]_{ij}$ can be obtained within linear response theory
by calculating the spin susceptibility, $\chi^{(s)}_{ij}(\qq,\omega)$, and then taking
the limit $\omega=0$, $\qq\to0$: $[D_s^{(s)}]_{ij} = \lim_{\qq\to 0}\chi^{(s)}_{ij}(\qq, \omega=0)$.
Starting from a microscopic model, it is straightforward to see that 
the expression of $\chi^{(s)}(\qq,\omega=0)$ up to order $\qq^2$,
involves the first and second derivatives of the Hamiltonian with respect to the momentum $\kk$. 
For single-band systems, such derivatives lead only to the appearance 
of derivatives with respect to $\kk$ of the energy eigenvalues, similarly to the first term of  Eq.(\ref{simplecurrentequation}). 
However, for multi-band systems, the second term in \cref{simplecurrentequation} appears, involving the quantum geometry
of the Bloch states. 

In superconductors, the superfluid stiffness is directly proportional to the superfluid weight
and so it can be directly probed by measuring the current response to an external vector field.
For ferromagnetic states, the most straightforward way to probe QGT's effects is by
probing the dispersion of the low-energy spin-waves, \ie, the Goldstone modes associated
to the continuous symmetry spontaneously broken by the ground state, something that is not straightforward to
do for superconductors~\cite{xiao2024} also due to the Anderson-Higgs mechanism.
For 2D XY ferromagnets, the effect of the quantum geometry can potentially also be inferred
indirectly by measuring $T_{BKT}$, as discussed in the case of superconductors.

So far the role of quantum geometry in ferromagnetic systems -- and especially in realistic experimental systems has not received much attention.
Recent works have investigated the connection between $[D^{(s)}_s]_{ij}$ 
for specific systems~\cite{wu2020c,bernevig2021f,khalaf2021a,kitamura2024,kang2024}.
One can obtain an exact solution of the ferromagnetic ground state and its excitations of a flat band subject to a repulsive interaction in the condition that makes the projected orbital occupation the same \cite{herzog-arbeitman2022b} (analogous to the uniform pairing condition for superconductors). In this case, the single particle charge excitations are flat. However, the spin wave spectrum can be solved exactly and it can be shown, in this class of models, that the spin stiffness is the same as the integrated minimal quantum metric~\cite{herzog-arbeitman2022b}. In moir\'e systems, projected Hamiltonians~\cite{Kang2018TBGFragile,song2021b}
do not satisfy the uniform pairing condition, and as such even the single-particle dispersion on top of the ferromagnetic state at integer fillings involves the quantum distance \cite{bernevig2021f}.
To exemplify the effect of the quantum metric in ferromagnets in \cref{app:spin_stiffness} we
describe the key results for 2D moir\'e systems~\cite{wu2020c,bernevig2021f}
and saturated ferromagnetism~\cite{kang2024}.

\subsection{Bose-Einstein condensation}\label{BEC}

Superconductivity is closely related to the physics of Bose-Einstein condensation (BEC) of electron pairs, highlighted by the smooth BCS-BEC crossover and a common mean-field ground state for both regimes. Nevertheless, when it comes to the role of quantum geometry, the BEC limit may show
quite a different phenomenology from that of superconductors.  Quantum geometry describes how the properties of quantum states vary throughout the Brillouin zone. This raises the question: Does quantum geometry have any impact on a (BEC) that occupies a single quantum state? For a non-interacting BEC at equilibrium, quantum geometry is indeed irrelevant. However, when interactions are introduced and excitations are considered, quantum geometry begins to play a significant role. 
Another natural question is: what is the bosonic counterpart of superconductivity in a flat band? Specifically, where would bosons condense in a flat band where all energies are degenerate? Once again, interactions change the scenario. Due to Hartree-type renormalization of the bands, certain momenta can acquire slightly lower energies, making them favorable sites for condensation~\cite{Huber2010BEC,You2012_BECkagome}. This leads to an important question: under what general conditions are such condensates stable? Given that the energies are essentially degenerate, even minimal interactions might excite particles to arbitrary momenta, potentially destabilizing the condensate.

Quantum geometry also plays a crucial role in Bose-Einstein Condensates (BECs). In a weakly interacting BEC within a flat band, the speed of sound—which must be positive to ensure superfluidity—is proportional to the interaction energy $U$ and the square root of a generalized quantum metric~\cite{Torma2021FlatBandBEC,Julku2023BECrevisited}.
Note again the linear dependence on the interaction energy $U$, typical for flat band phenomena: this is an immediate consequence of the existence of only one energy scale. This should be contrasted to the case of a usual dispersive band where the speed of sound is proportional to $\sqrt{U}$. The stability of a BEC can be also determined by calculating the fraction of excitations, due to weak interactions, 
that result in a finite particle density outside
the condensate state, $n_{\rm ex}(\bsl{k})$. This is also called the quantum depletion and was found~\cite{Torma2021FlatBandBEC} to be given by the condensate \textit{quantum distance} $\tilde{d}_c(\bsl{q})$ (similar to the Hilbert-Schmidt quantum distance $d_{HS}$ defined in the introduction), and, in the limit of vanishing interaction, is related to  $n_{\rm ex}(\bsl{k})$ via the equation
\begin{align}
\lim_{U\rightarrow 0} n_{ex}(\bsl{k}) = \frac{1-\tilde{d}_c(\bsl{q})}{2\tilde{d}_c(\bsl{q})},
\label{excitationfraction}
\end{align}
where $\bsl{q} = \bsl{k} - \bsl{k_{\rm c}}$ and $\tilde{d}_c(\bsl{q})$ includes overlaps of the Bloch state at the condensate momentum $\bsl{k_{\rm c}}$ with states at other momenta. The physical intuition is that depletion of the condensate to excitations is limited not by energetic reasons as in dispersive bands, but by a finite quantum distance between the initial (ground) and the excited state. 
The result~\eqref{excitationfraction} also implies that quantum excitations on top of the mean-field condensate do not vanish in the limit of small interactions; flat bands are thus an ideal platform for studying the beyond-mean-field physics of condensates. The quantum distance appears instead of the quantum metric because the quantum depletion includes finite momentum excitations. The quantum metric on the other hand is an infinitesimal measure and corresponds to long-wavelength limit quantities such as the speed of sound, and supercurrent in the case of superconductors. Quantum geometry manifests also in the superfluid weight of BEC~\cite{Julku2021BECExcitations,Julku2023BECrevisited,Iskin2023Bogoliubov,lukin2023unconventional}. 

\subsection{Exciton condensates}
An exciton is a bosonic quasiparticle formed by an electron ($e$), bound to a hole ($h$).
At low temperatures, a gas of excitons can form an exciton condensate (EC)~\cite{keldysh1965,halperin1968}.
Due to the effective interaction among excitons, resulting from the Coulomb interaction,
an EC will exhibit superfluidity. An EC can be regarded as superfluid BEC, see Sec[\ref{BEC}].
However, it is also analogous to a superconducting state, or a ferromagnetic state, as we 
discuss in this section.

Shortly after the proposal that electron-hole pairs could form an exciton condensate (EC), it was suggested that spatially separating the electrons and holes would enhance the stability of the EC by reducing the rate of electron-hole 
recombination~\cite{lozovik1975}.
This can be realized in 2D systems formed by an e-doped
2D layer and h-doped 2D layer separated by a high-quality, thin, 
dielectric film~\cite{lozovik1975,eisenstein2004}.
In these double-layer structures, when the doping is sufficiently low, and gates sufficiently far
away, so that screening effects are minimized~\cite{neilson2014,lu2016},
an EC can form when the carriers' intralayer distance $\approx \sqrt{1/n}$ is comparable to the interlayer distance $d$.
In such conditions, the layer degree of freedom can be treated as a pseudospin degree of freedom, 
or as the particle-hole degree of freedom of a superconductor.
In the first case the EC can be regarded as an easy-plane ferromagnet (see \cref{sec.spin-wave}) in the second case as a ``charge neutral superconductor''.

\begin{figure}[!!!t]
 \begin{center}
  \centering
  \includegraphics[width=0.49\textwidth]{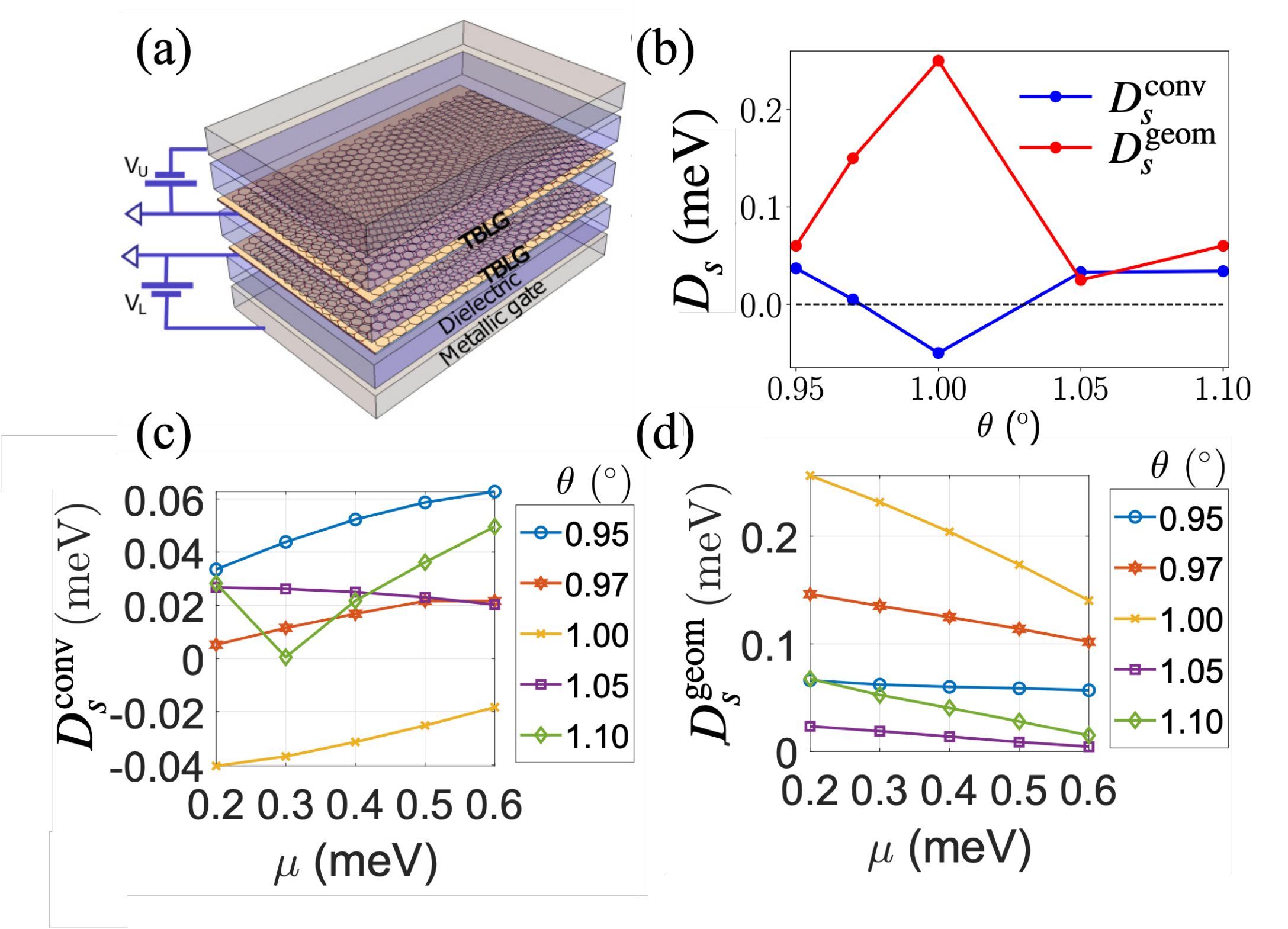}
  \caption{
           (a) Schematics of a double layer formed by two TBGs
           in which an EC state is expected to form when the chemical 
           potential in the top layer $\mu_T$ is equal and opposite to
           the chemical potential in the bottom layer $\mu_B$.
           (b) Dependence on twist angle of the conventional and geometric contributions to the superfluid weight $D_s$ of an EC formed in a double TBG, for fixed chemical potential $\mu\equiv\mu_T=-\mu_B=0.2$~mev. 
           (c), (d) Dependence on $\mu$, for different twist angles, of conventional and geometric contributions, respectively, to $D_s$.
           Adapted from~\cite{hu2022}.    
          } 
  \label{fig.EC}
 \end{center}
\end{figure} 

We can define the superfluid weight of an EC, in analogy to the definition introduced for a superconductor,
as the long-wavelength, zero-frequency, response of the system to a transverse vector field 
having opposite directions for electrons and holes. For an EC formed in a 2D double-layer 
this corresponds to having a vector field $\bsl{\mathcal{A}}$ in the top layer and a vector field $-\bsl{\mathcal{A}}$ in the bottom layer. 
It is then straightforward to derive the expression of $\left[D_s\right]_{ij}$ as done for the superconducting case, (roughly speaking) by reinterpreting
the particle-hole index \cite{peotta2015,liang2017,Huhtinen2022FlatBandSCQuantumMetric},
as the layer index~\cite{hu2022,rossi2021}.
In this analogy, the superconducting order parameter $\Delta$ corresponds to the mean-field order parameter
describing the EC,
$\Delta^{\rm EC}_{\alpha\alpha',\rr \rr'}\equiv \langle c^\dagger_{\alpha\rr T}V_{\rm TB}(\rr-\rr',d)c_{\alpha'\rr' B}\rangle$,
where $\alpha$ ($\alpha'$) is a general orbital degree of freedom, $T$, $B$ are the indices denoting the top and bottom layer, respectively,
and $V_{\rm TB}(\rr-\rr',d)$ is the effective interlayer Coulomb interaction with $d$ the distance between the layers.

Similar to superconductors, the critical temperature $T_c$  for forming an exciton condensate (EC) is enhanced in flat-band systems. As the bands become flatter, the conventional contribution to the EC's superfluid weight and stiffness is suppressed, reducing the neutral superfluid current and making the EC unstable to thermal and quantum fluctuations despite a high $T_c$. This changes when flat bands have nontrivial geometry: a geometric contribution to the superfluid weight \ds emerges, strengthening the stability of the EC \cite{hu2022,verma2023}.

This situation is most apparent in the proposal for an EC formed in a double layer formed by TMDs~\cite{verma2023}, or
by an e-doped TBG and h-doped TBG separated by a thin dielectric layer~\cite{hu2022},
as schematically shown in Fig.~\ref{fig.EC}~(a).
In this case, for certain twist angles, the conventional contribution to \ds can even
be negative, as shown in Fig.~\ref{fig.EC}~(b). 
However, for the same twist angle, the geometric
contribution is positive and very large guaranteeing the stability of the EC, Fig.~\ref{fig.EC}~(b)-(d).
We see that for an EC the effect of the quantum geometry of the bands can be even more significant than for superconductors.

\subsection{Electron-Phonon Coupling}

One main  interaction in solids is the electron-phonon coupling (EPC), which is crucial for various quantum phases, and in particular for  superconductivity~\cite{BCS1957SC,Migdal1958EPC,Eliashberg1960EPCSC}.
It is conceptually intriguing to ask if the EPC has any clear relation to quantum geometry, in particular in the generic case where the electrons have a Fermi surface - characteristic of the great majority of known superconductors. 
Uncovering this relationship could be crucial for identifying new superconductors, considering the vast array of topological materials~\cite{Bradlyn2017TQC,Po2017SymIndi,PhysRevX.7.041069,Bernevig2019TopoMat,Fang2019TopoMat,Wan2019TopoMat,Bernevig2020MTQCMat,Narang2021TopologyBands}. 

Recently, \refcite{Yu05032023GeometryEPC} revealed a direct connection between electron band geometry/topology and the bulk electron-phonon coupling (EPC). 
The study introduces a ``Gaussian approximation'' where this connection becomes explicit. 
Within this approximation, a quantum geometric contribution to the electron-phonon coupling constant $\lambda$ can be naturally distinguished from an energetic contribution. The EPC is the sum of the two (up to a cross-term).
(See \appref{app:EPC} for more details.)
Explicitly, the geometric contribution is supported by the quantum metric or an extended orbital-selective version of the quantum metric \cite{Yu05032023GeometryEPC,Torma2018SelectiveQuantumMetric}, and is bounded from below by the topological contributions over the electronic Fermi surface.

The Gaussian approximation can naturally be applied to graphene, where the short-range hopping and the symmetries make it exact, and its generalized version can naturally describe another well-known superconductor, MgB$_2$.
Combined with the \emph{ab initio} calculation, \refcite{Yu05032023GeometryEPC} finds that the quantum geometric (topological) contribution to $\lambda$ accounts for roughly 50\% (50\%) and 90\% (43\%) of the total EPC constant $\lambda$ in graphene and {\mgb}, respectively. 
The large contributions from quantum geometry to EPC can be intuitively understood: the quantum geometry affects the real-space localization of the electron Wannier functions, and then affects how the electron hopping changes under the motions of ions. This is an important part of the electron-phonon coupling (and in many cases ---such as graphene and MgB$_2$--- is the largest part of EPC).

The analysis for graphene in \refcite{Yu05032023GeometryEPC} can be tested by measuring the phonon linewidth by  Raman spectroscopy as well as measuring phonon frequencies by the inelastic x-ray scattering.
\refcite{Aris2024EPCPhotovaltaicCurrent} found that the quantum metric modifies the electron-phonon coupling by enhancing small-angle scattering. 
The formalism in \refcite{Yu05032023GeometryEPC} can in principle also be applied to other systems such as Weyl semimetals.
One major future direction is to develop a general framework that relates quantum geometry to the bulk EPC for realistic systems.
Such a study may provide new guidance for the future search for superconductors from the perspective of quantum geometry.

\subsection{Fractional Chern insulators}

Fractional Chern insulators (FCIs) are zero-magnetic-field analogs of the fractional quantum Hall effect.
By definition, FCIs should exhibit fractionally quantized Hall resistance and vanishing longitudinal resistance under zero external magnetic field.
FCIs were first proposed in toy models~\cite{neupert, sheng, regnault}, where fractionally filled nearly flat Chern bands~\cite{Sun2011,Tang11} (in zero magnetic field) and repulsive interactions are identified as the essential ingredients.

Recall that the fractional quantum Hall effect requires two ingredients: Landau levels (from an external magnetic field) and repulsive interactions.
As the repulsive interaction (\eg, Coulomb interaction) is ubiquitous, the special ingredient for the FQH is the  Landau level.
Therefore, one of the routes (but not the only one)  to realizing FCIs is to mimic Landau levels without external magnetic fields.

In this route, quantum geometry plays an important role in assessing how closely a realistic set of bands approximates the Landau levels.
Besides its exact flatness and nonzero Chern numbers, the $n$th Landau level ($n=0,1,2,3,...$) is characterized by the following three geometric properties: (i) uniform (in momentum space) quantum metric, (ii) uniform (in momentum space) Berry curvature, and (iii) the trace of the quantum metric equals $(2n+1)$ times the absolute value of the Berry curvature.
Therefore, the best system for FCIs is the one that hosts nearly flat Chern bands that nearly satisfy those three geometric properties of the Landau levels.
\refcite{Jie2021IdealBands,Parker2023IdealBands,Valentin2023IdealBands,liu2024theorygeneralizedlandaulevels} suggest that a flat Chern band is already favorable to realize FCIs even as long as an integrated version of (iii) is satisfied, even if (i) and (ii) conditions are strongly violated. (See also \cite{roy2014band, claassen2015position, northe2022interplay,ji2024quantum,PhysRevB.109.245111}.) 
\refcite{Jie2021IdealBands,Parker2023IdealBands} promote a concept called ``ideal Chern bands'', which motivate the study of analogy in other topological bands, such as ideal Euler bands~\cite{BJY2024EulerBoundQG}.
However, the claims in \refcite{Jie2021IdealBands,Parker2023IdealBands} are made for special short-range interactions instead of generic repulsive interactions and for continuum rather than tight-binding models. 
In practice, as long as a realistic model hosts nearly flat Chern bands near the Fermi level, it is reasonable to consider the possibility of realizing FCIs in such a system, as shown in \appref{app:FCI}.
Besides the way of mimicking Landau levels, one may also carefully desire the interaction to realize FCI in bands with zero Chern number.~\cite{Read2015Interaction}

Moir\'e systems are natural platforms for FCIs, since the quantum interference owing to moir\'e superlattice can easily lead to nearly flat topological bands. Upon spontaneous magnetizing due to interactions, these bands become nearly flat Chern bands.
Remarkably, last year, FCIs were experimentally realized in twisted bilayer MoTe$_2$ at fillings $-2/3$ and $-3/5$, as well as integer Chern insulator at filling $-1$~\cite{cai2023signatures,zeng2023integer,park2023observation,Xu2023FCItMoTe2,Ji2024LocalProbetMoTe2,Young2024MagtMoTe2,Kang2024_tMoTe2_2.13,xu2024interplaytopologycorrelationssecond,park_Ferromagnetism_2024}.
Theoretically, the system indeed hosts nearly flat Chern bands that have relatively uniform quantum metric and Berry curvature in each spin subspace~\cite{xiao_coupled_2012,wu_topological_2019,pan_band_2020,PhysRevResearch.3.L032070,zhang_electronic_2021,devakul_magic_2021,wang_topological_2023,reddy_fractional_2023,dong_composite_2023,qiu_interaction-driven_2023,wang_topology_2023,reddy_toward_2023,wang_fractional_2024,yu_fractional_2024,xu_maximally_2024,abouelkomsan_band_2024,jia_moire_2024,zhang_polarization-driven_2024,PhysRevResearch.6.L032063,PhysRevB.109.245131}.
Upon spin polarization (Stoner magnetization), the appearance of FCIs follows heuristically from the connection to a single Landau level~\cite{wang_topological_2023,reddy_fractional_2023,dong_composite_2023,Yu2023FCI}.
However, as shown in \refcite{Yu2023FCI}, the understanding of the spin properties requires more bands to be considered, \ie, band mixing is essential.

Following the first discovery of FCIs in twisted bilayer MoTe$_2$, clear evidence of FCIs was later observed in rhombohedral multi-layer graphene-hexagonal boron nitride superlattice (at fractional electron fillings)~\cite{Lu2024fractional,xie_Even_2024, choi_Electric_2024, lu_Extended_2024}, which has almost-fully-connected conduction bands.
The theoretical understanding of experimental observations at fractional fillings in those systems requires careful study of various issues, such as the interaction scheme and the roles of temperature and disorder~\cite{park_Topological_2023,herzog-arbeitman_Moire_2024,kwan_Moire_2023,yu_Moire_2024,guo_Fractional_2024,zhou_Fractional_2024,dong_Anomalous_2024,soejima_Anomalous_2024,huang_Impurityinduced_2024,tan_Wavefunction_2024,dong_Theory_2024,huang_Selfconsistent_2024,dassarma_Thermal_2024,xie_Integer_2024,dong_Stability_2024,kudo_Quantum_2024,zhou_New_2024}.

\section{Physical Responses}

The quantum mechanical uncertainty in the position of electrons in solids, quantified by the QGT $Q_{ij}(\bsl{k})$ in \cref{eq:qgt} or its integrated version $ \mathcal{Q}_{ij}$in \cref{eq:groundstateqgt}, leads to physical responses, which will be discussed in this section.
For this section, we will resume $\hbar$ explicitly.

\subsection{Polarization fluctuations}
Following fluctuation-dissipation theorems \cite{Martin.Souza.1999,Resta.Resta.200670o}, the quantum fluctuations of a material's polarization lead to dissipation in the presence of an external field. 
The electric polarization in solids is obtained \cite{KingSmith1993} by the expectation value of the position operator $p_i=e\ev{r_i}$, which can be reduced using Eq.\eqref{eq:blount-matrix-element} to the integral over the single band Berry connection $\left[A_i(\bsl{k})\right]_{nn}$ in the Brillouin zone. The position fluctuations in the ground state captured by the QGT  are therefore associated with polarization fluctuations~\cite{Noack.Aebischer.2001}. They are also hence associated with dissipation in the presence of perturbations that couple with the dipole operator, i.e. in the presence of an applied electric field $\mathcal{E}_i(t)=\mathcal{E}_ie^{i\omega t}$, which modifies the polarization of the medium by the polarizability $p_i(t)=\sum_j\chi_{ij}(t)\E_j(t)$.

To draw the parallel between electric dipole fluctuations of the ground state and quantum geometry, it is convenient to introduce time-dependence to the integrated QGT \cref{eq:groundstateqgt}: $\mathcal{Q}_{ij}(t-t') =\langle r_i(t)(1-P)r_i(t')\rangle$~\cite{komissarov2024quantum,verma2024instantaneous}. This captures the fact that virtual interband (dipole) transitions leading to a nontrivial $\Q_{ij}$ are modified in the presence of $\E_{i}(t)$ by how much time the state populates the virtual bands and how much the position operator has evolved with the static Hamiltonian $\H$. At $t=0$ it reduces to the integrated QGT \eqref{eq:groundstateqgt}, but away from the time origin, it has the expression
\eqa{
\Q_{ij}(t) & =\mathcal{D}_{ij}(t)+\sum_{mn}\int_{\rm BZ} \frac{d^dk}{(2\pi)^d}\ f_{n\bsl{k}}(1-f_{m\bsl{k}})\\
& \quad \times(\left[\mathfrak{g}_{ij}(\bsl{k})\right]_{nm}+\ii \left[\mathfrak{b}_{ij}(\bsl{k})\right]_{nm})e^{i\omega_{mn\bsl{k}}t}\label{eq:tqgt}
}
with $\omega_{mn\bsl{k}}=(\epsilon_{m\bsl{k}}-\epsilon_{n\bsl{k}})$, $\epsilon_{n\bsl{k}}$ the dispersion of the $n$th band, and $f_{n\bsl{k}}$ the occupation factor of the $n$th band at $
\bsl{k}$.
Here
\eqa{
& \left[\mathfrak{g}_{ij}(\bsl{k})\right]_{nm} = \frac{1}{2}\left[ A_i(\bsl{k}) \right]_{nm}\left[ A_j(\bsl{k}) \right]_{mn} + (i\leftrightarrow j) \\
& \left[\mathfrak{b}_{ij}(\bsl{k})\right]_{nm} = \frac{1}{2 \ii}\left[ A_i(\bsl{k}) \right]_{nm}\left[ A_j(\bsl{k}) \right]_{mn} - (i\leftrightarrow j) \ ,
}
are the interband metric and curvature matrix elements.
It also contains a Fermi surface contribution, $\mathcal{D}_{ij}(t)$, from the second term in the position operator of \cref{eq:blount-matrix-element}. The Fermi surface contribution is normally single-band and only present for metals $\D_{ij}(t)=F_{ij}+itD_{ij}$, containing the Drude weight $D_{ij}=\int_{\rm BZ} \frac{d^dk}{(2\pi)^d} f_{n\bsl{k}}\partial_{k_i}\partial_{k_j}\epsilon_{n\bsl{k}}$ from the dispersion curvature at the Fermi surface; as well as a divergent piece due to the discontinuity at the Fermi surface $F_{ij}=\int_{\rm BZ} \frac{d^dk}{(2\pi)^d}(f_{n\bsl{k}}')^2(\partial_{k_i}\epsilon_{n\bsl{k}})(\partial_{k_j}\epsilon_{n\bsl{k}})$. The latter term appears in the real and symmetric part of the geometric tensor, and explains the divergence of the metric for metals, even in single-band metals\cite{Resta.Notes}. The singularities coming from the Fermi surface get regularized by the introduction of a scattering time $\tau$.
Finally we note that $\Q_{ij}(t)$ in \cref{eq:tqgt} has Hermitian and anti-Hermitian components, $\Q_{ij}(t)=\Q^{s}_{ij}(t)+i\Q^{as}_{ij}(t)$, which can in principle be independently measured \cite{verma2024step}.

\subsection{Nondissipative geometric response}
The relationship between $\Q_{ij}$ and response functions such as the electric susceptibility $\chi_{ij}(\omega)$ or the electric conductivity $\sigma_{ij}(\omega)$ follows naturally in the geometric picture. Namely, the antisymmetric (in both $ij$ and $\pm t$) $\Q^{\rm as}_{ij}(t) = (\Q_{ij}(t) - \Q_{ji}(-t))/(\ii)$ is directly related to the polarizability, $\chi(t)=(\pi e^2)\Theta(t)\Q^{as}(t)$~\cite{Resta.Resta.2006.FDT,Queiroz.Verma.2024.step}. Noticing that $J_i(t)=\partial_t p_i(t)$, it follows that the conductivity can be written compactly as~\cite{Queiroz.Verma.2024.instantaneous}
\begin{align}
    \sigma_{ij}(t) &=  {\pi e^2\over \hbar} \Theta(t)\partial_t \Q^{\rm as}_{ij}(t). \label{eq:sigma-Q-relation-t}
\end{align}
which fully reproduces the Kubo formula\cite{Queiroz.Komissarov.2024}. We introduce back $\hbar$ explicitly.
The two response functions are simply related in the frequency domain by $\sigma_{ij}(\omega)=-i\omega\chi_{ij}(\omega)$. 
It becomes particularly apparent that for insulators without a Fermi surface, $D_{ij}=0$, the response at frequencies below the gap, and therefore non-dissipative, is strictly geometric, containing both longitudinal contributions with origin in the interband quantum metric matrix elements, $g_{ij}$, and Hall contributions from $B_{ij}$. 
To make it more apparent, we can expand Eq.\eqref{eq:sigma-Q-relation-t} for ingap frequencies, \cite{verma2024step} 
\eqa{
\sigma_{ij}(\omega) & = \frac{e^2}{i\hbar } \int_{\rm BZ} \frac{d^dk}{(2\pi)^d}\sum_{mn} f_{mn\bsl{k}} (\left[\mathfrak{g}_{ij}(\bsl{k})\right]_{nm}+ \ii \left[\mathfrak{b}_{ij}(\bsl{k}) \right]_{nm})\\
& \quad \times (1+{\omega\over\omega_{mn\bsl{k}}}+\cdots)\ . \label{eq:cond-omega-Taylor}
}
Since $f_{mn\bsl{k}}=(f_{n\bsl{k}}-f_{m\bsl{k}})$ is anti-symmetric in band indices, it is apparent that each order in frequency it selects either the symmetric (quantum metric) or antisymmetric (Berry curvature) geometric matrix elements. 
Namely, in the DC limit, only the Berry curvature contributes. In two dimensions, we obtain the celebrated Thouless-Kohmoto-Nightingale-Nijs (TKNN) formula $\sigma_{ij}=\epsilon_{ij} e^2C/h $ \cite{TKNN} with $\epsilon_{ij}$ the Levi-Civita symbol. At linear order in $\omega$, the non-dissipative geometric response -- the static polarizability, or capacitance of the insulator, $\chi_{ij}$ -- emerges from the matrix element of the quantum metric weighted by the inverse energy gaps \cite{komissarov2024quantum}:
\begin{equation}
    \sigma_{ij}(\omega) = \dfrac{e^2}{h} \big( C\epsilon_{ij} + i \omega\chi_{ij} +\cdots\big).
\end{equation}

\subsection{Sum rules of dissipative response}

In this part, we discuss several sum rule for band insulators.
First note that, integrating the conductivity tensor $\sigma_{ij}(\omega)$ over all frequencies (up to $
\omega$ dependent factors) can realize the instantaneous, $t=0$, response.
This realization leads to a sum rule that relates the integrated quantum metric 
\eq{
\mathcal{G}_{ij} = \int_{\rm BZ} \frac{d^dk}{(2\pi)^d} g_{ij}(\bsl{k}) = \frac{1}{2}\left[ \Q_{ij} + \mathcal{Q}_{ji} \right]
}
to the integrated real part of the longitudinal conductivity  as  
\eq{
\Tr \left[\mathcal{G} \right] = \frac{\hbar}{\pi e^2}\int_0^\infty d\omega\sum_{i}\frac{\Re\left[\sigma_{ii}(\omega)\right]}{\omega}\ ,
}
which is now recogonized as the Souza-Wilkens-Martin (SWM) sum rule~\cite{Martin.Souza.1999}.
The quantum geometry can also be found in the static structure factor\cite{Resta.Resta.2006.FDT,Yu.Tam.2024,Fu.Onishi.2024.SF} \begin{align}
S(\bsl{q}) = \frac{1}{2} \sum_{ij}q_i q_j \mathcal{G}_{ij} + ...   
\end{align}
Importantly, dissipation also occurs from the magnetic dipole moment of the medium, leading to Hall response. The simplest example is the Hall counterpart of the SWM sum rule, which is exactly the Kramers-Kr\"onig counterpart of the DC Hall conductivity~\cite{Fu.Onishi.2023}
or the dichroic sum rule~\cite{Vanderbilt.Souza.2008} which captures the orbital magnetic moment of bound electrons permitted by the Berry curvature\cite{Resta.Thonhauser.2005,Niu.Xiao.2005}.
Here $\sigma^H_{ij}(\omega) = [\sigma_{ij}(\omega) - \sigma_{ji}(\omega)]/2$.

Let us here present the generalization of these results by utilizing Eq.\eqref{eq:sigma-Q-relation-t} in Ref.~\cite{verma2024instantaneous}. Different sum rules can be constructed by weighting different powers $\eta$ of frequency. Each $\eta$ captures an instantaneous property of the medium characterizing a given moment of the zero point motion of the ground state, or the various time derivatives of $\Q_{ij}(t)$,
\begin{equation}
   \mathcal{S}_{ij}^\eta =\int\limits_0^\infty d\omega \; \dfrac{ \sigma^{\rm abs}_{ij}(\omega) }{ \omega^{1-\eta} }= {\pi e^2\over \hbar} \left. (-i\partial_t)^\eta \mathcal{Q}_{ij}(t) \right|_{t=0} . \label{eq:sumrules-main-result}
\end{equation}
The absorptive (or Hermitian) part of the conductivity $\sigma^{\rm abs}_{ij}=\Re\sigma^L_{ij}(\omega)+i\Im\sigma^H_{ij}(\omega)$ (with $\sigma^L_{ij}(\omega) = [\sigma_{ij}(\omega) + \sigma_{ji}(\omega)]/2$) contains a symmetric and real part due to coupling of $\E_i$ aligned with the dipole moment of the dielectric medium, but also a Hall contribution from the coupling with the magnetic dipole, perpendicular to the field.
Importantly in the sum rules $\S^\eta$, the entire $\Q_{ij}$ tensor Eq.\eqref{eq:tqgt} appears, not only $\Q^{\rm as}_{ij}$, and therefore, sum rules are sensitive to geometric quantities absent in nondissipative linear response, such as the integrated quantum metric\cite{Martin.Souza.1999}. 

Sum rules naturally divide into longitudinal and Hall contributions, where each $\eta$-time moment of $\Q_{ij}$ corresponds to a convolution with $\omega_{mn\bsl{k}}^\eta$. In insulators, all moments are given exclusively in terms of geometric matrix elements and explicitly given by 
\eq{
\S^\eta_{L,ij}= {\pi e^2 \over\hbar }\sum_{nm}\int_{\rm BZ} \frac{d^dk}{(2\pi)^d} f_{n\bk}(1-f_{m\bk}) \left[\mathfrak{g}_{ij}(\bsl{k})\right]_{nm} \omega_{mn\bsl{k}}^{\eta}\label{eq:sumrule}
}
and 
\eq{
\S^\eta_{H,ij}=  {\pi e^2 \over\hbar } \sum_{nm} \int_{\rm BZ} \frac{d^dk}{(2\pi)^d} f_{n\bk}(1-f_{m\bk})   \left[\mathfrak{b}_{ij}(\bsl{k})\right]_{nm} \omega_{mn\bsl{k}}^{\eta}\ .
}
These fluctuation moments reflect various instantaneous properties of bound electrons in periodic lattices \cite{Queiroz.Verma.2024.instantaneous} and have been dubbed quantum weights in \refcite{Fu.Onishi.2024}.
Let us now focus on 2D ($d=2$), where $\G_{ij}$ has no units.
Starting with the zeroth moment of longitudinal fluctuations $\S^0_{L,ij}= \pi e^2/\hbar \mathcal{G}_{ij} $, it captures exactly the integrated quantum metric, which is exactly the SWM sum rule~\cite{Martin.Souza.1999}. In Chern insulators or Landau levels, where projected of the position operators in orthogonal directions do not commute, the zeroth moment of Hall fluctuations is nonzero $\S^0_{H,ij}=-(\pi e^2/2h)C \epsilon^{ij}$. At $\eta=1$, quantifying the speed of the polarization fluctuations is the $f$-sum rule, defining the total spectral weight, $\S^1_L=\pi e^2 n/ 2m$; and the dichroic sum rule\cite{Souza2008}, defining the orbital magnetic moment $\S^1_H=\mu_M$, measured in Ref.\cite{Weitenberg.Asteria.2019}. Shot noise, that is zero temperature current fluctuations appear in the second fluctuation moment $\S^2_L$ and $\S^2_H$\cite{Mudry.Neupert.2013}. Intriguingly, and also noticed early on \cite{Pines.Nozières.1958}, metals and insulators don't show remarkably different behavior in current fluctuations $\S^\eta$ with $\eta>0$. However, $\S^0_L$, proportional to the integrated quantum metric, can qualitatively distinguish the two, completing the effort of Kohn to unambiguously distinguish a metal from an insulator~\cite{Kohn.Kohn.1964}. 

\subsection{Spectral transfer and optical bounds}

In metals not all $\S^\eta$ are well defined, but we can focus on the well-behaved first moment of the longitudinal response, the $f$-sum rule $\S^1_L$. This sum rule relates to nondissipative response by 
expanding the conductivity Eq.~\eqref{eq:sigma-Q-relation-t} to frequencies far above optical transitions, $\sigma_{ij}(\omega)\sim i\omega_p^2/\omega$. Here $\omega_p^2={4\pi n e^2}/ m$~\cite{Pines.Nozières.1958,Pines.Nozières.19587nn}, containing the electron density $n$ and optical mass $m$. The optical mass is defined by $\S^1_L$, which has a Fermi surface contribution given by the Drude weight $D_{ij}$, the linear in $t$ part of the $\Q_{ij}(t)$ \eqnref{eq:tqgt}, and a geometric component from the oscillations in $\Q_{ij}(t)$. Therefore, it is natural to separate the spectral weight into the charge stiffness $n/m^*$ obtained by the dispersion curvature at the Fermi level $\pi e^2n/m^*_{ij}=D_{ij}$ and a geometric contribution from interband optical transitions, $\pi e^2n/2m^g_{ij}\equiv\sum'\omega_{mn\bsl{k}}\left[\mathfrak{g}_{ij}(\bsl{k})\right]_{nm}$ \cite{Randeria.Hazra.2019,Queiroz.Verma.2024.instantaneous}, where $\sum'$ is a shorthand for the sum and prefactor of Eq.\eqref{eq:sumrule}.  Therefore, we have that the optical mass of electrons, defined by the sum rule and therefore indicating the instantaneous mass of electrons (usually the bare electron mass\cite{Austern.Sachs.1951}) relates to the Fermi surface electron mass and the geometric mass by
\begin{align}
{1\over m_{ij}} = {1\over m^*_{ij}}+{1\over m^g_{ij}}\label{eq:old-main-result}
\end{align}
The geometric contribution generally makes the electrons lighter at short time scales~\cite{IskinPRA2019}. In the extreme example of flat bands with nontrivial quantum geometry, in which quantum interference creates a band with $m^*=\infty$, the mass is purely geometric. This means that although semiclassical transport would be dictated by infinitely heavy electrons that do not conduct, at short time scales the electrons behave as if they have their original mass before the quenching of the band dispersion. A consequence, as discussed extensively above, is that electrons may still form a superconducting or excitonic condensate. 

\begin{figure}
    \centering
    \includegraphics[width=1\linewidth]{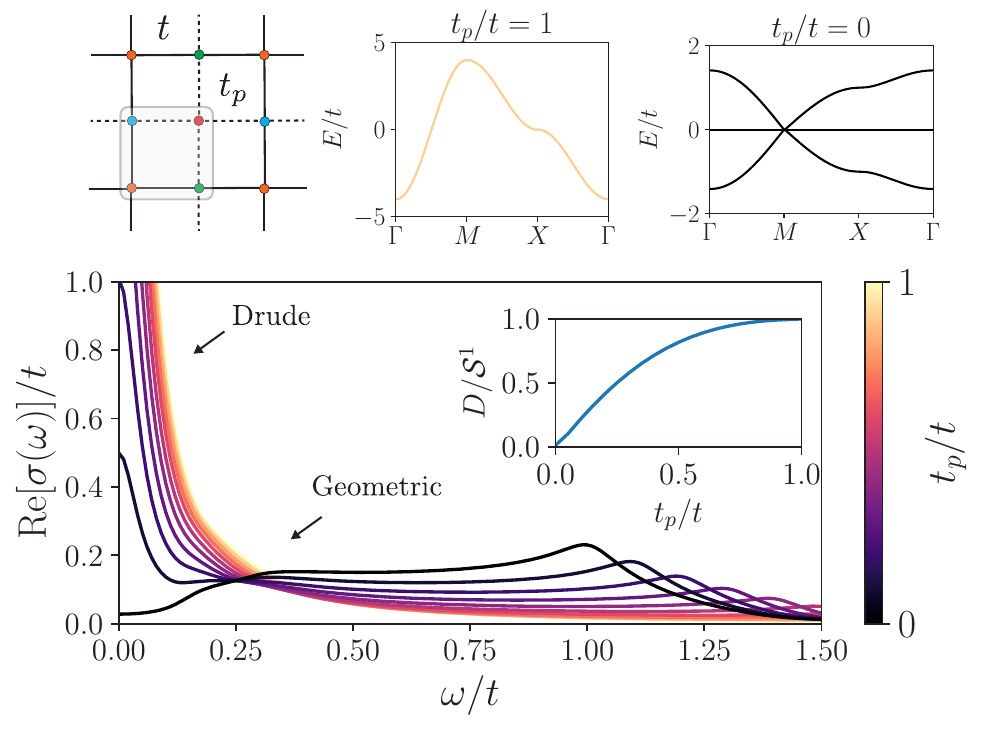}
    \caption{Transfer of optical spectral weight ${\rm Re}~\sigma_{xx}(\omega)$ from the Fermi surface (Drude weight) to high frequencies from destructive interference in a frustrated hopping lattice (Lieb). After cutting the hopping $t_p$, the resulting Hamiltonian has a flat band with no Drude weight, which was distributed to higher energies in the form of dipolar transitions between the flat and dispersive bands.}
    \label{fig:weighttransfer}
\end{figure}

The transfer of weight from the Fermi surface mass $m^*$ to the geometric mass $m^g$ can be best appreciated in Fig.\ref{fig:weighttransfer} where we consider a square lattice tight-binding model and show the evolution of the total spectral weight and optical mass $\S^1_L$. The hoppings are tuned such that the bands evolve smoothly into a Lieb lattice. In this process, the Drude peak gets progressively reduced while the geometric mass is built up to the point that exactly in the Lieb limit, the band is perfectly flat and contributes only to the geometric weight.

By looking at the different geometric responses within a unified framework, some identities become apparent. First, an insulator has a spectral gap $E_g$, which means the energy differences are bounded by $\omega_{mn\bsl{k}}\!>\!E_g/\hbar$. It follows that for $\eta>0$, $S^{\eta}_{L,ij}=\sum'\omega^\eta_{mn\bsl{k}} \left[\mathfrak{g}_{ij}(\bsl{k})\right]_{nm}\ge (\pi e^2/\hbar^2 V)E^\eta_g  \mathcal{G}_{ij}$. Similarly $S^{\eta}_H\ge (4\pi^2 e^2/\hbar^2)E^\eta_g C$. The signs are reversed for the negative powers of $\eta$. Focusing on the $f-$sum rule $\eta=1$, we have $\pi e^2 n d /m\ge (\pi e^2
/\hbar^2)\Tr\mathcal{G} E_g$, which gives $\hbar^2 n d/m\ge \Tr\mathcal{G} E_g$, where $d$ is the spacial dimension.
This fact was pointed out by Kivelson in 1982~\cite{Kivelson1982} for 1D insulators.

A similar relation was obtained for the electric susceptibility\cite{Martin.Martin.2004,Queiroz.Komissarov.2024,Fu.Onishi.2024.Dielectric}. Combining this result with the trace condition Eq.\eqref{eq.det.in} it can be observed that in the presence of a Chern number, it also holds that the energy gap is bounded by the inverse Chern number $E_g \leq {2 \pi n}/(m|C|)$~\cite{Fu.Onishi.2023}. In addition, relevant optical upper and lower bounds can be established~\cite{Martin.Martin.2004,Noack.Aebischer.2001,Queiroz.Verma.2024.instantaneous,Fu.Onishi.2024,Stengel.Souza.2024} utilizing sum rule inequalities well established in atomic physics \cite{Traini.Traini.1996}.

Let us conclude this discussion with the example of free electrons in two dimensions under a magnetic field. From Kohn's theorem~\cite{Kohn.Kohn.1961},  Galilean invariance requires that all optical transitions happen exclusively across consecutive Landau levels separated by the cyclotron frequency $\omega_c$. In this case, all sum rules become saturated \cite{onishi2024universal}. In fact, they can be compactly expressed by $\S^\eta_{ij}={\pi e^2\over \hbar}C\omega_c^\eta (\delta_{ij}+2\pi\epsilon_{ij})$ with $\delta_{ij}$ the Kronecker delta and $\varepsilon_{ij}$ the Levi-Civita symbol, which saturates all the bounds~\cite{verma2024instantaneous}. In this case, all responses are quantized, including DC conductivity, capacitance, or magnetic moment.
The quantum geometric effect on the optical response in the the presence of correlations was also studied in \refcite{PhysRevLett.133.196501,mao2024lowenergyopticalabsorptioncorrelated}.

\section{Landau levels}

	Semiclassical quantization of electronic states into Landau levels (LLs) under a magnetic field can be described by the generalized Onsager's rule:
	\begin{align}
		S_0(\epsilon) = 2\pi e B\left( n + \frac{1}{2} - \frac{\gamma_{\epsilon,B}}{2\pi}\right), 
		\label{eq:Onsager}
	\end{align}
	where $S_0(\epsilon)$ is the momentum space area of the closed semiclassical orbit at the energy $\epsilon$, $B$ is magnetic field, $-e$ is the electron charge, $n$ is a nonnegative integer, and $\gamma_{\epsilon,B}$ is the quantum correction due to Berry phase, magnetic susceptibility, and other band properties~\cite{Onsager1952interpretation,Roth1966semiclassical,Mikitik1999manifestation,Gao2017zero,Fuchs2018landau}.
	The semiclassical approach can successfully describe the band energy and the geometric properties of Bloch states in metallic systems with energy dispersion. This includes free electron gas with parabolic dispersion and Dirac electrons with linear dispersion~\cite{Berry1984quantal, Xiao2010rmp}.
	However, when applied to dispersionless flat bands, the implication of \cref{eq:Onsager} is subtle since semiclassical orbits are ill-defined.
	Naively, one may expect vanishing LL spacings due to the infinite effective mass, and thus no response of flat bands to the magnetic field.

	However, when a flat band exists in multi-band systems with sizable interband coupling, this naive expectation completely breaks down. 
	In this section, we will first review how the LLs are affected by interband coupling~\cite{hwang2021geometric}. 
	In particular, we discuss the role of interband Berry connection and symmetry of the system at zero magnetic field on the LL spectra. 
	After that, we discuss the anomalous magnetic responses of singular flat bands in which the flat band has a quadratic band crossing with another parabolic band at which the Bloch wave function becomes singular~\cite{rhim2020quantum}. 
	The geometric idea to describe the LL of singular flat bands can be further generalized to describe the LL spectra of generic quadratic band crossing~\cite{jung2024quantum}.
	We will discuss the complication when the flat band is made to be weakly dispersive and explain how the geometric effect can be extracted.
	Based on it, we revisit the magnetic responses of the Bernal stacked bilayer graphene~\cite{oh2024revisiting}.

	\subsection{Isolated flat bands}
	The Landau level spread of isolated single flat bands can be described by using the \textit{modified semiclassical approach} developed by M.-C. Chang and Q. Niu~\cite{chang1996berry}.
	Contrary to  Onsager's approach, where the band structure at zero magnetic field $\epsilon_{n\bk}$ is used to define the closed semiclassical orbits and the corresponding area $S_0(\epsilon)$, the modified semiclassical approach~\cite{chang1996berry} employs the modified band structure given by 
	\ba
	E_{n,B}(\bk) = \epsilon_{n \bk} + \mu_n(\bk) B,
	\label{eq:EnB_def1}
	\ea
	where $\b B=B\hat{z}$ is the magnetic field, $n$ is the band index, and $\mu_n(\bk)$ is the orbital magnetic moment of the $n$-th magnetic band in the $z$-direction whose explicit form is
	\ba
	\mu_n(\bk) = e \, {\rm Im} \bra{\der_x u_n(\bk)} \left[ \epsilon_{n \bk} - H(\bk) \right] \ket{\der_y u_n(\bk)},
	\label{eq:moment_def}
	\ea
	where $H(\bk)$ is the matrix Hamiltonian in momentum space.
	Hence, the second term on the right-hand side of \cref{eq:EnB_def1} indicates the leading energy correction from the orbital magnetic moment coupled to the magnetic field.
	In usual dispersive bands, the $B$-linear quantum correction is negligibly small in weak magnetic field limit compared to the zero-field bandwidth.
	
\begin{figure}
		\begin{center}
			\includegraphics[width=\columnwidth]{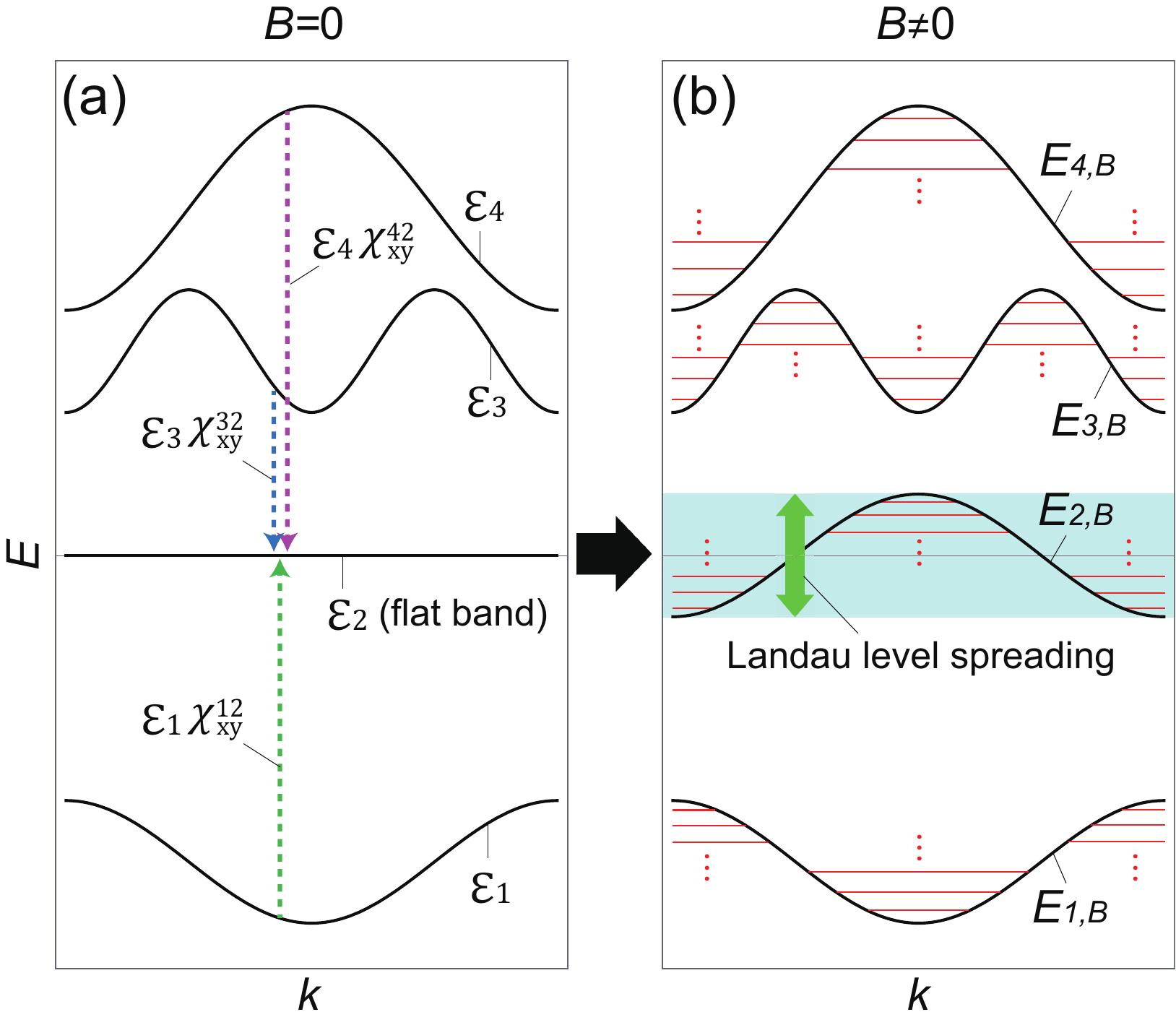}
		\end{center}
		\caption{
			\textbf{Landau level spread of an isolated flat band.}
			(a) The band structure of a 2D system in the absence of a magnetic field.
			The second band with the energy $\varepsilon_2=0$ corresponds to the isolated flat band.
			The inter-band coupling $\vep_m \chi_{xy}^{m2}$ of the isolated flat band with the other dispersive band of the energy $\varepsilon_m$ $(m=1,3,4)$ is indicated by a dashed vertical arrow.
			(b) The modified band dispersion $E_{m,B}$ $(m=1,\dots,4)$ in the presence of the magnetic field.
			The corresponding Landau levels are shown by red solid lines.
			The LL spread of the isolated flat band is represented by the green arrow. [Adapted from Ref.~\cite{hwang2021geometric}]
		}
		\label{LL_fig1}
	\end{figure}

	In the case of a flat band with zero bandwidth, on the other hand, the $B$-linear quantum correction always dominates the modified band structure $E_{n,B}(\bk)$ in \eqn{eq:EnB_def1} even in weak magnetic field limit.
	Moreover, the modified band dispersion of an isolated flat band is generally dispersive so that the relevant semiclassical orbits can be defined unambiguously.
	As a result, one can obtain the LL spreading of the isolated flat band in the adjacent gapped regions by applying the semiclassical quantization rule to $E_{n,B}(\bk)$, which naturally explains the LL spread of the isolated flat band.
	Especially, around the band edges of $E_{n,B}(\bk)$, one can define the effective mass $m^*$, which is inversely proportional to $B$, from which Onsager's scheme predicts Landau levels with a spacing $ eB/m^* \propto B^2$.
	The resulting LL spectrum is bounded by the upper and lower band edges of $E_{n,B}(\bk)$, and thus the total magnitude $\Delta$ of the LL spread is given by $\Delta = {\rm max}\,E_{n,B}(\bk) - {\rm min}\,E_{n,B}(\bk)$.
	This result is valid as long as the band gap $E_{\rm gap}$ between the isolated flat band and its neighboring band at zero magnetic field is large enough, \ie, $E_{\rm gap} \gg {\rm max}|E_{n,B}(\bk)|$.
	The generic behavior of an isolated flat band under a magnetic field is schematically described in \cref{LL_fig1} where one can clearly observe that the LL spread of the isolated flat band start filling the gaps at zero-field above and below the isolated flat band. 

	Interestingly, the LL spreading of isolated flat bands is a manifestation of the non-trivial wave function geometry of the flat band arising from inter-band couplings~\cite{hwang2021geometric}.
	One can show that the modified band dispersion of the isolated flat band is given by 
	\bg
	E_{n,B}(\bk) = -2\pi \frac{\phi}{\phi_0} \frac{1}{A_0} {\rm Im} \, \sum_{m\ne n} \epsilon_{m \bk}\ \chi^{nm}_{xy}(\bk),
	\label{eq:EnB_def2}
	\eg
	in which
	\ba
	\chi^{nm}_{ij}(\b k)&=\brk{\der_i u_n(\b k)}{u_m(\b k)} \brk{u_m(\b k)}{\der_j u_n(\b k)} \nonumber\\
	&= \left[ A_i(\b k)\right]_{mn}^* \left[ A_j(\b k)\right]_{mn},
	\label{eq:chi_def}
	\ea
	where $\phi_0=h/e$, $\phi=BA_0$ is the magnetic flux per unit cell, and $A_0$ is the unit cell area assumed to be $A_0=1$.
	Here, we assume that the $n$-th band is the isolated flat band at zero energy without loss of generality 
	so that  $\epsilon_{m \bk}$ in \cref{eq:EnB_def2} should be interpreted as the energy of the $m$-th band with respect to the flat band energy.
	We note that $\left[A_i(\b k)\right]_{nm} = \ii \brk{u_m(\b k)}{\der_i u_n(\b k)} $ indicates the cross-gap Berry connection between the $n$-th and $m$-th bands ($n\neq m$),
	and $\chi^{nm}_{ij}(\b k)$ is the corresponding fidelity tensor that describes the transition amplitude between the $n$-th and $m$-th bands. We note that $\left[A_i(\b k)\right]_{nm}$ is gauge-covariant while $\chi^{nm}_{ij}(\b k)$ is gauge-invariant, thus directly related to physical observables. 
	Hence, \eqn{eq:EnB_def2} indicates that the modified band dispersion of the isolated flat band is given by the summation of the transition amplitudes $\chi^{nm}_{xy}(\bk)$
	between the isolated flat band and the $m$-th band weighted by the energy $\epsilon_{m \bk}$ of the $m$-th band as illustrated in \cref{LL_fig1}.
	This means that the immobile carriers with infinite effective mass in an isolated flat band can respond to the external magnetic field through the inter-band coupling, characterized by the cross-gap Berry connection, to dispersive bands.

	The LL spread of an isolated flat band is strongly constrained by the symmetry group of the system~\cite{hwang2021geometric}.
	The $B$-linear correction to the modified band dispersion $E_{n,B}(\bk)$ vanishes when the system respects the chiral $C$ or space-time-inversion $I_{ST}$ symmetries in the zero magnetic flux.
	The LL spreading is proportional to $B^2$ for a flat-band system with $I_{ST}$ symmetry in the zero magnetic field, while the LL spreading is forbidden in the presence of chiral symmetry.
	Interestingly, although $I_{ST}$ symmetry would be broken as the magnetic field is turned on, the LL spreading is strongly constrained by $I_{ST}$ symmetry.
	We further find that ${\rm max}\,E_{n,B}(\bk)=-{\rm min}\,E_{n,B}(\bk)$ when the system respects time-reversal $T$ or reflection $R$ symmetry, at the zero magnetic field, thus the minimum and maximum values of the LL spreading have the same magnitude but with the opposite signs. 
    
	\begin{figure}
		\begin{center}
			\includegraphics[width=\columnwidth]{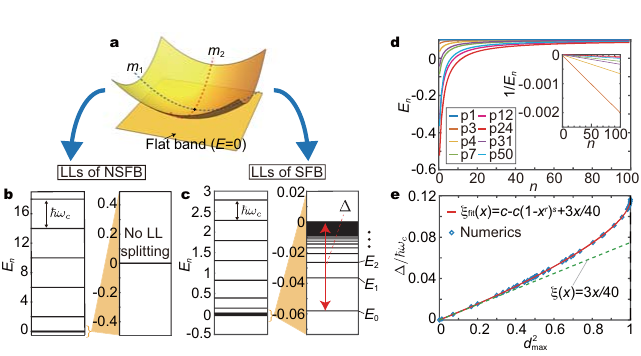}
		\end{center}
		\caption{
			\textbf{Landau level spectrum of a singular flat band.} 
			(a) The band structure of a flat band with a quadratic band-crossing. Here $m_1$ and $m_2$ are the maximum and minimum effective masses of the dispersive band. 
			(b) and (c) show the LL spreading for a non-singular flat band (NSFB) and singular flat band (SFB), respectively.  
			(d) LL spreading of various flat band models (denoted by the labeling [$p~m$] with an integer $m$ in the inset) as functions of the LL index $n$. Inset shows their $1/n$ dependences for $n\gg 1$. 
			(e) The universal relationship between $\Delta/\omega_c$ and $d_\mathrm{max}$. Numerical data (diamond symbols) are from the 50 flat band models.
			[Adapted from Ref.~\cite{rhim2020quantum}.]
		}
		\label{LL_fig2}
	\end{figure}

	\subsection{Singular flat band with quadratic touching}
	Next, let us consider the LL spectrum of singular flat bands in which a flat band has a band crossing with other dispersive bands at a momentum where the flat band wave function develops a singularity~\cite{rhim2019classification,rhim2021singular,hwang2021general}. As a minimal model of a singular flat band, we consider a two-band model describing a flat band crossing quadratically with a parabolic band.
	Explicitly, we consider the most general form of two-band continuum quadratic Hamiltonian given by
	\begin{align}\label{eq:twoband}
		\mathcal{H}_Q(\v k) \equiv f_0(\v k)\sigma_0 + \sum_{\alpha=x,y,z}  f_\alpha(\v k)\sigma_\alpha, 
	\end{align}
	where $\sigma_{\alpha}$'s are Pauli matrices and $\sigma_0$ is the $2\times2$ identity matrix. The quadratic functions $f_{\alpha=0,x,y,z}(\v k)$ take the form of $f_{\alpha}(\v k)=a_\alpha k_x^2 + b_\alpha k_x k_y + c_\alpha k_y^2$ with real coefficients $\{a_\alpha, b_\alpha, c_\alpha\}$. 
	A flat band touching with another parabolic band can be obtained by imposing the band flatness condition $\mathrm{det}\left[\mathcal{H}_Q(\v k)\right] =0$.
	If the resulting flat band wave function develops discontinuity at the band crossing point, we obtain a singular flat band. Otherwise, we have a non-singular flat band.

	The singular band crossing point can be characterized by the canting structure of the pseudospin $\v s(\v k) \equiv \sum_{\alpha=x,y,z} f_\alpha(\v k)/\sqrt{f_x(\v k)^2+f_y(\v k)^2+f_z(\v k)^2} \hat{\alpha}$ around the band crossing point.
	The canting structure arises due to the singularity at the band crossing point where the pseudospin direction cannot be uniquely determined. 
	The strength of the singularity can be characterized by the maximum canting angle $\Delta\theta_\mathrm{max}$ of the pseudospin around the band crossing point.
	Interestingly, the canting angle between two pseudospins at the momenta $\v k,\v k^\prime$ is related to the Hilbert-Schmidt quantum distance $d_\mathrm{HS}(\v k,\v k^\prime)\equiv\sqrt{1-\left|\langle u_{\bsl{k}} | u_{\bsl{k}^\prime} \rangle \right|^2}$ 
    of the perodic parts of the relevant Bloch states $\ket{u_{\bsl{k}}},~\ket{u_{\bsl{k}^\prime}}$ as $\Delta\theta(\v k,\v k^\prime) = 2\sin^{-1}\left(d_\mathrm{HS}(\v k,\v k^\prime)\right)$. 
	Denoting the maximum value of $d_\mathrm{HS}$ as $d_\mathrm{max}$ gives $\Delta\theta_\mathrm{max} = 2\sin^{-1}\left(d_\mathrm{max}\right)$.
	So, either $\Delta\theta_\mathrm{max}$ or $d_\mathrm{max}$ can equivalently measure the strength of the singularity~\cite{rhim2020quantum}.
	In the perspective of the quantum distance, the singularity at the band crossing point prevents Bloch wave functions from getting close to each other even in the limit $\v k \rightarrow 0$ yielding nonzero $d_\mathrm{max}$.

	A singular flat band under a magnetic field develops an anomalous LL structure, which directly manifests the quantum geometry of the wave function associated with the singularity at the band crossing point~\cite{rhim2020quantum,jung2024quantum,oh2024revisiting}.
	\cref{LL_fig2} b, c show the generic LL spectra of a non-singular flat band and a singular flat band, obtained by solving \cref{eq:twoband} under magnetic field.
	One can see that the non-singular flat band does not respond to the magnetic field, and all of its LL states are located at the same energy (that of the zero field flat band) without any spread. 
	On the other hand, the singular flat band develops its LL spreading in the {\it empty region} (\ie, with energy below that of the flat band).
	In both cases, the parabolic band develops a conventional LL structure with equal energy spacing $\omega_c$.

	 One can define the total LL spread $\Delta$ of the singular flat band as the difference between the energy of the singular flat band and that of the lowest LL ($E_0$) assuming that the flat band has lower energy than the parabolic band. 
	One striking observation~\cite{rhim2020quantum} is that there is a universal relationship between $\Delta/\omega_c$ and $d_\mathrm{max}$, independent of model parameters used to define the quadratic Hamiltonian in \cref{eq:twoband}, given by
	\begin{align}
		\frac{\Delta}{\omega_c} = \xi\left(d_\mathrm{max}^2 \right),\label{eq:total_LL_spacing}
	\end{align}
	where $\xi(x)$ is a monotonically increasing curve shown in \cref{LL_fig2}e.
	The existence of the universal function $\xi(x)$ can be proved analytically as well as numerically.
	Therefore, measuring $\Delta/\omega_c$, $d_\mathrm{max}$ of the singular flat band can be experimentally extracted.
	\cref{eq:total_LL_spacing} implies that $\Delta/\omega_c$ is determined solely by $d_\mathrm{max}$, and the LL spreading of singular flat bands is characterized by two distinct energy scales $\Delta$ and $\omega_c$, contrary to the case of non-singular flat band with just one energy scale $\omega_c$.
	Describing the LL generation of singular flat bands in the empty region is completely beyond the scope of semiclassical analysis, arising from the level repulsion between the LL spreading from the singular flat band and those from the parabolic band, which is encoded in the maximum quantum distance.

	\subsection{Quadratic band crossing and bilayer graphene}
	
	Now we relax the flat band condition and consider the LL spectrum of a general two-band quadratic band crossing Hamiltonian $\mathcal{H}_Q(\v k)$ in Eq.~(\ref{eq:twoband}) with $f_{\alpha=0,x,y,z}(\v k)=a_\alpha k_x^2 + b_\alpha k_x k_y + c_\alpha k_y^2$.
	After a series of unitary transformations, the Hamiltonian can be transformed to
	\ba
	H_{Q}(\bk)=&[q_1(k_x^2+k_y^2)+q_2(k_x^2-k_y^2)+q_3(2k_xk_y)]\sigma_0 \nn \\
	+&[b_2(k_x^2-k_y^2)+b_3(2k_xk_y)]\sigma_1+[c_3(2k_xk_y)]\sigma_2 \nn \\
	+&[a_1(k_x^2+k_y^2)+a_2(k_x^2-k_y^2)+a_3(2k_xk_y)]\sigma_3, 
	\label{eq:quad_H}
	\ea
	thus, the number of the Hamiltonian parameters reduced from twelve to nine~\cite{rhim2019classification}. 
	Interestingly, among the nine parameters, six correspond to the mass tensors of the two dispersive quadratic bands, while the other three describe the interband coupling~\cite{rhim2020quantum}. 
	In particular, considering that the wave function geometry of $\mathcal{H}_Q(\v k)$ appears in the form of an elliptic shape on the Bloch sphere, the three interband coupling parameters determine the major $d_1$ and minor $d_2$ diameters of the elliptic curve and the orientation of the ellipse represented by an angular variable $\phi$~\cite{jung2024quantum, hwang2021wave}. 
	When the flat band condition is additionally imposed in a way that one of the two bands becomes flat since it generates five constraint equations among the nine parameters, only four of them become independent and correspond to the three mass tensors of the parabolic band and one interband coupling parameter, which is nothing but the maximum quantum distance~\cite{rhim2020quantum}. 
	
	Moreover, one can show that the flat band condition is nothing but the condition that the energy eigenvalues of the two-band Hamiltonian have a quadratic analytic form~\cite{jung2024quantum}. 
	Under such quadratic form condition, the generic two-band Hamiltonian has only one interband coupling parameter, which is equivalent to the maximum quantum distance, and the corresponding wave function trajectory on the Bloch sphere is a circle whose diameter is equal to the maximum quantum distance~\cite{jung2024quantum, hwang2021wave}.

	The LL spectrum of generic two-band quadratic band crossing Hamiltonians is also significantly influenced by the three interband coupling parameters.
	One can reveal the role of the interband coupling in the LL spectrum by comparing two quadratic bands with identical mass tensors but different interband couplings~\cite{jung2024quantum}. 
To demonstrate the role of the interband coupling in the LL spectrum of quadratic band crossing Hamiltonian, let us consider the following model Hamiltonian~\cite{oh2024revisiting},
	    \begin{eqnarray}
		\mathcal{H}_{0}({\bm{k}}) = \sum_{\alpha=0,x,y,z } f_\alpha ({\bm{k}}) \sigma_\alpha , \label{eq:Ham}
	\end{eqnarray}
	where $f_{z} ({\bm{k}}) = {-d\sqrt{1-d^2}} k_y^2,~f_{y} ({\bm{k}}) = d\ k_x k_y,~f_{x} ({\bm{k}}) = k_x^2/2+(1-2d^2)k_y^2/2$, and $f_{0} ({\bm{k}}) = 0$. 
	Here, the parameter $d$ is defined as $d=\xi d_\mathrm{max}$ in which $\xi =\pm 1$, and $d_\mathrm{max}$ is the maximum quantum distance. 
	The energy eigenvalues of $\mathcal{H}_{0}({\bm{k}})$ remain fixed to be $\epsilon_{\pm,\bm{k}} = \pm\frac{1}{2}(k_x^2+k_y^2)$ regardless of the value of \( d \) within the range \( -1 \leq d \leq 1 \).
	When $d_\mathrm{max}=1$, the Hamiltonian in \cref{eq:Ham} corresponds to the low energy Hamiltonian of the Bernal stacked bilayer graphene, and $\xi=\pm1$ denotes the valley index.
	
	\begin{figure}[t]
		\includegraphics[width=85mm]{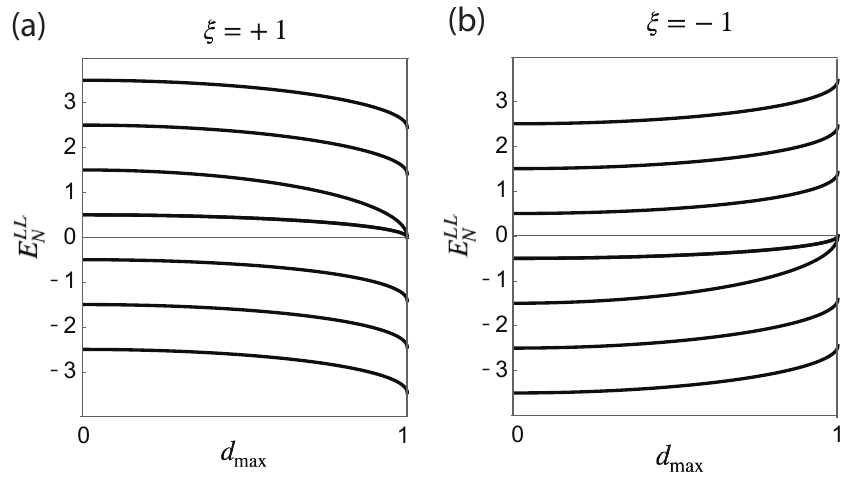} 
		\caption{\label{LL_fig3}
			The evolution of the Landau levels $E_N^{LL}$ of a quadratic band crossing Hamiltonian in \cref{eq:Ham} as a function of the maximum quantum distance $d_\mathrm{max}$ 
			for (a) $\xi=+1$ and (b) $\xi=-1$, respectively, with $\omega=1$. Here, $\xi=\pm1$ is related to the valley index of graphene. [Adapted from Ref.~\cite{oh2024revisiting}.]
		} 
	\end{figure}
	
In \cref{LL_fig3}, the $d_\mathrm{max}$-dependence of LLs is depicted~\cite{oh2024revisiting}. 
When $d_\mathrm{max}=0$, the LLs $E^{LL}_N$ are equivalent to those of the conventional parabolic bands $\epsilon^{\mathrm{conv}}_N=\pm\omega(N+\frac{1}{2})$ with the cyclotron frequency $\omega=eB/m$ 
but start to deviate from $\epsilon_N^{\mathrm{conv}}$ as $d_\mathrm{max}$ increases. 
When $d_\mathrm{max}$ reaches one, they become $\epsilon^{\mathrm{bilayer}}_N=\pm \omega\sqrt{N(N-1)}$, the LL of the bilayer graphene. 
The degeneracy of LLs between $E^{LL}_0(d_\mathrm{max}=1)$ and $E^{LL}_1(d_\mathrm{max}=1)$ leads to the absence of a zero energy plateau in the quantum Hall effect.
Since such degeneracy of LLs occurs only when $d_\mathrm{max}=1$, the zero energy plateau is absent only when $d_\mathrm{max}=1$ while it exists in other cases with $d_\mathrm{max}\neq1$~\cite{oh2024revisiting}.

One can verify that the degeneracy at $d_\mathrm{max}=1$ exists for both $\xi =\pm 1$. 
However, depending on $\xi$, the origin of zero LLs is different~\cite{oh2024revisiting}. For $\xi=+1$, the two zero energy levels come from the upper band, while for $\xi=-1$, the two zero energy levels come from the lower band. 
Furthermore, when $d_\mathrm{max}=1$ or $d_\mathrm{max}=0$, the LLs are symmetric with respect to $E=0$, as shown in \cref{LL_fig3}(a,b). 
This result arises from chiral symmetry, represented by the operator $\sigma_z$, which satisfies $\sigma_z H_0(\bm{k})\sigma_z=-H_0(\bm{k})$, exclusively when $d_\mathrm{max}=1$ or $d_\mathrm{max}=0$. 
This symmetry holds even in the presence of a magnetic field. 
In fact, the chiral symmetry is crucial for the degeneracy observed at $d_\mathrm{max}=1$. 
Thus, the presence of the zero energy plateau necessitates chiral symmetry as well as $d_\mathrm{max}=1$.
It is worth noting that although chiral symmetry is not crystalline but approximate symmetry~\cite{mccann2013electronic}, it is an excellent symmetry of bilayer graphene, and thus the absence of the zero energy plateau can be experimentally demonstrated~\cite{novoselov2006unconventional}.
	
\subsection{Quantum geometry of Landau levels and ideal bands}
Both the quantum metric and Berry curvature of LLs are uniform in momentum space~\cite{peotta2015, ozawa2021relations}, and satisfy the equalities in Eq.~(\ref{eq.det.in}).
This momentum independence of the LL quantum geometry enabled detailed analytical
understanding of the LL physics even in the presence of strong
electron-electron interactions~\cite{haldane1983fractional}. 

Recently, there is a surge of activities to understand the interplay of the band geometry and interaction effect in periodic lattice systems with nonuniform quantum geometry by generalizing the band geometry of the lowest LL~\cite{roy2014band, claassen2015position, wang2021exact, northe2022interplay, ledwith2023vortexability,ji2024quantum}.
For instance, Ref.~\onlinecite{wang2021exact} proposed ideal flatbands with nonzero Chern number in which the Berry curvature is positive definite and fluctuates in sync with the quantum metric, thus the inequalities in Eq.~(\ref{eq.det.in}) are saturated. 
Due to an exact correspondence between an ideal flatband and the lowest LL, the electron-electron interaction in an ideal
flat band can be exactly mapped to the case of interacting lowested LL.
Extending the idea of ideal flat bands to more general symmetry classes, such as ideal Euler bands~\cite{BJY2024EulerBoundQG}, would be an interesting direction to explore the interplay of electron-electron interaction and band geometry in correlated flat bands.

\section{Outlook}

The exploration of quantum geometry in condensed matter physics is in its early stages, with many avenues for further research (see \refcite{torma2023}). While the equilibrium behavior of quantum geometric superconductivity under assumed attractive interactions is fairly well understood, simple analytical results are mainly confined to the isolated flat band limit. A significant challenge is that interactions in real materials are predominantly repulsive due to Coulomb forces, making ferromagnetism the primary instability in flat bands with quantum geometry. This is evident in kagome metals~\cite{santiagoblancocanosa}, where most systems become ferromagnetic despite the presence of flat bands at the Fermi level, with only one known low-temperature superconductor.

Further research is needed to identify conditions (if any) where flat-band superconductivity is favored over flat-band ferromagnetism, extending beyond current models. Numerical studies show that superconductivity with purely attractive interactions is enhanced by nearby bands adjacent to flat bands~\cite{Julku2016,Huhtinen2022FlatBandSCQuantumMetric,Chan2022BandTouching,jiang2024superfluidweightcrossovercritical}. Analytical insights into non-isolated flat bands and especially repulsive interactions would be highly beneficial. Additionally, understanding how flat bands and quantum geometry affect various superconducting phenomena---such as critical currents, coherence lengths, vortex properties, Josephson junctions, Andreev reflections, and non-equilibrium responses---is crucial. While research in these areas is still emerging~\cite{Iskin2023GL,Chen2024Ginzburg-Landau,hu2024anomalouscoherencelengthsuperconductors,thumin2024correlation,iskin2024pairsizequantumgeometry,li2024flatbandjosephsonjunctions,virtanen2024}, early indications suggest that flat-band quantum geometric superconductivity could offer new functionalities like devices with suppressed quasiparticle currents~\cite{Pyykkonen2023Non-Eq}. However, definitive experiments and proposals are still lacking, and most current evidence remains circumstantial.

The exploration of quantum geometry in non-equilibrium phenomena should extend beyond superconductivity to include non-Hermitian systems. Collaborations between condensed matter physics and photonic or artificial quantum systems are particularly promising ~\cite{Solnyshkov2021,Liao2021ExceptionalPoint,Amelio2024Lasing,tesfaye2024}. Several experimental observations of the QGT have been reported in the latter systems~\cite{YuNatSciRev2019,gianfrate2020,ZhengChinPhysLett2022,Yi2023,Cuerda2024}.
Connecting current knowledge on quantum geometric effects in superconductivity and Bose-Einstein condensates via BCS-BEC crossover studies~\cite{iskin2024coherencelengthquantumgeometry} would be valuable. To fully grasp the significance of quantum geometry on correlated states, additional experiments and thorough theoretical analyses using realistic models are necessary. Notably, recent experiments have begun exploring quantum geometric effects in twisted bilayer graphene superconductivity~\cite{tian2023,tanaka2024kinetic}, although the extent to which flat band quantum geometry or other, renormalized bands contribute to the physics is unknown. 
Exploring the effect of quantum geometry in unconventional magentism is also a worthy direction~\cite{PhysRevLett.133.106701}.
In fractional Chern insulators
~\cite{cai2023signatures,zeng2023integer,park2023observation,Xu2023FCItMoTe2,Ji2024LocalProbetMoTe2,Young2024MagtMoTe2,Kang2024_tMoTe2_2.13,xu2024interplaytopologycorrelationssecond,park_Ferromagnetism_2024,Lu2024fractional,xie_Even_2024, choi_Electric_2024, lu_Extended_2024}, it is expected that both the Berry curvature and quantum geometry will lead to new effects, but a clear theoretical demonstration of this and an experimental observation are missing. 
Additionally, the first direct observations of quantum geometric quantities in solid-state systems are emerging~\cite{bałut2024quantumentanglementquantumgeometry,Kang2024QGT,Kim2024QGT} (albeit~\cite{bałut2024quantumentanglementquantumgeometry} in an ionic system).

To expand the impact of quantum geometry in materials research, a key challenge is to define, extract, and apply quantum geometric concepts in ab initio calculations and machine learning methods. These concepts are already utilized in the Wannier approach to electronic structures, serving as indicators for Wannier spread. However, dealing with entangled bands requires addressing issues like divergences in the QGT due to band touchings and extracting QGT from numerical Green's functions. Expanding the range of quantum geometric tools beyond the QGT---such as incorporating quantum distance or the real-space-local quantum metric~\cite{Torma2018SelectiveQuantumMetric,Marrazzo2019,Yu05032023GeometryEPC}—may be necessary.

It is important to note that the QGT is a basis-dependent quantity~\cite{simon2020}; in the case of Bloch states, it depends on Fourier transformation conventions. While some physical observables are basis-dependent and can be proportional to the QGT, others are basis-independent and cannot. For instance, the superfluid weight is proportional to the \emph{minimal} quantum metric~\cite{Huhtinen2022FlatBandSCQuantumMetric}, which resolves this issue. Basis-independent measures like quantum distance and the real-space-local quantum metric~\cite{Torma2018SelectiveQuantumMetric,Marrazzo2019,Yu05032023GeometryEPC} suggest that new, physically relevant quantum geometric quantities remain to be discovered.

This review highlighted the role of the single-particle QGT in condensed matter physics. Remarkably, properties of many-body systems—--even those that are strongly interacting---can often be described approximately, and sometimes exactly~\cite{Julku2016,Tovmasyan2018,Huhtinen2022FlatBandSCQuantumMetric,Herzog2022ManyBodySCFlatBand,han2024quantumgeometricnestingsolvable}, using the easily computed single-particle QGT. Looking forward, it is promising to explore the physical significance of many-body quantum geometry. For example, the many-body QGT has been predicted to provide a bound on the Drude weight in gapped systems~\cite{Salerno2023} and to characterize many-body localization~\cite{faugno2023geometriccharacterizationbodylocalization}. Recent works demonstrate the relation between many-body quantum geometry and entanglement~\cite{tam2024quantum,wu2024corner}, which means many-body QGT could become valuable in characterizing highly correlated quantum materials where mean-field approaches fail, where the many-body QGT is calculated for many-body states by reinterprating $\bsl{k}$ in \cref{eq:qgt_multiband} as the phase of the twisted boundary condition. In this context, calculating the many-body QGT and other quantum geometric quantities using current and future quantum computers~\cite{Niedermeier2024,Chen2024directprobetopologygeometry} is important.
In summary, quantum geometry has garnered significant attention~\cite{PhysRevA.92.063627,PhysRevB.94.134423,PhysRevLett.133.186601,yogendra2024fractionalwannierorbitalstightbinding,gong2024nonlineartransporttheoryorder,bouhon2023quantumgeometryprojectivesingle}, and its study is expected to yield more interesting results in the future.

\section{Acknowledgements} 
B.A.B. was supported by the Gordon and Betty Moore Foundation through Grant No. GBMF8685 towards the Princeton theory program, the Gordon and Betty Moore Foundation’s EPiQS Initiative (Grant No. GBMF11070), the Office of Naval Research (ONR Grant No. N00014-20-1-2303), the Global Collaborative Network Grant at Princeton University, the Simons Investigator Grant No. 404513, the BSF Israel US foundation No. 2018226, the NSF-MERSEC (Grant No. MERSEC DMR 2011750), by a grant from the Simons Foundation (SFI-MPS-NFS-00006741-01, B.A.B.) in the Simons Collaboration on New Frontiers in Superconductivity, and the Schmidt Foundation at the Princeton University. 
P.T. was supported by the Research Council of Finland under project number 339313 and by Jane and Aatos Erkko Foundation, Keele Foundation, and Magnus Ehrnrooth Foundation as part of the SuperC collaboration, as well as by a grant from the Simons Foundation (SFI-MPS-NFS-00006741-12, P.T.) in the Simons Collaboration on New Frontiers in Superconductivity. 
B.-J.Y. was supported by Samsung Science and Technology Foundation under Project No.  SSTF-BA2002-06, National Research Foundation of Korea  (NRF) grants funded by the government of Korea (MSIT)  (Grants No. NRF-2021R1A5A1032996), and GRDC(Global  Research Development Center) Cooperative Hub Program through the National Research Foundation of Korea(NRF)  funded by the Ministry of Science and ICT(MSIT) (RS-2023 00258359)).
E.R. was supported by the US Department of Energy, Office of Basic Energy Sciences, via Award DE-SC0022245, and thanks
the Kavli Institute for Theoretical Physics (KITP), supported in part by NSF grant PHY-2309135, and the Aspen Center for Physics, supported by NSF Grant  PHY-2210452, where part of this work was performed.
R.Q. was supported by the National Science Foundation under Award No. DMR-2340394, and the NSF-MRSEC (DMR-2011738), and the U.S. Department of Energy (DOE), Office of Science, Basic Energy Sciences (BES), under award DE-SC0019443
J. Y.'s work at Princeton University is supported by the Gordon and Betty Moore Foundation through Grant No. GBMF8685 towards the Princeton theory program.
J. Y.'s work at University of Florida is supported by startup funds at University of Florida.

\clearpage


\begin{thebibliography}{334}%
\makeatletter
\providecommand \@ifxundefined [1]{%
 \@ifx{#1\undefined}
}%
\providecommand \@ifnum [1]{%
 \ifnum #1\expandafter \@firstoftwo
 \else \expandafter \@secondoftwo
 \fi
}%
\providecommand \@ifx [1]{%
 \ifx #1\expandafter \@firstoftwo
 \else \expandafter \@secondoftwo
 \fi
}%
\providecommand \natexlab [1]{#1}%
\providecommand \enquote  [1]{``#1''}%
\providecommand \bibnamefont  [1]{#1}%
\providecommand \bibfnamefont [1]{#1}%
\providecommand \citenamefont [1]{#1}%
\providecommand \href@noop [0]{\@secondoftwo}%
\providecommand \href [0]{\begingroup \@sanitize@url \@href}%
\providecommand \@href[1]{\@@startlink{#1}\@@href}%
\providecommand \@@href[1]{\endgroup#1\@@endlink}%
\providecommand \@sanitize@url [0]{\catcode `\\12\catcode `\$12\catcode `\&12\catcode `\#12\catcode `\^12\catcode `\_12\catcode `\%12\relax}%
\providecommand \@@startlink[1]{}%
\providecommand \@@endlink[0]{}%
\providecommand \url  [0]{\begingroup\@sanitize@url \@url }%
\providecommand \@url [1]{\endgroup\@href {#1}{\urlprefix }}%
\providecommand \urlprefix  [0]{URL }%
\providecommand \Eprint [0]{\href }%
\providecommand \doibase [0]{https://doi.org/}%
\providecommand \selectlanguage [0]{\@gobble}%
\providecommand \bibinfo  [0]{\@secondoftwo}%
\providecommand \bibfield  [0]{\@secondoftwo}%
\providecommand \translation [1]{[#1]}%
\providecommand \BibitemOpen [0]{}%
\providecommand \bibitemStop [0]{}%
\providecommand \bibitemNoStop [0]{.\EOS\space}%
\providecommand \EOS [0]{\spacefactor3000\relax}%
\providecommand \BibitemShut  [1]{\csname bibitem#1\endcsname}%
\let\auto@bib@innerbib\@empty
\bibitem [{\citenamefont {Thouless}\ \emph {et~al.}(1982{\natexlab{a}})\citenamefont {Thouless}, \citenamefont {Kohmoto}, \citenamefont {Nightingale},\ and\ \citenamefont {{den Nijs}}}]{thouless1982b}%
  \BibitemOpen
  \bibfield  {author} {\bibinfo {author} {\bibfnamefont {D.~J.}\ \bibnamefont {Thouless}}, \bibinfo {author} {\bibfnamefont {M.}~\bibnamefont {Kohmoto}}, \bibinfo {author} {\bibfnamefont {M.~P.}\ \bibnamefont {Nightingale}},\ and\ \bibinfo {author} {\bibfnamefont {M.}~\bibnamefont {{den Nijs}}},\ }\bibfield  {title} {\bibinfo {title} {Quantized {{Hall Conductance}} in a {{Two-Dimensional Periodic Potential}}},\ }\href {https://doi.org/10.1103/PhysRevLett.49.405} {\bibfield  {journal} {\bibinfo  {journal} {Physical Review Letters}\ }\textbf {\bibinfo {volume} {49}},\ \bibinfo {pages} {405} (\bibinfo {year} {1982}{\natexlab{a}})}\BibitemShut {NoStop}%
\bibitem [{\citenamefont {Provost}\ and\ \citenamefont {Vallee}(1980)}]{provost1980}%
  \BibitemOpen
  \bibfield  {author} {\bibinfo {author} {\bibfnamefont {J.~P.}\ \bibnamefont {Provost}}\ and\ \bibinfo {author} {\bibfnamefont {G.}~\bibnamefont {Vallee}},\ }\bibfield  {title} {\bibinfo {title} {Riemannian structure on manifolds of quantum states},\ }\href {https://doi.org/10.1007/BF02193559} {\bibfield  {journal} {\bibinfo  {journal} {Communications in Mathematical Physics}\ }\textbf {\bibinfo {volume} {76}},\ \bibinfo {pages} {289} (\bibinfo {year} {1980})}\BibitemShut {NoStop}%
\bibitem [{\citenamefont {Fubini}(1904)}]{Fubini1904}%
  \BibitemOpen
  \bibfield  {author} {\bibinfo {author} {\bibfnamefont {G.}~\bibnamefont {Fubini}},\ }\bibfield  {title} {\bibinfo {title} {Sulle metriche definite da una forma hermitiana},\ }\href@noop {} {\bibfield  {journal} {\bibinfo  {journal} {Atti del Reale istituto Veneto di Scienze, Lettere ed Arti}\ }\textbf {\bibinfo {volume} {63}},\ \bibinfo {pages} {502} (\bibinfo {year} {1904})}\BibitemShut {NoStop}%
\bibitem [{\citenamefont {Study}(1905)}]{Study1905}%
  \BibitemOpen
  \bibfield  {author} {\bibinfo {author} {\bibfnamefont {E.}~\bibnamefont {Study}},\ }\bibfield  {title} {\bibinfo {title} {K{\"u}rzeste {{Wege}} im komplexen {{Gebiet}}},\ }\href {https://doi.org/10.1007/BF01457616} {\bibfield  {journal} {\bibinfo  {journal} {Mathematische Annalen}\ }\textbf {\bibinfo {volume} {60}},\ \bibinfo {pages} {321} (\bibinfo {year} {1905})}\BibitemShut {NoStop}%
\bibitem [{\citenamefont {Berry}(1984{\natexlab{a}})}]{Berry1984}%
  \BibitemOpen
  \bibfield  {author} {\bibinfo {author} {\bibfnamefont {M.~V.}\ \bibnamefont {Berry}},\ }\bibfield  {title} {\bibinfo {title} {Quantal phase factors accompanying adiabatic changes},\ }\href {https://doi.org/10.1098/rspa.1984.0023} {\bibfield  {journal} {\bibinfo  {journal} {Proceedings of the Royal Society of London. A. Mathematical and Physical Sciences}\ }\textbf {\bibinfo {volume} {392}},\ \bibinfo {pages} {45} (\bibinfo {year} {1984}{\natexlab{a}})}\BibitemShut {NoStop}%
\bibitem [{\citenamefont {Simon}(1983)}]{Simon1983}%
  \BibitemOpen
  \bibfield  {author} {\bibinfo {author} {\bibfnamefont {B.}~\bibnamefont {Simon}},\ }\bibfield  {title} {\bibinfo {title} {Holonomy, the {{Quantum Adiabatic Theorem}}, and {{Berry}}'s {{Phase}}},\ }\href {https://doi.org/10.1103/PhysRevLett.51.2167} {\bibfield  {journal} {\bibinfo  {journal} {Physical Review Letters}\ }\textbf {\bibinfo {volume} {51}},\ \bibinfo {pages} {2167} (\bibinfo {year} {1983})}\BibitemShut {NoStop}%
\bibitem [{\citenamefont {Aharonov}\ and\ \citenamefont {Anandan}(1987)}]{Aharonov1987}%
  \BibitemOpen
  \bibfield  {author} {\bibinfo {author} {\bibfnamefont {Y.}~\bibnamefont {Aharonov}}\ and\ \bibinfo {author} {\bibfnamefont {J.}~\bibnamefont {Anandan}},\ }\bibfield  {title} {\bibinfo {title} {Phase change during a cyclic quantum evolution},\ }\href {https://doi.org/10.1103/PhysRevLett.58.1593} {\bibfield  {journal} {\bibinfo  {journal} {Physical Review Letters}\ }\textbf {\bibinfo {volume} {58}},\ \bibinfo {pages} {1593} (\bibinfo {year} {1987})}\BibitemShut {NoStop}%
\bibitem [{\citenamefont {Roy}(2014{\natexlab{a}})}]{roy2014}%
  \BibitemOpen
  \bibfield  {author} {\bibinfo {author} {\bibfnamefont {R.}~\bibnamefont {Roy}},\ }\bibfield  {title} {\bibinfo {title} {Band geometry of fractional topological insulators},\ }\href {https://doi.org/10.1103/PhysRevB.90.165139} {\bibfield  {journal} {\bibinfo  {journal} {Physical Review B}\ }\textbf {\bibinfo {volume} {90}},\ \bibinfo {pages} {165139} (\bibinfo {year} {2014}{\natexlab{a}})}\BibitemShut {NoStop}%
\bibitem [{\citenamefont {Xie}\ \emph {et~al.}(2020{\natexlab{a}})\citenamefont {Xie}, \citenamefont {Song}, \citenamefont {Lian},\ and\ \citenamefont {Bernevig}}]{Xie2020TopologyBoundSCTBG}%
  \BibitemOpen
  \bibfield  {author} {\bibinfo {author} {\bibfnamefont {F.}~\bibnamefont {Xie}}, \bibinfo {author} {\bibfnamefont {Z.}~\bibnamefont {Song}}, \bibinfo {author} {\bibfnamefont {B.}~\bibnamefont {Lian}},\ and\ \bibinfo {author} {\bibfnamefont {B.~A.}\ \bibnamefont {Bernevig}},\ }\bibfield  {title} {\bibinfo {title} {Topology-bounded superfluid weight in twisted bilayer graphene},\ }\href {https://doi.org/10.1103/PhysRevLett.124.167002} {\bibfield  {journal} {\bibinfo  {journal} {Phys. Rev. Lett.}\ }\textbf {\bibinfo {volume} {124}},\ \bibinfo {pages} {167002} (\bibinfo {year} {2020}{\natexlab{a}})}\BibitemShut {NoStop}%
\bibitem [{\citenamefont {Yu}\ \emph {et~al.}(2023{\natexlab{a}})\citenamefont {Yu}, \citenamefont {Xie}, \citenamefont {Wu},\ and\ \citenamefont {Das~Sarma}}]{Yu2022EOCPTBG}%
  \BibitemOpen
  \bibfield  {author} {\bibinfo {author} {\bibfnamefont {J.}~\bibnamefont {Yu}}, \bibinfo {author} {\bibfnamefont {M.}~\bibnamefont {Xie}}, \bibinfo {author} {\bibfnamefont {F.}~\bibnamefont {Wu}},\ and\ \bibinfo {author} {\bibfnamefont {S.}~\bibnamefont {Das~Sarma}},\ }\bibfield  {title} {\bibinfo {title} {Euler-obstructed nematic nodal superconductivity in twisted bilayer graphene},\ }\href {https://doi.org/10.1103/PhysRevB.107.L201106} {\bibfield  {journal} {\bibinfo  {journal} {Phys. Rev. B}\ }\textbf {\bibinfo {volume} {107}},\ \bibinfo {pages} {L201106} (\bibinfo {year} {2023}{\natexlab{a}})}\BibitemShut {NoStop}%
\bibitem [{\citenamefont {Kwon}\ and\ \citenamefont {Yang}(2024)}]{BJY2024EulerBoundQG}%
  \BibitemOpen
  \bibfield  {author} {\bibinfo {author} {\bibfnamefont {S.}~\bibnamefont {Kwon}}\ and\ \bibinfo {author} {\bibfnamefont {B.-J.}\ \bibnamefont {Yang}},\ }\bibfield  {title} {\bibinfo {title} {Quantum geometric bound and ideal condition for euler band topology},\ }\href {https://doi.org/10.1103/PhysRevB.109.L161111} {\bibfield  {journal} {\bibinfo  {journal} {Phys. Rev. B}\ }\textbf {\bibinfo {volume} {109}},\ \bibinfo {pages} {L161111} (\bibinfo {year} {2024})}\BibitemShut {NoStop}%
\bibitem [{\citenamefont {Jankowski}\ \emph {et~al.}(2024)\citenamefont {Jankowski}, \citenamefont {Morris}, \citenamefont {Bouhon}, \citenamefont {Ünal},\ and\ \citenamefont {Slager}}]{Slager2024EulerOptical}%
  \BibitemOpen
  \bibfield  {author} {\bibinfo {author} {\bibfnamefont {W.~J.}\ \bibnamefont {Jankowski}}, \bibinfo {author} {\bibfnamefont {A.~S.}\ \bibnamefont {Morris}}, \bibinfo {author} {\bibfnamefont {A.}~\bibnamefont {Bouhon}}, \bibinfo {author} {\bibfnamefont {F.~N.}\ \bibnamefont {Ünal}},\ and\ \bibinfo {author} {\bibfnamefont {R.-J.}\ \bibnamefont {Slager}},\ }\href {https://arxiv.org/abs/2311.07545} {\bibinfo {title} {Optical manifestations of topological euler class}} (\bibinfo {year} {2024}),\ \Eprint {https://arxiv.org/abs/2311.07545} {arXiv:2311.07545 [cond-mat.mes-hall]} \BibitemShut {NoStop}%
\bibitem [{\citenamefont {Herzog-Arbeitman}\ \emph {et~al.}(2022{\natexlab{a}})\citenamefont {Herzog-Arbeitman}, \citenamefont {Peri}, \citenamefont {Schindler}, \citenamefont {Huber},\ and\ \citenamefont {Bernevig}}]{Herzog-Arbeitman2021}%
  \BibitemOpen
  \bibfield  {author} {\bibinfo {author} {\bibfnamefont {J.}~\bibnamefont {Herzog-Arbeitman}}, \bibinfo {author} {\bibfnamefont {V.}~\bibnamefont {Peri}}, \bibinfo {author} {\bibfnamefont {F.}~\bibnamefont {Schindler}}, \bibinfo {author} {\bibfnamefont {S.~D.}\ \bibnamefont {Huber}},\ and\ \bibinfo {author} {\bibfnamefont {B.~A.}\ \bibnamefont {Bernevig}},\ }\bibfield  {title} {\bibinfo {title} {Superfluid weight bounds from symmetry and quantum geometry in flat bands},\ }\href {https://doi.org/10.1103/PhysRevLett.128.087002} {\bibfield  {journal} {\bibinfo  {journal} {Phys. Rev. Lett.}\ }\textbf {\bibinfo {volume} {128}},\ \bibinfo {pages} {087002} (\bibinfo {year} {2022}{\natexlab{a}})}\BibitemShut {NoStop}%
\bibitem [{\citenamefont {Yu}\ \emph {et~al.}()\citenamefont {Yu}, \citenamefont {Herzog-Arbeitman}, ,\ and\ \citenamefont {Bernevig}}]{Yu2024Z2boundQG}%
  \BibitemOpen
  \bibfield  {author} {\bibinfo {author} {\bibfnamefont {J.}~\bibnamefont {Yu}}, \bibinfo {author} {\bibfnamefont {J.}~\bibnamefont {Herzog-Arbeitman}}, \ and\ \bibinfo {author} {\bibfnamefont {B.~A.}\ \bibnamefont {Bernevig}},\ }\href@noop {} {\bibinfo  {journal} {To appear}\ }\BibitemShut {NoStop}%
\bibitem [{\citenamefont {Kane}\ and\ \citenamefont {Mele}(2005{\natexlab{a}})}]{Kane2005Z2}%
  \BibitemOpen
\bibfield  {journal} {  }\bibfield  {author} {\bibinfo {author} {\bibfnamefont {C.~L.}\ \bibnamefont {Kane}}\ and\ \bibinfo {author} {\bibfnamefont {E.~J.}\ \bibnamefont {Mele}},\ }\bibfield  {title} {\bibinfo {title} {${Z}_{2}$ topological order and the quantum spin hall effect},\ }\href {https://doi.org/10.1103/PhysRevLett.95.146802} {\bibfield  {journal} {\bibinfo  {journal} {Phys. Rev. Lett.}\ }\textbf {\bibinfo {volume} {95}},\ \bibinfo {pages} {146802} (\bibinfo {year} {2005}{\natexlab{a}})}\BibitemShut {NoStop}%
\bibitem [{\citenamefont {Bernevig}\ and\ \citenamefont {Zhang}(2006)}]{Zhang2006QSH}%
  \BibitemOpen
  \bibfield  {author} {\bibinfo {author} {\bibfnamefont {B.~A.}\ \bibnamefont {Bernevig}}\ and\ \bibinfo {author} {\bibfnamefont {S.-C.}\ \bibnamefont {Zhang}},\ }\bibfield  {title} {\bibinfo {title} {Quantum spin hall effect},\ }\href {https://doi.org/10.1103/PhysRevLett.96.106802} {\bibfield  {journal} {\bibinfo  {journal} {Phys. Rev. Lett.}\ }\textbf {\bibinfo {volume} {96}},\ \bibinfo {pages} {106802} (\bibinfo {year} {2006})}\BibitemShut {NoStop}%
\bibitem [{\citenamefont {Kane}\ and\ \citenamefont {Mele}(2005{\natexlab{b}})}]{Kane2005QSH}%
  \BibitemOpen
  \bibfield  {author} {\bibinfo {author} {\bibfnamefont {C.~L.}\ \bibnamefont {Kane}}\ and\ \bibinfo {author} {\bibfnamefont {E.~J.}\ \bibnamefont {Mele}},\ }\bibfield  {title} {\bibinfo {title} {Quantum spin hall effect in graphene},\ }\href {https://doi.org/10.1103/PhysRevLett.95.226801} {\bibfield  {journal} {\bibinfo  {journal} {Phys. Rev. Lett.}\ }\textbf {\bibinfo {volume} {95}},\ \bibinfo {pages} {226801} (\bibinfo {year} {2005}{\natexlab{b}})}\BibitemShut {NoStop}%
\bibitem [{\citenamefont {Bernevig}\ \emph {et~al.}(2006)\citenamefont {Bernevig}, \citenamefont {Hughes},\ and\ \citenamefont {Zhang}}]{Bernevig2006BHZ}%
  \BibitemOpen
  \bibfield  {author} {\bibinfo {author} {\bibfnamefont {B.~A.}\ \bibnamefont {Bernevig}}, \bibinfo {author} {\bibfnamefont {T.~L.}\ \bibnamefont {Hughes}},\ and\ \bibinfo {author} {\bibfnamefont {S.-C.}\ \bibnamefont {Zhang}},\ }\bibfield  {title} {\bibinfo {title} {Quantum spin hall effect and topological phase transition in hgte quantum wells},\ }\href {https://doi.org/10.1126/science.1133734} {\bibfield  {journal} {\bibinfo  {journal} {Science}\ }\textbf {\bibinfo {volume} {314}},\ \bibinfo {pages} {1757} (\bibinfo {year} {2006})}\BibitemShut {NoStop}%
\bibitem [{\citenamefont {Hasan}\ and\ \citenamefont {Kane}(2010)}]{hasan2010a}%
  \BibitemOpen
  \bibfield  {author} {\bibinfo {author} {\bibfnamefont {M.~Z.}\ \bibnamefont {Hasan}}\ and\ \bibinfo {author} {\bibfnamefont {C.~L.}\ \bibnamefont {Kane}},\ }\bibfield  {title} {\bibinfo {title} {{\emph{Colloquium}} : {{Topological}} insulators},\ }\href {https://doi.org/10.1103/RevModPhys.82.3045} {\bibfield  {journal} {\bibinfo  {journal} {Reviews of Modern Physics}\ }\textbf {\bibinfo {volume} {82}},\ \bibinfo {pages} {3045} (\bibinfo {year} {2010})}\BibitemShut {NoStop}%
\bibitem [{\citenamefont {Qi}\ and\ \citenamefont {Zhang}(2011)}]{qi2011a}%
  \BibitemOpen
  \bibfield  {author} {\bibinfo {author} {\bibfnamefont {X.-L.}\ \bibnamefont {Qi}}\ and\ \bibinfo {author} {\bibfnamefont {S.-C.}\ \bibnamefont {Zhang}},\ }\bibfield  {title} {\bibinfo {title} {Topological insulators and superconductors},\ }\href {https://doi.org/10.1103/RevModPhys.83.1057} {\bibfield  {journal} {\bibinfo  {journal} {Reviews of Modern Physics}\ }\textbf {\bibinfo {volume} {83}},\ \bibinfo {pages} {1057} (\bibinfo {year} {2011})}\BibitemShut {NoStop}%
\bibitem [{\citenamefont {Chiu}\ \emph {et~al.}(2016)\citenamefont {Chiu}, \citenamefont {Teo}, \citenamefont {Schnyder},\ and\ \citenamefont {Ryu}}]{chiu2016}%
  \BibitemOpen
  \bibfield  {author} {\bibinfo {author} {\bibfnamefont {C.-K.}\ \bibnamefont {Chiu}}, \bibinfo {author} {\bibfnamefont {J.~C.~Y.}\ \bibnamefont {Teo}}, \bibinfo {author} {\bibfnamefont {A.~P.}\ \bibnamefont {Schnyder}},\ and\ \bibinfo {author} {\bibfnamefont {S.}~\bibnamefont {Ryu}},\ }\bibfield  {title} {\bibinfo {title} {Classification of topological quantum matter with symmetries},\ }\href {https://doi.org/10.1103/RevModPhys.88.035005} {\bibfield  {journal} {\bibinfo  {journal} {Reviews of Modern Physics}\ }\textbf {\bibinfo {volume} {88}},\ \bibinfo {pages} {035005} (\bibinfo {year} {2016})}\BibitemShut {NoStop}%
\bibitem [{\citenamefont {Vafek}\ and\ \citenamefont {Vishwanath}(2014)}]{Vafek2014}%
  \BibitemOpen
  \bibfield  {author} {\bibinfo {author} {\bibfnamefont {O.}~\bibnamefont {Vafek}}\ and\ \bibinfo {author} {\bibfnamefont {A.}~\bibnamefont {Vishwanath}},\ }\bibfield  {title} {\bibinfo {title} {Dirac {{Fermions}} in {{Solids}}: {{From High-T}} c {{Cuprates}} and {{Graphene}} to {{Topological Insulators}} and {{Weyl Semimetals}}},\ }\href {https://doi.org/10.1146/annurev-conmatphys-031113-133841} {\bibfield  {journal} {\bibinfo  {journal} {Annual Review of Condensed Matter Physics}\ }\textbf {\bibinfo {volume} {5}},\ \bibinfo {pages} {83} (\bibinfo {year} {2014})}\BibitemShut {NoStop}%
\bibitem [{\citenamefont {Wehling}\ \emph {et~al.}(2014)\citenamefont {Wehling}, \citenamefont {{Black-Schaffer}},\ and\ \citenamefont {Balatsky}}]{Wehling2014}%
  \BibitemOpen
  \bibfield  {author} {\bibinfo {author} {\bibfnamefont {T.}~\bibnamefont {Wehling}}, \bibinfo {author} {\bibfnamefont {A.}~\bibnamefont {{Black-Schaffer}}},\ and\ \bibinfo {author} {\bibfnamefont {A.}~\bibnamefont {Balatsky}},\ }\bibfield  {title} {\bibinfo {title} {Dirac materials},\ }\href {https://doi.org/10.1080/00018732.2014.927109} {\bibfield  {journal} {\bibinfo  {journal} {Advances in Physics}\ }\textbf {\bibinfo {volume} {63}},\ \bibinfo {pages} {1} (\bibinfo {year} {2014})}\BibitemShut {NoStop}%
\bibitem [{\citenamefont {Burkov}(2018)}]{Burkov2018}%
  \BibitemOpen
  \bibfield  {author} {\bibinfo {author} {\bibfnamefont {A.}~\bibnamefont {Burkov}},\ }\bibfield  {title} {\bibinfo {title} {Weyl {{Metals}}},\ }\href {https://doi.org/10.1146/annurev-conmatphys-033117-054129} {\bibfield  {journal} {\bibinfo  {journal} {Annual Review of Condensed Matter Physics}\ }\textbf {\bibinfo {volume} {9}},\ \bibinfo {pages} {359} (\bibinfo {year} {2018})}\BibitemShut {NoStop}%
\bibitem [{\citenamefont {Benalcazar}\ \emph {et~al.}(2017)\citenamefont {Benalcazar}, \citenamefont {Bernevig},\ and\ \citenamefont {Hughes}}]{Benalcazar2017a}%
  \BibitemOpen
  \bibfield  {author} {\bibinfo {author} {\bibfnamefont {W.~A.}\ \bibnamefont {Benalcazar}}, \bibinfo {author} {\bibfnamefont {B.~A.}\ \bibnamefont {Bernevig}},\ and\ \bibinfo {author} {\bibfnamefont {T.~L.}\ \bibnamefont {Hughes}},\ }\bibfield  {title} {\bibinfo {title} {Quantized electric multipole insulators},\ }\href {https://doi.org/10.1126/science.aah6442} {\bibfield  {journal} {\bibinfo  {journal} {Science}\ }\textbf {\bibinfo {volume} {357}},\ \bibinfo {pages} {61} (\bibinfo {year} {2017})}\BibitemShut {NoStop}%
\bibitem [{\citenamefont {Schindler}\ \emph {et~al.}(2018)\citenamefont {Schindler}, \citenamefont {Cook}, \citenamefont {Vergniory}, \citenamefont {Wang}, \citenamefont {Parkin}, \citenamefont {Bernevig},\ and\ \citenamefont {Neupert}}]{schindler2018a}%
  \BibitemOpen
  \bibfield  {author} {\bibinfo {author} {\bibfnamefont {F.}~\bibnamefont {Schindler}}, \bibinfo {author} {\bibfnamefont {A.~M.}\ \bibnamefont {Cook}}, \bibinfo {author} {\bibfnamefont {M.~G.}\ \bibnamefont {Vergniory}}, \bibinfo {author} {\bibfnamefont {Z.}~\bibnamefont {Wang}}, \bibinfo {author} {\bibfnamefont {S.~S.~P.}\ \bibnamefont {Parkin}}, \bibinfo {author} {\bibfnamefont {B.~A.}\ \bibnamefont {Bernevig}},\ and\ \bibinfo {author} {\bibfnamefont {T.}~\bibnamefont {Neupert}},\ }\bibfield  {title} {\bibinfo {title} {Higher-order topological insulators},\ }\href@noop {} {\bibfield  {journal} {\bibinfo  {journal} {SCIENCE ADVANCES}\ } (\bibinfo {year} {2018})}\BibitemShut {NoStop}%
\bibitem [{\citenamefont {Song}\ \emph {et~al.}(2017)\citenamefont {Song}, \citenamefont {Fang},\ and\ \citenamefont {Fang}}]{Song2017}%
  \BibitemOpen
  \bibfield  {author} {\bibinfo {author} {\bibfnamefont {Z.}~\bibnamefont {Song}}, \bibinfo {author} {\bibfnamefont {Z.}~\bibnamefont {Fang}},\ and\ \bibinfo {author} {\bibfnamefont {C.}~\bibnamefont {Fang}},\ }\bibfield  {title} {\bibinfo {title} {Dimensional edge states of rotation symmetry protected topological states},\ }\href {https://doi.org/10.1103/PhysRevLett.119.246402} {\bibfield  {journal} {\bibinfo  {journal} {Physical Review Letters}\ }\textbf {\bibinfo {volume} {119}},\ \bibinfo {pages} {246402} (\bibinfo {year} {2017})}\BibitemShut {NoStop}%
\bibitem [{\citenamefont {Khalaf}(2018)}]{khalaf2018a}%
  \BibitemOpen
  \bibfield  {author} {\bibinfo {author} {\bibfnamefont {E.}~\bibnamefont {Khalaf}},\ }\bibfield  {title} {\bibinfo {title} {Higher-order topological insulators and superconductors protected by inversion symmetry},\ }\href {https://doi.org/10.1103/PhysRevB.97.205136} {\bibfield  {journal} {\bibinfo  {journal} {Physical Review B}\ }\textbf {\bibinfo {volume} {97}},\ \bibinfo {pages} {205136} (\bibinfo {year} {2018})}\BibitemShut {NoStop}%
\bibitem [{\citenamefont {Ezawa}(2018)}]{Ezawa2018}%
  \BibitemOpen
  \bibfield  {author} {\bibinfo {author} {\bibfnamefont {M.}~\bibnamefont {Ezawa}},\ }\bibfield  {title} {\bibinfo {title} {Minimal models for {{Wannier-type}} higher-order topological insulators and phosphorene},\ }\href {https://doi.org/10.1103/PhysRevB.98.045125} {\bibfield  {journal} {\bibinfo  {journal} {Physical Review B}\ }\textbf {\bibinfo {volume} {98}},\ \bibinfo {pages} {045125} (\bibinfo {year} {2018})}\BibitemShut {NoStop}%
\bibitem [{\citenamefont {Trifunovic}\ and\ \citenamefont {Brouwer}(2019)}]{Trifunovic2019}%
  \BibitemOpen
  \bibfield  {author} {\bibinfo {author} {\bibfnamefont {L.}~\bibnamefont {Trifunovic}}\ and\ \bibinfo {author} {\bibfnamefont {P.~W.}\ \bibnamefont {Brouwer}},\ }\bibfield  {title} {\bibinfo {title} {Higher-{{Order Bulk-Boundary Correspondence}} for {{Topological Crystalline Phases}}},\ }\href {https://doi.org/10.1103/PhysRevX.9.011012} {\bibfield  {journal} {\bibinfo  {journal} {Physical Review X}\ }\textbf {\bibinfo {volume} {9}},\ \bibinfo {pages} {011012} (\bibinfo {year} {2019})},\ \bibinfo {note} {cited By 67}\BibitemShut {NoStop}%
\bibitem [{\citenamefont {Avron}\ \emph {et~al.}(1995)\citenamefont {Avron}, \citenamefont {Seiler},\ and\ \citenamefont {Zograf}}]{Avron1995}%
  \BibitemOpen
  \bibfield  {author} {\bibinfo {author} {\bibfnamefont {J.~E.}\ \bibnamefont {Avron}}, \bibinfo {author} {\bibfnamefont {R.}~\bibnamefont {Seiler}},\ and\ \bibinfo {author} {\bibfnamefont {P.~G.}\ \bibnamefont {Zograf}},\ }\bibfield  {title} {\bibinfo {title} {Viscosity of {{Quantum Hall Fluids}}},\ }\href {https://doi.org/10.1103/PhysRevLett.75.697} {\bibfield  {journal} {\bibinfo  {journal} {Physical Review Letters}\ }\textbf {\bibinfo {volume} {75}},\ \bibinfo {pages} {697} (\bibinfo {year} {1995})}\BibitemShut {NoStop}%
\bibitem [{\citenamefont {Read}(2009)}]{Read2009}%
  \BibitemOpen
  \bibfield  {author} {\bibinfo {author} {\bibfnamefont {N.}~\bibnamefont {Read}},\ }\bibfield  {title} {\bibinfo {title} {Non-{{Abelian}} adiabatic statistics and {{Hall}} viscosity in quantum {{Hall}} states and p x + i p y paired superfluids},\ }\href {https://doi.org/10.1103/PhysRevB.79.045308} {\bibfield  {journal} {\bibinfo  {journal} {Physical Review B}\ }\textbf {\bibinfo {volume} {79}},\ \bibinfo {pages} {045308} (\bibinfo {year} {2009})}\BibitemShut {NoStop}%
\bibitem [{\citenamefont {Haldane}(2009)}]{haldane2009b}%
  \BibitemOpen
  \bibfield  {author} {\bibinfo {author} {\bibfnamefont {F.~D.~M.}\ \bibnamefont {Haldane}},\ }\href@noop {} {\bibinfo {title} {"{{Hall}} viscosity" and intrinsic metric of incompressible fractional {{Hall}} fluids}} (\bibinfo {year} {2009}),\ \Eprint {https://arxiv.org/abs/0906.1854} {arXiv:0906.1854 [cond-mat, physics:hep-th]} \BibitemShut {NoStop}%
\bibitem [{\citenamefont {Read}\ and\ \citenamefont {Rezayi}(2011)}]{Read2011}%
  \BibitemOpen
  \bibfield  {author} {\bibinfo {author} {\bibfnamefont {N.}~\bibnamefont {Read}}\ and\ \bibinfo {author} {\bibfnamefont {E.~H.}\ \bibnamefont {Rezayi}},\ }\bibfield  {title} {\bibinfo {title} {Hall viscosity, orbital spin, and geometry: {{Paired}} superfluids and quantum {{Hall}} systems},\ }\href {https://doi.org/10.1103/PhysRevB.84.085316} {\bibfield  {journal} {\bibinfo  {journal} {Physical Review B}\ }\textbf {\bibinfo {volume} {84}},\ \bibinfo {pages} {085316} (\bibinfo {year} {2011})}\BibitemShut {NoStop}%
\bibitem [{\citenamefont {Hoyos}\ and\ \citenamefont {Son}(2012)}]{Hoyos2012}%
  \BibitemOpen
  \bibfield  {author} {\bibinfo {author} {\bibfnamefont {C.}~\bibnamefont {Hoyos}}\ and\ \bibinfo {author} {\bibfnamefont {D.~T.}\ \bibnamefont {Son}},\ }\bibfield  {title} {\bibinfo {title} {Hall {{Viscosity}} and {{Electromagnetic Response}}},\ }\href {https://doi.org/10.1103/PhysRevLett.108.066805} {\bibfield  {journal} {\bibinfo  {journal} {Physical Review Letters}\ }\textbf {\bibinfo {volume} {108}},\ \bibinfo {pages} {066805} (\bibinfo {year} {2012})}\BibitemShut {NoStop}%
\bibitem [{\citenamefont {Bradlyn}\ \emph {et~al.}(2012)\citenamefont {Bradlyn}, \citenamefont {Goldstein},\ and\ \citenamefont {Read}}]{bradlyn2012b}%
  \BibitemOpen
  \bibfield  {author} {\bibinfo {author} {\bibfnamefont {B.}~\bibnamefont {Bradlyn}}, \bibinfo {author} {\bibfnamefont {M.}~\bibnamefont {Goldstein}},\ and\ \bibinfo {author} {\bibfnamefont {N.}~\bibnamefont {Read}},\ }\bibfield  {title} {\bibinfo {title} {Kubo formulas for viscosity: {{Hall}} viscosity, {{Ward}} identities, and the relation with conductivity},\ }\href {https://doi.org/10.1103/PhysRevB.86.245309} {\bibfield  {journal} {\bibinfo  {journal} {Physical Review B}\ }\textbf {\bibinfo {volume} {86}},\ \bibinfo {pages} {245309} (\bibinfo {year} {2012})}\BibitemShut {NoStop}%
\bibitem [{\citenamefont {Haldane}\ and\ \citenamefont {Shen}(2015)}]{Haldane2015}%
  \BibitemOpen
  \bibfield  {author} {\bibinfo {author} {\bibfnamefont {F.~D.~M.}\ \bibnamefont {Haldane}}\ and\ \bibinfo {author} {\bibfnamefont {Y.}~\bibnamefont {Shen}},\ }\bibfield  {title} {\bibinfo {title} {Geometry of {{Landau}} orbits in the absence of rotational symmetry},\ }\href@noop {} {\  (\bibinfo {year} {2015})},\ \Eprint {https://arxiv.org/abs/1512.04502} {arXiv:1512.04502} \BibitemShut {NoStop}%
\bibitem [{\citenamefont {Shapourian}\ \emph {et~al.}(2015)\citenamefont {Shapourian}, \citenamefont {Hughes},\ and\ \citenamefont {Ryu}}]{Shapourian2015}%
  \BibitemOpen
  \bibfield  {author} {\bibinfo {author} {\bibfnamefont {H.}~\bibnamefont {Shapourian}}, \bibinfo {author} {\bibfnamefont {T.~L.}\ \bibnamefont {Hughes}},\ and\ \bibinfo {author} {\bibfnamefont {S.}~\bibnamefont {Ryu}},\ }\bibfield  {title} {\bibinfo {title} {Viscoelastic response of topological tight-binding models in two and three dimensions},\ }\href {https://doi.org/10.1103/PhysRevB.92.165131} {\bibfield  {journal} {\bibinfo  {journal} {Physical Review B}\ }\textbf {\bibinfo {volume} {92}},\ \bibinfo {pages} {165131} (\bibinfo {year} {2015})}\BibitemShut {NoStop}%
\bibitem [{\citenamefont {Holder}\ \emph {et~al.}(2019)\citenamefont {Holder}, \citenamefont {Queiroz},\ and\ \citenamefont {Stern}}]{holder2019}%
  \BibitemOpen
  \bibfield  {author} {\bibinfo {author} {\bibfnamefont {T.}~\bibnamefont {Holder}}, \bibinfo {author} {\bibfnamefont {R.}~\bibnamefont {Queiroz}},\ and\ \bibinfo {author} {\bibfnamefont {A.}~\bibnamefont {Stern}},\ }\bibfield  {title} {\bibinfo {title} {Unified {{Description}} of the {{Classical Hall Viscosity}}},\ }\href {https://doi.org/10.1103/PhysRevLett.123.106801} {\bibfield  {journal} {\bibinfo  {journal} {Physical Review Letters}\ }\textbf {\bibinfo {volume} {123}},\ \bibinfo {pages} {106801} (\bibinfo {year} {2019})}\BibitemShut {NoStop}%
\bibitem [{\citenamefont {Resta}(2011{\natexlab{a}})}]{Resta2011b}%
  \BibitemOpen
  \bibfield  {author} {\bibinfo {author} {\bibfnamefont {R.}~\bibnamefont {Resta}},\ }\bibfield  {title} {\bibinfo {title} {The insulating state of matter: A geometrical theory},\ }\href {https://doi.org/10.1140/epjb/e2010-10874-4} {\bibfield  {journal} {\bibinfo  {journal} {The European Physical Journal B}\ }\textbf {\bibinfo {volume} {79}},\ \bibinfo {pages} {121} (\bibinfo {year} {2011}{\natexlab{a}})}\BibitemShut {NoStop}%
\bibitem [{\citenamefont {Gao}\ \emph {et~al.}(2015)\citenamefont {Gao}, \citenamefont {Yang},\ and\ \citenamefont {Niu}}]{gao2015}%
  \BibitemOpen
  \bibfield  {author} {\bibinfo {author} {\bibfnamefont {Y.}~\bibnamefont {Gao}}, \bibinfo {author} {\bibfnamefont {S.~A.}\ \bibnamefont {Yang}},\ and\ \bibinfo {author} {\bibfnamefont {Q.}~\bibnamefont {Niu}},\ }\bibfield  {title} {\bibinfo {title} {Geometrical effects in orbital magnetic susceptibility},\ }\href {https://doi.org/10.1103/PhysRevB.91.214405} {\bibfield  {journal} {\bibinfo  {journal} {Physical Review B}\ }\textbf {\bibinfo {volume} {91}},\ \bibinfo {pages} {214405} (\bibinfo {year} {2015})}\BibitemShut {NoStop}%
\bibitem [{\citenamefont {Resta}(2018)}]{resta2018}%
  \BibitemOpen
  \bibfield  {author} {\bibinfo {author} {\bibfnamefont {R.}~\bibnamefont {Resta}},\ }\bibfield  {title} {\bibinfo {title} {Drude weight and superconducting weight},\ }\href {https://doi.org/10.1088/1361-648X/aade19} {\bibfield  {journal} {\bibinfo  {journal} {Journal of Physics: Condensed Matter}\ }\textbf {\bibinfo {volume} {30}},\ \bibinfo {pages} {414001} (\bibinfo {year} {2018})}\BibitemShut {NoStop}%
\bibitem [{\citenamefont {Marrazzo}\ and\ \citenamefont {Resta}(2019)}]{Marrazzo2019}%
  \BibitemOpen
  \bibfield  {author} {\bibinfo {author} {\bibfnamefont {A.}~\bibnamefont {Marrazzo}}\ and\ \bibinfo {author} {\bibfnamefont {R.}~\bibnamefont {Resta}},\ }\bibfield  {title} {\bibinfo {title} {Local theory of the insulating state},\ }\href {https://doi.org/10.1103/PhysRevLett.122.166602} {\bibfield  {journal} {\bibinfo  {journal} {Phys. Rev. Lett.}\ }\textbf {\bibinfo {volume} {122}},\ \bibinfo {pages} {166602} (\bibinfo {year} {2019})}\BibitemShut {NoStop}%
\bibitem [{\citenamefont {Bellomia}\ and\ \citenamefont {Resta}(2020)}]{Bellomia2020}%
  \BibitemOpen
  \bibfield  {author} {\bibinfo {author} {\bibfnamefont {G.}~\bibnamefont {Bellomia}}\ and\ \bibinfo {author} {\bibfnamefont {R.}~\bibnamefont {Resta}},\ }\bibfield  {title} {\bibinfo {title} {Drude weight in systems with open boundary conditions},\ }\href {https://doi.org/10.1103/PhysRevB.102.205123} {\bibfield  {journal} {\bibinfo  {journal} {Physical Review B}\ }\textbf {\bibinfo {volume} {102}},\ \bibinfo {pages} {205123} (\bibinfo {year} {2020})}\BibitemShut {NoStop}%
\bibitem [{\citenamefont {Gao}\ \emph {et~al.}(2014)\citenamefont {Gao}, \citenamefont {Yang},\ and\ \citenamefont {Niu}}]{gao2014}%
  \BibitemOpen
  \bibfield  {author} {\bibinfo {author} {\bibfnamefont {Y.}~\bibnamefont {Gao}}, \bibinfo {author} {\bibfnamefont {S.~A.}\ \bibnamefont {Yang}},\ and\ \bibinfo {author} {\bibfnamefont {Q.}~\bibnamefont {Niu}},\ }\bibfield  {title} {\bibinfo {title} {Field {{Induced Positional Shift}} of {{Bloch Electrons}} and {{Its Dynamical Implications}}},\ }\href {https://doi.org/10.1103/PhysRevLett.112.166601} {\bibfield  {journal} {\bibinfo  {journal} {Physical Review Letters}\ }\textbf {\bibinfo {volume} {112}},\ \bibinfo {pages} {166601} (\bibinfo {year} {2014})}\BibitemShut {NoStop}%
\bibitem [{\citenamefont {Morimoto}\ and\ \citenamefont {Nagaosa}(2016)}]{morimoto2016b}%
  \BibitemOpen
  \bibfield  {author} {\bibinfo {author} {\bibfnamefont {T.}~\bibnamefont {Morimoto}}\ and\ \bibinfo {author} {\bibfnamefont {N.}~\bibnamefont {Nagaosa}},\ }\bibfield  {title} {\bibinfo {title} {Topological nature of nonlinear optical effects in solids},\ }\href {https://doi.org/10.1126/sciadv.1501524} {\bibfield  {journal} {\bibinfo  {journal} {Science Advances}\ }\textbf {\bibinfo {volume} {2}},\ \bibinfo {pages} {e1501524} (\bibinfo {year} {2016})}\BibitemShut {NoStop}%
\bibitem [{\citenamefont {Nagaosa}\ and\ \citenamefont {Morimoto}(2017)}]{nagaosa2017a}%
  \BibitemOpen
  \bibfield  {author} {\bibinfo {author} {\bibfnamefont {N.}~\bibnamefont {Nagaosa}}\ and\ \bibinfo {author} {\bibfnamefont {T.}~\bibnamefont {Morimoto}},\ }\bibfield  {title} {\bibinfo {title} {Concept of {{Quantum Geometry}} in {{Optoelectronic Processes}} in {{Solids}}: {{Application}} to {{Solar Cells}}},\ }\href {https://doi.org/10.1002/adma.201603345} {\bibfield  {journal} {\bibinfo  {journal} {Advanced Materials}\ }\textbf {\bibinfo {volume} {29}},\ \bibinfo {pages} {1603345} (\bibinfo {year} {2017})}\BibitemShut {NoStop}%
\bibitem [{\citenamefont {Kolodrubetz}\ \emph {et~al.}(2017)\citenamefont {Kolodrubetz}, \citenamefont {Sels}, \citenamefont {Mehta},\ and\ \citenamefont {Polkovnikov}}]{kolodrubetz2017}%
  \BibitemOpen
  \bibfield  {author} {\bibinfo {author} {\bibfnamefont {M.}~\bibnamefont {Kolodrubetz}}, \bibinfo {author} {\bibfnamefont {D.}~\bibnamefont {Sels}}, \bibinfo {author} {\bibfnamefont {P.}~\bibnamefont {Mehta}},\ and\ \bibinfo {author} {\bibfnamefont {A.}~\bibnamefont {Polkovnikov}},\ }\bibfield  {title} {\bibinfo {title} {Geometry and non-adiabatic response in quantum and classical systems},\ }\href {https://doi.org/10.1016/j.physrep.2017.07.001} {\bibfield  {journal} {\bibinfo  {journal} {Physics Reports}\ }\textbf {\bibinfo {volume} {697}},\ \bibinfo {pages} {1} (\bibinfo {year} {2017})}\BibitemShut {NoStop}%
\bibitem [{\citenamefont {Ahn}\ \emph {et~al.}(2020)\citenamefont {Ahn}, \citenamefont {Guo},\ and\ \citenamefont {Nagaosa}}]{ahn2020b}%
  \BibitemOpen
  \bibfield  {author} {\bibinfo {author} {\bibfnamefont {J.}~\bibnamefont {Ahn}}, \bibinfo {author} {\bibfnamefont {G.-Y.}\ \bibnamefont {Guo}},\ and\ \bibinfo {author} {\bibfnamefont {N.}~\bibnamefont {Nagaosa}},\ }\bibfield  {title} {\bibinfo {title} {Low-{{Frequency Divergence}} and {{Quantum Geometry}} of the {{Bulk Photovoltaic Effect}} in {{Topological Semimetals}}},\ }\href {https://doi.org/10.1103/PhysRevX.10.041041} {\bibfield  {journal} {\bibinfo  {journal} {Physical Review X}\ }\textbf {\bibinfo {volume} {10}},\ \bibinfo {pages} {041041} (\bibinfo {year} {2020})}\BibitemShut {NoStop}%
\bibitem [{\citenamefont {Ahn}\ \emph {et~al.}(2022)\citenamefont {Ahn}, \citenamefont {Guo}, \citenamefont {Nagaosa},\ and\ \citenamefont {Vishwanath}}]{ahn2022a}%
  \BibitemOpen
  \bibfield  {author} {\bibinfo {author} {\bibfnamefont {J.}~\bibnamefont {Ahn}}, \bibinfo {author} {\bibfnamefont {G.-Y.}\ \bibnamefont {Guo}}, \bibinfo {author} {\bibfnamefont {N.}~\bibnamefont {Nagaosa}},\ and\ \bibinfo {author} {\bibfnamefont {A.}~\bibnamefont {Vishwanath}},\ }\bibfield  {title} {\bibinfo {title} {Riemannian geometry of resonant optical responses},\ }\href {https://doi.org/10.1038/s41567-021-01465-z} {\bibfield  {journal} {\bibinfo  {journal} {Nature Physics}\ }\textbf {\bibinfo {volume} {18}},\ \bibinfo {pages} {290} (\bibinfo {year} {2022})}\BibitemShut {NoStop}%
\bibitem [{\citenamefont {Wang}\ \emph {et~al.}(2021{\natexlab{a}})\citenamefont {Wang}, \citenamefont {Gao},\ and\ \citenamefont {Xiao}}]{wang2021a}%
  \BibitemOpen
  \bibfield  {author} {\bibinfo {author} {\bibfnamefont {C.}~\bibnamefont {Wang}}, \bibinfo {author} {\bibfnamefont {Y.}~\bibnamefont {Gao}},\ and\ \bibinfo {author} {\bibfnamefont {D.}~\bibnamefont {Xiao}},\ }\bibfield  {title} {\bibinfo {title} {Intrinsic {{Nonlinear Hall Effect}} in {{Antiferromagnetic Tetragonal CuMnAs}}},\ }\href {https://doi.org/10.1103/PhysRevLett.127.277201} {\bibfield  {journal} {\bibinfo  {journal} {Physical Review Letters}\ }\textbf {\bibinfo {volume} {127}},\ \bibinfo {pages} {277201} (\bibinfo {year} {2021}{\natexlab{a}})}\BibitemShut {NoStop}%
\bibitem [{\citenamefont {Ma}\ \emph {et~al.}(2023)\citenamefont {Ma}, \citenamefont {Arora}, \citenamefont {Vignale},\ and\ \citenamefont {Song}}]{ma2023}%
  \BibitemOpen
  \bibfield  {author} {\bibinfo {author} {\bibfnamefont {D.}~\bibnamefont {Ma}}, \bibinfo {author} {\bibfnamefont {A.}~\bibnamefont {Arora}}, \bibinfo {author} {\bibfnamefont {G.}~\bibnamefont {Vignale}},\ and\ \bibinfo {author} {\bibfnamefont {J.~C.~W.}\ \bibnamefont {Song}},\ }\bibfield  {title} {\bibinfo {title} {Anomalous {{Skew-Scattering Nonlinear Hall Effect}} and {{Chiral Photocurrents}} in \${\textbackslash}mathcal\{\vphantom\}{{PT}}\vphantom\{\}\$-{{Symmetric Antiferromagnets}}},\ }\href {https://doi.org/10.1103/PhysRevLett.131.076601} {\bibfield  {journal} {\bibinfo  {journal} {Physical Review Letters}\ }\textbf {\bibinfo {volume} {131}},\ \bibinfo {pages} {076601} (\bibinfo {year} {2023})}\BibitemShut {NoStop}%
\bibitem [{\citenamefont {Liu}\ \emph {et~al.}(2024{\natexlab{a}})\citenamefont {Liu}, \citenamefont {Qiang}, \citenamefont {Lu},\ and\ \citenamefont {Xie}}]{Liu_2024QGreview}%
  \BibitemOpen
  \bibfield  {author} {\bibinfo {author} {\bibfnamefont {T.}~\bibnamefont {Liu}}, \bibinfo {author} {\bibfnamefont {X.-B.}\ \bibnamefont {Qiang}}, \bibinfo {author} {\bibfnamefont {H.-Z.}\ \bibnamefont {Lu}},\ and\ \bibinfo {author} {\bibfnamefont {X.~C.}\ \bibnamefont {Xie}},\ }\bibfield  {title} {\bibinfo {title} {Quantum geometry in condensed matter},\ }\bibfield  {journal} {\bibinfo  {journal} {National Science Review}\ }\href {https://doi.org/10.1093/nsr/nwae334} {10.1093/nsr/nwae334} (\bibinfo {year} {2024}{\natexlab{a}})\BibitemShut {NoStop}%
\bibitem [{\citenamefont {Avdoshkin}\ \emph {et~al.}(2024)\citenamefont {Avdoshkin}, \citenamefont {Mitscherling},\ and\ \citenamefont {Moore}}]{avdoshkin2024}%
  \BibitemOpen
  \bibfield  {author} {\bibinfo {author} {\bibfnamefont {A.}~\bibnamefont {Avdoshkin}}, \bibinfo {author} {\bibfnamefont {J.}~\bibnamefont {Mitscherling}},\ and\ \bibinfo {author} {\bibfnamefont {J.~E.}\ \bibnamefont {Moore}},\ }\href {https://doi.org/10.48550/arXiv.2409.16358} {\bibinfo {title} {The multi-state geometry of shift current and polarization}} (\bibinfo {year} {2024}),\ \Eprint {https://arxiv.org/abs/2409.16358} {arXiv:2409.16358} \BibitemShut {NoStop}%
\bibitem [{\citenamefont {Sala}\ \emph {et~al.}(2024)\citenamefont {Sala}, \citenamefont {Mercaldo}, \citenamefont {Domi}, \citenamefont {Gariglio}, \citenamefont {Cuoco}, \citenamefont {Ortix},\ and\ \citenamefont {Caviglia}}]{sala2024}%
  \BibitemOpen
  \bibfield  {author} {\bibinfo {author} {\bibfnamefont {G.}~\bibnamefont {Sala}}, \bibinfo {author} {\bibfnamefont {M.~T.}\ \bibnamefont {Mercaldo}}, \bibinfo {author} {\bibfnamefont {K.}~\bibnamefont {Domi}}, \bibinfo {author} {\bibfnamefont {S.}~\bibnamefont {Gariglio}}, \bibinfo {author} {\bibfnamefont {M.}~\bibnamefont {Cuoco}}, \bibinfo {author} {\bibfnamefont {C.}~\bibnamefont {Ortix}},\ and\ \bibinfo {author} {\bibfnamefont {A.~D.}\ \bibnamefont {Caviglia}},\ }\href {https://arxiv.org/abs/2407.06659} {\bibinfo {title} {The quantum metric of electrons with spin-momentum locking}} (\bibinfo {year} {2024}),\ \Eprint {https://arxiv.org/abs/2407.06659} {arXiv:2407.06659 [cond-mat.mes-hall]} \BibitemShut {NoStop}%
\bibitem [{\citenamefont {Peotta}\ and\ \citenamefont {T{\"o}rm{\"a}}(2015)}]{peotta2015}%
  \BibitemOpen
  \bibfield  {author} {\bibinfo {author} {\bibfnamefont {S.}~\bibnamefont {Peotta}}\ and\ \bibinfo {author} {\bibfnamefont {P.}~\bibnamefont {T{\"o}rm{\"a}}},\ }\bibfield  {title} {\bibinfo {title} {Superfluidity in topologically nontrivial flat bands},\ }\href {https://doi.org/10.1038/ncomms9944} {\bibfield  {journal} {\bibinfo  {journal} {Nature Communications}\ }\textbf {\bibinfo {volume} {6}},\ \bibinfo {pages} {8944} (\bibinfo {year} {2015})}\BibitemShut {NoStop}%
\bibitem [{\citenamefont {Julku}\ \emph {et~al.}(2016)\citenamefont {Julku}, \citenamefont {Peotta}, \citenamefont {Vanhala}, \citenamefont {Kim},\ and\ \citenamefont {T{\"o}rm{\"a}}}]{Julku2016}%
  \BibitemOpen
  \bibfield  {author} {\bibinfo {author} {\bibfnamefont {A.}~\bibnamefont {Julku}}, \bibinfo {author} {\bibfnamefont {S.}~\bibnamefont {Peotta}}, \bibinfo {author} {\bibfnamefont {T.~I.}\ \bibnamefont {Vanhala}}, \bibinfo {author} {\bibfnamefont {D.-H.}\ \bibnamefont {Kim}},\ and\ \bibinfo {author} {\bibfnamefont {P.}~\bibnamefont {T{\"o}rm{\"a}}},\ }\bibfield  {title} {\bibinfo {title} {Geometric {{Origin}} of {{Superfluidity}} in the {{Lieb-Lattice Flat Band}}},\ }\href {https://doi.org/10.1103/PhysRevLett.117.045303} {\bibfield  {journal} {\bibinfo  {journal} {Physical Review Letters}\ }\textbf {\bibinfo {volume} {117}},\ \bibinfo {pages} {045303} (\bibinfo {year} {2016})}\BibitemShut {NoStop}%
\bibitem [{\citenamefont {Liang}\ \emph {et~al.}(2017)\citenamefont {Liang}, \citenamefont {Vanhala}, \citenamefont {Peotta}, \citenamefont {Siro}, \citenamefont {Harju},\ and\ \citenamefont {T{\"o}rm{\"a}}}]{liang2017}%
  \BibitemOpen
  \bibfield  {author} {\bibinfo {author} {\bibfnamefont {L.}~\bibnamefont {Liang}}, \bibinfo {author} {\bibfnamefont {T.~I.}\ \bibnamefont {Vanhala}}, \bibinfo {author} {\bibfnamefont {S.}~\bibnamefont {Peotta}}, \bibinfo {author} {\bibfnamefont {T.}~\bibnamefont {Siro}}, \bibinfo {author} {\bibfnamefont {A.}~\bibnamefont {Harju}},\ and\ \bibinfo {author} {\bibfnamefont {P.}~\bibnamefont {T{\"o}rm{\"a}}},\ }\bibfield  {title} {\bibinfo {title} {Band geometry, {{Berry}} curvature, and superfluid weight},\ }\href {https://doi.org/10.1103/PhysRevB.95.024515} {\bibfield  {journal} {\bibinfo  {journal} {Physical Review B}\ }\textbf {\bibinfo {volume} {95}},\ \bibinfo {pages} {1} (\bibinfo {year} {2017})}\BibitemShut {NoStop}%
\bibitem [{\citenamefont {Hu}\ \emph {et~al.}(2019)\citenamefont {Hu}, \citenamefont {Hyart}, \citenamefont {Pikulin},\ and\ \citenamefont {Rossi}}]{Hu2019}%
  \BibitemOpen
  \bibfield  {author} {\bibinfo {author} {\bibfnamefont {X.}~\bibnamefont {Hu}}, \bibinfo {author} {\bibfnamefont {T.}~\bibnamefont {Hyart}}, \bibinfo {author} {\bibfnamefont {D.~I.}\ \bibnamefont {Pikulin}},\ and\ \bibinfo {author} {\bibfnamefont {E.}~\bibnamefont {Rossi}},\ }\bibfield  {title} {\bibinfo {title} {Geometric and {{Conventional Contribution}} to the {{Superfluid Weight}} in {{Twisted Bilayer Graphene}}},\ }\href {https://doi.org/10.1103/PhysRevLett.123.237002} {\bibfield  {journal} {\bibinfo  {journal} {Physical Review Letters}\ }\textbf {\bibinfo {volume} {123}},\ \bibinfo {pages} {237002} (\bibinfo {year} {2019})}\BibitemShut {NoStop}%
\bibitem [{\citenamefont {Xie}\ \emph {et~al.}(2020{\natexlab{b}})\citenamefont {Xie}, \citenamefont {Song}, \citenamefont {Lian},\ and\ \citenamefont {Bernevig}}]{xie2020}%
  \BibitemOpen
  \bibfield  {author} {\bibinfo {author} {\bibfnamefont {F.}~\bibnamefont {Xie}}, \bibinfo {author} {\bibfnamefont {Z.}~\bibnamefont {Song}}, \bibinfo {author} {\bibfnamefont {B.}~\bibnamefont {Lian}},\ and\ \bibinfo {author} {\bibfnamefont {B.~A.}\ \bibnamefont {Bernevig}},\ }\bibfield  {title} {\bibinfo {title} {Topology-{{Bounded Superfluid Weight}} in {{Twisted Bilayer Graphene}}},\ }\href {https://doi.org/10.1103/PhysRevLett.124.167002} {\bibfield  {journal} {\bibinfo  {journal} {Physical Review Letters}\ }\textbf {\bibinfo {volume} {124}},\ \bibinfo {pages} {167002} (\bibinfo {year} {2020}{\natexlab{b}})}\BibitemShut {NoStop}%
\bibitem [{\citenamefont {Julku}\ \emph {et~al.}(2020)\citenamefont {Julku}, \citenamefont {Peltonen}, \citenamefont {Liang}, \citenamefont {Heikkil{\"a}},\ and\ \citenamefont {T{\"o}rm{\"a}}}]{julku2020}%
  \BibitemOpen
  \bibfield  {author} {\bibinfo {author} {\bibfnamefont {A.}~\bibnamefont {Julku}}, \bibinfo {author} {\bibfnamefont {T.~J.}\ \bibnamefont {Peltonen}}, \bibinfo {author} {\bibfnamefont {L.}~\bibnamefont {Liang}}, \bibinfo {author} {\bibfnamefont {T.~T.}\ \bibnamefont {Heikkil{\"a}}},\ and\ \bibinfo {author} {\bibfnamefont {P.}~\bibnamefont {T{\"o}rm{\"a}}},\ }\bibfield  {title} {\bibinfo {title} {Superfluid weight and {{Berezinskii-Kosterlitz-Thouless}} transition temperature of twisted bilayer graphene},\ }\href {https://doi.org/10.1103/PhysRevB.101.060505} {\bibfield  {journal} {\bibinfo  {journal} {Physical Review B}\ }\textbf {\bibinfo {volume} {101}},\ \bibinfo {pages} {060505} (\bibinfo {year} {2020})}\BibitemShut {NoStop}%
\bibitem [{\citenamefont {Rossi}(2021{\natexlab{a}})}]{rossi2021}%
  \BibitemOpen
  \bibfield  {author} {\bibinfo {author} {\bibfnamefont {E.}~\bibnamefont {Rossi}},\ }\bibfield  {title} {\bibinfo {title} {Quantum metric and correlated states in two-dimensional systems},\ }\href {https://doi.org/10.1016/j.cossms.2021.100952} {\bibfield  {journal} {\bibinfo  {journal} {Current Opinion in Solid State and Materials Science}\ }\textbf {\bibinfo {volume} {25}},\ \bibinfo {pages} {100952} (\bibinfo {year} {2021}{\natexlab{a}})}\BibitemShut {NoStop}%
\bibitem [{\citenamefont {T{\"o}rm{\"a}}\ \emph {et~al.}(2022)\citenamefont {T{\"o}rm{\"a}}, \citenamefont {Peotta},\ and\ \citenamefont {Bernevig}}]{Torma2022ReviewQuantumGeometry}%
  \BibitemOpen
  \bibfield  {author} {\bibinfo {author} {\bibfnamefont {P.}~\bibnamefont {T{\"o}rm{\"a}}}, \bibinfo {author} {\bibfnamefont {S.}~\bibnamefont {Peotta}},\ and\ \bibinfo {author} {\bibfnamefont {B.~A.}\ \bibnamefont {Bernevig}},\ }\bibfield  {title} {\bibinfo {title} {Superconductivity, superfluidity and quantum geometry in twisted multilayer systems},\ }\href {https://doi.org/10.1038/s42254-022-00466-y} {\bibfield  {journal} {\bibinfo  {journal} {Nature Reviews Physics}\ }\textbf {\bibinfo {volume} {4}},\ \bibinfo {pages} {528} (\bibinfo {year} {2022})}\BibitemShut {NoStop}%
\bibitem [{\citenamefont {T{\"o}rm{\"a}}(2023)}]{torma2023}%
  \BibitemOpen
  \bibfield  {author} {\bibinfo {author} {\bibfnamefont {P.}~\bibnamefont {T{\"o}rm{\"a}}},\ }\bibfield  {title} {\bibinfo {title} {Essay: {{Where Can Quantum Geometry Lead Us}}?},\ }\href {https://doi.org/10.1103/PhysRevLett.131.240001} {\bibfield  {journal} {\bibinfo  {journal} {Physical Review Letters}\ }\textbf {\bibinfo {volume} {131}},\ \bibinfo {pages} {240001} (\bibinfo {year} {2023})}\BibitemShut {NoStop}%
\bibitem [{\citenamefont {Tian}\ \emph {et~al.}(2023)\citenamefont {Tian}, \citenamefont {Gao}, \citenamefont {Zhang}, \citenamefont {Che}, \citenamefont {Xu}, \citenamefont {Cheung}, \citenamefont {Watanabe}, \citenamefont {Taniguchi}, \citenamefont {Randeria}, \citenamefont {Zhang}, \citenamefont {Lau},\ and\ \citenamefont {Bockrath}}]{tian2023}%
  \BibitemOpen
  \bibfield  {author} {\bibinfo {author} {\bibfnamefont {H.}~\bibnamefont {Tian}}, \bibinfo {author} {\bibfnamefont {X.}~\bibnamefont {Gao}}, \bibinfo {author} {\bibfnamefont {Y.}~\bibnamefont {Zhang}}, \bibinfo {author} {\bibfnamefont {S.}~\bibnamefont {Che}}, \bibinfo {author} {\bibfnamefont {T.}~\bibnamefont {Xu}}, \bibinfo {author} {\bibfnamefont {P.}~\bibnamefont {Cheung}}, \bibinfo {author} {\bibfnamefont {K.}~\bibnamefont {Watanabe}}, \bibinfo {author} {\bibfnamefont {T.}~\bibnamefont {Taniguchi}}, \bibinfo {author} {\bibfnamefont {M.}~\bibnamefont {Randeria}}, \bibinfo {author} {\bibfnamefont {F.}~\bibnamefont {Zhang}}, \bibinfo {author} {\bibfnamefont {C.~N.}\ \bibnamefont {Lau}},\ and\ \bibinfo {author} {\bibfnamefont {M.~W.}\ \bibnamefont {Bockrath}},\ }\bibfield  {title} {\bibinfo {title} {Evidence for {{Dirac}} flat band superconductivity enabled by quantum geometry},\ }\href {https://doi.org/10.1038/s41586-022-05576-2} {\bibfield  {journal} {\bibinfo  {journal} {Nature}\ }\textbf {\bibinfo
  {volume} {614}},\ \bibinfo {pages} {440} (\bibinfo {year} {2023})}\BibitemShut {NoStop}%
\bibitem [{\citenamefont {Cao}\ \emph {et~al.}(2018{\natexlab{a}})\citenamefont {Cao}, \citenamefont {Fatemi}, \citenamefont {Demir}, \citenamefont {Fang}, \citenamefont {Tomarken}, \citenamefont {Luo}, \citenamefont {{Sanchez-Yamagishi}}, \citenamefont {Watanabe}, \citenamefont {Taniguchi}, \citenamefont {Kaxiras}, \citenamefont {Ashoori},\ and\ \citenamefont {{Jarillo-Herrero}}}]{Cao2018c}%
  \BibitemOpen
  \bibfield  {author} {\bibinfo {author} {\bibfnamefont {Y.}~\bibnamefont {Cao}}, \bibinfo {author} {\bibfnamefont {V.}~\bibnamefont {Fatemi}}, \bibinfo {author} {\bibfnamefont {A.}~\bibnamefont {Demir}}, \bibinfo {author} {\bibfnamefont {S.}~\bibnamefont {Fang}}, \bibinfo {author} {\bibfnamefont {S.}~\bibnamefont {Tomarken}}, \bibinfo {author} {\bibfnamefont {J.}~\bibnamefont {Luo}}, \bibinfo {author} {\bibfnamefont {J.}~\bibnamefont {{Sanchez-Yamagishi}}}, \bibinfo {author} {\bibfnamefont {K.}~\bibnamefont {Watanabe}}, \bibinfo {author} {\bibfnamefont {T.}~\bibnamefont {Taniguchi}}, \bibinfo {author} {\bibfnamefont {E.}~\bibnamefont {Kaxiras}}, \bibinfo {author} {\bibfnamefont {R.}~\bibnamefont {Ashoori}},\ and\ \bibinfo {author} {\bibfnamefont {P.}~\bibnamefont {{Jarillo-Herrero}}},\ }\bibfield  {title} {\bibinfo {title} {Correlated insulator behaviour at half-filling in magic-angle graphene superlattices},\ }\href {https://doi.org/10.1038/nature26154} {\bibfield  {journal} {\bibinfo  {journal} {Nature}\
  }\textbf {\bibinfo {volume} {556}},\ \bibinfo {pages} {80} (\bibinfo {year} {2018}{\natexlab{a}})}\BibitemShut {NoStop}%
\bibitem [{\citenamefont {Cao}\ \emph {et~al.}(2018{\natexlab{b}})\citenamefont {Cao}, \citenamefont {Fatemi}, \citenamefont {Fang}, \citenamefont {Watanabe}, \citenamefont {Taniguchi}, \citenamefont {Kaxiras},\ and\ \citenamefont {{Jarillo-Herrero}}}]{Cao2018a}%
  \BibitemOpen
  \bibfield  {author} {\bibinfo {author} {\bibfnamefont {Y.}~\bibnamefont {Cao}}, \bibinfo {author} {\bibfnamefont {V.}~\bibnamefont {Fatemi}}, \bibinfo {author} {\bibfnamefont {S.}~\bibnamefont {Fang}}, \bibinfo {author} {\bibfnamefont {K.}~\bibnamefont {Watanabe}}, \bibinfo {author} {\bibfnamefont {T.}~\bibnamefont {Taniguchi}}, \bibinfo {author} {\bibfnamefont {E.}~\bibnamefont {Kaxiras}},\ and\ \bibinfo {author} {\bibfnamefont {P.}~\bibnamefont {{Jarillo-Herrero}}},\ }\bibfield  {title} {\bibinfo {title} {Unconventional superconductivity in magic-angle graphene superlattices},\ }\href {https://doi.org/10.1038/nature26160} {\bibfield  {journal} {\bibinfo  {journal} {Nature}\ }\textbf {\bibinfo {volume} {556}},\ \bibinfo {pages} {43} (\bibinfo {year} {2018}{\natexlab{b}})}\BibitemShut {NoStop}%
\bibitem [{\citenamefont {Yankowitz}\ \emph {et~al.}(2019)\citenamefont {Yankowitz}, \citenamefont {Chen}, \citenamefont {Polshyn}, \citenamefont {Zhang}, \citenamefont {Watanabe}, \citenamefont {Taniguchi}, \citenamefont {Graf}, \citenamefont {Young},\ and\ \citenamefont {Dean}}]{Yankowitz2019}%
  \BibitemOpen
  \bibfield  {author} {\bibinfo {author} {\bibfnamefont {M.}~\bibnamefont {Yankowitz}}, \bibinfo {author} {\bibfnamefont {S.}~\bibnamefont {Chen}}, \bibinfo {author} {\bibfnamefont {H.}~\bibnamefont {Polshyn}}, \bibinfo {author} {\bibfnamefont {Y.}~\bibnamefont {Zhang}}, \bibinfo {author} {\bibfnamefont {K.}~\bibnamefont {Watanabe}}, \bibinfo {author} {\bibfnamefont {T.}~\bibnamefont {Taniguchi}}, \bibinfo {author} {\bibfnamefont {D.}~\bibnamefont {Graf}}, \bibinfo {author} {\bibfnamefont {A.~F.}\ \bibnamefont {Young}},\ and\ \bibinfo {author} {\bibfnamefont {C.~R.}\ \bibnamefont {Dean}},\ }\bibfield  {title} {\bibinfo {title} {Tuning superconductivity in twisted bilayer graphene},\ }\href {https://doi.org/10.1126/science.aav1910} {\bibfield  {journal} {\bibinfo  {journal} {Science}\ }\textbf {\bibinfo {volume} {363}},\ \bibinfo {pages} {1059} (\bibinfo {year} {2019})}\BibitemShut {NoStop}%
\bibitem [{\citenamefont {Chen}\ \emph {et~al.}(2019)\citenamefont {Chen}, \citenamefont {Sharpe}, \citenamefont {Gallagher}, \citenamefont {Rosen}, \citenamefont {Fox}, \citenamefont {Jiang}, \citenamefont {Lyu}, \citenamefont {Li}, \citenamefont {Watanabe}, \citenamefont {Taniguchi}, \citenamefont {Zhang},\ and\ \citenamefont {Wang}}]{Chen2019}%
  \BibitemOpen
  \bibfield  {author} {\bibinfo {author} {\bibfnamefont {G.}~\bibnamefont {Chen}}, \bibinfo {author} {\bibfnamefont {A.}~\bibnamefont {Sharpe}}, \bibinfo {author} {\bibfnamefont {P.}~\bibnamefont {Gallagher}}, \bibinfo {author} {\bibfnamefont {I.}~\bibnamefont {Rosen}}, \bibinfo {author} {\bibfnamefont {E.}~\bibnamefont {Fox}}, \bibinfo {author} {\bibfnamefont {L.}~\bibnamefont {Jiang}}, \bibinfo {author} {\bibfnamefont {B.}~\bibnamefont {Lyu}}, \bibinfo {author} {\bibfnamefont {H.}~\bibnamefont {Li}}, \bibinfo {author} {\bibfnamefont {K.}~\bibnamefont {Watanabe}}, \bibinfo {author} {\bibfnamefont {T.}~\bibnamefont {Taniguchi}}, \bibinfo {author} {\bibfnamefont {Y.}~\bibnamefont {Zhang}},\ and\ \bibinfo {author} {\bibfnamefont {F.}~\bibnamefont {Wang}},\ }\bibfield  {title} {\bibinfo {title} {Signatures of tunable superconductivity in a trilayer graphene moir{\'e} superlattice},\ }\href {https://doi.org/10.1038/s41586-019-1393-y} {\bibfield  {journal} {\bibinfo  {journal} {Nature}\ }\textbf {\bibinfo {volume}
  {572}},\ \bibinfo {pages} {215} (\bibinfo {year} {2019})}\BibitemShut {NoStop}%
\bibitem [{\citenamefont {Lu}\ \emph {et~al.}(2019)\citenamefont {Lu}, \citenamefont {Stepanov}, \citenamefont {Yang}, \citenamefont {Xie}, \citenamefont {Aamir}, \citenamefont {Das}, \citenamefont {Urgell}, \citenamefont {Watanabe}, \citenamefont {Taniguchi}, \citenamefont {Zhang}, \citenamefont {Bachtold}, \citenamefont {MacDonald},\ and\ \citenamefont {Efetov}}]{lu2019a}%
  \BibitemOpen
  \bibfield  {author} {\bibinfo {author} {\bibfnamefont {X.}~\bibnamefont {Lu}}, \bibinfo {author} {\bibfnamefont {P.}~\bibnamefont {Stepanov}}, \bibinfo {author} {\bibfnamefont {W.}~\bibnamefont {Yang}}, \bibinfo {author} {\bibfnamefont {M.}~\bibnamefont {Xie}}, \bibinfo {author} {\bibfnamefont {M.~A.}\ \bibnamefont {Aamir}}, \bibinfo {author} {\bibfnamefont {I.}~\bibnamefont {Das}}, \bibinfo {author} {\bibfnamefont {C.}~\bibnamefont {Urgell}}, \bibinfo {author} {\bibfnamefont {K.}~\bibnamefont {Watanabe}}, \bibinfo {author} {\bibfnamefont {T.}~\bibnamefont {Taniguchi}}, \bibinfo {author} {\bibfnamefont {G.}~\bibnamefont {Zhang}}, \bibinfo {author} {\bibfnamefont {A.}~\bibnamefont {Bachtold}}, \bibinfo {author} {\bibfnamefont {A.~H.}\ \bibnamefont {MacDonald}},\ and\ \bibinfo {author} {\bibfnamefont {D.~K.}\ \bibnamefont {Efetov}},\ }\bibfield  {title} {\bibinfo {title} {Superconductors, orbital magnets and correlated states in magic-angle bilayer graphene},\ }\href
  {https://doi.org/10.1038/s41586-019-1695-0} {\bibfield  {journal} {\bibinfo  {journal} {Nature}\ }\textbf {\bibinfo {volume} {574}},\ \bibinfo {pages} {653} (\bibinfo {year} {2019})}\BibitemShut {NoStop}%
\bibitem [{\citenamefont {Choi}\ \emph {et~al.}(2019)\citenamefont {Choi}, \citenamefont {Kemmer}, \citenamefont {Peng}, \citenamefont {Thomson}, \citenamefont {Arora}, \citenamefont {Polski}, \citenamefont {Zhang}, \citenamefont {Ren}, \citenamefont {Alicea}, \citenamefont {Refael}, \citenamefont {Taniguchi},\ and\ \citenamefont {{Nadj-Perge}}}]{Choi2019}%
  \BibitemOpen
  \bibfield  {author} {\bibinfo {author} {\bibfnamefont {Y.}~\bibnamefont {Choi}}, \bibinfo {author} {\bibfnamefont {J.}~\bibnamefont {Kemmer}}, \bibinfo {author} {\bibfnamefont {Y.}~\bibnamefont {Peng}}, \bibinfo {author} {\bibfnamefont {A.}~\bibnamefont {Thomson}}, \bibinfo {author} {\bibfnamefont {H.}~\bibnamefont {Arora}}, \bibinfo {author} {\bibfnamefont {R.}~\bibnamefont {Polski}}, \bibinfo {author} {\bibfnamefont {Y.}~\bibnamefont {Zhang}}, \bibinfo {author} {\bibfnamefont {H.}~\bibnamefont {Ren}}, \bibinfo {author} {\bibfnamefont {J.}~\bibnamefont {Alicea}}, \bibinfo {author} {\bibfnamefont {G.}~\bibnamefont {Refael}}, \bibinfo {author} {\bibfnamefont {T.}~\bibnamefont {Taniguchi}},\ and\ \bibinfo {author} {\bibfnamefont {S.}~\bibnamefont {{Nadj-Perge}}},\ }\bibfield  {title} {\bibinfo {title} {Electronic correlations in twisted bilayer graphene near the magic angle},\ }\href {https://doi.org/10.1038/s41567-019-0606-5} {\bibfield  {journal} {\bibinfo  {journal} {Nature Physics}\ }\textbf {\bibinfo
  {volume} {15}},\ \bibinfo {pages} {1174} (\bibinfo {year} {2019})}\BibitemShut {NoStop}%
\bibitem [{\citenamefont {Sharpe}\ \emph {et~al.}(2019)\citenamefont {Sharpe}, \citenamefont {Fox}, \citenamefont {Barnard}, \citenamefont {Finney}, \citenamefont {Watanabe}, \citenamefont {Taniguchi}, \citenamefont {Kastner},\ and\ \citenamefont {{Goldhaber-Gordon}}}]{sharpe2019}%
  \BibitemOpen
  \bibfield  {author} {\bibinfo {author} {\bibfnamefont {A.~L.}\ \bibnamefont {Sharpe}}, \bibinfo {author} {\bibfnamefont {E.~J.}\ \bibnamefont {Fox}}, \bibinfo {author} {\bibfnamefont {A.~W.}\ \bibnamefont {Barnard}}, \bibinfo {author} {\bibfnamefont {J.}~\bibnamefont {Finney}}, \bibinfo {author} {\bibfnamefont {K.}~\bibnamefont {Watanabe}}, \bibinfo {author} {\bibfnamefont {T.}~\bibnamefont {Taniguchi}}, \bibinfo {author} {\bibfnamefont {M.~A.}\ \bibnamefont {Kastner}},\ and\ \bibinfo {author} {\bibfnamefont {D.}~\bibnamefont {{Goldhaber-Gordon}}},\ }\bibfield  {title} {\bibinfo {title} {Emergent ferromagnetism near three-quarters filling in twisted bilayer graphene},\ }\href {https://doi.org/10.1126/science.aaw3780} {\bibfield  {journal} {\bibinfo  {journal} {Science}\ }\textbf {\bibinfo {volume} {365}},\ \bibinfo {pages} {605} (\bibinfo {year} {2019})}\BibitemShut {NoStop}%
\bibitem [{\citenamefont {Polshyn}\ \emph {et~al.}(2019)\citenamefont {Polshyn}, \citenamefont {Yankowitz}, \citenamefont {Chen}, \citenamefont {Zhang}, \citenamefont {Watanabe}, \citenamefont {Taniguchi}, \citenamefont {Dean},\ and\ \citenamefont {Young}}]{Polshyn2019}%
  \BibitemOpen
  \bibfield  {author} {\bibinfo {author} {\bibfnamefont {H.}~\bibnamefont {Polshyn}}, \bibinfo {author} {\bibfnamefont {M.}~\bibnamefont {Yankowitz}}, \bibinfo {author} {\bibfnamefont {S.}~\bibnamefont {Chen}}, \bibinfo {author} {\bibfnamefont {Y.}~\bibnamefont {Zhang}}, \bibinfo {author} {\bibfnamefont {K.}~\bibnamefont {Watanabe}}, \bibinfo {author} {\bibfnamefont {T.}~\bibnamefont {Taniguchi}}, \bibinfo {author} {\bibfnamefont {C.~R.}\ \bibnamefont {Dean}},\ and\ \bibinfo {author} {\bibfnamefont {A.~F.}\ \bibnamefont {Young}},\ }\bibfield  {title} {\bibinfo {title} {Large linear-in-temperature resistivity in twisted bilayer graphene},\ }\href {https://doi.org/10.1038/s41567-019-0596-3} {\bibfield  {journal} {\bibinfo  {journal} {Nature Physics}\ }\textbf {\bibinfo {volume} {15}},\ \bibinfo {pages} {1011} (\bibinfo {year} {2019})}\BibitemShut {NoStop}%
\bibitem [{\citenamefont {Codecido}\ \emph {et~al.}(2019)\citenamefont {Codecido}, \citenamefont {Wang}, \citenamefont {Koester}, \citenamefont {Che}, \citenamefont {Tian}, \citenamefont {Lv}, \citenamefont {Tran}, \citenamefont {Watanabe}, \citenamefont {Taniguchi}, \citenamefont {Zhang}, \citenamefont {Bockrath},\ and\ \citenamefont {Lau}}]{Codecido2019}%
  \BibitemOpen
  \bibfield  {author} {\bibinfo {author} {\bibfnamefont {E.}~\bibnamefont {Codecido}}, \bibinfo {author} {\bibfnamefont {Q.}~\bibnamefont {Wang}}, \bibinfo {author} {\bibfnamefont {R.}~\bibnamefont {Koester}}, \bibinfo {author} {\bibfnamefont {S.}~\bibnamefont {Che}}, \bibinfo {author} {\bibfnamefont {H.}~\bibnamefont {Tian}}, \bibinfo {author} {\bibfnamefont {R.}~\bibnamefont {Lv}}, \bibinfo {author} {\bibfnamefont {S.}~\bibnamefont {Tran}}, \bibinfo {author} {\bibfnamefont {K.}~\bibnamefont {Watanabe}}, \bibinfo {author} {\bibfnamefont {T.}~\bibnamefont {Taniguchi}}, \bibinfo {author} {\bibfnamefont {F.}~\bibnamefont {Zhang}}, \bibinfo {author} {\bibfnamefont {M.}~\bibnamefont {Bockrath}},\ and\ \bibinfo {author} {\bibfnamefont {C.~N.}\ \bibnamefont {Lau}},\ }\bibfield  {title} {\bibinfo {title} {Correlated insulating and superconducting states in twisted bilayer graphene below the magic angle},\ }\href {https://doi.org/10.1126/sciadv.aaw9770} {\bibfield  {journal} {\bibinfo  {journal} {Science Advances}\
  }\textbf {\bibinfo {volume} {5}},\ \bibinfo {pages} {eaaw9770} (\bibinfo {year} {2019})}\BibitemShut {NoStop}%
\bibitem [{\citenamefont {Liu}\ \emph {et~al.}(2020)\citenamefont {Liu}, \citenamefont {Hao}, \citenamefont {Khalaf}, \citenamefont {Lee}, \citenamefont {Ronen}, \citenamefont {Yoo}, \citenamefont {Haei~Najafabadi}, \citenamefont {Watanabe}, \citenamefont {Taniguchi}, \citenamefont {Vishwanath},\ and\ \citenamefont {Kim}}]{Liu2020a}%
  \BibitemOpen
  \bibfield  {author} {\bibinfo {author} {\bibfnamefont {X.}~\bibnamefont {Liu}}, \bibinfo {author} {\bibfnamefont {Z.}~\bibnamefont {Hao}}, \bibinfo {author} {\bibfnamefont {E.}~\bibnamefont {Khalaf}}, \bibinfo {author} {\bibfnamefont {J.~Y.}\ \bibnamefont {Lee}}, \bibinfo {author} {\bibfnamefont {Y.}~\bibnamefont {Ronen}}, \bibinfo {author} {\bibfnamefont {H.}~\bibnamefont {Yoo}}, \bibinfo {author} {\bibfnamefont {D.}~\bibnamefont {Haei~Najafabadi}}, \bibinfo {author} {\bibfnamefont {K.}~\bibnamefont {Watanabe}}, \bibinfo {author} {\bibfnamefont {T.}~\bibnamefont {Taniguchi}}, \bibinfo {author} {\bibfnamefont {A.}~\bibnamefont {Vishwanath}},\ and\ \bibinfo {author} {\bibfnamefont {P.}~\bibnamefont {Kim}},\ }\bibfield  {title} {\bibinfo {title} {Tunable spin-polarized correlated states in twisted double bilayer graphene},\ }\href {https://doi.org/10.1038/s41586-020-2458-7} {\bibfield  {journal} {\bibinfo  {journal} {Nature}\ }\textbf {\bibinfo {volume} {583}},\ \bibinfo {pages} {221} (\bibinfo {year}
  {2020})}\BibitemShut {NoStop}%
\bibitem [{\citenamefont {Shen}\ \emph {et~al.}(2020)\citenamefont {Shen}, \citenamefont {Chu}, \citenamefont {Wu}, \citenamefont {Li}, \citenamefont {Wang}, \citenamefont {Zhao}, \citenamefont {Tang}, \citenamefont {Liu}, \citenamefont {Tian}, \citenamefont {Watanabe}, \citenamefont {Taniguchi}, \citenamefont {Yang}, \citenamefont {Meng}, \citenamefont {Shi}, \citenamefont {Yazyev},\ and\ \citenamefont {Zhang}}]{Shen2020}%
  \BibitemOpen
  \bibfield  {author} {\bibinfo {author} {\bibfnamefont {C.}~\bibnamefont {Shen}}, \bibinfo {author} {\bibfnamefont {Y.}~\bibnamefont {Chu}}, \bibinfo {author} {\bibfnamefont {Q.~S.}\ \bibnamefont {Wu}}, \bibinfo {author} {\bibfnamefont {N.}~\bibnamefont {Li}}, \bibinfo {author} {\bibfnamefont {S.}~\bibnamefont {Wang}}, \bibinfo {author} {\bibfnamefont {Y.}~\bibnamefont {Zhao}}, \bibinfo {author} {\bibfnamefont {J.}~\bibnamefont {Tang}}, \bibinfo {author} {\bibfnamefont {J.}~\bibnamefont {Liu}}, \bibinfo {author} {\bibfnamefont {J.}~\bibnamefont {Tian}}, \bibinfo {author} {\bibfnamefont {K.}~\bibnamefont {Watanabe}}, \bibinfo {author} {\bibfnamefont {T.}~\bibnamefont {Taniguchi}}, \bibinfo {author} {\bibfnamefont {R.}~\bibnamefont {Yang}}, \bibinfo {author} {\bibfnamefont {Z.~Y.}\ \bibnamefont {Meng}}, \bibinfo {author} {\bibfnamefont {D.}~\bibnamefont {Shi}}, \bibinfo {author} {\bibfnamefont {O.~V.}\ \bibnamefont {Yazyev}},\ and\ \bibinfo {author} {\bibfnamefont {G.}~\bibnamefont {Zhang}},\ }\bibfield
  {title} {\bibinfo {title} {Correlated states in twisted double bilayer graphene},\ }\href {https://doi.org/10.1038/s41567-020-0825-9} {\bibfield  {journal} {\bibinfo  {journal} {Nature Physics}\ }\textbf {\bibinfo {volume} {16}},\ \bibinfo {pages} {520} (\bibinfo {year} {2020})}\BibitemShut {NoStop}%
\bibitem [{\citenamefont {Chen}\ \emph {et~al.}(2020)\citenamefont {Chen}, \citenamefont {Sharpe}, \citenamefont {Fox}, \citenamefont {Zhang}, \citenamefont {Wang}, \citenamefont {Jiang}, \citenamefont {Lyu}, \citenamefont {Li}, \citenamefont {Watanabe}, \citenamefont {Taniguchi}, \citenamefont {Shi}, \citenamefont {Senthil}, \citenamefont {{Goldhaber-Gordon}}, \citenamefont {Zhang},\ and\ \citenamefont {Wang}}]{Chen2020}%
  \BibitemOpen
  \bibfield  {author} {\bibinfo {author} {\bibfnamefont {G.}~\bibnamefont {Chen}}, \bibinfo {author} {\bibfnamefont {A.~L.}\ \bibnamefont {Sharpe}}, \bibinfo {author} {\bibfnamefont {E.~J.}\ \bibnamefont {Fox}}, \bibinfo {author} {\bibfnamefont {Y.-H.}\ \bibnamefont {Zhang}}, \bibinfo {author} {\bibfnamefont {S.}~\bibnamefont {Wang}}, \bibinfo {author} {\bibfnamefont {L.}~\bibnamefont {Jiang}}, \bibinfo {author} {\bibfnamefont {B.}~\bibnamefont {Lyu}}, \bibinfo {author} {\bibfnamefont {H.}~\bibnamefont {Li}}, \bibinfo {author} {\bibfnamefont {K.}~\bibnamefont {Watanabe}}, \bibinfo {author} {\bibfnamefont {T.}~\bibnamefont {Taniguchi}}, \bibinfo {author} {\bibfnamefont {Z.}~\bibnamefont {Shi}}, \bibinfo {author} {\bibfnamefont {T.}~\bibnamefont {Senthil}}, \bibinfo {author} {\bibfnamefont {D.}~\bibnamefont {{Goldhaber-Gordon}}}, \bibinfo {author} {\bibfnamefont {Y.}~\bibnamefont {Zhang}},\ and\ \bibinfo {author} {\bibfnamefont {F.}~\bibnamefont {Wang}},\ }\bibfield  {title} {\bibinfo {title} {Tunable
  correlated {{Chern}} insulator and ferromagnetism in a moir{\'e} superlattice},\ }\href {https://doi.org/10.1038/s41586-020-2049-7} {\bibfield  {journal} {\bibinfo  {journal} {Nature}\ }\textbf {\bibinfo {volume} {579}},\ \bibinfo {pages} {56} (\bibinfo {year} {2020})}\BibitemShut {NoStop}%
\bibitem [{\citenamefont {Cao}\ \emph {et~al.}(2020)\citenamefont {Cao}, \citenamefont {{Rodan-Legrain}}, \citenamefont {{Rubies-Bigorda}}, \citenamefont {Park}, \citenamefont {Watanabe}, \citenamefont {Taniguchi},\ and\ \citenamefont {{Jarillo-Herrero}}}]{Cao2020}%
  \BibitemOpen
  \bibfield  {author} {\bibinfo {author} {\bibfnamefont {Y.}~\bibnamefont {Cao}}, \bibinfo {author} {\bibfnamefont {D.}~\bibnamefont {{Rodan-Legrain}}}, \bibinfo {author} {\bibfnamefont {O.}~\bibnamefont {{Rubies-Bigorda}}}, \bibinfo {author} {\bibfnamefont {J.~M.}\ \bibnamefont {Park}}, \bibinfo {author} {\bibfnamefont {K.}~\bibnamefont {Watanabe}}, \bibinfo {author} {\bibfnamefont {T.}~\bibnamefont {Taniguchi}},\ and\ \bibinfo {author} {\bibfnamefont {P.}~\bibnamefont {{Jarillo-Herrero}}},\ }\bibfield  {title} {\bibinfo {title} {Tunable correlated states and spin-polarized phases in twisted bilayer--bilayer graphene},\ }\href {https://doi.org/10.1038/s41586-020-2260-6} {\bibfield  {journal} {\bibinfo  {journal} {Nature}\ }\textbf {\bibinfo {volume} {583}},\ \bibinfo {pages} {215} (\bibinfo {year} {2020})}\BibitemShut {NoStop}%
\bibitem [{\citenamefont {Andrei}\ and\ \citenamefont {MacDonald}(2020)}]{andrei2020}%
  \BibitemOpen
  \bibfield  {author} {\bibinfo {author} {\bibfnamefont {E.~Y.}\ \bibnamefont {Andrei}}\ and\ \bibinfo {author} {\bibfnamefont {A.~H.}\ \bibnamefont {MacDonald}},\ }\bibfield  {title} {\bibinfo {title} {Graphene bilayers with a twist},\ }\href {https://doi.org/10.1038/s41563-020-00840-0} {\bibfield  {journal} {\bibinfo  {journal} {Nature Materials}\ }\textbf {\bibinfo {volume} {19}},\ \bibinfo {pages} {1265} (\bibinfo {year} {2020})}\BibitemShut {NoStop}%
\bibitem [{\citenamefont {{Mao}}\ \emph {et~al.}(2023)\citenamefont {{Mao}}, \citenamefont {{Xu}}, \citenamefont {{Li}}, \citenamefont {{Bao}}, \citenamefont {{Liu}}, \citenamefont {{Xu}}, \citenamefont {{Felser}}, \citenamefont {{Fu}},\ and\ \citenamefont {{Zhang}}}]{Zhang2023MoTe2}%
  \BibitemOpen
  \bibfield  {author} {\bibinfo {author} {\bibfnamefont {N.}~\bibnamefont {{Mao}}}, \bibinfo {author} {\bibfnamefont {C.}~\bibnamefont {{Xu}}}, \bibinfo {author} {\bibfnamefont {J.}~\bibnamefont {{Li}}}, \bibinfo {author} {\bibfnamefont {T.}~\bibnamefont {{Bao}}}, \bibinfo {author} {\bibfnamefont {P.}~\bibnamefont {{Liu}}}, \bibinfo {author} {\bibfnamefont {Y.}~\bibnamefont {{Xu}}}, \bibinfo {author} {\bibfnamefont {C.}~\bibnamefont {{Felser}}}, \bibinfo {author} {\bibfnamefont {L.}~\bibnamefont {{Fu}}},\ and\ \bibinfo {author} {\bibfnamefont {Y.}~\bibnamefont {{Zhang}}},\ }\bibfield  {title} {\bibinfo {title} {{Lattice relaxation, electronic structure and continuum model for twisted bilayer MoTe$_2$}},\ }\href {https://doi.org/10.48550/arXiv.2311.07533} {\bibfield  {journal} {\bibinfo  {journal} {arXiv e-prints}\ ,\ \bibinfo {eid} {arXiv:2311.07533}} (\bibinfo {year} {2023})},\ \Eprint {https://arxiv.org/abs/2311.07533} {arXiv:2311.07533 [cond-mat.str-el]} \BibitemShut {NoStop}%
\bibitem [{\citenamefont {Dong}\ \emph {et~al.}(2023{\natexlab{a}})\citenamefont {Dong}, \citenamefont {Wang}, \citenamefont {Ledwith}, \citenamefont {Vishwanath},\ and\ \citenamefont {Parker}}]{Dong2023CFLtMoTe2}%
  \BibitemOpen
  \bibfield  {author} {\bibinfo {author} {\bibfnamefont {J.}~\bibnamefont {Dong}}, \bibinfo {author} {\bibfnamefont {J.}~\bibnamefont {Wang}}, \bibinfo {author} {\bibfnamefont {P.~J.}\ \bibnamefont {Ledwith}}, \bibinfo {author} {\bibfnamefont {A.}~\bibnamefont {Vishwanath}},\ and\ \bibinfo {author} {\bibfnamefont {D.~E.}\ \bibnamefont {Parker}},\ }\bibfield  {title} {\bibinfo {title} {Composite fermi liquid at zero magnetic field in twisted mote $ \_2$},\ }\href@noop {} {\bibfield  {journal} {\bibinfo  {journal} {arXiv preprint arXiv:2306.01719}\ } (\bibinfo {year} {2023}{\natexlab{a}})}\BibitemShut {NoStop}%
\bibitem [{\citenamefont {{Redekop}}\ \emph {et~al.}(2024)\citenamefont {{Redekop}}, \citenamefont {{Zhang}}, \citenamefont {{Park}}, \citenamefont {{Cai}}, \citenamefont {{Anderson}}, \citenamefont {{Sheekey}}, \citenamefont {{Arp}}, \citenamefont {{Babikyan}}, \citenamefont {{Salters}}, \citenamefont {{Watanabe}}, \citenamefont {{Taniguchi}}, \citenamefont {{Xu}},\ and\ \citenamefont {{Young}}}]{Young2024MagtMoTe2}%
  \BibitemOpen
  \bibfield  {author} {\bibinfo {author} {\bibfnamefont {E.}~\bibnamefont {{Redekop}}}, \bibinfo {author} {\bibfnamefont {C.}~\bibnamefont {{Zhang}}}, \bibinfo {author} {\bibfnamefont {H.}~\bibnamefont {{Park}}}, \bibinfo {author} {\bibfnamefont {J.}~\bibnamefont {{Cai}}}, \bibinfo {author} {\bibfnamefont {E.}~\bibnamefont {{Anderson}}}, \bibinfo {author} {\bibfnamefont {O.}~\bibnamefont {{Sheekey}}}, \bibinfo {author} {\bibfnamefont {T.}~\bibnamefont {{Arp}}}, \bibinfo {author} {\bibfnamefont {G.}~\bibnamefont {{Babikyan}}}, \bibinfo {author} {\bibfnamefont {S.}~\bibnamefont {{Salters}}}, \bibinfo {author} {\bibfnamefont {K.}~\bibnamefont {{Watanabe}}}, \bibinfo {author} {\bibfnamefont {T.}~\bibnamefont {{Taniguchi}}}, \bibinfo {author} {\bibfnamefont {X.}~\bibnamefont {{Xu}}},\ and\ \bibinfo {author} {\bibfnamefont {A.~F.}\ \bibnamefont {{Young}}},\ }\bibfield  {title} {\bibinfo {title} {{Direct magnetic imaging of fractional Chern insulators in twisted MoTe$_2$ with a superconducting sensor}},\ }\href
  {https://doi.org/10.48550/arXiv.2405.10269} {\bibfield  {journal} {\bibinfo  {journal} {arXiv e-prints}\ ,\ \bibinfo {eid} {arXiv:2405.10269}} (\bibinfo {year} {2024})},\ \Eprint {https://arxiv.org/abs/2405.10269} {arXiv:2405.10269 [cond-mat.mes-hall]} \BibitemShut {NoStop}%
\bibitem [{\citenamefont {Helstrom}(1976)}]{helstrom1976}%
  \BibitemOpen
  \bibfield  {author} {\bibinfo {author} {\bibfnamefont {C.~W.}\ \bibnamefont {Helstrom}},\ }\href@noop {} {\emph {\bibinfo {title} {Quantum Detection and Estimation Theory}}},\ \bibinfo {series} {Mathematics in Science and Engineering}\ No.\ \bibinfo {number} {v. 123}\ (\bibinfo  {publisher} {Academic Press},\ \bibinfo {address} {New York},\ \bibinfo {year} {1976})\BibitemShut {NoStop}%
\bibitem [{\citenamefont {Pezz{\'e}}\ and\ \citenamefont {Smerzi}(2009)}]{pezze2009}%
  \BibitemOpen
  \bibfield  {author} {\bibinfo {author} {\bibfnamefont {L.}~\bibnamefont {Pezz{\'e}}}\ and\ \bibinfo {author} {\bibfnamefont {A.}~\bibnamefont {Smerzi}},\ }\bibfield  {title} {\bibinfo {title} {Entanglement, {{Nonlinear Dynamics}}, and the {{Heisenberg Limit}}},\ }\href {https://doi.org/10.1103/PhysRevLett.102.100401} {\bibfield  {journal} {\bibinfo  {journal} {Physical Review Letters}\ }\textbf {\bibinfo {volume} {102}},\ \bibinfo {pages} {100401} (\bibinfo {year} {2009})}\BibitemShut {NoStop}%
\bibitem [{\citenamefont {Paris}(2009)}]{paris2009b}%
  \BibitemOpen
  \bibfield  {author} {\bibinfo {author} {\bibfnamefont {M.~G.~A.}\ \bibnamefont {Paris}},\ }\bibfield  {title} {\bibinfo {title} {Quantum {{Estimation}} for {{Quantum Technology}}},\ }\href {https://doi.org/10.1142/S0219749909004839} {\bibfield  {journal} {\bibinfo  {journal} {International Journal of Quantum Information}\ }\textbf {\bibinfo {volume} {07}},\ \bibinfo {pages} {125} (\bibinfo {year} {2009})}\BibitemShut {NoStop}%
\bibitem [{\citenamefont {Bernevig}\ and\ \citenamefont {Hughes}(2013)}]{bernevig2013}%
  \BibitemOpen
  \bibfield  {author} {\bibinfo {author} {\bibfnamefont {B.~A.}\ \bibnamefont {Bernevig}}\ and\ \bibinfo {author} {\bibfnamefont {T.~L.}\ \bibnamefont {Hughes}},\ }\href@noop {} {\emph {\bibinfo {title} {Topological Insulators and Topological Superconductors}}}\ (\bibinfo  {publisher} {Princeton University Press},\ \bibinfo {address} {Princeton},\ \bibinfo {year} {2013})\BibitemShut {NoStop}%
\bibitem [{\citenamefont {Neupert}\ \emph {et~al.}(2013{\natexlab{a}})\citenamefont {Neupert}, \citenamefont {Chamon},\ and\ \citenamefont {Mudry}}]{neupert2013}%
  \BibitemOpen
  \bibfield  {author} {\bibinfo {author} {\bibfnamefont {T.}~\bibnamefont {Neupert}}, \bibinfo {author} {\bibfnamefont {C.}~\bibnamefont {Chamon}},\ and\ \bibinfo {author} {\bibfnamefont {C.}~\bibnamefont {Mudry}},\ }\bibfield  {title} {\bibinfo {title} {Measuring the quantum geometry of {{Bloch}} bands with current noise},\ }\href {https://doi.org/10.1103/PhysRevB.87.245103} {\bibfield  {journal} {\bibinfo  {journal} {Physical Review B}\ }\textbf {\bibinfo {volume} {87}},\ \bibinfo {pages} {245103} (\bibinfo {year} {2013}{\natexlab{a}})}\BibitemShut {NoStop}%
\bibitem [{\citenamefont {Fertig}\ and\ \citenamefont {Brey}(2024)}]{fertig2024}%
  \BibitemOpen
  \bibfield  {author} {\bibinfo {author} {\bibfnamefont {H.~A.}\ \bibnamefont {Fertig}}\ and\ \bibinfo {author} {\bibfnamefont {L.}~\bibnamefont {Brey}},\ }\href {https://doi.org/10.48550/arXiv.2406.12089} {\bibinfo {title} {Many-{{Body Quantum Geometric Dipole}}}} (\bibinfo {year} {2024}),\ \Eprint {https://arxiv.org/abs/2406.12089} {arXiv:2406.12089} \BibitemShut {NoStop}%
\bibitem [{\citenamefont {Blount}(1962)}]{Blount1962}%
  \BibitemOpen
  \bibfield  {author} {\bibinfo {author} {\bibfnamefont {E.}~\bibnamefont {Blount}},\ }\bibfield  {title} {\bibinfo {title} {Formalisms of band theory}\ }(\bibinfo  {publisher} {Academic Press},\ \bibinfo {year} {1962})\ pp.\ \bibinfo {pages} {305 -- 373}\BibitemShut {NoStop}%
\bibitem [{\citenamefont {Resta}(2011{\natexlab{b}})}]{Resta.Resta.2011}%
  \BibitemOpen
  \bibfield  {author} {\bibinfo {author} {\bibfnamefont {R.}~\bibnamefont {Resta}},\ }\bibfield  {title} {\bibinfo {title} {{The insulating state of matter: a geometrical theory}},\ }\href {https://doi.org/10.1140/epjb/e2010-10874-4} {\bibfield  {journal} {\bibinfo  {journal} {The European Physical Journal B}\ }\textbf {\bibinfo {volume} {79}},\ \bibinfo {pages} {121} (\bibinfo {year} {2011}{\natexlab{b}})},\ \Eprint {https://arxiv.org/abs/1012.5776} {1012.5776} \BibitemShut {NoStop}%
\bibitem [{\citenamefont {Kohn}(1964)}]{Kohn.Kohn.1964}%
  \BibitemOpen
  \bibfield  {author} {\bibinfo {author} {\bibfnamefont {W.}~\bibnamefont {Kohn}},\ }\bibfield  {title} {\bibinfo {title} {{Theory of the Insulating State}},\ }\href {https://doi.org/10.1103/physrev.133.a171} {\bibfield  {journal} {\bibinfo  {journal} {Physical Review}\ }\textbf {\bibinfo {volume} {133}},\ \bibinfo {pages} {A171} (\bibinfo {year} {1964})}\BibitemShut {NoStop}%
\bibitem [{\citenamefont {Resta}\ and\ \citenamefont {Sorella}(1999)}]{Sorella.Resta.1999}%
  \BibitemOpen
  \bibfield  {author} {\bibinfo {author} {\bibfnamefont {R.}~\bibnamefont {Resta}}\ and\ \bibinfo {author} {\bibfnamefont {S.}~\bibnamefont {Sorella}},\ }\bibfield  {title} {\bibinfo {title} {{Electron Localization in the Insulating State}},\ }\href {https://doi.org/10.1103/physrevlett.82.370} {\bibfield  {journal} {\bibinfo  {journal} {Physical Review Letters}\ }\textbf {\bibinfo {volume} {82}},\ \bibinfo {pages} {370} (\bibinfo {year} {1999})},\ \Eprint {https://arxiv.org/abs/cond-mat/9808151} {cond-mat/9808151} \BibitemShut {NoStop}%
\bibitem [{\citenamefont {Souza}\ \emph {et~al.}(1999)\citenamefont {Souza}, \citenamefont {Wilkens},\ and\ \citenamefont {Martin}}]{Martin.Souza.1999}%
  \BibitemOpen
  \bibfield  {author} {\bibinfo {author} {\bibfnamefont {I.}~\bibnamefont {Souza}}, \bibinfo {author} {\bibfnamefont {T.}~\bibnamefont {Wilkens}},\ and\ \bibinfo {author} {\bibfnamefont {R.~M.}\ \bibnamefont {Martin}},\ }\bibfield  {title} {\bibinfo {title} {{Polarization and localization in insulators: Generating function approach}},\ }\href {https://doi.org/10.1103/physrevb.62.1666} {\bibfield  {journal} {\bibinfo  {journal} {Physical Review B}\ }\textbf {\bibinfo {volume} {62}},\ \bibinfo {pages} {1666} (\bibinfo {year} {1999})},\ \Eprint {https://arxiv.org/abs/cond-mat/9911007} {cond-mat/9911007} \BibitemShut {NoStop}%
\bibitem [{\citenamefont {Marzari}\ \emph {et~al.}(2012{\natexlab{a}})\citenamefont {Marzari}, \citenamefont {Mostofi}, \citenamefont {Yates}, \citenamefont {Souza},\ and\ \citenamefont {Vanderbilt}}]{MarzariRMP2012}%
  \BibitemOpen
  \bibfield  {author} {\bibinfo {author} {\bibfnamefont {N.}~\bibnamefont {Marzari}}, \bibinfo {author} {\bibfnamefont {A.~A.}\ \bibnamefont {Mostofi}}, \bibinfo {author} {\bibfnamefont {J.~R.}\ \bibnamefont {Yates}}, \bibinfo {author} {\bibfnamefont {I.}~\bibnamefont {Souza}},\ and\ \bibinfo {author} {\bibfnamefont {D.}~\bibnamefont {Vanderbilt}},\ }\bibfield  {title} {\bibinfo {title} {Maximally localized wannier functions: Theory and applications},\ }\href {https://doi.org/10.1103/RevModPhys.84.1419} {\bibfield  {journal} {\bibinfo  {journal} {Rev. Mod. Phys.}\ }\textbf {\bibinfo {volume} {84}},\ \bibinfo {pages} {1419} (\bibinfo {year} {2012}{\natexlab{a}})}\BibitemShut {NoStop}%
\bibitem [{\citenamefont {Marzari}\ and\ \citenamefont {Vanderbilt}(1997)}]{Vanderbilt.Marzari.1997}%
  \BibitemOpen
  \bibfield  {author} {\bibinfo {author} {\bibfnamefont {N.}~\bibnamefont {Marzari}}\ and\ \bibinfo {author} {\bibfnamefont {D.}~\bibnamefont {Vanderbilt}},\ }\bibfield  {title} {\bibinfo {title} {{Maximally localized generalized Wannier functions for composite energy bands}},\ }\href {https://doi.org/10.1103/physrevb.56.12847} {\bibfield  {journal} {\bibinfo  {journal} {Physical Review B}\ }\textbf {\bibinfo {volume} {56}},\ \bibinfo {pages} {12847} (\bibinfo {year} {1997})},\ \Eprint {https://arxiv.org/abs/cond-mat/9707145} {cond-mat/9707145} \BibitemShut {NoStop}%
\bibitem [{\citenamefont {Brouder}\ \emph {et~al.}(2007{\natexlab{a}})\citenamefont {Brouder}, \citenamefont {Panati}, \citenamefont {Calandra}, \citenamefont {Mourougane},\ and\ \citenamefont {Marzari}}]{Marzari.Brouder.2007}%
  \BibitemOpen
  \bibfield  {author} {\bibinfo {author} {\bibfnamefont {C.}~\bibnamefont {Brouder}}, \bibinfo {author} {\bibfnamefont {G.}~\bibnamefont {Panati}}, \bibinfo {author} {\bibfnamefont {M.}~\bibnamefont {Calandra}}, \bibinfo {author} {\bibfnamefont {C.}~\bibnamefont {Mourougane}},\ and\ \bibinfo {author} {\bibfnamefont {N.}~\bibnamefont {Marzari}},\ }\bibfield  {title} {\bibinfo {title} {{Exponential Localization of Wannier Functions in Insulators}},\ }\href {https://doi.org/10.1103/physrevlett.98.046402} {\bibfield  {journal} {\bibinfo  {journal} {Physical Review Letters}\ }\textbf {\bibinfo {volume} {98}},\ \bibinfo {pages} {046402} (\bibinfo {year} {2007}{\natexlab{a}})},\ \Eprint {https://arxiv.org/abs/cond-mat/0606726} {cond-mat/0606726} \BibitemShut {NoStop}%
\bibitem [{\citenamefont {Scalapino}\ \emph {et~al.}(1993)\citenamefont {Scalapino}, \citenamefont {White},\ and\ \citenamefont {Zhang}}]{Scalapino1993}%
  \BibitemOpen
  \bibfield  {author} {\bibinfo {author} {\bibfnamefont {D.~J.}\ \bibnamefont {Scalapino}}, \bibinfo {author} {\bibfnamefont {S.~R.}\ \bibnamefont {White}},\ and\ \bibinfo {author} {\bibfnamefont {S.}~\bibnamefont {Zhang}},\ }\bibfield  {title} {\bibinfo {title} {Insulator, metal, or superconductor: The criteria},\ }\href {https://doi.org/10.1103/PhysRevB.47.7995} {\bibfield  {journal} {\bibinfo  {journal} {Phys. Rev. B}\ }\textbf {\bibinfo {volume} {47}},\ \bibinfo {pages} {7995} (\bibinfo {year} {1993})}\BibitemShut {NoStop}%
\bibitem [{\citenamefont {Huhtinen}\ \emph {et~al.}(2022)\citenamefont {Huhtinen}, \citenamefont {Herzog-Arbeitman}, \citenamefont {Chew}, \citenamefont {Bernevig},\ and\ \citenamefont {T\"orm\"a}}]{Huhtinen2022FlatBandSCQuantumMetric}%
  \BibitemOpen
  \bibfield  {author} {\bibinfo {author} {\bibfnamefont {K.-E.}\ \bibnamefont {Huhtinen}}, \bibinfo {author} {\bibfnamefont {J.}~\bibnamefont {Herzog-Arbeitman}}, \bibinfo {author} {\bibfnamefont {A.}~\bibnamefont {Chew}}, \bibinfo {author} {\bibfnamefont {B.~A.}\ \bibnamefont {Bernevig}},\ and\ \bibinfo {author} {\bibfnamefont {P.}~\bibnamefont {T\"orm\"a}},\ }\bibfield  {title} {\bibinfo {title} {Revisiting flat band superconductivity: Dependence on minimal quantum metric and band touchings},\ }\href {https://doi.org/10.1103/PhysRevB.106.014518} {\bibfield  {journal} {\bibinfo  {journal} {Phys. Rev. B}\ }\textbf {\bibinfo {volume} {106}},\ \bibinfo {pages} {014518} (\bibinfo {year} {2022})}\BibitemShut {NoStop}%
\bibitem [{\citenamefont {Simon}\ and\ \citenamefont {Rudner}(2020)}]{simon2020}%
  \BibitemOpen
  \bibfield  {author} {\bibinfo {author} {\bibfnamefont {S.~H.}\ \bibnamefont {Simon}}\ and\ \bibinfo {author} {\bibfnamefont {M.~S.}\ \bibnamefont {Rudner}},\ }\bibfield  {title} {\bibinfo {title} {Contrasting lattice geometry dependent versus independent quantities: {Ramifications} for {Berry} curvature, energy gaps, and dynamics},\ }\href {https://doi.org/10.1103/PhysRevB.102.165148} {\bibfield  {journal} {\bibinfo  {journal} {Phys. Rev. B}\ }\textbf {\bibinfo {volume} {102}},\ \bibinfo {pages} {165148} (\bibinfo {year} {2020})}\BibitemShut {NoStop}%
\bibitem [{\citenamefont {Tam}\ and\ \citenamefont {Peotta}(2024)}]{Tam2024GeomIndependence}%
  \BibitemOpen
  \bibfield  {author} {\bibinfo {author} {\bibfnamefont {M.}~\bibnamefont {Tam}}\ and\ \bibinfo {author} {\bibfnamefont {S.}~\bibnamefont {Peotta}},\ }\bibfield  {title} {\bibinfo {title} {Geometry-independent superfluid weight in multiorbital lattices from the generalized random phase approximation},\ }\href {https://doi.org/10.1103/PhysRevResearch.6.013256} {\bibfield  {journal} {\bibinfo  {journal} {Phys. Rev. Res.}\ }\textbf {\bibinfo {volume} {6}},\ \bibinfo {pages} {013256} (\bibinfo {year} {2024})}\BibitemShut {NoStop}%
\bibitem [{\citenamefont {Tovmasyan}\ \emph {et~al.}(2016)\citenamefont {Tovmasyan}, \citenamefont {Peotta}, \citenamefont {Törmä},\ and\ \citenamefont {Huber}}]{tovmasyan2016}%
  \BibitemOpen
  \bibfield  {author} {\bibinfo {author} {\bibfnamefont {M.}~\bibnamefont {Tovmasyan}}, \bibinfo {author} {\bibfnamefont {S.}~\bibnamefont {Peotta}}, \bibinfo {author} {\bibfnamefont {P.}~\bibnamefont {Törmä}},\ and\ \bibinfo {author} {\bibfnamefont {S.~D.}\ \bibnamefont {Huber}},\ }\bibfield  {title} {\bibinfo {title} {Effective theory and emergent {SU}(2) symmetry in the flat bands of attractive {H}ubbard models},\ }\href {https://doi.org/10.1103/PhysRevB.94.245149} {\bibfield  {journal} {\bibinfo  {journal} {Phys. Rev. B}\ }\textbf {\bibinfo {volume} {94}},\ \bibinfo {pages} {245149} (\bibinfo {year} {2016})}\BibitemShut {NoStop}%
\bibitem [{\citenamefont {Mondaini}\ \emph {et~al.}(2018)\citenamefont {Mondaini}, \citenamefont {Batrouni},\ and\ \citenamefont {Gr{\'e}maud}}]{Mondaini2018}%
  \BibitemOpen
  \bibfield  {author} {\bibinfo {author} {\bibfnamefont {R.}~\bibnamefont {Mondaini}}, \bibinfo {author} {\bibfnamefont {G.~G.}\ \bibnamefont {Batrouni}},\ and\ \bibinfo {author} {\bibfnamefont {B.}~\bibnamefont {Gr{\'e}maud}},\ }\bibfield  {title} {\bibinfo {title} {Pairing and superconductivity in the flat band: Creutz lattice},\ }\href {https://doi.org/10.1103/PhysRevB.98.155142} {\bibfield  {journal} {\bibinfo  {journal} {Physical Review B}\ }\textbf {\bibinfo {volume} {98}},\ \bibinfo {pages} {155142} (\bibinfo {year} {2018})}\BibitemShut {NoStop}%
\bibitem [{\citenamefont {Hofmann}\ \emph {et~al.}(2020)\citenamefont {Hofmann}, \citenamefont {Berg},\ and\ \citenamefont {Chowdhury}}]{Hofmann2020_SC}%
  \BibitemOpen
  \bibfield  {author} {\bibinfo {author} {\bibfnamefont {J.~S.}\ \bibnamefont {Hofmann}}, \bibinfo {author} {\bibfnamefont {E.}~\bibnamefont {Berg}},\ and\ \bibinfo {author} {\bibfnamefont {D.}~\bibnamefont {Chowdhury}},\ }\bibfield  {title} {\bibinfo {title} {Superconductivity, pseudogap, and phase separation in topological flat bands},\ }\href {https://doi.org/10.1103/PhysRevB.102.201112} {\bibfield  {journal} {\bibinfo  {journal} {Physical Review B}\ }\textbf {\bibinfo {volume} {102}},\ \bibinfo {pages} {201112} (\bibinfo {year} {2020})}\BibitemShut {NoStop}%
\bibitem [{\citenamefont {Peri}\ \emph {et~al.}(2021)\citenamefont {Peri}, \citenamefont {Song}, \citenamefont {Bernevig},\ and\ \citenamefont {Huber}}]{Peri2021TBGFragileAndSC}%
  \BibitemOpen
  \bibfield  {author} {\bibinfo {author} {\bibfnamefont {V.}~\bibnamefont {Peri}}, \bibinfo {author} {\bibfnamefont {Z.-D.}\ \bibnamefont {Song}}, \bibinfo {author} {\bibfnamefont {B.~A.}\ \bibnamefont {Bernevig}},\ and\ \bibinfo {author} {\bibfnamefont {S.~D.}\ \bibnamefont {Huber}},\ }\bibfield  {title} {\bibinfo {title} {Fragile topology and flat-band superconductivity in the strong-coupling regime},\ }\href {https://doi.org/10.1103/PhysRevLett.126.027002} {\bibfield  {journal} {\bibinfo  {journal} {Phys. Rev. Lett.}\ }\textbf {\bibinfo {volume} {126}},\ \bibinfo {pages} {027002} (\bibinfo {year} {2021})}\BibitemShut {NoStop}%
\bibitem [{\citenamefont {Herzog-Arbeitman}\ \emph {et~al.}(2022{\natexlab{b}})\citenamefont {Herzog-Arbeitman}, \citenamefont {Chew}, \citenamefont {Huhtinen}, \citenamefont {T{\"o}rm{\"a}},\ and\ \citenamefont {Bernevig}}]{Herzog2022ManyBodySCFlatBand}%
  \BibitemOpen
  \bibfield  {author} {\bibinfo {author} {\bibfnamefont {J.}~\bibnamefont {Herzog-Arbeitman}}, \bibinfo {author} {\bibfnamefont {A.}~\bibnamefont {Chew}}, \bibinfo {author} {\bibfnamefont {K.-E.}\ \bibnamefont {Huhtinen}}, \bibinfo {author} {\bibfnamefont {P.}~\bibnamefont {T{\"o}rm{\"a}}},\ and\ \bibinfo {author} {\bibfnamefont {B.~A.}\ \bibnamefont {Bernevig}},\ }\bibfield  {title} {\bibinfo {title} {Many-body superconductivity in topological flat bands},\ }\href@noop {} {\bibfield  {journal} {\bibinfo  {journal} {arXiv preprint arXiv:2209.00007}\ } (\bibinfo {year} {2022}{\natexlab{b}})}\BibitemShut {NoStop}%
\bibitem [{\citenamefont {Rossi}(2021{\natexlab{b}})}]{Rossi2021review}%
  \BibitemOpen
  \bibfield  {author} {\bibinfo {author} {\bibfnamefont {E.}~\bibnamefont {Rossi}},\ }\bibfield  {title} {\bibinfo {title} {Quantum metric and correlated states in two-dimensional systems},\ }\href {https://doi.org/10.1016/j.cossms.2021.100952} {\bibfield  {journal} {\bibinfo  {journal} {Current Opinion in Solid State and Materials Science}\ }\textbf {\bibinfo {volume} {25}},\ \bibinfo {pages} {100952} (\bibinfo {year} {2021}{\natexlab{b}})}\BibitemShut {NoStop}%
\bibitem [{\citenamefont {Peotta}\ \emph {et~al.}(2023)\citenamefont {Peotta}, \citenamefont {Huhtinen},\ and\ \citenamefont {T\"orm\"a}}]{Peotta2023review}%
  \BibitemOpen
  \bibfield  {author} {\bibinfo {author} {\bibfnamefont {S.}~\bibnamefont {Peotta}}, \bibinfo {author} {\bibfnamefont {K.-E.}\ \bibnamefont {Huhtinen}},\ and\ \bibinfo {author} {\bibfnamefont {P.}~\bibnamefont {T\"orm\"a}},\ }\href@noop {} {\bibinfo {title} {Quantum geometry in superfluidity and superconductivity}} (\bibinfo {year} {2023}),\ \Eprint {https://arxiv.org/abs/2308.08248} {arXiv:2308.08248 [cond-mat.quant-gas]} \BibitemShut {NoStop}%
\bibitem [{\citenamefont {Bouzerar}\ and\ \citenamefont {Thumin}(2024)}]{bouzerar2024hidden}%
  \BibitemOpen
  \bibfield  {author} {\bibinfo {author} {\bibfnamefont {G.}~\bibnamefont {Bouzerar}}\ and\ \bibinfo {author} {\bibfnamefont {M.}~\bibnamefont {Thumin}},\ }\href@noop {} {\bibinfo {title} {Hidden symmetry of {Bogoliubov de Gennes} quasi-particle eigenstates and universal relations in flat band superconducting bipartite lattices}} (\bibinfo {year} {2024}),\ \Eprint {https://arxiv.org/abs/2310.06589} {arXiv:2310.06589 [cond-mat.supr-con]} \BibitemShut {NoStop}%
\bibitem [{\citenamefont {Penttilä}\ \emph {et~al.}(2024)\citenamefont {Penttilä}, \citenamefont {Huhtinen},\ and\ \citenamefont {Törmä}}]{penttila2024flatband}%
  \BibitemOpen
  \bibfield  {author} {\bibinfo {author} {\bibfnamefont {R.~P.~S.}\ \bibnamefont {Penttilä}}, \bibinfo {author} {\bibfnamefont {K.-E.}\ \bibnamefont {Huhtinen}},\ and\ \bibinfo {author} {\bibfnamefont {P.}~\bibnamefont {Törmä}},\ }\href@noop {} {\bibinfo {title} {Flat-band ratio and quantum metric in the superconductivity of modified {L}ieb lattices}} (\bibinfo {year} {2024}),\ \Eprint {https://arxiv.org/abs/2404.12993} {arXiv:2404.12993 [cond-mat.supr-con]} \BibitemShut {NoStop}%
\bibitem [{\citenamefont {Jiang}\ \emph {et~al.}(2024)\citenamefont {Jiang}, \citenamefont {Törmä},\ and\ \citenamefont {Barlas}}]{jiang2024superfluidweightcrossovercritical}%
  \BibitemOpen
  \bibfield  {author} {\bibinfo {author} {\bibfnamefont {G.}~\bibnamefont {Jiang}}, \bibinfo {author} {\bibfnamefont {P.}~\bibnamefont {Törmä}},\ and\ \bibinfo {author} {\bibfnamefont {Y.}~\bibnamefont {Barlas}},\ }\href {https://arxiv.org/abs/2407.14919} {\bibinfo {title} {Superfluid weight crossover and critical temperature enhancement in singular flat bands}} (\bibinfo {year} {2024}),\ \Eprint {https://arxiv.org/abs/2407.14919} {arXiv:2407.14919 [cond-mat.supr-con]} \BibitemShut {NoStop}%
\bibitem [{\citenamefont {Cooper}(1956)}]{Cooper1956}%
  \BibitemOpen
  \bibfield  {author} {\bibinfo {author} {\bibfnamefont {L.~N.}\ \bibnamefont {Cooper}},\ }\bibfield  {title} {\bibinfo {title} {Bound {{Electron Pairs}} in a {{Degenerate Fermi Gas}}},\ }\href {https://doi.org/10.1103/PhysRev.104.1189} {\bibfield  {journal} {\bibinfo  {journal} {Physical Review}\ }\textbf {\bibinfo {volume} {104}},\ \bibinfo {pages} {1189} (\bibinfo {year} {1956})}\BibitemShut {NoStop}%
\bibitem [{\citenamefont {T\"orm\"a}\ \emph {et~al.}(2018)\citenamefont {T\"orm\"a}, \citenamefont {Liang},\ and\ \citenamefont {Peotta}}]{Torma2018SelectiveQuantumMetric}%
  \BibitemOpen
  \bibfield  {author} {\bibinfo {author} {\bibfnamefont {P.}~\bibnamefont {T\"orm\"a}}, \bibinfo {author} {\bibfnamefont {L.}~\bibnamefont {Liang}},\ and\ \bibinfo {author} {\bibfnamefont {S.}~\bibnamefont {Peotta}},\ }\bibfield  {title} {\bibinfo {title} {Quantum metric and effective mass of a two-body bound state in a flat band},\ }\href {https://doi.org/10.1103/PhysRevB.98.220511} {\bibfield  {journal} {\bibinfo  {journal} {Phys. Rev. B}\ }\textbf {\bibinfo {volume} {98}},\ \bibinfo {pages} {220511} (\bibinfo {year} {2018})}\BibitemShut {NoStop}%
\bibitem [{\citenamefont {Marzari}\ \emph {et~al.}(2012{\natexlab{b}})\citenamefont {Marzari}, \citenamefont {Mostofi}, \citenamefont {Yates}, \citenamefont {Souza},\ and\ \citenamefont {Vanderbilt}}]{marzari2011}%
  \BibitemOpen
  \bibfield  {author} {\bibinfo {author} {\bibfnamefont {N.}~\bibnamefont {Marzari}}, \bibinfo {author} {\bibfnamefont {A.~A.}\ \bibnamefont {Mostofi}}, \bibinfo {author} {\bibfnamefont {J.~R.}\ \bibnamefont {Yates}}, \bibinfo {author} {\bibfnamefont {I.}~\bibnamefont {Souza}},\ and\ \bibinfo {author} {\bibfnamefont {D.}~\bibnamefont {Vanderbilt}},\ }\bibfield  {title} {\bibinfo {title} {Maximally localized wannier functions: Theory and applications},\ }\href {https://doi.org/10.1103/RevModPhys.84.1419} {\bibfield  {journal} {\bibinfo  {journal} {Rev. Mod. Phys.}\ }\textbf {\bibinfo {volume} {84}},\ \bibinfo {pages} {1419} (\bibinfo {year} {2012}{\natexlab{b}})}\BibitemShut {NoStop}%
\bibitem [{\citenamefont {Brouder}\ \emph {et~al.}(2007{\natexlab{b}})\citenamefont {Brouder}, \citenamefont {Panati}, \citenamefont {Calandra}, \citenamefont {Mourougane},\ and\ \citenamefont {Marzari}}]{Brouder2007Wannier}%
  \BibitemOpen
  \bibfield  {author} {\bibinfo {author} {\bibfnamefont {C.}~\bibnamefont {Brouder}}, \bibinfo {author} {\bibfnamefont {G.}~\bibnamefont {Panati}}, \bibinfo {author} {\bibfnamefont {M.}~\bibnamefont {Calandra}}, \bibinfo {author} {\bibfnamefont {C.}~\bibnamefont {Mourougane}},\ and\ \bibinfo {author} {\bibfnamefont {N.}~\bibnamefont {Marzari}},\ }\bibfield  {title} {\bibinfo {title} {Exponential localization of wannier functions in insulators},\ }\href {https://doi.org/10.1103/PhysRevLett.98.046402} {\bibfield  {journal} {\bibinfo  {journal} {Phys. Rev. Lett.}\ }\textbf {\bibinfo {volume} {98}},\ \bibinfo {pages} {046402} (\bibinfo {year} {2007}{\natexlab{b}})}\BibitemShut {NoStop}%
\bibitem [{\citenamefont {Verma}\ \emph {et~al.}(2021)\citenamefont {Verma}, \citenamefont {Hazra},\ and\ \citenamefont {Randeria}}]{Verma2021FlatBandSC}%
  \BibitemOpen
  \bibfield  {author} {\bibinfo {author} {\bibfnamefont {N.}~\bibnamefont {Verma}}, \bibinfo {author} {\bibfnamefont {T.}~\bibnamefont {Hazra}},\ and\ \bibinfo {author} {\bibfnamefont {M.}~\bibnamefont {Randeria}},\ }\bibfield  {title} {\bibinfo {title} {Optical spectral weight, phase stiffness, and {$T_c$} bounds for trivial and topological flat band superconductors},\ }\href {https://doi.org/10.1073/pnas.2106744118} {\bibfield  {journal} {\bibinfo  {journal} {Proceedings of the National Academy of Sciences}\ }\textbf {\bibinfo {volume} {118}},\ \bibinfo {pages} {e2106744118} (\bibinfo {year} {2021})},\ \Eprint {https://arxiv.org/abs/https://www.pnas.org/doi/pdf/10.1073/pnas.2106744118} {https://www.pnas.org/doi/pdf/10.1073/pnas.2106744118} \BibitemShut {NoStop}%
\bibitem [{\citenamefont {Iskin}(2023{\natexlab{a}})}]{Iskin2023GL}%
  \BibitemOpen
  \bibfield  {author} {\bibinfo {author} {\bibfnamefont {M.}~\bibnamefont {Iskin}},\ }\bibfield  {title} {\bibinfo {title} {Extracting quantum-geometric effects from {Ginzburg-Landau} theory in a multiband hubbard model},\ }\href {https://doi.org/10.1103/PhysRevB.107.224505} {\bibfield  {journal} {\bibinfo  {journal} {Phys. Rev. B}\ }\textbf {\bibinfo {volume} {107}},\ \bibinfo {pages} {224505} (\bibinfo {year} {2023}{\natexlab{a}})}\BibitemShut {NoStop}%
\bibitem [{\citenamefont {Chen}\ and\ \citenamefont {Law}(2024)}]{Chen2024Ginzburg-Landau}%
  \BibitemOpen
  \bibfield  {author} {\bibinfo {author} {\bibfnamefont {S.~A.}\ \bibnamefont {Chen}}\ and\ \bibinfo {author} {\bibfnamefont {K.~T.}\ \bibnamefont {Law}},\ }\bibfield  {title} {\bibinfo {title} {{Ginzburg-Landau} theory of flat-band superconductors with quantum metric},\ }\href {https://doi.org/10.1103/PhysRevLett.132.026002} {\bibfield  {journal} {\bibinfo  {journal} {Phys. Rev. Lett.}\ }\textbf {\bibinfo {volume} {132}},\ \bibinfo {pages} {026002} (\bibinfo {year} {2024})}\BibitemShut {NoStop}%
\bibitem [{\citenamefont {Hu}\ \emph {et~al.}(2024)\citenamefont {Hu}, \citenamefont {Chen},\ and\ \citenamefont {Law}}]{hu2024anomalouscoherencelengthsuperconductors}%
  \BibitemOpen
  \bibfield  {author} {\bibinfo {author} {\bibfnamefont {J.-X.}\ \bibnamefont {Hu}}, \bibinfo {author} {\bibfnamefont {S.~A.}\ \bibnamefont {Chen}},\ and\ \bibinfo {author} {\bibfnamefont {K.~T.}\ \bibnamefont {Law}},\ }\href {https://arxiv.org/abs/2308.05686} {\bibinfo {title} {Anomalous coherence length in superconductors with quantum metric}} (\bibinfo {year} {2024}),\ \Eprint {https://arxiv.org/abs/2308.05686} {arXiv:2308.05686 [cond-mat.supr-con]} \BibitemShut {NoStop}%
\bibitem [{\citenamefont {Iskin}(2024{\natexlab{a}})}]{iskin2024coherencelengthquantumgeometry}%
  \BibitemOpen
  \bibfield  {author} {\bibinfo {author} {\bibfnamefont {M.}~\bibnamefont {Iskin}},\ }\href {https://arxiv.org/abs/2407.08449} {\bibinfo {title} {Coherence length and quantum geometry in a dilute flat-band superconductor}} (\bibinfo {year} {2024}{\natexlab{a}}),\ \Eprint {https://arxiv.org/abs/2407.08449} {arXiv:2407.08449 [cond-mat.supr-con]} \BibitemShut {NoStop}%
\bibitem [{\citenamefont {Iskin}(2024{\natexlab{b}})}]{iskin2024pairsizequantumgeometry}%
  \BibitemOpen
  \bibfield  {author} {\bibinfo {author} {\bibfnamefont {M.}~\bibnamefont {Iskin}},\ }\href {https://arxiv.org/abs/2409.14921} {\bibinfo {title} {Pair size and quantum geometry in a multiband hubbard model}} (\bibinfo {year} {2024}{\natexlab{b}}),\ \Eprint {https://arxiv.org/abs/2409.14921} {arXiv:2409.14921 [cond-mat.supr-con]} \BibitemShut {NoStop}%
\bibitem [{\citenamefont {Thumin}\ and\ \citenamefont {Bouzerar}(2024)}]{thumin2024correlation}%
  \BibitemOpen
  \bibfield  {author} {\bibinfo {author} {\bibfnamefont {M.}~\bibnamefont {Thumin}}\ and\ \bibinfo {author} {\bibfnamefont {G.}~\bibnamefont {Bouzerar}},\ }\href@noop {} {\bibinfo {title} {Correlation functions and characteristic lengthscales in flat band superconductors}} (\bibinfo {year} {2024}),\ \Eprint {https://arxiv.org/abs/2405.06215} {arXiv:2405.06215 [cond-mat.supr-con]} \BibitemShut {NoStop}%
\bibitem [{\citenamefont {Virtanen}\ \emph {et~al.}(2024)\citenamefont {Virtanen}, \citenamefont {Penttilä}, \citenamefont {Törmä}, \citenamefont {Díez-Carlón}, \citenamefont {Efetov},\ and\ \citenamefont {Heikkilä}}]{virtanen2024}%
  \BibitemOpen
  \bibfield  {author} {\bibinfo {author} {\bibfnamefont {P.}~\bibnamefont {Virtanen}}, \bibinfo {author} {\bibfnamefont {R.~P.~S.}\ \bibnamefont {Penttilä}}, \bibinfo {author} {\bibfnamefont {P.}~\bibnamefont {Törmä}}, \bibinfo {author} {\bibfnamefont {A.}~\bibnamefont {Díez-Carlón}}, \bibinfo {author} {\bibfnamefont {D.~K.}\ \bibnamefont {Efetov}},\ and\ \bibinfo {author} {\bibfnamefont {T.~T.}\ \bibnamefont {Heikkilä}},\ }\href {https://arxiv.org/abs/2410.23121} {\bibinfo {title} {Superconducting junctions with flat bands}} (\bibinfo {year} {2024}),\ \Eprint {https://arxiv.org/abs/2410.23121} {arXiv:2410.23121 [cond-mat.supr-con]} \BibitemShut {NoStop}%
\bibitem [{\citenamefont {Xu}\ \emph {et~al.}(2021)\citenamefont {Xu}, \citenamefont {Elcoro}, \citenamefont {Li}, \citenamefont {Song}, \citenamefont {Regnault}, \citenamefont {Yang}, \citenamefont {Sun}, \citenamefont {Parkin}, \citenamefont {Felser},\ and\ \citenamefont {Bernevig}}]{yuanfengOAI}%
  \BibitemOpen
  \bibfield  {author} {\bibinfo {author} {\bibfnamefont {Y.}~\bibnamefont {Xu}}, \bibinfo {author} {\bibfnamefont {L.}~\bibnamefont {Elcoro}}, \bibinfo {author} {\bibfnamefont {G.}~\bibnamefont {Li}}, \bibinfo {author} {\bibfnamefont {Z.-D.}\ \bibnamefont {Song}}, \bibinfo {author} {\bibfnamefont {N.}~\bibnamefont {Regnault}}, \bibinfo {author} {\bibfnamefont {Q.}~\bibnamefont {Yang}}, \bibinfo {author} {\bibfnamefont {Y.}~\bibnamefont {Sun}}, \bibinfo {author} {\bibfnamefont {S.}~\bibnamefont {Parkin}}, \bibinfo {author} {\bibfnamefont {C.}~\bibnamefont {Felser}},\ and\ \bibinfo {author} {\bibfnamefont {B.~A.}\ \bibnamefont {Bernevig}},\ }\href {https://arxiv.org/abs/2111.02433} {\bibinfo {title} {Three-dimensional real space invariants, obstructed atomic insulators and a new principle for active catalytic sites}} (\bibinfo {year} {2021}),\ \Eprint {https://arxiv.org/abs/2111.02433} {arXiv:2111.02433 [cond-mat.mtrl-sci]} \BibitemShut {NoStop}%
\bibitem [{\citenamefont {Lau}\ \emph {et~al.}(2022)\citenamefont {Lau}, \citenamefont {Peotta}, \citenamefont {Pikulin}, \citenamefont {Rossi},\ and\ \citenamefont {Hyart}}]{Lau2022Disorder}%
  \BibitemOpen
  \bibfield  {author} {\bibinfo {author} {\bibfnamefont {A.}~\bibnamefont {Lau}}, \bibinfo {author} {\bibfnamefont {S.}~\bibnamefont {Peotta}}, \bibinfo {author} {\bibfnamefont {D.~I.}\ \bibnamefont {Pikulin}}, \bibinfo {author} {\bibfnamefont {E.}~\bibnamefont {Rossi}},\ and\ \bibinfo {author} {\bibfnamefont {T.}~\bibnamefont {Hyart}},\ }\bibfield  {title} {\bibinfo {title} {{Universal suppression of superfluid weight by non-magnetic disorder in $s$-wave superconductors independent of quantum geometry and band dispersion}},\ }\href {https://doi.org/10.21468/SciPostPhys.13.4.086} {\bibfield  {journal} {\bibinfo  {journal} {SciPost Phys.}\ }\textbf {\bibinfo {volume} {13}},\ \bibinfo {pages} {086} (\bibinfo {year} {2022})}\BibitemShut {NoStop}%
\bibitem [{\citenamefont {Marsal}\ and\ \citenamefont {Black-Schaffer}(2024)}]{Marsal2024}%
  \BibitemOpen
  \bibfield  {author} {\bibinfo {author} {\bibfnamefont {Q.}~\bibnamefont {Marsal}}\ and\ \bibinfo {author} {\bibfnamefont {A.~M.}\ \bibnamefont {Black-Schaffer}},\ }\bibfield  {title} {\bibinfo {title} {Enhanced quantum metric due to vacancies in graphene},\ }\href {https://doi.org/10.1103/PhysRevLett.133.026002} {\bibfield  {journal} {\bibinfo  {journal} {Phys. Rev. Lett.}\ }\textbf {\bibinfo {volume} {133}},\ \bibinfo {pages} {026002} (\bibinfo {year} {2024})}\BibitemShut {NoStop}%
\bibitem [{\citenamefont {Romeral}\ \emph {et~al.}(2024)\citenamefont {Romeral}, \citenamefont {Cummings},\ and\ \citenamefont {Roche}}]{romeral2024}%
  \BibitemOpen
  \bibfield  {author} {\bibinfo {author} {\bibfnamefont {J.~M.}\ \bibnamefont {Romeral}}, \bibinfo {author} {\bibfnamefont {A.~W.}\ \bibnamefont {Cummings}},\ and\ \bibinfo {author} {\bibfnamefont {S.}~\bibnamefont {Roche}},\ }\href {https://arxiv.org/abs/2406.12677} {\bibinfo {title} {Scaling of the integrated quantum metric in disordered topological phases}} (\bibinfo {year} {2024}),\ \Eprint {https://arxiv.org/abs/2406.12677} {arXiv:2406.12677 [cond-mat.dis-nn]} \BibitemShut {NoStop}%
\bibitem [{\citenamefont {Pyykk\"onen}\ \emph {et~al.}(2023)\citenamefont {Pyykk\"onen}, \citenamefont {Peotta},\ and\ \citenamefont {T\"orm\"a}}]{Pyykkonen2023Non-Eq}%
  \BibitemOpen
  \bibfield  {author} {\bibinfo {author} {\bibfnamefont {V.~A.~J.}\ \bibnamefont {Pyykk\"onen}}, \bibinfo {author} {\bibfnamefont {S.}~\bibnamefont {Peotta}},\ and\ \bibinfo {author} {\bibfnamefont {P.}~\bibnamefont {T\"orm\"a}},\ }\bibfield  {title} {\bibinfo {title} {Suppression of nonequilibrium quasiparticle transport in flat-band superconductors},\ }\href {https://doi.org/10.1103/PhysRevLett.130.216003} {\bibfield  {journal} {\bibinfo  {journal} {Phys. Rev. Lett.}\ }\textbf {\bibinfo {volume} {130}},\ \bibinfo {pages} {216003} (\bibinfo {year} {2023})}\BibitemShut {NoStop}%
\bibitem [{\citenamefont {Hofmann}\ \emph {et~al.}(2023)\citenamefont {Hofmann}, \citenamefont {Berg},\ and\ \citenamefont {Chowdhury}}]{Hofmann2023SC_CWD}%
  \BibitemOpen
  \bibfield  {author} {\bibinfo {author} {\bibfnamefont {J.~S.}\ \bibnamefont {Hofmann}}, \bibinfo {author} {\bibfnamefont {E.}~\bibnamefont {Berg}},\ and\ \bibinfo {author} {\bibfnamefont {D.}~\bibnamefont {Chowdhury}},\ }\bibfield  {title} {\bibinfo {title} {Superconductivity, charge density wave, and supersolidity in flat bands with a tunable quantum metric},\ }\href {https://doi.org/10.1103/PhysRevLett.130.226001} {\bibfield  {journal} {\bibinfo  {journal} {Phys. Rev. Lett.}\ }\textbf {\bibinfo {volume} {130}},\ \bibinfo {pages} {226001} (\bibinfo {year} {2023})}\BibitemShut {NoStop}%
\bibitem [{\citenamefont {Jiang}\ and\ \citenamefont {Barlas}(2023)}]{Jiang2023PWD}%
  \BibitemOpen
  \bibfield  {author} {\bibinfo {author} {\bibfnamefont {G.}~\bibnamefont {Jiang}}\ and\ \bibinfo {author} {\bibfnamefont {Y.}~\bibnamefont {Barlas}},\ }\bibfield  {title} {\bibinfo {title} {Pair density waves from local band geometry},\ }\href {https://doi.org/10.1103/PhysRevLett.131.016002} {\bibfield  {journal} {\bibinfo  {journal} {Phys. Rev. Lett.}\ }\textbf {\bibinfo {volume} {131}},\ \bibinfo {pages} {016002} (\bibinfo {year} {2023})}\BibitemShut {NoStop}%
\bibitem [{\citenamefont {Chen}\ and\ \citenamefont {Huang}(2023)}]{Chen2023PWD}%
  \BibitemOpen
  \bibfield  {author} {\bibinfo {author} {\bibfnamefont {W.}~\bibnamefont {Chen}}\ and\ \bibinfo {author} {\bibfnamefont {W.}~\bibnamefont {Huang}},\ }\bibfield  {title} {\bibinfo {title} {Pair density wave facilitated by bloch quantum geometry in nearly flat band multiorbital superconductors},\ }\href {https://doi.org/10.1007s11433-023-2122-4} {\bibfield  {journal} {\bibinfo  {journal} {Science China Physics, Mechanics \& Astronomy}\ }\textbf {\bibinfo {volume} {66}},\ \bibinfo {pages} {287212} (\bibinfo {year} {2023})}\BibitemShut {NoStop}%
\bibitem [{\citenamefont {Shavit}\ and\ \citenamefont {Alicea}(2024)}]{shavit2024}%
  \BibitemOpen
  \bibfield  {author} {\bibinfo {author} {\bibfnamefont {G.}~\bibnamefont {Shavit}}\ and\ \bibinfo {author} {\bibfnamefont {J.}~\bibnamefont {Alicea}},\ }\href {https://arxiv.org/abs/2411.05071} {\bibinfo {title} {Quantum geometric unconventional superconductivity}} (\bibinfo {year} {2024}),\ \Eprint {https://arxiv.org/abs/2411.05071} {arXiv:2411.05071 [cond-mat.supr-con]} \BibitemShut {NoStop}%
\bibitem [{\citenamefont {Jahin}\ and\ \citenamefont {Lin}(2024)}]{jahin2024}%
  \BibitemOpen
  \bibfield  {author} {\bibinfo {author} {\bibfnamefont {A.}~\bibnamefont {Jahin}}\ and\ \bibinfo {author} {\bibfnamefont {S.-Z.}\ \bibnamefont {Lin}},\ }\href {https://arxiv.org/abs/2411.09664} {\bibinfo {title} {Enhanced kohn-luttinger topological superconductivity in bands with nontrivial geometry}} (\bibinfo {year} {2024}),\ \Eprint {https://arxiv.org/abs/2411.09664} {arXiv:2411.09664 [cond-mat.supr-con]} \BibitemShut {NoStop}%
\bibitem [{\citenamefont {Roddick}\ and\ \citenamefont {Stroud}(1995)}]{roddick1995}%
  \BibitemOpen
  \bibfield  {author} {\bibinfo {author} {\bibfnamefont {E.}~\bibnamefont {Roddick}}\ and\ \bibinfo {author} {\bibfnamefont {D.}~\bibnamefont {Stroud}},\ }\bibfield  {title} {\bibinfo {title} {Effect of {{Phase Fluctuations}} on the {{Low-Temperature Penetration Depth}} of {{High- T}} c {{Superconductors}}},\ }\href {https://doi.org/10.1103/PhysRevLett.74.1430} {\bibfield  {journal} {\bibinfo  {journal} {Physical Review Letters}\ }\textbf {\bibinfo {volume} {74}},\ \bibinfo {pages} {1430} (\bibinfo {year} {1995})}\BibitemShut {NoStop}%
\bibitem [{\citenamefont {Emery}\ and\ \citenamefont {Kivelson}(1995)}]{emery1995}%
  \BibitemOpen
  \bibfield  {author} {\bibinfo {author} {\bibfnamefont {V.~J.}\ \bibnamefont {Emery}}\ and\ \bibinfo {author} {\bibfnamefont {S.~A.}\ \bibnamefont {Kivelson}},\ }\bibfield  {title} {\bibinfo {title} {Superconductivity in {{Bad Metals}}},\ }\href {https://doi.org/10.1103/PhysRevLett.74.3253} {\bibfield  {journal} {\bibinfo  {journal} {Physical Review Letters}\ }\textbf {\bibinfo {volume} {74}},\ \bibinfo {pages} {3253} (\bibinfo {year} {1995})}\BibitemShut {NoStop}%
\bibitem [{\citenamefont {Xiao}\ and\ \citenamefont {Hao}(2024)}]{xiao2024}%
  \BibitemOpen
  \bibfield  {author} {\bibinfo {author} {\bibfnamefont {Y.}~\bibnamefont {Xiao}}\ and\ \bibinfo {author} {\bibfnamefont {N.}~\bibnamefont {Hao}},\ }\href {https://doi.org/10.48550/arXiv.2409.10891} {\bibinfo {title} {Quantum {{Geometric Effects}} on the {{Higgs Mode}} in {{Flat-band Superconductors}}}} (\bibinfo {year} {2024}),\ \Eprint {https://arxiv.org/abs/2409.10891} {arXiv:2409.10891} \BibitemShut {NoStop}%
\bibitem [{\citenamefont {Wu}\ and\ \citenamefont {Das~Sarma}(2020)}]{wu2020c}%
  \BibitemOpen
  \bibfield  {author} {\bibinfo {author} {\bibfnamefont {F.}~\bibnamefont {Wu}}\ and\ \bibinfo {author} {\bibfnamefont {S.}~\bibnamefont {Das~Sarma}},\ }\bibfield  {title} {\bibinfo {title} {Quantum geometry and stability of moir{\'e} flatband ferromagnetism},\ }\href {https://doi.org/10.1103/PhysRevB.102.165118} {\bibfield  {journal} {\bibinfo  {journal} {Physical Review B}\ }\textbf {\bibinfo {volume} {102}},\ \bibinfo {pages} {165118} (\bibinfo {year} {2020})}\BibitemShut {NoStop}%
\bibitem [{\citenamefont {Bernevig}\ \emph {et~al.}(2021{\natexlab{a}})\citenamefont {Bernevig}, \citenamefont {Lian}, \citenamefont {Cowsik}, \citenamefont {Xie}, \citenamefont {Regnault},\ and\ \citenamefont {Song}}]{bernevig2021f}%
  \BibitemOpen
  \bibfield  {author} {\bibinfo {author} {\bibfnamefont {B.~A.}\ \bibnamefont {Bernevig}}, \bibinfo {author} {\bibfnamefont {B.}~\bibnamefont {Lian}}, \bibinfo {author} {\bibfnamefont {A.}~\bibnamefont {Cowsik}}, \bibinfo {author} {\bibfnamefont {F.}~\bibnamefont {Xie}}, \bibinfo {author} {\bibfnamefont {N.}~\bibnamefont {Regnault}},\ and\ \bibinfo {author} {\bibfnamefont {Z.-D.}\ \bibnamefont {Song}},\ }\bibfield  {title} {\bibinfo {title} {Twisted bilayer graphene. {{V}}. {{Exact}} analytic many-body excitations in {{Coulomb Hamiltonians}}: {{Charge}} gap, {{Goldstone}} modes, and absence of {{Cooper}} pairing},\ }\href {https://doi.org/10.1103/PhysRevB.103.205415} {\bibfield  {journal} {\bibinfo  {journal} {Physical Review B}\ }\textbf {\bibinfo {volume} {103}},\ \bibinfo {pages} {205415} (\bibinfo {year} {2021}{\natexlab{a}})}\BibitemShut {NoStop}%
\bibitem [{\citenamefont {Khalaf}\ \emph {et~al.}(2021)\citenamefont {Khalaf}, \citenamefont {Chatterjee}, \citenamefont {Bultinck}, \citenamefont {Zaletel},\ and\ \citenamefont {Vishwanath}}]{khalaf2021a}%
  \BibitemOpen
  \bibfield  {author} {\bibinfo {author} {\bibfnamefont {E.}~\bibnamefont {Khalaf}}, \bibinfo {author} {\bibfnamefont {S.}~\bibnamefont {Chatterjee}}, \bibinfo {author} {\bibfnamefont {N.}~\bibnamefont {Bultinck}}, \bibinfo {author} {\bibfnamefont {M.~P.}\ \bibnamefont {Zaletel}},\ and\ \bibinfo {author} {\bibfnamefont {A.}~\bibnamefont {Vishwanath}},\ }\bibfield  {title} {\bibinfo {title} {Charged skyrmions and topological origin of superconductivity in magic-angle graphene},\ }\href {https://doi.org/10.1126/sciadv.abf5299} {\bibfield  {journal} {\bibinfo  {journal} {Science Advances}\ }\textbf {\bibinfo {volume} {7}},\ \bibinfo {pages} {eabf5299} (\bibinfo {year} {2021})}\BibitemShut {NoStop}%
\bibitem [{\citenamefont {Kitamura}\ \emph {et~al.}(2024)\citenamefont {Kitamura}, \citenamefont {Daido},\ and\ \citenamefont {Yanase}}]{kitamura2024}%
  \BibitemOpen
  \bibfield  {author} {\bibinfo {author} {\bibfnamefont {T.}~\bibnamefont {Kitamura}}, \bibinfo {author} {\bibfnamefont {A.}~\bibnamefont {Daido}},\ and\ \bibinfo {author} {\bibfnamefont {Y.}~\bibnamefont {Yanase}},\ }\bibfield  {title} {\bibinfo {title} {Spin-{{Triplet Superconductivity}} from {{Quantum-Geometry-Induced Ferromagnetic Fluctuation}}},\ }\href {https://doi.org/10.1103/PhysRevLett.132.036001} {\bibfield  {journal} {\bibinfo  {journal} {Physical Review Letters}\ }\textbf {\bibinfo {volume} {132}},\ \bibinfo {pages} {036001} (\bibinfo {year} {2024})}\BibitemShut {NoStop}%
\bibitem [{\citenamefont {Kang}\ \emph {et~al.}(2024{\natexlab{a}})\citenamefont {Kang}, \citenamefont {Oh}, \citenamefont {Lee},\ and\ \citenamefont {Yang}}]{kang2024}%
  \BibitemOpen
  \bibfield  {author} {\bibinfo {author} {\bibfnamefont {J.}~\bibnamefont {Kang}}, \bibinfo {author} {\bibfnamefont {T.}~\bibnamefont {Oh}}, \bibinfo {author} {\bibfnamefont {J.}~\bibnamefont {Lee}},\ and\ \bibinfo {author} {\bibfnamefont {B.-J.}\ \bibnamefont {Yang}},\ }\href@noop {} {\bibinfo {title} {Quantum geometric bound for saturated ferromagnetism}} (\bibinfo {year} {2024}{\natexlab{a}}),\ \Eprint {https://arxiv.org/abs/2402.07171} {arXiv:2402.07171 [cond-mat]} \BibitemShut {NoStop}%
\bibitem [{\citenamefont {{Herzog-Arbeitman}}\ \emph {et~al.}(2022)\citenamefont {{Herzog-Arbeitman}}, \citenamefont {Chew}, \citenamefont {Huhtinen}, \citenamefont {T{\"o}rm{\"a}},\ and\ \citenamefont {Bernevig}}]{herzog-arbeitman2022b}%
  \BibitemOpen
  \bibfield  {author} {\bibinfo {author} {\bibfnamefont {J.}~\bibnamefont {{Herzog-Arbeitman}}}, \bibinfo {author} {\bibfnamefont {A.}~\bibnamefont {Chew}}, \bibinfo {author} {\bibfnamefont {K.-E.}\ \bibnamefont {Huhtinen}}, \bibinfo {author} {\bibfnamefont {P.}~\bibnamefont {T{\"o}rm{\"a}}},\ and\ \bibinfo {author} {\bibfnamefont {B.~A.}\ \bibnamefont {Bernevig}},\ }\bibfield  {title} {\bibinfo {title} {Many-{{Body Superconductivity}} in {{Topological Flat Bands}}},\ }\href {10.48550/arXiv.2209.00007} {\bibfield  {journal} {\bibinfo  {journal} {{arXiv}}\ ,\ \bibinfo {pages} {2209.00007}} (\bibinfo {year} {2022})},\ \Eprint {https://arxiv.org/abs/2209.00007} {arxiv:2209.00007 [cond-mat]} \BibitemShut {NoStop}%
\bibitem [{\citenamefont {Kang}\ and\ \citenamefont {Vafek}(2018)}]{Kang2018TBGFragile}%
  \BibitemOpen
  \bibfield  {author} {\bibinfo {author} {\bibfnamefont {J.}~\bibnamefont {Kang}}\ and\ \bibinfo {author} {\bibfnamefont {O.}~\bibnamefont {Vafek}},\ }\bibfield  {title} {\bibinfo {title} {Symmetry, maximally localized wannier states, and a low-energy model for twisted bilayer graphene narrow bands},\ }\href {https://doi.org/10.1103/PhysRevX.8.031088} {\bibfield  {journal} {\bibinfo  {journal} {Phys. Rev. X}\ }\textbf {\bibinfo {volume} {8}},\ \bibinfo {pages} {031088} (\bibinfo {year} {2018})}\BibitemShut {NoStop}%
\bibitem [{\citenamefont {Song}\ \emph {et~al.}(2021)\citenamefont {Song}, \citenamefont {Lian}, \citenamefont {Regnault},\ and\ \citenamefont {Bernevig}}]{song2021b}%
  \BibitemOpen
  \bibfield  {author} {\bibinfo {author} {\bibfnamefont {Z.-D.}\ \bibnamefont {Song}}, \bibinfo {author} {\bibfnamefont {B.}~\bibnamefont {Lian}}, \bibinfo {author} {\bibfnamefont {N.}~\bibnamefont {Regnault}},\ and\ \bibinfo {author} {\bibfnamefont {B.~A.}\ \bibnamefont {Bernevig}},\ }\bibfield  {title} {\bibinfo {title} {Twisted bilayer graphene. {{II}}. {{Stable}} symmetry anomaly},\ }\href {https://doi.org/10.1103/PhysRevB.103.205412} {\bibfield  {journal} {\bibinfo  {journal} {Physical Review B}\ }\textbf {\bibinfo {volume} {103}},\ \bibinfo {pages} {205412} (\bibinfo {year} {2021})}\BibitemShut {NoStop}%
\bibitem [{\citenamefont {Huber}\ and\ \citenamefont {Altman}(2010)}]{Huber2010BEC}%
  \BibitemOpen
  \bibfield  {author} {\bibinfo {author} {\bibfnamefont {S.~D.}\ \bibnamefont {Huber}}\ and\ \bibinfo {author} {\bibfnamefont {E.}~\bibnamefont {Altman}},\ }\bibfield  {title} {\bibinfo {title} {Bose condensation in flat bands},\ }\href {https://doi.org/10.1103/PhysRevB.82.184502} {\bibfield  {journal} {\bibinfo  {journal} {Phys. Rev. B}\ }\textbf {\bibinfo {volume} {82}},\ \bibinfo {pages} {184502} (\bibinfo {year} {2010})}\BibitemShut {NoStop}%
\bibitem [{\citenamefont {You}\ \emph {et~al.}(2012)\citenamefont {You}, \citenamefont {Chen}, \citenamefont {Sun},\ and\ \citenamefont {Zhai}}]{You2012_BECkagome}%
  \BibitemOpen
  \bibfield  {author} {\bibinfo {author} {\bibfnamefont {Y.-Z.}\ \bibnamefont {You}}, \bibinfo {author} {\bibfnamefont {Z.}~\bibnamefont {Chen}}, \bibinfo {author} {\bibfnamefont {X.-Q.}\ \bibnamefont {Sun}},\ and\ \bibinfo {author} {\bibfnamefont {H.}~\bibnamefont {Zhai}},\ }\bibfield  {title} {\bibinfo {title} {Superfluidity of bosons in kagome lattices with frustration},\ }\href {https://doi.org/10.1103/PhysRevLett.109.265302} {\bibfield  {journal} {\bibinfo  {journal} {Phys. Rev. Lett.}\ }\textbf {\bibinfo {volume} {109}},\ \bibinfo {pages} {265302} (\bibinfo {year} {2012})}\BibitemShut {NoStop}%
\bibitem [{\citenamefont {Julku}\ \emph {et~al.}(2021{\natexlab{a}})\citenamefont {Julku}, \citenamefont {Bruun},\ and\ \citenamefont {T\"orm\"a}}]{Torma2021FlatBandBEC}%
  \BibitemOpen
  \bibfield  {author} {\bibinfo {author} {\bibfnamefont {A.}~\bibnamefont {Julku}}, \bibinfo {author} {\bibfnamefont {G.~M.}\ \bibnamefont {Bruun}},\ and\ \bibinfo {author} {\bibfnamefont {P.}~\bibnamefont {T\"orm\"a}},\ }\bibfield  {title} {\bibinfo {title} {Quantum geometry and flat band {Bose-Einstein} condensation},\ }\href {https://doi.org/10.1103/PhysRevLett.127.170404} {\bibfield  {journal} {\bibinfo  {journal} {Phys. Rev. Lett.}\ }\textbf {\bibinfo {volume} {127}},\ \bibinfo {pages} {170404} (\bibinfo {year} {2021}{\natexlab{a}})}\BibitemShut {NoStop}%
\bibitem [{\citenamefont {Julku}\ \emph {et~al.}(2023)\citenamefont {Julku}, \citenamefont {Salerno},\ and\ \citenamefont {Törmä}}]{Julku2023BECrevisited}%
  \BibitemOpen
  \bibfield  {author} {\bibinfo {author} {\bibfnamefont {A.}~\bibnamefont {Julku}}, \bibinfo {author} {\bibfnamefont {G.}~\bibnamefont {Salerno}},\ and\ \bibinfo {author} {\bibfnamefont {P.}~\bibnamefont {Törmä}},\ }\bibfield  {title} {\bibinfo {title} {{Superfluidity of flat band {Bose–Einstein} condensates revisited}},\ }\href {https://doi.org/10.1063/10.0019426} {\bibfield  {journal} {\bibinfo  {journal} {Low Temperature Physics}\ }\textbf {\bibinfo {volume} {49}},\ \bibinfo {pages} {701} (\bibinfo {year} {2023})}\BibitemShut {NoStop}%
\bibitem [{\citenamefont {Julku}\ \emph {et~al.}(2021{\natexlab{b}})\citenamefont {Julku}, \citenamefont {Bruun},\ and\ \citenamefont {T\"orm\"a}}]{Julku2021BECExcitations}%
  \BibitemOpen
  \bibfield  {author} {\bibinfo {author} {\bibfnamefont {A.}~\bibnamefont {Julku}}, \bibinfo {author} {\bibfnamefont {G.~M.}\ \bibnamefont {Bruun}},\ and\ \bibinfo {author} {\bibfnamefont {P.}~\bibnamefont {T\"orm\"a}},\ }\bibfield  {title} {\bibinfo {title} {Excitations of a {Bose-Einstein} condensate and the quantum geometry of a flat band},\ }\href {https://doi.org/10.1103/PhysRevB.104.144507} {\bibfield  {journal} {\bibinfo  {journal} {Phys. Rev. B}\ }\textbf {\bibinfo {volume} {104}},\ \bibinfo {pages} {144507} (\bibinfo {year} {2021}{\natexlab{b}})}\BibitemShut {NoStop}%
\bibitem [{\citenamefont {Iskin}(2023{\natexlab{b}})}]{Iskin2023Bogoliubov}%
  \BibitemOpen
  \bibfield  {author} {\bibinfo {author} {\bibfnamefont {M.}~\bibnamefont {Iskin}},\ }\bibfield  {title} {\bibinfo {title} {Quantum-geometric contribution to the {Bogoliubov} modes in a two-band {Bose-Einstein} condensate},\ }\href {https://doi.org/10.1103/PhysRevA.107.023313} {\bibfield  {journal} {\bibinfo  {journal} {Phys. Rev. A}\ }\textbf {\bibinfo {volume} {107}},\ \bibinfo {pages} {023313} (\bibinfo {year} {2023}{\natexlab{b}})}\BibitemShut {NoStop}%
\bibitem [{\citenamefont {Lukin}\ \emph {et~al.}(2023)\citenamefont {Lukin}, \citenamefont {Sotnikov},\ and\ \citenamefont {Kruchkov}}]{lukin2023unconventional}%
  \BibitemOpen
  \bibfield  {author} {\bibinfo {author} {\bibfnamefont {I.}~\bibnamefont {Lukin}}, \bibinfo {author} {\bibfnamefont {A.}~\bibnamefont {Sotnikov}},\ and\ \bibinfo {author} {\bibfnamefont {A.}~\bibnamefont {Kruchkov}},\ }\href@noop {} {\bibinfo {title} {Unconventional superfluidity and quantum geometry of topological bosons}} (\bibinfo {year} {2023}),\ \Eprint {https://arxiv.org/abs/2307.08748} {arXiv:2307.08748 [cond-mat.quant-gas]} \BibitemShut {NoStop}%
\bibitem [{\citenamefont {Keldysh}\ and\ \citenamefont {Kopaev}(1965)}]{keldysh1965}%
  \BibitemOpen
  \bibfield  {author} {\bibinfo {author} {\bibfnamefont {L.~V.}\ \bibnamefont {Keldysh}}\ and\ \bibinfo {author} {\bibfnamefont {Y.~V.}\ \bibnamefont {Kopaev}},\ }\bibfield  {title} {\bibinfo {title} {Possible instability of semimetallic state toward coulomb interaction},\ }\href@noop {} {\bibfield  {journal} {\bibinfo  {journal} {Soviet Phys. Solid State,ussr}\ }\textbf {\bibinfo {volume} {6}},\ \bibinfo {pages} {2219} (\bibinfo {year} {1965})}\BibitemShut {NoStop}%
\bibitem [{\citenamefont {Halperin}\ and\ \citenamefont {Rice}(1968)}]{halperin1968}%
  \BibitemOpen
  \bibfield  {author} {\bibinfo {author} {\bibfnamefont {B.~I.}\ \bibnamefont {Halperin}}\ and\ \bibinfo {author} {\bibfnamefont {T.~M.}\ \bibnamefont {Rice}},\ }\bibfield  {title} {\bibinfo {title} {The {{Excitonic State}} at the {{Semiconductor-Semimetal Transition}}*},\ }in\ \href {https://doi.org/10.1016/S0081-1947(08)60740-7} {\emph {\bibinfo {booktitle} {Solid {{State Physics}}}}},\ Vol.~\bibinfo {volume} {21},\ \bibinfo {editor} {edited by\ \bibinfo {editor} {\bibfnamefont {F.}~\bibnamefont {Seitz}}, \bibinfo {editor} {\bibfnamefont {D.}~\bibnamefont {Turnbull}},\ and\ \bibinfo {editor} {\bibfnamefont {H.}~\bibnamefont {Ehrenreich}}}\ (\bibinfo  {publisher} {Academic Press},\ \bibinfo {year} {1968})\ pp.\ \bibinfo {pages} {115--192}\BibitemShut {NoStop}%
\bibitem [{\citenamefont {Lozovik}\ and\ \citenamefont {Yudson}(1975)}]{lozovik1975}%
  \BibitemOpen
  \bibfield  {author} {\bibinfo {author} {\bibfnamefont {Y.~E.}\ \bibnamefont {Lozovik}}\ and\ \bibinfo {author} {\bibfnamefont {V.~I.}\ \bibnamefont {Yudson}},\ }\bibfield  {title} {\bibinfo {title} {Feasibility of superfluidity of paired spatially separated electrons and holes - new superconductivity mechanism},\ }\href@noop {} {\bibfield  {journal} {\bibinfo  {journal} {Jetp Lett.}\ }\textbf {\bibinfo {volume} {22}},\ \bibinfo {pages} {274} (\bibinfo {year} {1975})}\BibitemShut {NoStop}%
\bibitem [{\citenamefont {Eisenstein}\ and\ \citenamefont {MacDonald}(2004)}]{eisenstein2004}%
  \BibitemOpen
  \bibfield  {author} {\bibinfo {author} {\bibfnamefont {J.~P.}\ \bibnamefont {Eisenstein}}\ and\ \bibinfo {author} {\bibfnamefont {A.~H.}\ \bibnamefont {MacDonald}},\ }\bibfield  {title} {\bibinfo {title} {Bose-{{Einstein}} condensation of excitons in bilayer electron systems},\ }\href@noop {} {\bibfield  {journal} {\bibinfo  {journal} {Nature}\ }\textbf {\bibinfo {volume} {432}},\ \bibinfo {pages} {691} (\bibinfo {year} {2004})}\BibitemShut {NoStop}%
\bibitem [{\citenamefont {Neilson}\ \emph {et~al.}(2014)\citenamefont {Neilson}, \citenamefont {Perali},\ and\ \citenamefont {Hamilton}}]{neilson2014}%
  \BibitemOpen
  \bibfield  {author} {\bibinfo {author} {\bibfnamefont {D.}~\bibnamefont {Neilson}}, \bibinfo {author} {\bibfnamefont {A.}~\bibnamefont {Perali}},\ and\ \bibinfo {author} {\bibfnamefont {A.~R.}\ \bibnamefont {Hamilton}},\ }\bibfield  {title} {\bibinfo {title} {Excitonic superfluidity and screening in electron-hole bilayer systems},\ }\href {https://doi.org/10.1103/PhysRevB.89.060502} {\bibfield  {journal} {\bibinfo  {journal} {Physical Review B}\ }\textbf {\bibinfo {volume} {89}},\ \bibinfo {pages} {060502} (\bibinfo {year} {2014})}\BibitemShut {NoStop}%
\bibitem [{\citenamefont {Lu}\ \emph {et~al.}(2016)\citenamefont {Lu}, \citenamefont {{Rodriguez-Vega}}, \citenamefont {Li}, \citenamefont {{Luican-Mayer}}, \citenamefont {Watanabe}, \citenamefont {Taniguchi}, \citenamefont {Rossi},\ and\ \citenamefont {Andrei}}]{lu2016}%
  \BibitemOpen
  \bibfield  {author} {\bibinfo {author} {\bibfnamefont {C.-P.}\ \bibnamefont {Lu}}, \bibinfo {author} {\bibfnamefont {M.}~\bibnamefont {{Rodriguez-Vega}}}, \bibinfo {author} {\bibfnamefont {G.}~\bibnamefont {Li}}, \bibinfo {author} {\bibfnamefont {A.}~\bibnamefont {{Luican-Mayer}}}, \bibinfo {author} {\bibfnamefont {K.}~\bibnamefont {Watanabe}}, \bibinfo {author} {\bibfnamefont {T.}~\bibnamefont {Taniguchi}}, \bibinfo {author} {\bibfnamefont {E.}~\bibnamefont {Rossi}},\ and\ \bibinfo {author} {\bibfnamefont {E.~Y.}\ \bibnamefont {Andrei}},\ }\bibfield  {title} {\bibinfo {title} {Local, global, and nonlinear screening in twisted double-layer graphene},\ }\href {https://doi.org/10.1073/pnas.1606278113} {\bibfield  {journal} {\bibinfo  {journal} {Proceedings of the National Academy of Sciences}\ }\textbf {\bibinfo {volume} {113}},\ \bibinfo {pages} {6623} (\bibinfo {year} {2016})}\BibitemShut {NoStop}%
\bibitem [{\citenamefont {Hu}\ \emph {et~al.}(2022)\citenamefont {Hu}, \citenamefont {Hyart}, \citenamefont {Pikulin},\ and\ \citenamefont {Rossi}}]{hu2022}%
  \BibitemOpen
  \bibfield  {author} {\bibinfo {author} {\bibfnamefont {X.}~\bibnamefont {Hu}}, \bibinfo {author} {\bibfnamefont {T.}~\bibnamefont {Hyart}}, \bibinfo {author} {\bibfnamefont {D.~I.}\ \bibnamefont {Pikulin}},\ and\ \bibinfo {author} {\bibfnamefont {E.}~\bibnamefont {Rossi}},\ }\bibfield  {title} {\bibinfo {title} {Quantum-metric-enabled exciton condensate in double twisted bilayer graphene},\ }\href {https://doi.org/10.1103/PhysRevB.105.L140506} {\bibfield  {journal} {\bibinfo  {journal} {Physical Review B}\ }\textbf {\bibinfo {volume} {105}},\ \bibinfo {pages} {L140506} (\bibinfo {year} {2022})}\BibitemShut {NoStop}%
\bibitem [{\citenamefont {Verma}\ \emph {et~al.}(2024)\citenamefont {Verma}, \citenamefont {Guerci},\ and\ \citenamefont {Queiroz}}]{verma2023}%
  \BibitemOpen
  \bibfield  {author} {\bibinfo {author} {\bibfnamefont {N.}~\bibnamefont {Verma}}, \bibinfo {author} {\bibfnamefont {D.}~\bibnamefont {Guerci}},\ and\ \bibinfo {author} {\bibfnamefont {R.}~\bibnamefont {Queiroz}},\ }\bibfield  {title} {\bibinfo {title} {{Geometric Stiffness in Interlayer Exciton Condensates}},\ }\href {https://doi.org/10.1103/physrevlett.132.236001} {\bibfield  {journal} {\bibinfo  {journal} {Physical Review Letters}\ }\textbf {\bibinfo {volume} {132}},\ \bibinfo {pages} {236001} (\bibinfo {year} {2024})},\ \Eprint {https://arxiv.org/abs/2307.01253} {2307.01253} \BibitemShut {NoStop}%
\bibitem [{\citenamefont {Bardeen}\ \emph {et~al.}(1957)\citenamefont {Bardeen}, \citenamefont {Cooper},\ and\ \citenamefont {Schrieffer}}]{BCS1957SC}%
  \BibitemOpen
  \bibfield  {author} {\bibinfo {author} {\bibfnamefont {J.}~\bibnamefont {Bardeen}}, \bibinfo {author} {\bibfnamefont {L.~N.}\ \bibnamefont {Cooper}},\ and\ \bibinfo {author} {\bibfnamefont {J.~R.}\ \bibnamefont {Schrieffer}},\ }\bibfield  {title} {\bibinfo {title} {Theory of superconductivity},\ }\href {https://doi.org/10.1103/PhysRev.108.1175} {\bibfield  {journal} {\bibinfo  {journal} {Phys. Rev.}\ }\textbf {\bibinfo {volume} {108}},\ \bibinfo {pages} {1175} (\bibinfo {year} {1957})}\BibitemShut {NoStop}%
\bibitem [{\citenamefont {Migdal}(1958)}]{Migdal1958EPC}%
  \BibitemOpen
  \bibfield  {author} {\bibinfo {author} {\bibfnamefont {A.}~\bibnamefont {Migdal}},\ }\bibfield  {title} {\bibinfo {title} {Interaction between electrons and lattice vibrations in a normal metal},\ }\href@noop {} {\bibfield  {journal} {\bibinfo  {journal} {Sov. Phys. JETP}\ }\textbf {\bibinfo {volume} {7}},\ \bibinfo {pages} {996} (\bibinfo {year} {1958})}\BibitemShut {NoStop}%
\bibitem [{\citenamefont {Eliashberg}(1960)}]{Eliashberg1960EPCSC}%
  \BibitemOpen
  \bibfield  {author} {\bibinfo {author} {\bibfnamefont {G.}~\bibnamefont {Eliashberg}},\ }\bibfield  {title} {\bibinfo {title} {Interactions between electrons and lattice vibrations in a superconductor},\ }\href@noop {} {\bibfield  {journal} {\bibinfo  {journal} {Sov. Phys. JETP}\ }\textbf {\bibinfo {volume} {11}},\ \bibinfo {pages} {696} (\bibinfo {year} {1960})}\BibitemShut {NoStop}%
\bibitem [{\citenamefont {Bradlyn}\ \emph {et~al.}(2017)\citenamefont {Bradlyn}, \citenamefont {Elcoro}, \citenamefont {Cano}, \citenamefont {Vergniory}, \citenamefont {Wang}, \citenamefont {Felser}, \citenamefont {Aroyo},\ and\ \citenamefont {Bernevig}}]{Bradlyn2017TQC}%
  \BibitemOpen
  \bibfield  {author} {\bibinfo {author} {\bibfnamefont {B.}~\bibnamefont {Bradlyn}}, \bibinfo {author} {\bibfnamefont {L.}~\bibnamefont {Elcoro}}, \bibinfo {author} {\bibfnamefont {J.}~\bibnamefont {Cano}}, \bibinfo {author} {\bibfnamefont {M.}~\bibnamefont {Vergniory}}, \bibinfo {author} {\bibfnamefont {Z.}~\bibnamefont {Wang}}, \bibinfo {author} {\bibfnamefont {C.}~\bibnamefont {Felser}}, \bibinfo {author} {\bibfnamefont {M.}~\bibnamefont {Aroyo}},\ and\ \bibinfo {author} {\bibfnamefont {B.~A.}\ \bibnamefont {Bernevig}},\ }\bibfield  {title} {\bibinfo {title} {Topological quantum chemistry},\ }\href {https://doi.org/10.1038/nature23268} {\bibfield  {journal} {\bibinfo  {journal} {Nature}\ }\textbf {\bibinfo {volume} {547}},\ \bibinfo {pages} {298} (\bibinfo {year} {2017})}\BibitemShut {NoStop}%
\bibitem [{\citenamefont {Po}\ \emph {et~al.}(2017)\citenamefont {Po}, \citenamefont {Vishwanath},\ and\ \citenamefont {Watanabe}}]{Po2017SymIndi}%
  \BibitemOpen
  \bibfield  {author} {\bibinfo {author} {\bibfnamefont {H.~C.}\ \bibnamefont {Po}}, \bibinfo {author} {\bibfnamefont {A.}~\bibnamefont {Vishwanath}},\ and\ \bibinfo {author} {\bibfnamefont {H.}~\bibnamefont {Watanabe}},\ }\bibfield  {title} {\bibinfo {title} {Symmetry-based indicators of band topology in the 230 space groups},\ }\href {https://doi.org/10.1038/s41467-017-00133-2} {\bibfield  {journal} {\bibinfo  {journal} {Nature communications}\ }\textbf {\bibinfo {volume} {8}},\ \bibinfo {pages} {50} (\bibinfo {year} {2017})}\BibitemShut {NoStop}%
\bibitem [{\citenamefont {Vergniory}\ \emph {et~al.}(2019)\citenamefont {Vergniory}, \citenamefont {Elcoro}, \citenamefont {Felser}, \citenamefont {Regnault}, \citenamefont {Bernevig},\ and\ \citenamefont {Wang}}]{Bernevig2019TopoMat}%
  \BibitemOpen
  \bibfield  {author} {\bibinfo {author} {\bibfnamefont {M.~G.}\ \bibnamefont {Vergniory}}, \bibinfo {author} {\bibfnamefont {L.}~\bibnamefont {Elcoro}}, \bibinfo {author} {\bibfnamefont {C.}~\bibnamefont {Felser}}, \bibinfo {author} {\bibfnamefont {N.}~\bibnamefont {Regnault}}, \bibinfo {author} {\bibfnamefont {B.~A.}\ \bibnamefont {Bernevig}},\ and\ \bibinfo {author} {\bibfnamefont {Z.}~\bibnamefont {Wang}},\ }\bibfield  {title} {\bibinfo {title} {A complete catalogue of high-quality topological materials},\ }\href {https://doi.org/10.1038/s41586-019-0954-4} {\bibfield  {journal} {\bibinfo  {journal} {Nature}\ }\textbf {\bibinfo {volume} {566}},\ \bibinfo {pages} {480} (\bibinfo {year} {2019})}\BibitemShut {NoStop}%
\bibitem [{\citenamefont {Zhang}\ \emph {et~al.}(2019)\citenamefont {Zhang}, \citenamefont {Jiang}, \citenamefont {Song}, \citenamefont {Huang}, \citenamefont {He}, \citenamefont {Fang}, \citenamefont {Weng},\ and\ \citenamefont {Fang}}]{Fang2019TopoMat}%
  \BibitemOpen
  \bibfield  {author} {\bibinfo {author} {\bibfnamefont {T.}~\bibnamefont {Zhang}}, \bibinfo {author} {\bibfnamefont {Y.}~\bibnamefont {Jiang}}, \bibinfo {author} {\bibfnamefont {Z.}~\bibnamefont {Song}}, \bibinfo {author} {\bibfnamefont {H.}~\bibnamefont {Huang}}, \bibinfo {author} {\bibfnamefont {Y.}~\bibnamefont {He}}, \bibinfo {author} {\bibfnamefont {Z.}~\bibnamefont {Fang}}, \bibinfo {author} {\bibfnamefont {H.}~\bibnamefont {Weng}},\ and\ \bibinfo {author} {\bibfnamefont {C.}~\bibnamefont {Fang}},\ }\bibfield  {title} {\bibinfo {title} {Catalogue of topological electronic materials},\ }\href {https://doi.org/10.1038/s41586-019-0944-6} {\bibfield  {journal} {\bibinfo  {journal} {Nature}\ }\textbf {\bibinfo {volume} {566}},\ \bibinfo {pages} {475} (\bibinfo {year} {2019})}\BibitemShut {NoStop}%
\bibitem [{\citenamefont {Tang}\ \emph {et~al.}(2019)\citenamefont {Tang}, \citenamefont {Po}, \citenamefont {Vishwanath},\ and\ \citenamefont {Wan}}]{Wan2019TopoMat}%
  \BibitemOpen
  \bibfield  {author} {\bibinfo {author} {\bibfnamefont {F.}~\bibnamefont {Tang}}, \bibinfo {author} {\bibfnamefont {H.~C.}\ \bibnamefont {Po}}, \bibinfo {author} {\bibfnamefont {A.}~\bibnamefont {Vishwanath}},\ and\ \bibinfo {author} {\bibfnamefont {X.}~\bibnamefont {Wan}},\ }\bibfield  {title} {\bibinfo {title} {Comprehensive search for topological materials using symmetry indicators},\ }\href {https://doi.org/10.1038/s41586-019-0937-5} {\bibfield  {journal} {\bibinfo  {journal} {Nature}\ }\textbf {\bibinfo {volume} {566}},\ \bibinfo {pages} {486} (\bibinfo {year} {2019})}\BibitemShut {NoStop}%
\bibitem [{\citenamefont {Xu}\ \emph {et~al.}(2020)\citenamefont {Xu}, \citenamefont {Elcoro}, \citenamefont {Song}, \citenamefont {Wieder}, \citenamefont {Vergniory}, \citenamefont {Regnault}, \citenamefont {Chen}, \citenamefont {Felser},\ and\ \citenamefont {Bernevig}}]{Bernevig2020MTQCMat}%
  \BibitemOpen
  \bibfield  {author} {\bibinfo {author} {\bibfnamefont {Y.}~\bibnamefont {Xu}}, \bibinfo {author} {\bibfnamefont {L.}~\bibnamefont {Elcoro}}, \bibinfo {author} {\bibfnamefont {Z.}~\bibnamefont {Song}}, \bibinfo {author} {\bibfnamefont {B.~J.}\ \bibnamefont {Wieder}}, \bibinfo {author} {\bibfnamefont {M.}~\bibnamefont {Vergniory}}, \bibinfo {author} {\bibfnamefont {N.}~\bibnamefont {Regnault}}, \bibinfo {author} {\bibfnamefont {Y.}~\bibnamefont {Chen}}, \bibinfo {author} {\bibfnamefont {C.}~\bibnamefont {Felser}},\ and\ \bibinfo {author} {\bibfnamefont {B.~A.}\ \bibnamefont {Bernevig}},\ }\bibfield  {title} {\bibinfo {title} {High-throughput calculations of antiferromagnetic topological materials from magnetic topological quantum chemistry},\ }\href {https://arxiv.org/abs/2003.00012} {\bibfield  {journal} {\bibinfo  {journal} {arXiv:2003.00012}\ } (\bibinfo {year} {2020})}\BibitemShut {NoStop}%
\bibitem [{\citenamefont {Narang}\ \emph {et~al.}(2021)\citenamefont {Narang}, \citenamefont {Garcia},\ and\ \citenamefont {Felser}}]{Narang2021TopologyBands}%
  \BibitemOpen
  \bibfield  {author} {\bibinfo {author} {\bibfnamefont {P.}~\bibnamefont {Narang}}, \bibinfo {author} {\bibfnamefont {C.~A.~C.}\ \bibnamefont {Garcia}},\ and\ \bibinfo {author} {\bibfnamefont {C.}~\bibnamefont {Felser}},\ }\bibfield  {title} {\bibinfo {title} {The topology of electronic band structures},\ }\href {https://doi.org/10.1038/s41563-020-00820-4} {\bibfield  {journal} {\bibinfo  {journal} {Nature Materials}\ }\textbf {\bibinfo {volume} {20}},\ \bibinfo {pages} {293} (\bibinfo {year} {2021})}\BibitemShut {NoStop}%
\bibitem [{\citenamefont {Yu}\ \emph {et~al.}(2023{\natexlab{b}})\citenamefont {Yu}, \citenamefont {Ciccarino}, \citenamefont {Bianco}, \citenamefont {Errea}, \citenamefont {Narang},\ and\ \citenamefont {Bernevig}}]{Yu05032023GeometryEPC}%
  \BibitemOpen
  \bibfield  {author} {\bibinfo {author} {\bibfnamefont {J.}~\bibnamefont {Yu}}, \bibinfo {author} {\bibfnamefont {C.~J.}\ \bibnamefont {Ciccarino}}, \bibinfo {author} {\bibfnamefont {R.}~\bibnamefont {Bianco}}, \bibinfo {author} {\bibfnamefont {I.}~\bibnamefont {Errea}}, \bibinfo {author} {\bibfnamefont {P.}~\bibnamefont {Narang}},\ and\ \bibinfo {author} {\bibfnamefont {B.~A.}\ \bibnamefont {Bernevig}},\ }\bibfield  {title} {\bibinfo {title} {Nontrivial quantum geometry and the strength of electron-phonon coupling},\ }\href {https://arxiv.org/abs/2305.02340} {\bibfield  {journal} {\bibinfo  {journal} {arXiv:2305.02340}\ } (\bibinfo {year} {2023}{\natexlab{b}})}\BibitemShut {NoStop}%
\bibitem [{\citenamefont {Zhu}\ and\ \citenamefont {Alexandradinata}(2024)}]{Aris2024EPCPhotovaltaicCurrent}%
  \BibitemOpen
  \bibfield  {author} {\bibinfo {author} {\bibfnamefont {P.}~\bibnamefont {Zhu}}\ and\ \bibinfo {author} {\bibfnamefont {A.}~\bibnamefont {Alexandradinata}},\ }\bibfield  {title} {\bibinfo {title} {Anomalous shift and optical vorticity in the steady photovoltaic current},\ }\href {https://doi.org/10.1103/PhysRevB.110.115108} {\bibfield  {journal} {\bibinfo  {journal} {Phys. Rev. B}\ }\textbf {\bibinfo {volume} {110}},\ \bibinfo {pages} {115108} (\bibinfo {year} {2024})}\BibitemShut {NoStop}%
\bibitem [{\citenamefont {Neupert}\ \emph {et~al.}(2011)\citenamefont {Neupert}, \citenamefont {Santos}, \citenamefont {Chamon},\ and\ \citenamefont {Mudry}}]{neupert}%
  \BibitemOpen
  \bibfield  {author} {\bibinfo {author} {\bibfnamefont {T.}~\bibnamefont {Neupert}}, \bibinfo {author} {\bibfnamefont {L.}~\bibnamefont {Santos}}, \bibinfo {author} {\bibfnamefont {C.}~\bibnamefont {Chamon}},\ and\ \bibinfo {author} {\bibfnamefont {C.}~\bibnamefont {Mudry}},\ }\bibfield  {title} {\bibinfo {title} {Fractional quantum hall states at zero magnetic field},\ }\href {https://doi.org/10.1103/PhysRevLett.106.236804} {\bibfield  {journal} {\bibinfo  {journal} {Phys. Rev. Lett.}\ }\textbf {\bibinfo {volume} {106}},\ \bibinfo {pages} {236804} (\bibinfo {year} {2011})}\BibitemShut {NoStop}%
\bibitem [{\citenamefont {{Sheng}}\ \emph {et~al.}(2011)\citenamefont {{Sheng}}, \citenamefont {{Gu}}, \citenamefont {{Sun}},\ and\ \citenamefont {{Sheng}}}]{sheng}%
  \BibitemOpen
  \bibfield  {author} {\bibinfo {author} {\bibfnamefont {D.~N.}\ \bibnamefont {{Sheng}}}, \bibinfo {author} {\bibfnamefont {Z.-C.}\ \bibnamefont {{Gu}}}, \bibinfo {author} {\bibfnamefont {K.}~\bibnamefont {{Sun}}},\ and\ \bibinfo {author} {\bibfnamefont {L.}~\bibnamefont {{Sheng}}},\ }\bibfield  {title} {\bibinfo {title} {{Fractional quantum Hall effect in the absence of Landau levels}},\ }\href {https://doi.org/10.1038/ncomms1380} {\bibfield  {journal} {\bibinfo  {journal} {Nature Communications}\ }\textbf {\bibinfo {volume} {2}},\ \bibinfo {eid} {389} (\bibinfo {year} {2011})}\BibitemShut {NoStop}%
\bibitem [{\citenamefont {Regnault}\ and\ \citenamefont {Bernevig}(2011{\natexlab{a}})}]{regnault}%
  \BibitemOpen
  \bibfield  {author} {\bibinfo {author} {\bibfnamefont {N.}~\bibnamefont {Regnault}}\ and\ \bibinfo {author} {\bibfnamefont {B.~A.}\ \bibnamefont {Bernevig}},\ }\bibfield  {title} {\bibinfo {title} {Fractional chern insulator},\ }\href {https://doi.org/10.1103/PhysRevX.1.021014} {\bibfield  {journal} {\bibinfo  {journal} {Phys. Rev. X}\ }\textbf {\bibinfo {volume} {1}},\ \bibinfo {pages} {021014} (\bibinfo {year} {2011}{\natexlab{a}})}\BibitemShut {NoStop}%
\bibitem [{\citenamefont {Sun}\ \emph {et~al.}(2011)\citenamefont {Sun}, \citenamefont {Gu}, \citenamefont {Katsura},\ and\ \citenamefont {Das~Sarma}}]{Sun2011}%
  \BibitemOpen
  \bibfield  {author} {\bibinfo {author} {\bibfnamefont {K.}~\bibnamefont {Sun}}, \bibinfo {author} {\bibfnamefont {Z.}~\bibnamefont {Gu}}, \bibinfo {author} {\bibfnamefont {H.}~\bibnamefont {Katsura}},\ and\ \bibinfo {author} {\bibfnamefont {S.}~\bibnamefont {Das~Sarma}},\ }\bibfield  {title} {\bibinfo {title} {Nearly flatbands with nontrivial topology},\ }\href {https://doi.org/10.1103/PhysRevLett.106.236803} {\bibfield  {journal} {\bibinfo  {journal} {Phys. Rev. Lett.}\ }\textbf {\bibinfo {volume} {106}},\ \bibinfo {pages} {236803} (\bibinfo {year} {2011})}\BibitemShut {NoStop}%
\bibitem [{\citenamefont {Tang}\ \emph {et~al.}(2011)\citenamefont {Tang}, \citenamefont {Mei},\ and\ \citenamefont {Wen}}]{Tang11}%
  \BibitemOpen
  \bibfield  {author} {\bibinfo {author} {\bibfnamefont {E.}~\bibnamefont {Tang}}, \bibinfo {author} {\bibfnamefont {J.-W.}\ \bibnamefont {Mei}},\ and\ \bibinfo {author} {\bibfnamefont {X.-G.}\ \bibnamefont {Wen}},\ }\bibfield  {title} {\bibinfo {title} {High-temperature fractional quantum hall states},\ }\href {https://doi.org/10.1103/PhysRevLett.106.236802} {\bibfield  {journal} {\bibinfo  {journal} {Phys. Rev. Lett.}\ }\textbf {\bibinfo {volume} {106}},\ \bibinfo {pages} {236802} (\bibinfo {year} {2011})}\BibitemShut {NoStop}%
\bibitem [{\citenamefont {Wang}\ \emph {et~al.}(2021{\natexlab{b}})\citenamefont {Wang}, \citenamefont {Cano}, \citenamefont {Millis}, \citenamefont {Liu},\ and\ \citenamefont {Yang}}]{Jie2021IdealBands}%
  \BibitemOpen
  \bibfield  {author} {\bibinfo {author} {\bibfnamefont {J.}~\bibnamefont {Wang}}, \bibinfo {author} {\bibfnamefont {J.}~\bibnamefont {Cano}}, \bibinfo {author} {\bibfnamefont {A.~J.}\ \bibnamefont {Millis}}, \bibinfo {author} {\bibfnamefont {Z.}~\bibnamefont {Liu}},\ and\ \bibinfo {author} {\bibfnamefont {B.}~\bibnamefont {Yang}},\ }\bibfield  {title} {\bibinfo {title} {Exact landau level description of geometry and interaction in a flatband},\ }\href {https://doi.org/10.1103/PhysRevLett.127.246403} {\bibfield  {journal} {\bibinfo  {journal} {Phys. Rev. Lett.}\ }\textbf {\bibinfo {volume} {127}},\ \bibinfo {pages} {246403} (\bibinfo {year} {2021}{\natexlab{b}})}\BibitemShut {NoStop}%
\bibitem [{\citenamefont {Ledwith}\ \emph {et~al.}(2023{\natexlab{a}})\citenamefont {Ledwith}, \citenamefont {Vishwanath},\ and\ \citenamefont {Parker}}]{Parker2023IdealBands}%
  \BibitemOpen
  \bibfield  {author} {\bibinfo {author} {\bibfnamefont {P.~J.}\ \bibnamefont {Ledwith}}, \bibinfo {author} {\bibfnamefont {A.}~\bibnamefont {Vishwanath}},\ and\ \bibinfo {author} {\bibfnamefont {D.~E.}\ \bibnamefont {Parker}},\ }\bibfield  {title} {\bibinfo {title} {Vortexability: A unifying criterion for ideal fractional chern insulators},\ }\href {https://doi.org/10.1103/PhysRevB.108.205144} {\bibfield  {journal} {\bibinfo  {journal} {Phys. Rev. B}\ }\textbf {\bibinfo {volume} {108}},\ \bibinfo {pages} {205144} (\bibinfo {year} {2023}{\natexlab{a}})}\BibitemShut {NoStop}%
\bibitem [{\citenamefont {Estienne}\ \emph {et~al.}(2023)\citenamefont {Estienne}, \citenamefont {Regnault},\ and\ \citenamefont {Cr\'epel}}]{Valentin2023IdealBands}%
  \BibitemOpen
  \bibfield  {author} {\bibinfo {author} {\bibfnamefont {B.}~\bibnamefont {Estienne}}, \bibinfo {author} {\bibfnamefont {N.}~\bibnamefont {Regnault}},\ and\ \bibinfo {author} {\bibfnamefont {V.}~\bibnamefont {Cr\'epel}},\ }\bibfield  {title} {\bibinfo {title} {Ideal chern bands as landau levels in curved space},\ }\href {https://doi.org/10.1103/PhysRevResearch.5.L032048} {\bibfield  {journal} {\bibinfo  {journal} {Phys. Rev. Res.}\ }\textbf {\bibinfo {volume} {5}},\ \bibinfo {pages} {L032048} (\bibinfo {year} {2023})}\BibitemShut {NoStop}%
\bibitem [{\citenamefont {Liu}\ \emph {et~al.}(2024{\natexlab{b}})\citenamefont {Liu}, \citenamefont {Mera}, \citenamefont {Fujimoto}, \citenamefont {Ozawa},\ and\ \citenamefont {Wang}}]{liu2024theorygeneralizedlandaulevels}%
  \BibitemOpen
  \bibfield  {author} {\bibinfo {author} {\bibfnamefont {Z.}~\bibnamefont {Liu}}, \bibinfo {author} {\bibfnamefont {B.}~\bibnamefont {Mera}}, \bibinfo {author} {\bibfnamefont {M.}~\bibnamefont {Fujimoto}}, \bibinfo {author} {\bibfnamefont {T.}~\bibnamefont {Ozawa}},\ and\ \bibinfo {author} {\bibfnamefont {J.}~\bibnamefont {Wang}},\ }\href {https://arxiv.org/abs/2405.14479} {\bibinfo {title} {Theory of generalized landau levels and implication for non-abelian states}} (\bibinfo {year} {2024}{\natexlab{b}}),\ \Eprint {https://arxiv.org/abs/2405.14479} {arXiv:2405.14479 [cond-mat.mes-hall]} \BibitemShut {NoStop}%
\bibitem [{\citenamefont {Roy}(2014{\natexlab{b}})}]{roy2014band}%
  \BibitemOpen
  \bibfield  {author} {\bibinfo {author} {\bibfnamefont {R.}~\bibnamefont {Roy}},\ }\bibfield  {title} {\bibinfo {title} {Band geometry of fractional topological insulators},\ }\href@noop {} {\bibfield  {journal} {\bibinfo  {journal} {Physical Review B}\ }\textbf {\bibinfo {volume} {90}},\ \bibinfo {pages} {165139} (\bibinfo {year} {2014}{\natexlab{b}})}\BibitemShut {NoStop}%
\bibitem [{\citenamefont {Claassen}\ \emph {et~al.}(2015)\citenamefont {Claassen}, \citenamefont {Lee}, \citenamefont {Thomale}, \citenamefont {Qi},\ and\ \citenamefont {Devereaux}}]{claassen2015position}%
  \BibitemOpen
  \bibfield  {author} {\bibinfo {author} {\bibfnamefont {M.}~\bibnamefont {Claassen}}, \bibinfo {author} {\bibfnamefont {C.~H.}\ \bibnamefont {Lee}}, \bibinfo {author} {\bibfnamefont {R.}~\bibnamefont {Thomale}}, \bibinfo {author} {\bibfnamefont {X.-L.}\ \bibnamefont {Qi}},\ and\ \bibinfo {author} {\bibfnamefont {T.~P.}\ \bibnamefont {Devereaux}},\ }\bibfield  {title} {\bibinfo {title} {Position-momentum duality and fractional quantum hall effect in chern insulators},\ }\href@noop {} {\bibfield  {journal} {\bibinfo  {journal} {Physical review letters}\ }\textbf {\bibinfo {volume} {114}},\ \bibinfo {pages} {236802} (\bibinfo {year} {2015})}\BibitemShut {NoStop}%
\bibitem [{\citenamefont {Northe}\ \emph {et~al.}(2022)\citenamefont {Northe}, \citenamefont {Palumbo}, \citenamefont {Sturm}, \citenamefont {Tutschku},\ and\ \citenamefont {Hankiewicz}}]{northe2022interplay}%
  \BibitemOpen
  \bibfield  {author} {\bibinfo {author} {\bibfnamefont {C.}~\bibnamefont {Northe}}, \bibinfo {author} {\bibfnamefont {G.}~\bibnamefont {Palumbo}}, \bibinfo {author} {\bibfnamefont {J.}~\bibnamefont {Sturm}}, \bibinfo {author} {\bibfnamefont {C.}~\bibnamefont {Tutschku}},\ and\ \bibinfo {author} {\bibfnamefont {E.~M.}\ \bibnamefont {Hankiewicz}},\ }\bibfield  {title} {\bibinfo {title} {Interplay of band geometry and topology in ideal chern insulators in the presence of external electromagnetic fields},\ }\href@noop {} {\bibfield  {journal} {\bibinfo  {journal} {Physical Review B}\ }\textbf {\bibinfo {volume} {105}},\ \bibinfo {pages} {155410} (\bibinfo {year} {2022})}\BibitemShut {NoStop}%
\bibitem [{\citenamefont {Ji}\ and\ \citenamefont {Yang}(2024)}]{ji2024quantum}%
  \BibitemOpen
  \bibfield  {author} {\bibinfo {author} {\bibfnamefont {G.}~\bibnamefont {Ji}}\ and\ \bibinfo {author} {\bibfnamefont {B.}~\bibnamefont {Yang}},\ }\bibfield  {title} {\bibinfo {title} {Quantum metric induced hole dispersion and emergent particle-hole symmetry in topological flat bands},\ }\href@noop {} {\bibfield  {journal} {\bibinfo  {journal} {arXiv preprint arXiv:2409.08324}\ } (\bibinfo {year} {2024})}\BibitemShut {NoStop}%
\bibitem [{\citenamefont {Simon}\ \emph {et~al.}(2015)\citenamefont {Simon}, \citenamefont {Harper},\ and\ \citenamefont {Read}}]{Read2015Interaction}%
  \BibitemOpen
  \bibfield  {author} {\bibinfo {author} {\bibfnamefont {S.~H.}\ \bibnamefont {Simon}}, \bibinfo {author} {\bibfnamefont {F.}~\bibnamefont {Harper}},\ and\ \bibinfo {author} {\bibfnamefont {N.}~\bibnamefont {Read}},\ }\bibfield  {title} {\bibinfo {title} {Fractional chern insulators in bands with zero berry curvature},\ }\href {https://doi.org/10.1103/PhysRevB.92.195104} {\bibfield  {journal} {\bibinfo  {journal} {Phys. Rev. B}\ }\textbf {\bibinfo {volume} {92}},\ \bibinfo {pages} {195104} (\bibinfo {year} {2015})}\BibitemShut {NoStop}%
\bibitem [{\citenamefont {Cai}\ \emph {et~al.}(2023)\citenamefont {Cai}, \citenamefont {Anderson}, \citenamefont {Wang}, \citenamefont {Zhang}, \citenamefont {Liu}, \citenamefont {Holtzmann}, \citenamefont {Zhang}, \citenamefont {Fan}, \citenamefont {Taniguchi}, \citenamefont {Watanabe}, \citenamefont {Ran}, \citenamefont {Cao}, \citenamefont {Fu}, \citenamefont {Xiao}, \citenamefont {Yao},\ and\ \citenamefont {Xu}}]{cai2023signatures}%
  \BibitemOpen
  \bibfield  {author} {\bibinfo {author} {\bibfnamefont {J.}~\bibnamefont {Cai}}, \bibinfo {author} {\bibfnamefont {E.}~\bibnamefont {Anderson}}, \bibinfo {author} {\bibfnamefont {C.}~\bibnamefont {Wang}}, \bibinfo {author} {\bibfnamefont {X.}~\bibnamefont {Zhang}}, \bibinfo {author} {\bibfnamefont {X.}~\bibnamefont {Liu}}, \bibinfo {author} {\bibfnamefont {W.}~\bibnamefont {Holtzmann}}, \bibinfo {author} {\bibfnamefont {Y.}~\bibnamefont {Zhang}}, \bibinfo {author} {\bibfnamefont {F.}~\bibnamefont {Fan}}, \bibinfo {author} {\bibfnamefont {T.}~\bibnamefont {Taniguchi}}, \bibinfo {author} {\bibfnamefont {K.}~\bibnamefont {Watanabe}}, \bibinfo {author} {\bibfnamefont {Y.}~\bibnamefont {Ran}}, \bibinfo {author} {\bibfnamefont {T.}~\bibnamefont {Cao}}, \bibinfo {author} {\bibfnamefont {L.}~\bibnamefont {Fu}}, \bibinfo {author} {\bibfnamefont {D.}~\bibnamefont {Xiao}}, \bibinfo {author} {\bibfnamefont {W.}~\bibnamefont {Yao}},\ and\ \bibinfo {author} {\bibfnamefont {X.}~\bibnamefont {Xu}},\ }\bibfield  {title}
  {\bibinfo {title} {Signatures of fractional quantum anomalous hall states in twisted mote2},\ }\bibfield  {journal} {\bibinfo  {journal} {Nature}\ }\href {https://doi.org/10.1038/s41586-023-06289-w} {10.1038/s41586-023-06289-w} (\bibinfo {year} {2023})\BibitemShut {NoStop}%
\bibitem [{\citenamefont {Zeng}\ \emph {et~al.}(2023)\citenamefont {Zeng}, \citenamefont {Xia}, \citenamefont {Kang}, \citenamefont {Zhu}, \citenamefont {Kn{\"u}ppel}, \citenamefont {Vaswani}, \citenamefont {Watanabe}, \citenamefont {Taniguchi}, \citenamefont {Mak},\ and\ \citenamefont {Shan}}]{zeng2023integer}%
  \BibitemOpen
  \bibfield  {author} {\bibinfo {author} {\bibfnamefont {Y.}~\bibnamefont {Zeng}}, \bibinfo {author} {\bibfnamefont {Z.}~\bibnamefont {Xia}}, \bibinfo {author} {\bibfnamefont {K.}~\bibnamefont {Kang}}, \bibinfo {author} {\bibfnamefont {J.}~\bibnamefont {Zhu}}, \bibinfo {author} {\bibfnamefont {P.}~\bibnamefont {Kn{\"u}ppel}}, \bibinfo {author} {\bibfnamefont {C.}~\bibnamefont {Vaswani}}, \bibinfo {author} {\bibfnamefont {K.}~\bibnamefont {Watanabe}}, \bibinfo {author} {\bibfnamefont {T.}~\bibnamefont {Taniguchi}}, \bibinfo {author} {\bibfnamefont {K.~F.}\ \bibnamefont {Mak}},\ and\ \bibinfo {author} {\bibfnamefont {J.}~\bibnamefont {Shan}},\ }\bibfield  {title} {\bibinfo {title} {Thermodynamic evidence of fractional chern insulator in moir{\'e} mote2},\ }\bibfield  {journal} {\bibinfo  {journal} {Nature}\ }\href {https://doi.org/10.1038/s41586-023-06452-3} {10.1038/s41586-023-06452-3} (\bibinfo {year} {2023})\BibitemShut {NoStop}%
\bibitem [{\citenamefont {{Park}}\ \emph {et~al.}(2023)\citenamefont {{Park}}, \citenamefont {{Cai}}, \citenamefont {{Anderson}}, \citenamefont {{Zhang}}, \citenamefont {{Zhu}}, \citenamefont {{Liu}}, \citenamefont {{Wang}}, \citenamefont {{Holtzmann}}, \citenamefont {{Hu}}, \citenamefont {{Liu}}, \citenamefont {{Taniguchi}}, \citenamefont {{Watanabe}}, \citenamefont {{Chu}}, \citenamefont {{Cao}}, \citenamefont {{Fu}}, \citenamefont {{Yao}}, \citenamefont {{Chang}}, \citenamefont {{Cobden}}, \citenamefont {{Xiao}},\ and\ \citenamefont {{Xu}}}]{park2023observation}%
  \BibitemOpen
  \bibfield  {author} {\bibinfo {author} {\bibfnamefont {H.}~\bibnamefont {{Park}}}, \bibinfo {author} {\bibfnamefont {J.}~\bibnamefont {{Cai}}}, \bibinfo {author} {\bibfnamefont {E.}~\bibnamefont {{Anderson}}}, \bibinfo {author} {\bibfnamefont {Y.}~\bibnamefont {{Zhang}}}, \bibinfo {author} {\bibfnamefont {J.}~\bibnamefont {{Zhu}}}, \bibinfo {author} {\bibfnamefont {X.}~\bibnamefont {{Liu}}}, \bibinfo {author} {\bibfnamefont {C.}~\bibnamefont {{Wang}}}, \bibinfo {author} {\bibfnamefont {W.}~\bibnamefont {{Holtzmann}}}, \bibinfo {author} {\bibfnamefont {C.}~\bibnamefont {{Hu}}}, \bibinfo {author} {\bibfnamefont {Z.}~\bibnamefont {{Liu}}}, \bibinfo {author} {\bibfnamefont {T.}~\bibnamefont {{Taniguchi}}}, \bibinfo {author} {\bibfnamefont {K.}~\bibnamefont {{Watanabe}}}, \bibinfo {author} {\bibfnamefont {J.-H.}\ \bibnamefont {{Chu}}}, \bibinfo {author} {\bibfnamefont {T.}~\bibnamefont {{Cao}}}, \bibinfo {author} {\bibfnamefont {L.}~\bibnamefont {{Fu}}}, \bibinfo {author} {\bibfnamefont {W.}~\bibnamefont
  {{Yao}}}, \bibinfo {author} {\bibfnamefont {C.-Z.}\ \bibnamefont {{Chang}}}, \bibinfo {author} {\bibfnamefont {D.}~\bibnamefont {{Cobden}}}, \bibinfo {author} {\bibfnamefont {D.}~\bibnamefont {{Xiao}}},\ and\ \bibinfo {author} {\bibfnamefont {X.}~\bibnamefont {{Xu}}},\ }\bibfield  {title} {\bibinfo {title} {{Observation of fractionally quantized anomalous Hall effect}},\ }\href {https://doi.org/10.1038/s41586-023-06536-0} {\bibfield  {journal} {\bibinfo  {journal} {\nat}\ }\textbf {\bibinfo {volume} {622}},\ \bibinfo {pages} {74} (\bibinfo {year} {2023})},\ \Eprint {https://arxiv.org/abs/2308.02657} {arXiv:2308.02657 [cond-mat.mes-hall]} \BibitemShut {NoStop}%
\bibitem [{\citenamefont {{Xu}}\ \emph {et~al.}(2023)\citenamefont {{Xu}}, \citenamefont {{Sun}}, \citenamefont {{Jia}}, \citenamefont {{Liu}}, \citenamefont {{Xu}}, \citenamefont {{Li}}, \citenamefont {{Gu}}, \citenamefont {{Watanabe}}, \citenamefont {{Taniguchi}}, \citenamefont {{Tong}}, \citenamefont {{Jia}}, \citenamefont {{Shi}}, \citenamefont {{Jiang}}, \citenamefont {{Zhang}}, \citenamefont {{Liu}},\ and\ \citenamefont {{Li}}}]{Xu2023FCItMoTe2}%
  \BibitemOpen
  \bibfield  {author} {\bibinfo {author} {\bibfnamefont {F.}~\bibnamefont {{Xu}}}, \bibinfo {author} {\bibfnamefont {Z.}~\bibnamefont {{Sun}}}, \bibinfo {author} {\bibfnamefont {T.}~\bibnamefont {{Jia}}}, \bibinfo {author} {\bibfnamefont {C.}~\bibnamefont {{Liu}}}, \bibinfo {author} {\bibfnamefont {C.}~\bibnamefont {{Xu}}}, \bibinfo {author} {\bibfnamefont {C.}~\bibnamefont {{Li}}}, \bibinfo {author} {\bibfnamefont {Y.}~\bibnamefont {{Gu}}}, \bibinfo {author} {\bibfnamefont {K.}~\bibnamefont {{Watanabe}}}, \bibinfo {author} {\bibfnamefont {T.}~\bibnamefont {{Taniguchi}}}, \bibinfo {author} {\bibfnamefont {B.}~\bibnamefont {{Tong}}}, \bibinfo {author} {\bibfnamefont {J.}~\bibnamefont {{Jia}}}, \bibinfo {author} {\bibfnamefont {Z.}~\bibnamefont {{Shi}}}, \bibinfo {author} {\bibfnamefont {S.}~\bibnamefont {{Jiang}}}, \bibinfo {author} {\bibfnamefont {Y.}~\bibnamefont {{Zhang}}}, \bibinfo {author} {\bibfnamefont {X.}~\bibnamefont {{Liu}}},\ and\ \bibinfo {author} {\bibfnamefont {T.}~\bibnamefont {{Li}}},\
  }\bibfield  {title} {\bibinfo {title} {{Observation of integer and fractional quantum anomalous Hall states in twisted bilayer MoTe2}},\ }\href {https://doi.org/10.48550/arXiv.2308.06177} {\bibfield  {journal} {\bibinfo  {journal} {arXiv e-prints}\ ,\ \bibinfo {eid} {arXiv:2308.06177}} (\bibinfo {year} {2023})},\ \Eprint {https://arxiv.org/abs/2308.06177} {arXiv:2308.06177 [cond-mat.mes-hall]} \BibitemShut {NoStop}%
\bibitem [{\citenamefont {{Ji}}\ \emph {et~al.}(2024)\citenamefont {{Ji}}, \citenamefont {{Park}}, \citenamefont {{Barber}}, \citenamefont {{Hu}}, \citenamefont {{Watanabe}}, \citenamefont {{Taniguchi}}, \citenamefont {{Chu}}, \citenamefont {{Xu}},\ and\ \citenamefont {{Shen}}}]{Ji2024LocalProbetMoTe2}%
  \BibitemOpen
  \bibfield  {author} {\bibinfo {author} {\bibfnamefont {Z.}~\bibnamefont {{Ji}}}, \bibinfo {author} {\bibfnamefont {H.}~\bibnamefont {{Park}}}, \bibinfo {author} {\bibfnamefont {M.~E.}\ \bibnamefont {{Barber}}}, \bibinfo {author} {\bibfnamefont {C.}~\bibnamefont {{Hu}}}, \bibinfo {author} {\bibfnamefont {K.}~\bibnamefont {{Watanabe}}}, \bibinfo {author} {\bibfnamefont {T.}~\bibnamefont {{Taniguchi}}}, \bibinfo {author} {\bibfnamefont {J.-H.}\ \bibnamefont {{Chu}}}, \bibinfo {author} {\bibfnamefont {X.}~\bibnamefont {{Xu}}},\ and\ \bibinfo {author} {\bibfnamefont {Z.-x.}\ \bibnamefont {{Shen}}},\ }\bibfield  {title} {\bibinfo {title} {{Local probe of bulk and edge states in a fractional Chern insulator}},\ }\href {https://doi.org/10.48550/arXiv.2404.07157} {\bibfield  {journal} {\bibinfo  {journal} {arXiv e-prints}\ ,\ \bibinfo {eid} {arXiv:2404.07157}} (\bibinfo {year} {2024})},\ \Eprint {https://arxiv.org/abs/2404.07157} {arXiv:2404.07157 [cond-mat.str-el]} \BibitemShut {NoStop}%
\bibitem [{\citenamefont {Kang}\ \emph {et~al.}(2024{\natexlab{b}})\citenamefont {Kang}, \citenamefont {Shen}, \citenamefont {Qiu}, \citenamefont {Zeng}, \citenamefont {Xia}, \citenamefont {Watanabe}, \citenamefont {Taniguchi}, \citenamefont {Shan},\ and\ \citenamefont {Mak}}]{Kang2024_tMoTe2_2.13}%
  \BibitemOpen
  \bibfield  {author} {\bibinfo {author} {\bibfnamefont {K.}~\bibnamefont {Kang}}, \bibinfo {author} {\bibfnamefont {B.}~\bibnamefont {Shen}}, \bibinfo {author} {\bibfnamefont {Y.}~\bibnamefont {Qiu}}, \bibinfo {author} {\bibfnamefont {Y.}~\bibnamefont {Zeng}}, \bibinfo {author} {\bibfnamefont {Z.}~\bibnamefont {Xia}}, \bibinfo {author} {\bibfnamefont {K.}~\bibnamefont {Watanabe}}, \bibinfo {author} {\bibfnamefont {T.}~\bibnamefont {Taniguchi}}, \bibinfo {author} {\bibfnamefont {J.}~\bibnamefont {Shan}},\ and\ \bibinfo {author} {\bibfnamefont {K.~F.}\ \bibnamefont {Mak}},\ }\bibfield  {title} {\bibinfo {title} {Evidence of the fractional quantum spin hall effect in moir{\'e} mote2},\ }\href {https://doi.org/10.1038/s41586-024-07214-5} {\bibfield  {journal} {\bibinfo  {journal} {Nature}\ }\textbf {\bibinfo {volume} {628}},\ \bibinfo {pages} {522} (\bibinfo {year} {2024}{\natexlab{b}})}\BibitemShut {NoStop}%
\bibitem [{\citenamefont {Xu}\ \emph {et~al.}(2024{\natexlab{a}})\citenamefont {Xu}, \citenamefont {Chang}, \citenamefont {Xiao}, \citenamefont {Zhang}, \citenamefont {Liu}, \citenamefont {Sun}, \citenamefont {Mao}, \citenamefont {Peshcherenko}, \citenamefont {Li}, \citenamefont {Watanabe}, \citenamefont {Taniguchi}, \citenamefont {Tong}, \citenamefont {Lu}, \citenamefont {Jia}, \citenamefont {Qian}, \citenamefont {Shi}, \citenamefont {Zhang}, \citenamefont {Liu}, \citenamefont {Jiang},\ and\ \citenamefont {Li}}]{xu2024interplaytopologycorrelationssecond}%
  \BibitemOpen
  \bibfield  {author} {\bibinfo {author} {\bibfnamefont {F.}~\bibnamefont {Xu}}, \bibinfo {author} {\bibfnamefont {X.}~\bibnamefont {Chang}}, \bibinfo {author} {\bibfnamefont {J.}~\bibnamefont {Xiao}}, \bibinfo {author} {\bibfnamefont {Y.}~\bibnamefont {Zhang}}, \bibinfo {author} {\bibfnamefont {F.}~\bibnamefont {Liu}}, \bibinfo {author} {\bibfnamefont {Z.}~\bibnamefont {Sun}}, \bibinfo {author} {\bibfnamefont {N.}~\bibnamefont {Mao}}, \bibinfo {author} {\bibfnamefont {N.}~\bibnamefont {Peshcherenko}}, \bibinfo {author} {\bibfnamefont {J.}~\bibnamefont {Li}}, \bibinfo {author} {\bibfnamefont {K.}~\bibnamefont {Watanabe}}, \bibinfo {author} {\bibfnamefont {T.}~\bibnamefont {Taniguchi}}, \bibinfo {author} {\bibfnamefont {B.}~\bibnamefont {Tong}}, \bibinfo {author} {\bibfnamefont {L.}~\bibnamefont {Lu}}, \bibinfo {author} {\bibfnamefont {J.}~\bibnamefont {Jia}}, \bibinfo {author} {\bibfnamefont {D.}~\bibnamefont {Qian}}, \bibinfo {author} {\bibfnamefont {Z.}~\bibnamefont {Shi}}, \bibinfo {author} {\bibfnamefont
  {Y.}~\bibnamefont {Zhang}}, \bibinfo {author} {\bibfnamefont {X.}~\bibnamefont {Liu}}, \bibinfo {author} {\bibfnamefont {S.}~\bibnamefont {Jiang}},\ and\ \bibinfo {author} {\bibfnamefont {T.}~\bibnamefont {Li}},\ }\href {https://arxiv.org/abs/2406.09687} {\bibinfo {title} {Interplay between topology and correlations in the second moir\'e band of twisted bilayer mote2}} (\bibinfo {year} {2024}{\natexlab{a}}),\ \Eprint {https://arxiv.org/abs/2406.09687} {arXiv:2406.09687 [cond-mat.mes-hall]} \BibitemShut {NoStop}%
\bibitem [{\citenamefont {Park}\ \emph {et~al.}(2024)\citenamefont {Park}, \citenamefont {Cai}, \citenamefont {Anderson}, \citenamefont {Zhang}, \citenamefont {Liu}, \citenamefont {Holtzmann}, \citenamefont {Li}, \citenamefont {Wang}, \citenamefont {Hu}, \citenamefont {Zhao}, \citenamefont {Taniguchi}, \citenamefont {Watanabe}, \citenamefont {Yang}, \citenamefont {Cobden}, \citenamefont {Chu}, \citenamefont {Regnault}, \citenamefont {Bernevig}, \citenamefont {Fu}, \citenamefont {Cao}, \citenamefont {Xiao},\ and\ \citenamefont {Xu}}]{park_Ferromagnetism_2024}%
  \BibitemOpen
  \bibfield  {author} {\bibinfo {author} {\bibfnamefont {H.}~\bibnamefont {Park}}, \bibinfo {author} {\bibfnamefont {J.}~\bibnamefont {Cai}}, \bibinfo {author} {\bibfnamefont {E.}~\bibnamefont {Anderson}}, \bibinfo {author} {\bibfnamefont {X.-W.}\ \bibnamefont {Zhang}}, \bibinfo {author} {\bibfnamefont {X.}~\bibnamefont {Liu}}, \bibinfo {author} {\bibfnamefont {W.}~\bibnamefont {Holtzmann}}, \bibinfo {author} {\bibfnamefont {W.}~\bibnamefont {Li}}, \bibinfo {author} {\bibfnamefont {C.}~\bibnamefont {Wang}}, \bibinfo {author} {\bibfnamefont {C.}~\bibnamefont {Hu}}, \bibinfo {author} {\bibfnamefont {Y.}~\bibnamefont {Zhao}}, \bibinfo {author} {\bibfnamefont {T.}~\bibnamefont {Taniguchi}}, \bibinfo {author} {\bibfnamefont {K.}~\bibnamefont {Watanabe}}, \bibinfo {author} {\bibfnamefont {J.}~\bibnamefont {Yang}}, \bibinfo {author} {\bibfnamefont {D.}~\bibnamefont {Cobden}}, \bibinfo {author} {\bibfnamefont {J.-H.}\ \bibnamefont {Chu}}, \bibinfo {author} {\bibfnamefont {N.}~\bibnamefont {Regnault}}, \bibinfo
  {author} {\bibfnamefont {B.~A.}\ \bibnamefont {Bernevig}}, \bibinfo {author} {\bibfnamefont {L.}~\bibnamefont {Fu}}, \bibinfo {author} {\bibfnamefont {T.}~\bibnamefont {Cao}}, \bibinfo {author} {\bibfnamefont {D.}~\bibnamefont {Xiao}},\ and\ \bibinfo {author} {\bibfnamefont {X.}~\bibnamefont {Xu}},\ }\href {https://doi.org/10.48550/arXiv.2406.09591} {\bibinfo {title} {Ferromagnetism and topology of the higher flat band in a fractional chern insulator}} (\bibinfo {year} {2024}),\ \Eprint {https://arxiv.org/abs/2406.09591} {arXiv:2406.09591 [cond-mat]} \BibitemShut {NoStop}%
\bibitem [{\citenamefont {Xiao}\ \emph {et~al.}(2012)\citenamefont {Xiao}, \citenamefont {Liu}, \citenamefont {Feng}, \citenamefont {Xu},\ and\ \citenamefont {Yao}}]{xiao_coupled_2012}%
  \BibitemOpen
  \bibfield  {author} {\bibinfo {author} {\bibfnamefont {D.}~\bibnamefont {Xiao}}, \bibinfo {author} {\bibfnamefont {G.-B.}\ \bibnamefont {Liu}}, \bibinfo {author} {\bibfnamefont {W.}~\bibnamefont {Feng}}, \bibinfo {author} {\bibfnamefont {X.}~\bibnamefont {Xu}},\ and\ \bibinfo {author} {\bibfnamefont {W.}~\bibnamefont {Yao}},\ }\bibfield  {title} {\bibinfo {title} {Coupled {Spin} and {Valley} {Physics} in {Monolayers} of \(\mathrm{MoS}_2\) and {Other} {Group}-{VI} {Dichalcogenides}},\ }\href {https://doi.org/10.1103/PhysRevLett.108.196802} {\bibfield  {journal} {\bibinfo  {journal} {Physical Review Letters}\ }\textbf {\bibinfo {volume} {108}},\ \bibinfo {pages} {196802} (\bibinfo {year} {2012})}\BibitemShut {NoStop}%
\bibitem [{\citenamefont {Wu}\ \emph {et~al.}(2019)\citenamefont {Wu}, \citenamefont {Lovorn}, \citenamefont {Tutuc}, \citenamefont {Martin},\ and\ \citenamefont {MacDonald}}]{wu_topological_2019}%
  \BibitemOpen
  \bibfield  {author} {\bibinfo {author} {\bibfnamefont {F.}~\bibnamefont {Wu}}, \bibinfo {author} {\bibfnamefont {T.}~\bibnamefont {Lovorn}}, \bibinfo {author} {\bibfnamefont {E.}~\bibnamefont {Tutuc}}, \bibinfo {author} {\bibfnamefont {I.}~\bibnamefont {Martin}},\ and\ \bibinfo {author} {\bibfnamefont {A.}~\bibnamefont {MacDonald}},\ }\bibfield  {title} {\bibinfo {title} {Topological {Insulators} in {Twisted} {Transition} {Metal} {Dichalcogenide} {Homobilayers}},\ }\href {https://doi.org/10.1103/PhysRevLett.122.086402} {\bibfield  {journal} {\bibinfo  {journal} {Physical Review Letters}\ }\textbf {\bibinfo {volume} {122}},\ \bibinfo {pages} {086402} (\bibinfo {year} {2019})}\BibitemShut {NoStop}%
\bibitem [{\citenamefont {Pan}\ \emph {et~al.}(2020)\citenamefont {Pan}, \citenamefont {Wu},\ and\ \citenamefont {Das~Sarma}}]{pan_band_2020}%
  \BibitemOpen
  \bibfield  {author} {\bibinfo {author} {\bibfnamefont {H.}~\bibnamefont {Pan}}, \bibinfo {author} {\bibfnamefont {F.}~\bibnamefont {Wu}},\ and\ \bibinfo {author} {\bibfnamefont {S.}~\bibnamefont {Das~Sarma}},\ }\bibfield  {title} {\bibinfo {title} {Band topology, {Hubbard} model, {Heisenberg} model, and {Dzyaloshinskii}-{Moriya} interaction in twisted bilayer \(\mathrm{WSe}_2\)},\ }\href {https://doi.org/10.1103/PhysRevResearch.2.033087} {\bibfield  {journal} {\bibinfo  {journal} {Physical Review Research}\ }\textbf {\bibinfo {volume} {2}},\ \bibinfo {pages} {033087} (\bibinfo {year} {2020})}\BibitemShut {NoStop}%
\bibitem [{\citenamefont {Zhang}\ \emph {et~al.}(2021)\citenamefont {Zhang}, \citenamefont {Liu},\ and\ \citenamefont {Fu}}]{zhang_electronic_2021}%
  \BibitemOpen
  \bibfield  {author} {\bibinfo {author} {\bibfnamefont {Y.}~\bibnamefont {Zhang}}, \bibinfo {author} {\bibfnamefont {T.}~\bibnamefont {Liu}},\ and\ \bibinfo {author} {\bibfnamefont {L.}~\bibnamefont {Fu}},\ }\bibfield  {title} {\bibinfo {title} {Electronic structures, charge transfer, and charge order in twisted transition metal dichalcogenide bilayers},\ }\href {https://doi.org/10.1103/PhysRevB.103.155142} {\bibfield  {journal} {\bibinfo  {journal} {Physical Review B}\ }\textbf {\bibinfo {volume} {103}},\ \bibinfo {pages} {155142} (\bibinfo {year} {2021})}\BibitemShut {NoStop}%
\bibitem [{\citenamefont {Devakul}\ \emph {et~al.}(2021)\citenamefont {Devakul}, \citenamefont {Crépel}, \citenamefont {Zhang},\ and\ \citenamefont {Fu}}]{devakul_magic_2021}%
  \BibitemOpen
  \bibfield  {author} {\bibinfo {author} {\bibfnamefont {T.}~\bibnamefont {Devakul}}, \bibinfo {author} {\bibfnamefont {V.}~\bibnamefont {Crépel}}, \bibinfo {author} {\bibfnamefont {Y.}~\bibnamefont {Zhang}},\ and\ \bibinfo {author} {\bibfnamefont {L.}~\bibnamefont {Fu}},\ }\bibfield  {title} {\bibinfo {title} {Magic in twisted transition metal dichalcogenide bilayers},\ }\href {https://doi.org/10.1038/s41467-021-27042-9} {\bibfield  {journal} {\bibinfo  {journal} {Nature Communications}\ }\textbf {\bibinfo {volume} {12}},\ \bibinfo {pages} {6730} (\bibinfo {year} {2021})}\BibitemShut {NoStop}%
\bibitem [{\citenamefont {Wang}\ \emph {et~al.}(2023{\natexlab{a}})\citenamefont {Wang}, \citenamefont {Devakul}, \citenamefont {Zaletel},\ and\ \citenamefont {Fu}}]{wang_topological_2023}%
  \BibitemOpen
  \bibfield  {author} {\bibinfo {author} {\bibfnamefont {T.}~\bibnamefont {Wang}}, \bibinfo {author} {\bibfnamefont {T.}~\bibnamefont {Devakul}}, \bibinfo {author} {\bibfnamefont {M.~P.}\ \bibnamefont {Zaletel}},\ and\ \bibinfo {author} {\bibfnamefont {L.}~\bibnamefont {Fu}},\ }\href {https://doi.org/10.48550/arXiv.2306.02501} {\bibinfo {title} {Topological magnets and magnons in twisted bilayer \(\mathrm{MoTe}_2\) and \(\mathrm{WSe}_2\)}} (\bibinfo {year} {2023}{\natexlab{a}}),\ \Eprint {https://arxiv.org/abs/2306.02501} {arXiv:2306.02501 [cond-mat]} \BibitemShut {NoStop}%
\bibitem [{\citenamefont {Reddy}\ \emph {et~al.}(2023)\citenamefont {Reddy}, \citenamefont {Alsallom}, \citenamefont {Zhang}, \citenamefont {Devakul},\ and\ \citenamefont {Fu}}]{reddy_fractional_2023}%
  \BibitemOpen
  \bibfield  {author} {\bibinfo {author} {\bibfnamefont {A.~P.}\ \bibnamefont {Reddy}}, \bibinfo {author} {\bibfnamefont {F.}~\bibnamefont {Alsallom}}, \bibinfo {author} {\bibfnamefont {Y.}~\bibnamefont {Zhang}}, \bibinfo {author} {\bibfnamefont {T.}~\bibnamefont {Devakul}},\ and\ \bibinfo {author} {\bibfnamefont {L.}~\bibnamefont {Fu}},\ }\bibfield  {title} {\bibinfo {title} {Fractional quantum anomalous {Hall} states in twisted bilayer \(\mathrm{MoTe}_2\) and \(\mathrm{WSe}_2\)},\ }\href {https://doi.org/10.1103/PhysRevB.108.085117} {\bibfield  {journal} {\bibinfo  {journal} {Physical Review B}\ }\textbf {\bibinfo {volume} {108}},\ \bibinfo {pages} {085117} (\bibinfo {year} {2023})}\BibitemShut {NoStop}%
\bibitem [{\citenamefont {Dong}\ \emph {et~al.}(2023{\natexlab{b}})\citenamefont {Dong}, \citenamefont {Wang}, \citenamefont {Ledwith}, \citenamefont {Vishwanath},\ and\ \citenamefont {Parker}}]{dong_composite_2023}%
  \BibitemOpen
  \bibfield  {author} {\bibinfo {author} {\bibfnamefont {J.}~\bibnamefont {Dong}}, \bibinfo {author} {\bibfnamefont {J.}~\bibnamefont {Wang}}, \bibinfo {author} {\bibfnamefont {P.~J.}\ \bibnamefont {Ledwith}}, \bibinfo {author} {\bibfnamefont {A.}~\bibnamefont {Vishwanath}},\ and\ \bibinfo {author} {\bibfnamefont {D.~E.}\ \bibnamefont {Parker}},\ }\bibfield  {title} {\bibinfo {title} {Composite {Fermi} {Liquid} at {Zero} {Magnetic} {Field} in {Twisted} \(\mathrm{MoTe}_2\)},\ }\href {https://doi.org/10.1103/PhysRevLett.131.136502} {\bibfield  {journal} {\bibinfo  {journal} {Physical Review Letters}\ }\textbf {\bibinfo {volume} {131}},\ \bibinfo {pages} {136502} (\bibinfo {year} {2023}{\natexlab{b}})}\BibitemShut {NoStop}%
\bibitem [{\citenamefont {Qiu}\ \emph {et~al.}(2023)\citenamefont {Qiu}, \citenamefont {Li}, \citenamefont {Luo},\ and\ \citenamefont {Wu}}]{qiu_interaction-driven_2023}%
  \BibitemOpen
  \bibfield  {author} {\bibinfo {author} {\bibfnamefont {W.-X.}\ \bibnamefont {Qiu}}, \bibinfo {author} {\bibfnamefont {B.}~\bibnamefont {Li}}, \bibinfo {author} {\bibfnamefont {X.-J.}\ \bibnamefont {Luo}},\ and\ \bibinfo {author} {\bibfnamefont {F.}~\bibnamefont {Wu}},\ }\bibfield  {title} {\bibinfo {title} {Interaction-{Driven} {Topological} {Phase} {Diagram} of {Twisted} {Bilayer} \(\mathrm{MoTe}_2\)},\ }\href {https://doi.org/10.1103/PhysRevX.13.041026} {\bibfield  {journal} {\bibinfo  {journal} {Physical Review X}\ }\textbf {\bibinfo {volume} {13}},\ \bibinfo {pages} {041026} (\bibinfo {year} {2023})}\BibitemShut {NoStop}%
\bibitem [{\citenamefont {Wang}\ \emph {et~al.}(2023{\natexlab{b}})\citenamefont {Wang}, \citenamefont {Wang}, \citenamefont {Kim}, \citenamefont {Louie}, \citenamefont {Fu},\ and\ \citenamefont {Zaletel}}]{wang_topology_2023}%
  \BibitemOpen
  \bibfield  {author} {\bibinfo {author} {\bibfnamefont {T.}~\bibnamefont {Wang}}, \bibinfo {author} {\bibfnamefont {M.}~\bibnamefont {Wang}}, \bibinfo {author} {\bibfnamefont {W.}~\bibnamefont {Kim}}, \bibinfo {author} {\bibfnamefont {S.~G.}\ \bibnamefont {Louie}}, \bibinfo {author} {\bibfnamefont {L.}~\bibnamefont {Fu}},\ and\ \bibinfo {author} {\bibfnamefont {M.~P.}\ \bibnamefont {Zaletel}},\ }\href {https://arxiv.org/abs/2312.12531v1} {\bibinfo {title} {Topology, magnetism and charge order in twisted \(\mathrm{MoTe}_2\) at higher integer hole fillings}} (\bibinfo {year} {2023}{\natexlab{b}})\BibitemShut {NoStop}%
\bibitem [{\citenamefont {Reddy}\ and\ \citenamefont {Fu}(2023)}]{reddy_toward_2023}%
  \BibitemOpen
  \bibfield  {author} {\bibinfo {author} {\bibfnamefont {A.~P.}\ \bibnamefont {Reddy}}\ and\ \bibinfo {author} {\bibfnamefont {L.}~\bibnamefont {Fu}},\ }\bibfield  {title} {\bibinfo {title} {Toward a global phase diagram of the fractional quantum anomalous {Hall} effect},\ }\href {https://doi.org/10.1103/PhysRevB.108.245159} {\bibfield  {journal} {\bibinfo  {journal} {Physical Review B}\ }\textbf {\bibinfo {volume} {108}},\ \bibinfo {pages} {245159} (\bibinfo {year} {2023})}\BibitemShut {NoStop}%
\bibitem [{\citenamefont {Wang}\ \emph {et~al.}(2024)\citenamefont {Wang}, \citenamefont {Zhang}, \citenamefont {Liu}, \citenamefont {He}, \citenamefont {Xu}, \citenamefont {Ran}, \citenamefont {Cao},\ and\ \citenamefont {Xiao}}]{wang_fractional_2024}%
  \BibitemOpen
  \bibfield  {author} {\bibinfo {author} {\bibfnamefont {C.}~\bibnamefont {Wang}}, \bibinfo {author} {\bibfnamefont {X.-W.}\ \bibnamefont {Zhang}}, \bibinfo {author} {\bibfnamefont {X.}~\bibnamefont {Liu}}, \bibinfo {author} {\bibfnamefont {Y.}~\bibnamefont {He}}, \bibinfo {author} {\bibfnamefont {X.}~\bibnamefont {Xu}}, \bibinfo {author} {\bibfnamefont {Y.}~\bibnamefont {Ran}}, \bibinfo {author} {\bibfnamefont {T.}~\bibnamefont {Cao}},\ and\ \bibinfo {author} {\bibfnamefont {D.}~\bibnamefont {Xiao}},\ }\bibfield  {title} {\bibinfo {title} {Fractional {Chern} {Insulator} in {Twisted} {Bilayer} \(\mathrm{MoTe}_2\)},\ }\href {https://doi.org/10.1103/PhysRevLett.132.036501} {\bibfield  {journal} {\bibinfo  {journal} {Physical Review Letters}\ }\textbf {\bibinfo {volume} {132}},\ \bibinfo {pages} {036501} (\bibinfo {year} {2024})}\BibitemShut {NoStop}%
\bibitem [{\citenamefont {Yu}\ \emph {et~al.}(2024{\natexlab{a}})\citenamefont {Yu}, \citenamefont {Herzog-Arbeitman}, \citenamefont {Wang}, \citenamefont {Vafek}, \citenamefont {Bernevig},\ and\ \citenamefont {Regnault}}]{yu_fractional_2024}%
  \BibitemOpen
  \bibfield  {author} {\bibinfo {author} {\bibfnamefont {J.}~\bibnamefont {Yu}}, \bibinfo {author} {\bibfnamefont {J.}~\bibnamefont {Herzog-Arbeitman}}, \bibinfo {author} {\bibfnamefont {M.}~\bibnamefont {Wang}}, \bibinfo {author} {\bibfnamefont {O.}~\bibnamefont {Vafek}}, \bibinfo {author} {\bibfnamefont {B.~A.}\ \bibnamefont {Bernevig}},\ and\ \bibinfo {author} {\bibfnamefont {N.}~\bibnamefont {Regnault}},\ }\bibfield  {title} {\bibinfo {title} {Fractional {Chern} insulators versus nonmagnetic states in twisted bilayer \(\mathrm{MoTe}_2\)},\ }\href {https://doi.org/10.1103/PhysRevB.109.045147} {\bibfield  {journal} {\bibinfo  {journal} {Physical Review B}\ }\textbf {\bibinfo {volume} {109}},\ \bibinfo {pages} {045147} (\bibinfo {year} {2024}{\natexlab{a}})}\BibitemShut {NoStop}%
\bibitem [{\citenamefont {Xu}\ \emph {et~al.}(2024{\natexlab{b}})\citenamefont {Xu}, \citenamefont {Li}, \citenamefont {Xu}, \citenamefont {Bi},\ and\ \citenamefont {Zhang}}]{xu_maximally_2024}%
  \BibitemOpen
  \bibfield  {author} {\bibinfo {author} {\bibfnamefont {C.}~\bibnamefont {Xu}}, \bibinfo {author} {\bibfnamefont {J.}~\bibnamefont {Li}}, \bibinfo {author} {\bibfnamefont {Y.}~\bibnamefont {Xu}}, \bibinfo {author} {\bibfnamefont {Z.}~\bibnamefont {Bi}},\ and\ \bibinfo {author} {\bibfnamefont {Y.}~\bibnamefont {Zhang}},\ }\bibfield  {title} {\bibinfo {title} {Maximally localized {Wannier} functions, interaction models, and fractional quantum anomalous {Hall} effect in twisted bilayer \(\mathrm{MoTe}_2\)},\ }\href {https://doi.org/10.1073/pnas.2316749121} {\bibfield  {journal} {\bibinfo  {journal} {Proceedings of the National Academy of Sciences}\ }\textbf {\bibinfo {volume} {121}},\ \bibinfo {pages} {e2316749121} (\bibinfo {year} {2024}{\natexlab{b}})}\BibitemShut {NoStop}%
\bibitem [{\citenamefont {Abouelkomsan}\ \emph {et~al.}(2024)\citenamefont {Abouelkomsan}, \citenamefont {Reddy}, \citenamefont {Fu},\ and\ \citenamefont {Bergholtz}}]{abouelkomsan_band_2024}%
  \BibitemOpen
  \bibfield  {author} {\bibinfo {author} {\bibfnamefont {A.}~\bibnamefont {Abouelkomsan}}, \bibinfo {author} {\bibfnamefont {A.~P.}\ \bibnamefont {Reddy}}, \bibinfo {author} {\bibfnamefont {L.}~\bibnamefont {Fu}},\ and\ \bibinfo {author} {\bibfnamefont {E.~J.}\ \bibnamefont {Bergholtz}},\ }\bibfield  {title} {\bibinfo {title} {Band mixing in the quantum anomalous {Hall} regime of twisted semiconductor bilayers},\ }\href {https://doi.org/10.1103/PhysRevB.109.L121107} {\bibfield  {journal} {\bibinfo  {journal} {Physical Review B}\ }\textbf {\bibinfo {volume} {109}},\ \bibinfo {pages} {L121107} (\bibinfo {year} {2024})}\BibitemShut {NoStop}%
\bibitem [{\citenamefont {Jia}\ \emph {et~al.}(2024)\citenamefont {Jia}, \citenamefont {Yu}, \citenamefont {Liu}, \citenamefont {Herzog-Arbeitman}, \citenamefont {Qi}, \citenamefont {Pi}, \citenamefont {Regnault}, \citenamefont {Weng}, \citenamefont {Bernevig},\ and\ \citenamefont {Wu}}]{jia_moire_2024}%
  \BibitemOpen
  \bibfield  {author} {\bibinfo {author} {\bibfnamefont {Y.}~\bibnamefont {Jia}}, \bibinfo {author} {\bibfnamefont {J.}~\bibnamefont {Yu}}, \bibinfo {author} {\bibfnamefont {J.}~\bibnamefont {Liu}}, \bibinfo {author} {\bibfnamefont {J.}~\bibnamefont {Herzog-Arbeitman}}, \bibinfo {author} {\bibfnamefont {Z.}~\bibnamefont {Qi}}, \bibinfo {author} {\bibfnamefont {H.}~\bibnamefont {Pi}}, \bibinfo {author} {\bibfnamefont {N.}~\bibnamefont {Regnault}}, \bibinfo {author} {\bibfnamefont {H.}~\bibnamefont {Weng}}, \bibinfo {author} {\bibfnamefont {B.~A.}\ \bibnamefont {Bernevig}},\ and\ \bibinfo {author} {\bibfnamefont {Q.}~\bibnamefont {Wu}},\ }\bibfield  {title} {\bibinfo {title} {Moir{\'e} fractional {Chern} insulators. {I}. {First}-principles calculations and continuum models of twisted bilayer \(\mathrm{MoTe}_2\)},\ }\href {https://doi.org/10.1103/PhysRevB.109.205121} {\bibfield  {journal} {\bibinfo  {journal} {Physical Review B}\ }\textbf {\bibinfo {volume} {109}},\ \bibinfo {pages} {205121} (\bibinfo {year}
  {2024})}\BibitemShut {NoStop}%
\bibitem [{\citenamefont {Zhang}\ \emph {et~al.}(2024)\citenamefont {Zhang}, \citenamefont {Wang}, \citenamefont {Liu}, \citenamefont {Fan}, \citenamefont {Cao},\ and\ \citenamefont {Xiao}}]{zhang_polarization-driven_2024}%
  \BibitemOpen
  \bibfield  {author} {\bibinfo {author} {\bibfnamefont {X.-W.}\ \bibnamefont {Zhang}}, \bibinfo {author} {\bibfnamefont {C.}~\bibnamefont {Wang}}, \bibinfo {author} {\bibfnamefont {X.}~\bibnamefont {Liu}}, \bibinfo {author} {\bibfnamefont {Y.}~\bibnamefont {Fan}}, \bibinfo {author} {\bibfnamefont {T.}~\bibnamefont {Cao}},\ and\ \bibinfo {author} {\bibfnamefont {D.}~\bibnamefont {Xiao}},\ }\bibfield  {title} {\bibinfo {title} {Polarization-driven band topology evolution in twisted \(\mathrm{MoTe}_2\) and \(\mathrm{WSe}_2\)},\ }\href {https://doi.org/10.1038/s41467-024-48511-x} {\bibfield  {journal} {\bibinfo  {journal} {Nature Communications}\ }\textbf {\bibinfo {volume} {15}},\ \bibinfo {pages} {4223} (\bibinfo {year} {2024})}\BibitemShut {NoStop}%
\bibitem [{\citenamefont {Yu}\ \emph {et~al.}(2023{\natexlab{c}})\citenamefont {Yu}, \citenamefont {Herzog-Arbeitman}, \citenamefont {Wang}, \citenamefont {Vafek}, \citenamefont {Bernevig},\ and\ \citenamefont {Regnault}}]{Yu2023FCI}%
  \BibitemOpen
  \bibfield  {author} {\bibinfo {author} {\bibfnamefont {J.}~\bibnamefont {Yu}}, \bibinfo {author} {\bibfnamefont {J.}~\bibnamefont {Herzog-Arbeitman}}, \bibinfo {author} {\bibfnamefont {M.}~\bibnamefont {Wang}}, \bibinfo {author} {\bibfnamefont {O.}~\bibnamefont {Vafek}}, \bibinfo {author} {\bibfnamefont {B.~A.}\ \bibnamefont {Bernevig}},\ and\ \bibinfo {author} {\bibfnamefont {N.}~\bibnamefont {Regnault}},\ }\href@noop {} {\bibinfo {title} {Fractional chern insulators vs. non-magnetic states in twisted bilayer mote$_2$}} (\bibinfo {year} {2023}{\natexlab{c}}),\ \Eprint {https://arxiv.org/abs/2309.14429} {arXiv:2309.14429 [cond-mat.mes-hall]} \BibitemShut {NoStop}%
\bibitem [{\citenamefont {Lu}\ \emph {et~al.}(2024{\natexlab{a}})\citenamefont {Lu}, \citenamefont {Han}, \citenamefont {Yao}, \citenamefont {Reddy}, \citenamefont {Yang}, \citenamefont {Seo}, \citenamefont {Watanabe}, \citenamefont {Taniguchi}, \citenamefont {Fu},\ and\ \citenamefont {Ju}}]{Lu2024fractional}%
  \BibitemOpen
  \bibfield  {author} {\bibinfo {author} {\bibfnamefont {Z.}~\bibnamefont {Lu}}, \bibinfo {author} {\bibfnamefont {T.}~\bibnamefont {Han}}, \bibinfo {author} {\bibfnamefont {Y.}~\bibnamefont {Yao}}, \bibinfo {author} {\bibfnamefont {A.~P.}\ \bibnamefont {Reddy}}, \bibinfo {author} {\bibfnamefont {J.}~\bibnamefont {Yang}}, \bibinfo {author} {\bibfnamefont {J.}~\bibnamefont {Seo}}, \bibinfo {author} {\bibfnamefont {K.}~\bibnamefont {Watanabe}}, \bibinfo {author} {\bibfnamefont {T.}~\bibnamefont {Taniguchi}}, \bibinfo {author} {\bibfnamefont {L.}~\bibnamefont {Fu}},\ and\ \bibinfo {author} {\bibfnamefont {L.}~\bibnamefont {Ju}},\ }\bibfield  {title} {\bibinfo {title} {Fractional quantum anomalous hall effect in multilayer graphene},\ }\href {https://doi.org/10.1038/s41586-023-07010-7} {\bibfield  {journal} {\bibinfo  {journal} {Nature}\ }\textbf {\bibinfo {volume} {626}},\ \bibinfo {pages} {759} (\bibinfo {year} {2024}{\natexlab{a}})}\BibitemShut {NoStop}%
\bibitem [{\citenamefont {Xie}\ \emph {et~al.}(2024)\citenamefont {Xie}, \citenamefont {Huo}, \citenamefont {Lu}, \citenamefont {Feng}, \citenamefont {Zhang}, \citenamefont {Wang}, \citenamefont {Yang}, \citenamefont {Watanabe}, \citenamefont {Taniguchi}, \citenamefont {Liu}, \citenamefont {Song}, \citenamefont {Xie}, \citenamefont {Liu},\ and\ \citenamefont {Lu}}]{xie_Even_2024}%
  \BibitemOpen
  \bibfield  {author} {\bibinfo {author} {\bibfnamefont {J.}~\bibnamefont {Xie}}, \bibinfo {author} {\bibfnamefont {Z.}~\bibnamefont {Huo}}, \bibinfo {author} {\bibfnamefont {X.}~\bibnamefont {Lu}}, \bibinfo {author} {\bibfnamefont {Z.}~\bibnamefont {Feng}}, \bibinfo {author} {\bibfnamefont {Z.}~\bibnamefont {Zhang}}, \bibinfo {author} {\bibfnamefont {W.}~\bibnamefont {Wang}}, \bibinfo {author} {\bibfnamefont {Q.}~\bibnamefont {Yang}}, \bibinfo {author} {\bibfnamefont {K.}~\bibnamefont {Watanabe}}, \bibinfo {author} {\bibfnamefont {T.}~\bibnamefont {Taniguchi}}, \bibinfo {author} {\bibfnamefont {K.}~\bibnamefont {Liu}}, \bibinfo {author} {\bibfnamefont {Z.}~\bibnamefont {Song}}, \bibinfo {author} {\bibfnamefont {X.~C.}\ \bibnamefont {Xie}}, \bibinfo {author} {\bibfnamefont {J.}~\bibnamefont {Liu}},\ and\ \bibinfo {author} {\bibfnamefont {X.}~\bibnamefont {Lu}},\ }\href {https://doi.org/10.48550/arXiv.2405.16944} {\bibinfo {title} {Even- and odd-denominator fractional quantum anomalous hall effect in graphene
  moir{\'e} superlattices}} (\bibinfo {year} {2024}),\ \Eprint {https://arxiv.org/abs/2405.16944} {arXiv:2405.16944 [cond-mat]} \BibitemShut {NoStop}%
\bibitem [{\citenamefont {Choi}\ \emph {et~al.}(2024)\citenamefont {Choi}, \citenamefont {Choi}, \citenamefont {Valentini}, \citenamefont {Patterson}, \citenamefont {Holleis}, \citenamefont {Sheekey}, \citenamefont {Stoyanov}, \citenamefont {Cheng}, \citenamefont {Taniguchi}, \citenamefont {Watanabe},\ and\ \citenamefont {Young}}]{choi_Electric_2024}%
  \BibitemOpen
  \bibfield  {author} {\bibinfo {author} {\bibfnamefont {Y.}~\bibnamefont {Choi}}, \bibinfo {author} {\bibfnamefont {Y.}~\bibnamefont {Choi}}, \bibinfo {author} {\bibfnamefont {M.}~\bibnamefont {Valentini}}, \bibinfo {author} {\bibfnamefont {C.~L.}\ \bibnamefont {Patterson}}, \bibinfo {author} {\bibfnamefont {L.~F.~W.}\ \bibnamefont {Holleis}}, \bibinfo {author} {\bibfnamefont {O.~I.}\ \bibnamefont {Sheekey}}, \bibinfo {author} {\bibfnamefont {H.}~\bibnamefont {Stoyanov}}, \bibinfo {author} {\bibfnamefont {X.}~\bibnamefont {Cheng}}, \bibinfo {author} {\bibfnamefont {T.}~\bibnamefont {Taniguchi}}, \bibinfo {author} {\bibfnamefont {K.}~\bibnamefont {Watanabe}},\ and\ \bibinfo {author} {\bibfnamefont {A.~F.}\ \bibnamefont {Young}},\ }\href {https://doi.org/10.48550/arXiv.2408.12584} {\bibinfo {title} {Electric field control of superconductivity and quantized anomalous hall effects in rhombohedral tetralayer graphene}} (\bibinfo {year} {2024}),\ \Eprint {https://arxiv.org/abs/2408.12584} {arXiv:2408.12584
  [cond-mat]} \BibitemShut {NoStop}%
\bibitem [{\citenamefont {Lu}\ \emph {et~al.}(2024{\natexlab{b}})\citenamefont {Lu}, \citenamefont {Han}, \citenamefont {Yao}, \citenamefont {Yang}, \citenamefont {Seo}, \citenamefont {Shi}, \citenamefont {Ye}, \citenamefont {Watanabe}, \citenamefont {Taniguchi},\ and\ \citenamefont {Ju}}]{lu_Extended_2024}%
  \BibitemOpen
  \bibfield  {author} {\bibinfo {author} {\bibfnamefont {Z.}~\bibnamefont {Lu}}, \bibinfo {author} {\bibfnamefont {T.}~\bibnamefont {Han}}, \bibinfo {author} {\bibfnamefont {Y.}~\bibnamefont {Yao}}, \bibinfo {author} {\bibfnamefont {J.}~\bibnamefont {Yang}}, \bibinfo {author} {\bibfnamefont {J.}~\bibnamefont {Seo}}, \bibinfo {author} {\bibfnamefont {L.}~\bibnamefont {Shi}}, \bibinfo {author} {\bibfnamefont {S.}~\bibnamefont {Ye}}, \bibinfo {author} {\bibfnamefont {K.}~\bibnamefont {Watanabe}}, \bibinfo {author} {\bibfnamefont {T.}~\bibnamefont {Taniguchi}},\ and\ \bibinfo {author} {\bibfnamefont {L.}~\bibnamefont {Ju}},\ }\href {https://doi.org/10.48550/arXiv.2408.10203} {\bibinfo {title} {Extended quantum anomalous hall states in graphene/\(\mathrm{hBN}\) moir{\'e} superlattices}} (\bibinfo {year} {2024}{\natexlab{b}}),\ \Eprint {https://arxiv.org/abs/2408.10203} {arXiv:2408.10203} \BibitemShut {NoStop}%
\bibitem [{\citenamefont {Park}\ \emph {et~al.}(2023)\citenamefont {Park}, \citenamefont {Kim}, \citenamefont {Chittari},\ and\ \citenamefont {Jung}}]{park_Topological_2023}%
  \BibitemOpen
  \bibfield  {author} {\bibinfo {author} {\bibfnamefont {Y.}~\bibnamefont {Park}}, \bibinfo {author} {\bibfnamefont {Y.}~\bibnamefont {Kim}}, \bibinfo {author} {\bibfnamefont {B.~L.}\ \bibnamefont {Chittari}},\ and\ \bibinfo {author} {\bibfnamefont {J.}~\bibnamefont {Jung}},\ }\bibfield  {title} {\bibinfo {title} {Topological flat bands in rhombohedral tetralayer and multilayer graphene on hexagonal boron nitride moir{\'e} superlattices},\ }\href {https://doi.org/10.1103/PhysRevB.108.155406} {\bibfield  {journal} {\bibinfo  {journal} {Physical Review B}\ }\textbf {\bibinfo {volume} {108}},\ \bibinfo {pages} {155406} (\bibinfo {year} {2023})}\BibitemShut {NoStop}%
\bibitem [{\citenamefont {{Herzog-Arbeitman}}\ \emph {et~al.}(2024)\citenamefont {{Herzog-Arbeitman}}, \citenamefont {Wang}, \citenamefont {Liu}, \citenamefont {Tam}, \citenamefont {Qi}, \citenamefont {Jia}, \citenamefont {Efetov}, \citenamefont {Vafek}, \citenamefont {Regnault}, \citenamefont {Weng}, \citenamefont {Wu}, \citenamefont {Bernevig},\ and\ \citenamefont {Yu}}]{herzog-arbeitman_Moire_2024}%
  \BibitemOpen
  \bibfield  {author} {\bibinfo {author} {\bibfnamefont {J.}~\bibnamefont {{Herzog-Arbeitman}}}, \bibinfo {author} {\bibfnamefont {Y.}~\bibnamefont {Wang}}, \bibinfo {author} {\bibfnamefont {J.}~\bibnamefont {Liu}}, \bibinfo {author} {\bibfnamefont {P.~M.}\ \bibnamefont {Tam}}, \bibinfo {author} {\bibfnamefont {Z.}~\bibnamefont {Qi}}, \bibinfo {author} {\bibfnamefont {Y.}~\bibnamefont {Jia}}, \bibinfo {author} {\bibfnamefont {D.~K.}\ \bibnamefont {Efetov}}, \bibinfo {author} {\bibfnamefont {O.}~\bibnamefont {Vafek}}, \bibinfo {author} {\bibfnamefont {N.}~\bibnamefont {Regnault}}, \bibinfo {author} {\bibfnamefont {H.}~\bibnamefont {Weng}}, \bibinfo {author} {\bibfnamefont {Q.}~\bibnamefont {Wu}}, \bibinfo {author} {\bibfnamefont {B.~A.}\ \bibnamefont {Bernevig}},\ and\ \bibinfo {author} {\bibfnamefont {J.}~\bibnamefont {Yu}},\ }\bibfield  {title} {\bibinfo {title} {Moir{\'e} fractional chern insulators. ii. first-principles calculations and continuum models of rhombohedral graphene superlattices},\ }\href
  {https://doi.org/10.1103/PhysRevB.109.205122} {\bibfield  {journal} {\bibinfo  {journal} {Physical Review B}\ }\textbf {\bibinfo {volume} {109}},\ \bibinfo {pages} {205122} (\bibinfo {year} {2024})}\BibitemShut {NoStop}%
\bibitem [{\citenamefont {Kwan}\ \emph {et~al.}(2023)\citenamefont {Kwan}, \citenamefont {Yu}, \citenamefont {{Herzog-Arbeitman}}, \citenamefont {Efetov}, \citenamefont {Regnault},\ and\ \citenamefont {Bernevig}}]{kwan_Moire_2023}%
  \BibitemOpen
  \bibfield  {author} {\bibinfo {author} {\bibfnamefont {Y.~H.}\ \bibnamefont {Kwan}}, \bibinfo {author} {\bibfnamefont {J.}~\bibnamefont {Yu}}, \bibinfo {author} {\bibfnamefont {J.}~\bibnamefont {{Herzog-Arbeitman}}}, \bibinfo {author} {\bibfnamefont {D.~K.}\ \bibnamefont {Efetov}}, \bibinfo {author} {\bibfnamefont {N.}~\bibnamefont {Regnault}},\ and\ \bibinfo {author} {\bibfnamefont {B.~A.}\ \bibnamefont {Bernevig}},\ }\href {https://doi.org/10.48550/arXiv.2312.11617} {\bibinfo {title} {Moir{\'e} fractional chern insulators iii: Hartree-fock phase diagram, magic angle regime for chern insulator states, the role of the moir{\'e} potential and goldstone gaps in rhombohedral graphene superlattices}} (\bibinfo {year} {2023}),\ \Eprint {https://arxiv.org/abs/2312.11617} {arXiv:2312.11617} \BibitemShut {NoStop}%
\bibitem [{\citenamefont {Yu}\ \emph {et~al.}(2024{\natexlab{b}})\citenamefont {Yu}, \citenamefont {{Herzog-Arbeitman}}, \citenamefont {Kwan}, \citenamefont {Regnault},\ and\ \citenamefont {Bernevig}}]{yu_Moire_2024}%
  \BibitemOpen
  \bibfield  {author} {\bibinfo {author} {\bibfnamefont {J.}~\bibnamefont {Yu}}, \bibinfo {author} {\bibfnamefont {J.}~\bibnamefont {{Herzog-Arbeitman}}}, \bibinfo {author} {\bibfnamefont {Y.~H.}\ \bibnamefont {Kwan}}, \bibinfo {author} {\bibfnamefont {N.}~\bibnamefont {Regnault}},\ and\ \bibinfo {author} {\bibfnamefont {B.~A.}\ \bibnamefont {Bernevig}},\ }\href {https://doi.org/10.48550/arXiv.2407.13770} {\bibinfo {title} {Moir{\'e} fractional chern insulators iv: Fluctuation-driven collapse of fcis in multi-band exact diagonalization calculations on rhombohedral graphene}} (\bibinfo {year} {2024}{\natexlab{b}}),\ \Eprint {https://arxiv.org/abs/2407.13770} {arXiv:2407.13770} \BibitemShut {NoStop}%
\bibitem [{\citenamefont {Guo}\ \emph {et~al.}(2024)\citenamefont {Guo}, \citenamefont {Lu}, \citenamefont {Xie},\ and\ \citenamefont {Liu}}]{guo_Fractional_2024}%
  \BibitemOpen
  \bibfield  {author} {\bibinfo {author} {\bibfnamefont {Z.}~\bibnamefont {Guo}}, \bibinfo {author} {\bibfnamefont {X.}~\bibnamefont {Lu}}, \bibinfo {author} {\bibfnamefont {B.}~\bibnamefont {Xie}},\ and\ \bibinfo {author} {\bibfnamefont {J.}~\bibnamefont {Liu}},\ }\bibfield  {title} {\bibinfo {title} {Fractional chern insulator states in multilayer graphene moir{\'e} superlattices},\ }\href {https://doi.org/10.1103/PhysRevB.110.075109} {\bibfield  {journal} {\bibinfo  {journal} {Physical Review B}\ }\textbf {\bibinfo {volume} {110}},\ \bibinfo {pages} {075109} (\bibinfo {year} {2024})}\BibitemShut {NoStop}%
\bibitem [{\citenamefont {Zhou}\ \emph {et~al.}(2024)\citenamefont {Zhou}, \citenamefont {Yang},\ and\ \citenamefont {Zhang}}]{zhou_Fractional_2024}%
  \BibitemOpen
  \bibfield  {author} {\bibinfo {author} {\bibfnamefont {B.}~\bibnamefont {Zhou}}, \bibinfo {author} {\bibfnamefont {H.}~\bibnamefont {Yang}},\ and\ \bibinfo {author} {\bibfnamefont {Y.-H.}\ \bibnamefont {Zhang}},\ }\href {https://doi.org/10.48550/arXiv.2311.04217} {\bibinfo {title} {Fractional quantum anomalous hall effects in rhombohedral multilayer graphene in the moir{\'e}less limit and in coulomb imprinted superlattice}} (\bibinfo {year} {2024}),\ \Eprint {https://arxiv.org/abs/2311.04217} {arXiv:2311.04217} \BibitemShut {NoStop}%
\bibitem [{\citenamefont {Dong}\ \emph {et~al.}(2024{\natexlab{a}})\citenamefont {Dong}, \citenamefont {Wang}, \citenamefont {Wang}, \citenamefont {Soejima}, \citenamefont {Zaletel}, \citenamefont {Vishwanath},\ and\ \citenamefont {Parker}}]{dong_Anomalous_2024}%
  \BibitemOpen
  \bibfield  {author} {\bibinfo {author} {\bibfnamefont {J.}~\bibnamefont {Dong}}, \bibinfo {author} {\bibfnamefont {T.}~\bibnamefont {Wang}}, \bibinfo {author} {\bibfnamefont {T.}~\bibnamefont {Wang}}, \bibinfo {author} {\bibfnamefont {T.}~\bibnamefont {Soejima}}, \bibinfo {author} {\bibfnamefont {M.~P.}\ \bibnamefont {Zaletel}}, \bibinfo {author} {\bibfnamefont {A.}~\bibnamefont {Vishwanath}},\ and\ \bibinfo {author} {\bibfnamefont {D.~E.}\ \bibnamefont {Parker}},\ }\href {https://doi.org/10.48550/arXiv.2311.05568} {\bibinfo {title} {Anomalous hall crystals in rhombohedral multilayer graphene i: Interaction-driven chern bands and fractional quantum hall states at zero magnetic field}} (\bibinfo {year} {2024}{\natexlab{a}}),\ \Eprint {https://arxiv.org/abs/2311.05568} {arXiv:2311.05568} \BibitemShut {NoStop}%
\bibitem [{\citenamefont {Soejima}\ \emph {et~al.}(2024)\citenamefont {Soejima}, \citenamefont {Dong}, \citenamefont {Wang}, \citenamefont {Wang}, \citenamefont {Zaletel}, \citenamefont {Vishwanath},\ and\ \citenamefont {Parker}}]{soejima_Anomalous_2024}%
  \BibitemOpen
  \bibfield  {author} {\bibinfo {author} {\bibfnamefont {T.}~\bibnamefont {Soejima}}, \bibinfo {author} {\bibfnamefont {J.}~\bibnamefont {Dong}}, \bibinfo {author} {\bibfnamefont {T.}~\bibnamefont {Wang}}, \bibinfo {author} {\bibfnamefont {T.}~\bibnamefont {Wang}}, \bibinfo {author} {\bibfnamefont {M.~P.}\ \bibnamefont {Zaletel}}, \bibinfo {author} {\bibfnamefont {A.}~\bibnamefont {Vishwanath}},\ and\ \bibinfo {author} {\bibfnamefont {D.~E.}\ \bibnamefont {Parker}},\ }\href {https://doi.org/10.48550/arXiv.2403.05522} {\bibinfo {title} {Anomalous hall crystals in rhombohedral multilayer graphene ii: General mechanism and a minimal model}} (\bibinfo {year} {2024}),\ \Eprint {https://arxiv.org/abs/2403.05522} {arXiv:2403.05522} \BibitemShut {NoStop}%
\bibitem [{\citenamefont {Huang}\ \emph {et~al.}(2024{\natexlab{a}})\citenamefont {Huang}, \citenamefont {Sarma},\ and\ \citenamefont {Li}}]{huang_Impurityinduced_2024}%
  \BibitemOpen
  \bibfield  {author} {\bibinfo {author} {\bibfnamefont {K.}~\bibnamefont {Huang}}, \bibinfo {author} {\bibfnamefont {S.~D.}\ \bibnamefont {Sarma}},\ and\ \bibinfo {author} {\bibfnamefont {X.}~\bibnamefont {Li}},\ }\href {https://doi.org/10.48550/arXiv.2409.04349} {\bibinfo {title} {Impurity-induced thermal crossover in fractional chern insulators}} (\bibinfo {year} {2024}{\natexlab{a}}),\ \Eprint {https://arxiv.org/abs/2409.04349} {arXiv:2409.04349} \BibitemShut {NoStop}%
\bibitem [{\citenamefont {Tan}\ \emph {et~al.}(2024)\citenamefont {Tan}, \citenamefont {{May-Mann}},\ and\ \citenamefont {Devakul}}]{tan_Wavefunction_2024}%
  \BibitemOpen
  \bibfield  {author} {\bibinfo {author} {\bibfnamefont {T.}~\bibnamefont {Tan}}, \bibinfo {author} {\bibfnamefont {J.}~\bibnamefont {{May-Mann}}},\ and\ \bibinfo {author} {\bibfnamefont {T.}~\bibnamefont {Devakul}},\ }\href {https://doi.org/10.48550/arXiv.2409.06775} {\bibinfo {title} {Wavefunction approach to the fractional anomalous hall crystal}} (\bibinfo {year} {2024}),\ \Eprint {https://arxiv.org/abs/2409.06775} {arXiv:2409.06775} \BibitemShut {NoStop}%
\bibitem [{\citenamefont {Dong}\ \emph {et~al.}(2024{\natexlab{b}})\citenamefont {Dong}, \citenamefont {Patri},\ and\ \citenamefont {Senthil}}]{dong_Theory_2024}%
  \BibitemOpen
  \bibfield  {author} {\bibinfo {author} {\bibfnamefont {Z.}~\bibnamefont {Dong}}, \bibinfo {author} {\bibfnamefont {A.~S.}\ \bibnamefont {Patri}},\ and\ \bibinfo {author} {\bibfnamefont {T.}~\bibnamefont {Senthil}},\ }\href {https://doi.org/10.48550/arXiv.2311.03445} {\bibinfo {title} {Theory of quantum anomalous hall phases in pentalayer rhombohedral graphene moir{\'e} structures}} (\bibinfo {year} {2024}{\natexlab{b}}),\ \Eprint {https://arxiv.org/abs/2311.03445} {arXiv:2311.03445} \BibitemShut {NoStop}%
\bibitem [{\citenamefont {Huang}\ \emph {et~al.}(2024{\natexlab{b}})\citenamefont {Huang}, \citenamefont {Li}, \citenamefont {Das~Sarma},\ and\ \citenamefont {Zhang}}]{huang_Selfconsistent_2024}%
  \BibitemOpen
  \bibfield  {author} {\bibinfo {author} {\bibfnamefont {K.}~\bibnamefont {Huang}}, \bibinfo {author} {\bibfnamefont {X.}~\bibnamefont {Li}}, \bibinfo {author} {\bibfnamefont {S.}~\bibnamefont {Das~Sarma}},\ and\ \bibinfo {author} {\bibfnamefont {F.}~\bibnamefont {Zhang}},\ }\bibfield  {title} {\bibinfo {title} {Self-consistent theory of fractional quantum anomalous hall states in rhombohedral graphene},\ }\href {https://doi.org/10.1103/PhysRevB.110.115146} {\bibfield  {journal} {\bibinfo  {journal} {Physical Review B}\ }\textbf {\bibinfo {volume} {110}},\ \bibinfo {pages} {115146} (\bibinfo {year} {2024}{\natexlab{b}})}\BibitemShut {NoStop}%
\bibitem [{\citenamefont {Das~Sarma}\ and\ \citenamefont {Xie}(2024)}]{dassarma_Thermal_2024}%
  \BibitemOpen
  \bibfield  {author} {\bibinfo {author} {\bibfnamefont {S.}~\bibnamefont {Das~Sarma}}\ and\ \bibinfo {author} {\bibfnamefont {M.}~\bibnamefont {Xie}},\ }\bibfield  {title} {\bibinfo {title} {Thermal crossover from a chern insulator to a fractional chern insulator in pentalayer graphene},\ }\href {https://doi.org/10.1103/PhysRevB.110.155148} {\bibfield  {journal} {\bibinfo  {journal} {Physical Review B}\ }\textbf {\bibinfo {volume} {110}},\ \bibinfo {pages} {155148} (\bibinfo {year} {2024})}\BibitemShut {NoStop}%
\bibitem [{\citenamefont {Xie}\ and\ \citenamefont {Das~Sarma}(2024)}]{xie_Integer_2024}%
  \BibitemOpen
  \bibfield  {author} {\bibinfo {author} {\bibfnamefont {M.}~\bibnamefont {Xie}}\ and\ \bibinfo {author} {\bibfnamefont {S.}~\bibnamefont {Das~Sarma}},\ }\bibfield  {title} {\bibinfo {title} {Integer and fractional quantum anomalous hall effects in pentalayer graphene},\ }\href {https://doi.org/10.1103/PhysRevB.109.L241115} {\bibfield  {journal} {\bibinfo  {journal} {Physical Review B}\ }\textbf {\bibinfo {volume} {109}},\ \bibinfo {pages} {L241115} (\bibinfo {year} {2024})}\BibitemShut {NoStop}%
\bibitem [{\citenamefont {Dong}\ \emph {et~al.}(2024{\natexlab{c}})\citenamefont {Dong}, \citenamefont {Patri},\ and\ \citenamefont {Senthil}}]{dong_Stability_2024}%
  \BibitemOpen
  \bibfield  {author} {\bibinfo {author} {\bibfnamefont {Z.}~\bibnamefont {Dong}}, \bibinfo {author} {\bibfnamefont {A.~S.}\ \bibnamefont {Patri}},\ and\ \bibinfo {author} {\bibfnamefont {T.}~\bibnamefont {Senthil}},\ }\href {https://doi.org/10.48550/arXiv.2403.07873} {\bibinfo {title} {Stability of anomalous hall crystals in multilayer rhombohedral graphene}} (\bibinfo {year} {2024}{\natexlab{c}}),\ \Eprint {https://arxiv.org/abs/2403.07873} {arXiv:2403.07873} \BibitemShut {NoStop}%
\bibitem [{\citenamefont {Kudo}\ \emph {et~al.}(2024)\citenamefont {Kudo}, \citenamefont {Nakai},\ and\ \citenamefont {Nomura}}]{kudo_Quantum_2024}%
  \BibitemOpen
  \bibfield  {author} {\bibinfo {author} {\bibfnamefont {K.}~\bibnamefont {Kudo}}, \bibinfo {author} {\bibfnamefont {R.}~\bibnamefont {Nakai}},\ and\ \bibinfo {author} {\bibfnamefont {K.}~\bibnamefont {Nomura}},\ }\href {https://doi.org/10.48550/arXiv.2406.14354} {\bibinfo {title} {Quantum anomalous, quantum spin, and quantum valley hall effects in pentalayer rhombohedral graphene}} (\bibinfo {year} {2024}),\ \Eprint {https://arxiv.org/abs/2406.14354} {arXiv:2406.14354} \BibitemShut {NoStop}%
\bibitem [{\citenamefont {Zhou}\ and\ \citenamefont {Zhang}(2024)}]{zhou_New_2024}%
  \BibitemOpen
  \bibfield  {author} {\bibinfo {author} {\bibfnamefont {B.}~\bibnamefont {Zhou}}\ and\ \bibinfo {author} {\bibfnamefont {Y.-H.}\ \bibnamefont {Zhang}},\ }\href {https://doi.org/10.48550/arXiv.2411.04174} {\bibinfo {title} {New classes of quantum anomalous hall crystals in multilayer graphene}} (\bibinfo {year} {2024}),\ \Eprint {https://arxiv.org/abs/2411.04174} {arXiv:2411.04174} \BibitemShut {NoStop}%
\bibitem [{\citenamefont {Resta}(2006{\natexlab{a}})}]{Resta.Resta.200670o}%
  \BibitemOpen
  \bibfield  {author} {\bibinfo {author} {\bibfnamefont {R.}~\bibnamefont {Resta}},\ }\bibfield  {title} {\bibinfo {title} {{Polarization Fluctuations in Insulators and Metals: New and Old Theories Merge}},\ }\href {https://doi.org/10.1103/physrevlett.96.137601} {\bibfield  {journal} {\bibinfo  {journal} {Physical Review Letters}\ }\textbf {\bibinfo {volume} {96}},\ \bibinfo {pages} {137601} (\bibinfo {year} {2006}{\natexlab{a}})},\ \Eprint {https://arxiv.org/abs/cond-mat/0512247} {cond-mat/0512247} \BibitemShut {NoStop}%
\bibitem [{\citenamefont {King-Smith}\ and\ \citenamefont {Vanderbilt}(1993)}]{KingSmith1993}%
  \BibitemOpen
  \bibfield  {author} {\bibinfo {author} {\bibfnamefont {R.~D.}\ \bibnamefont {King-Smith}}\ and\ \bibinfo {author} {\bibfnamefont {D.}~\bibnamefont {Vanderbilt}},\ }\bibfield  {title} {\bibinfo {title} {Theory of polarization of crystalline solids},\ }\href {https://doi.org/10.1103/PhysRevB.47.1651} {\bibfield  {journal} {\bibinfo  {journal} {Phys. Rev. B}\ }\textbf {\bibinfo {volume} {47}},\ \bibinfo {pages} {1651} (\bibinfo {year} {1993})}\BibitemShut {NoStop}%
\bibitem [{\citenamefont {Aebischer}\ \emph {et~al.}(2001)\citenamefont {Aebischer}, \citenamefont {Baeriswyl},\ and\ \citenamefont {Noack}}]{Noack.Aebischer.2001}%
  \BibitemOpen
  \bibfield  {author} {\bibinfo {author} {\bibfnamefont {C.}~\bibnamefont {Aebischer}}, \bibinfo {author} {\bibfnamefont {D.}~\bibnamefont {Baeriswyl}},\ and\ \bibinfo {author} {\bibfnamefont {R.~M.}\ \bibnamefont {Noack}},\ }\bibfield  {title} {\bibinfo {title} {{Dielectric Catastrophe at the Mott Transition}},\ }\href {https://doi.org/10.1103/physrevlett.86.468} {\bibfield  {journal} {\bibinfo  {journal} {Physical Review Letters}\ }\textbf {\bibinfo {volume} {86}},\ \bibinfo {pages} {468} (\bibinfo {year} {2001})},\ \Eprint {https://arxiv.org/abs/cond-mat/0006354} {cond-mat/0006354} \BibitemShut {NoStop}%
\bibitem [{\citenamefont {Komissarov}\ \emph {et~al.}(2024{\natexlab{a}})\citenamefont {Komissarov}, \citenamefont {Holder},\ and\ \citenamefont {Queiroz}}]{komissarov2024quantum}%
  \BibitemOpen
  \bibfield  {author} {\bibinfo {author} {\bibfnamefont {I.}~\bibnamefont {Komissarov}}, \bibinfo {author} {\bibfnamefont {T.}~\bibnamefont {Holder}},\ and\ \bibinfo {author} {\bibfnamefont {R.}~\bibnamefont {Queiroz}},\ }\bibfield  {title} {\bibinfo {title} {The quantum geometric origin of capacitance in insulators},\ }\href {https://www.nature.com/articles/s41467-024-48808-x} {\bibfield  {journal} {\bibinfo  {journal} {Nat. Commun.}\ }\textbf {\bibinfo {volume} {15}},\ \bibinfo {pages} {4621} (\bibinfo {year} {2024}{\natexlab{a}})}\BibitemShut {NoStop}%
\bibitem [{\citenamefont {Verma}\ and\ \citenamefont {Queiroz}(2024{\natexlab{a}})}]{verma2024instantaneous}%
  \BibitemOpen
  \bibfield  {author} {\bibinfo {author} {\bibfnamefont {N.}~\bibnamefont {Verma}}\ and\ \bibinfo {author} {\bibfnamefont {R.}~\bibnamefont {Queiroz}},\ }\bibfield  {title} {\bibinfo {title} {Instantaneous response and quantum geometry of insulators},\ }\href {https://arxiv.org/abs/2403.07052} {\bibfield  {journal} {\bibinfo  {journal} {arXiv:2403.07052}\ } (\bibinfo {year} {2024}{\natexlab{a}})}\BibitemShut {NoStop}%
\bibitem [{\citenamefont {Resta}(2022)}]{Resta.Notes}%
  \BibitemOpen
  \bibfield  {author} {\bibinfo {author} {\bibfnamefont {R.}~\bibnamefont {Resta}},\ }\bibfield  {title} {\bibinfo {title} {{Geometry and Topology in Electronic Structure Theory}},\ } {\bibfield  {journal} {\bibinfo  {journal} {Lecture Notes}\ } (\bibinfo {year} {2022})}\BibitemShut {NoStop}%
\bibitem [{\citenamefont {Verma}\ and\ \citenamefont {Queiroz}(2024{\natexlab{b}})}]{verma2024step}%
  \BibitemOpen
  \bibfield  {author} {\bibinfo {author} {\bibfnamefont {N.}~\bibnamefont {Verma}}\ and\ \bibinfo {author} {\bibfnamefont {R.}~\bibnamefont {Queiroz}},\ }\bibfield  {title} {\bibinfo {title} {Quantum metric in step response},\ }\href {https://arxiv.org/abs/2406.17845} {\bibfield  {journal} {\bibinfo  {journal} {arXiv preprint arXiv:2406.17845}\ } (\bibinfo {year} {2024}{\natexlab{b}})}\BibitemShut {NoStop}%
\bibitem [{\citenamefont {Resta}(2006{\natexlab{b}})}]{Resta.Resta.2006.FDT}%
  \BibitemOpen
  \bibfield  {author} {\bibinfo {author} {\bibfnamefont {R.}~\bibnamefont {Resta}},\ }\bibfield  {title} {\bibinfo {title} {{Polarization Fluctuations in Insulators and Metals: New and Old Theories Merge}},\ }\href {https://doi.org/10.1103/physrevlett.96.137601} {\bibfield  {journal} {\bibinfo  {journal} {Physical Review Letters}\ }\textbf {\bibinfo {volume} {96}},\ \bibinfo {pages} {137601} (\bibinfo {year} {2006}{\natexlab{b}})},\ \Eprint {https://arxiv.org/abs/cond-mat/0512247} {cond-mat/0512247} \BibitemShut {NoStop}%
\bibitem [{\citenamefont {Verma}\ and\ \citenamefont {Queiroz}(2024{\natexlab{c}})}]{Queiroz.Verma.2024.step}%
  \BibitemOpen
  \bibfield  {author} {\bibinfo {author} {\bibfnamefont {N.}~\bibnamefont {Verma}}\ and\ \bibinfo {author} {\bibfnamefont {R.}~\bibnamefont {Queiroz}},\ }\bibfield  {title} {\bibinfo {title} {{Quantum Metric in Step Response}},\ }\bibfield  {journal} {\bibinfo  {journal} {arXiv}\ }\href {https://doi.org/10.48550/arxiv.2406.17845} {10.48550/arxiv.2406.17845} (\bibinfo {year} {2024}{\natexlab{c}}),\ \Eprint {https://arxiv.org/abs/2406.17845} {2406.17845} \BibitemShut {NoStop}%
\bibitem [{\citenamefont {Verma}\ and\ \citenamefont {Queiroz}(2024{\natexlab{d}})}]{Queiroz.Verma.2024.instantaneous}%
  \BibitemOpen
  \bibfield  {author} {\bibinfo {author} {\bibfnamefont {N.}~\bibnamefont {Verma}}\ and\ \bibinfo {author} {\bibfnamefont {R.}~\bibnamefont {Queiroz}},\ }\bibfield  {title} {\bibinfo {title} {{Instantaneous Response and Quantum Geometry of Insulators}},\ }\bibfield  {journal} {\bibinfo  {journal} {arXiv}\ }\href {https://doi.org/10.48550/arxiv.2403.07052} {10.48550/arxiv.2403.07052} (\bibinfo {year} {2024}{\natexlab{d}}),\ \Eprint {https://arxiv.org/abs/2403.07052} {2403.07052} \BibitemShut {NoStop}%
\bibitem [{\citenamefont {Komissarov}\ \emph {et~al.}(2024{\natexlab{b}})\citenamefont {Komissarov}, \citenamefont {Holder},\ and\ \citenamefont {Queiroz}}]{Queiroz.Komissarov.2024}%
  \BibitemOpen
  \bibfield  {author} {\bibinfo {author} {\bibfnamefont {I.}~\bibnamefont {Komissarov}}, \bibinfo {author} {\bibfnamefont {T.}~\bibnamefont {Holder}},\ and\ \bibinfo {author} {\bibfnamefont {R.}~\bibnamefont {Queiroz}},\ }\bibfield  {title} {\bibinfo {title} {{The quantum geometric origin of capacitance in insulators}},\ }\href {https://doi.org/10.1038/s41467-024-48808-x} {\bibfield  {journal} {\bibinfo  {journal} {Nature Communications}\ }\textbf {\bibinfo {volume} {15}},\ \bibinfo {pages} {4621} (\bibinfo {year} {2024}{\natexlab{b}})},\ \Eprint {https://arxiv.org/abs/2306.08035} {2306.08035} \BibitemShut {NoStop}%
\bibitem [{\citenamefont {Thouless}\ \emph {et~al.}(1982{\natexlab{b}})\citenamefont {Thouless}, \citenamefont {Kohmoto}, \citenamefont {Nightingale},\ and\ \citenamefont {den Nijs}}]{TKNN}%
  \BibitemOpen
  \bibfield  {author} {\bibinfo {author} {\bibfnamefont {D.~J.}\ \bibnamefont {Thouless}}, \bibinfo {author} {\bibfnamefont {M.}~\bibnamefont {Kohmoto}}, \bibinfo {author} {\bibfnamefont {M.~P.}\ \bibnamefont {Nightingale}},\ and\ \bibinfo {author} {\bibfnamefont {M.}~\bibnamefont {den Nijs}},\ }\bibfield  {title} {\bibinfo {title} {Quantized hall conductance in a two-dimensional periodic potential},\ }\href {https://doi.org/10.1103/PhysRevLett.49.405} {\bibfield  {journal} {\bibinfo  {journal} {Phys. Rev. Lett.}\ }\textbf {\bibinfo {volume} {49}},\ \bibinfo {pages} {405} (\bibinfo {year} {1982}{\natexlab{b}})}\BibitemShut {NoStop}%
\bibitem [{\citenamefont {Tam}\ \emph {et~al.}(2024{\natexlab{a}})\citenamefont {Tam}, \citenamefont {Herzog-Arbeitman},\ and\ \citenamefont {Yu}}]{Yu.Tam.2024}%
  \BibitemOpen
  \bibfield  {author} {\bibinfo {author} {\bibfnamefont {P.~M.}\ \bibnamefont {Tam}}, \bibinfo {author} {\bibfnamefont {J.}~\bibnamefont {Herzog-Arbeitman}},\ and\ \bibinfo {author} {\bibfnamefont {J.}~\bibnamefont {Yu}},\ }\bibfield  {title} {\bibinfo {title} {{Corner Charge Fluctuation as an Observable for Quantum Geometry and Entanglement in Two-dimensional Insulators}},\ }\bibfield  {journal} {\bibinfo  {journal} {arXiv}\ }\href {https://doi.org/10.48550/arxiv.2406.17023} {10.48550/arxiv.2406.17023} (\bibinfo {year} {2024}{\natexlab{a}}),\ \Eprint {https://arxiv.org/abs/2406.17023} {2406.17023} \BibitemShut {NoStop}%
\bibitem [{\citenamefont {Onishi}\ and\ \citenamefont {Fu}(2024{\natexlab{a}})}]{Fu.Onishi.2024.SF}%
  \BibitemOpen
  \bibfield  {author} {\bibinfo {author} {\bibfnamefont {Y.}~\bibnamefont {Onishi}}\ and\ \bibinfo {author} {\bibfnamefont {L.}~\bibnamefont {Fu}},\ }\bibfield  {title} {\bibinfo {title} {{Topological bound on structure factor}},\ }\bibfield  {journal} {\bibinfo  {journal} {arXiv}\ }\href {https://doi.org/10.48550/arxiv.2406.18654} {10.48550/arxiv.2406.18654} (\bibinfo {year} {2024}{\natexlab{a}}),\ \Eprint {https://arxiv.org/abs/2406.18654} {2406.18654} \BibitemShut {NoStop}%
\bibitem [{\citenamefont {Onishi}\ and\ \citenamefont {Fu}(2023)}]{Fu.Onishi.2023}%
  \BibitemOpen
  \bibfield  {author} {\bibinfo {author} {\bibfnamefont {Y.}~\bibnamefont {Onishi}}\ and\ \bibinfo {author} {\bibfnamefont {L.}~\bibnamefont {Fu}},\ }\bibfield  {title} {\bibinfo {title} {{Fundamental bound on topological gap}},\ }\bibfield  {journal} {\bibinfo  {journal} {arXiv}\ }\href {https://doi.org/10.48550/arxiv.2306.00078} {10.48550/arxiv.2306.00078} (\bibinfo {year} {2023}),\ \Eprint {https://arxiv.org/abs/2306.00078} {2306.00078} \BibitemShut {NoStop}%
\bibitem [{\citenamefont {Souza}\ and\ \citenamefont {Vanderbilt}(2008{\natexlab{a}})}]{Vanderbilt.Souza.2008}%
  \BibitemOpen
  \bibfield  {author} {\bibinfo {author} {\bibfnamefont {I.}~\bibnamefont {Souza}}\ and\ \bibinfo {author} {\bibfnamefont {D.}~\bibnamefont {Vanderbilt}},\ }\bibfield  {title} {\bibinfo {title} {{Dichroic f-sum rule and the orbital magnetization of crystals}},\ }\href {https://doi.org/10.1103/physrevb.77.054438} {\bibfield  {journal} {\bibinfo  {journal} {Physical Review B}\ }\textbf {\bibinfo {volume} {77}},\ \bibinfo {pages} {054438} (\bibinfo {year} {2008}{\natexlab{a}})},\ \Eprint {https://arxiv.org/abs/0709.2389} {0709.2389} \BibitemShut {NoStop}%
\bibitem [{\citenamefont {Thonhauser}\ \emph {et~al.}(2005)\citenamefont {Thonhauser}, \citenamefont {Ceresoli}, \citenamefont {Vanderbilt},\ and\ \citenamefont {Resta}}]{Resta.Thonhauser.2005}%
  \BibitemOpen
  \bibfield  {author} {\bibinfo {author} {\bibfnamefont {T.}~\bibnamefont {Thonhauser}}, \bibinfo {author} {\bibfnamefont {D.}~\bibnamefont {Ceresoli}}, \bibinfo {author} {\bibfnamefont {D.}~\bibnamefont {Vanderbilt}},\ and\ \bibinfo {author} {\bibfnamefont {R.}~\bibnamefont {Resta}},\ }\bibfield  {title} {\bibinfo {title} {{Orbital Magnetization in Periodic Insulators}},\ }\href {https://doi.org/10.1103/physrevlett.95.137205} {\bibfield  {journal} {\bibinfo  {journal} {Physical Review Letters}\ }\textbf {\bibinfo {volume} {95}},\ \bibinfo {pages} {137205} (\bibinfo {year} {2005})},\ \Eprint {https://arxiv.org/abs/cond-mat/0505518} {cond-mat/0505518} \BibitemShut {NoStop}%
\bibitem [{\citenamefont {Xiao}\ \emph {et~al.}(2005)\citenamefont {Xiao}, \citenamefont {Shi},\ and\ \citenamefont {Niu}}]{Niu.Xiao.2005}%
  \BibitemOpen
  \bibfield  {author} {\bibinfo {author} {\bibfnamefont {D.}~\bibnamefont {Xiao}}, \bibinfo {author} {\bibfnamefont {J.}~\bibnamefont {Shi}},\ and\ \bibinfo {author} {\bibfnamefont {Q.}~\bibnamefont {Niu}},\ }\bibfield  {title} {\bibinfo {title} {{Berry Phase Correction to Electron Density of States in Solids}},\ }\href {https://doi.org/10.1103/physrevlett.95.137204} {\bibfield  {journal} {\bibinfo  {journal} {Phys. Rev. Lett.}\ }\textbf {\bibinfo {volume} {95}},\ \bibinfo {pages} {137204} (\bibinfo {year} {2005})},\ \Eprint {https://arxiv.org/abs/cond-mat/0502340} {cond-mat/0502340} \BibitemShut {NoStop}%
\bibitem [{\citenamefont {Onishi}\ and\ \citenamefont {Fu}(2024{\natexlab{b}})}]{Fu.Onishi.2024}%
  \BibitemOpen
  \bibfield  {author} {\bibinfo {author} {\bibfnamefont {Y.}~\bibnamefont {Onishi}}\ and\ \bibinfo {author} {\bibfnamefont {L.}~\bibnamefont {Fu}},\ }\bibfield  {title} {\bibinfo {title} {{Quantum weight}},\ }\bibfield  {journal} {\bibinfo  {journal} {arXiv}\ }\href {https://doi.org/10.48550/arxiv.2406.06783} {10.48550/arxiv.2406.06783} (\bibinfo {year} {2024}{\natexlab{b}}),\ \Eprint {https://arxiv.org/abs/2406.06783} {2406.06783} \BibitemShut {NoStop}%
\bibitem [{\citenamefont {Souza}\ and\ \citenamefont {Vanderbilt}(2008{\natexlab{b}})}]{Souza2008}%
  \BibitemOpen
  \bibfield  {author} {\bibinfo {author} {\bibfnamefont {I.}~\bibnamefont {Souza}}\ and\ \bibinfo {author} {\bibfnamefont {D.}~\bibnamefont {Vanderbilt}},\ }\bibfield  {title} {\bibinfo {title} {Dichroic $f$-sum rule and the orbital magnetization of crystals},\ }\href {https://doi.org/10.1103/PhysRevB.77.054438} {\bibfield  {journal} {\bibinfo  {journal} {Phys. Rev. B}\ }\textbf {\bibinfo {volume} {77}},\ \bibinfo {pages} {054438} (\bibinfo {year} {2008}{\natexlab{b}})}\BibitemShut {NoStop}%
\bibitem [{\citenamefont {Asteria}\ \emph {et~al.}(2019)\citenamefont {Asteria}, \citenamefont {Tran}, \citenamefont {Ozawa}, \citenamefont {Tarnowski}, \citenamefont {Rem}, \citenamefont {Fläschner}, \citenamefont {Sengstock}, \citenamefont {Goldman},\ and\ \citenamefont {Weitenberg}}]{Weitenberg.Asteria.2019}%
  \BibitemOpen
  \bibfield  {author} {\bibinfo {author} {\bibfnamefont {L.}~\bibnamefont {Asteria}}, \bibinfo {author} {\bibfnamefont {D.~T.}\ \bibnamefont {Tran}}, \bibinfo {author} {\bibfnamefont {T.}~\bibnamefont {Ozawa}}, \bibinfo {author} {\bibfnamefont {M.}~\bibnamefont {Tarnowski}}, \bibinfo {author} {\bibfnamefont {B.~S.}\ \bibnamefont {Rem}}, \bibinfo {author} {\bibfnamefont {N.}~\bibnamefont {Fläschner}}, \bibinfo {author} {\bibfnamefont {K.}~\bibnamefont {Sengstock}}, \bibinfo {author} {\bibfnamefont {N.}~\bibnamefont {Goldman}},\ and\ \bibinfo {author} {\bibfnamefont {C.}~\bibnamefont {Weitenberg}},\ }\bibfield  {title} {\bibinfo {title} {{Measuring quantized circular dichroism in ultracold topological matter}},\ }\href {https://doi.org/10.1038/s41567-019-0417-8} {\bibfield  {journal} {\bibinfo  {journal} {Nature Physics}\ }\textbf {\bibinfo {volume} {15}},\ \bibinfo {pages} {449} (\bibinfo {year} {2019})},\ \Eprint {https://arxiv.org/abs/1805.11077} {1805.11077} \BibitemShut {NoStop}%
\bibitem [{\citenamefont {Neupert}\ \emph {et~al.}(2013{\natexlab{b}})\citenamefont {Neupert}, \citenamefont {Chamon},\ and\ \citenamefont {Mudry}}]{Mudry.Neupert.2013}%
  \BibitemOpen
  \bibfield  {author} {\bibinfo {author} {\bibfnamefont {T.}~\bibnamefont {Neupert}}, \bibinfo {author} {\bibfnamefont {C.}~\bibnamefont {Chamon}},\ and\ \bibinfo {author} {\bibfnamefont {C.}~\bibnamefont {Mudry}},\ }\bibfield  {title} {\bibinfo {title} {{Measuring the quantum geometry of Bloch bands with current noise}},\ }\href {https://doi.org/10.1103/physrevb.87.245103} {\bibfield  {journal} {\bibinfo  {journal} {Physical Review B}\ }\textbf {\bibinfo {volume} {87}},\ \bibinfo {pages} {245103} (\bibinfo {year} {2013}{\natexlab{b}})},\ \Eprint {https://arxiv.org/abs/1303.4643} {1303.4643} \BibitemShut {NoStop}%
\bibitem [{\citenamefont {Nozières}\ and\ \citenamefont {Pines}(1958{\natexlab{a}})}]{Pines.Nozières.1958}%
  \BibitemOpen
  \bibfield  {author} {\bibinfo {author} {\bibfnamefont {P.}~\bibnamefont {Nozières}}\ and\ \bibinfo {author} {\bibfnamefont {D.}~\bibnamefont {Pines}},\ }\bibfield  {title} {\bibinfo {title} {{Electron Interaction in Solids. General Formulation}},\ }\href {https://doi.org/10.1103/physrev.109.741} {\bibfield  {journal} {\bibinfo  {journal} {Physical Review}\ }\textbf {\bibinfo {volume} {109}},\ \bibinfo {pages} {741} (\bibinfo {year} {1958}{\natexlab{a}})}\BibitemShut {NoStop}%
\bibitem [{\citenamefont {Nozières}\ and\ \citenamefont {Pines}(1958{\natexlab{b}})}]{Pines.Nozières.19587nn}%
  \BibitemOpen
  \bibfield  {author} {\bibinfo {author} {\bibfnamefont {P.}~\bibnamefont {Nozières}}\ and\ \bibinfo {author} {\bibfnamefont {D.}~\bibnamefont {Pines}},\ }\bibfield  {title} {\bibinfo {title} {{A dielectric formulation of the many body problem: Application to the free electron gas}},\ }\href {https://doi.org/10.1007/bf02725103} {\bibfield  {journal} {\bibinfo  {journal} {Il Nuovo Cimento (1955-1965)}\ }\textbf {\bibinfo {volume} {9}},\ \bibinfo {pages} {470} (\bibinfo {year} {1958}{\natexlab{b}})}\BibitemShut {NoStop}%
\bibitem [{\citenamefont {Hazra}\ \emph {et~al.}(2019)\citenamefont {Hazra}, \citenamefont {Verma},\ and\ \citenamefont {Randeria}}]{Randeria.Hazra.2019}%
  \BibitemOpen
  \bibfield  {author} {\bibinfo {author} {\bibfnamefont {T.}~\bibnamefont {Hazra}}, \bibinfo {author} {\bibfnamefont {N.}~\bibnamefont {Verma}},\ and\ \bibinfo {author} {\bibfnamefont {M.}~\bibnamefont {Randeria}},\ }\bibfield  {title} {\bibinfo {title} {{Bounds on the Superconducting Transition Temperature: Applications to Twisted Bilayer Graphene and Cold Atoms}},\ }\href {https://doi.org/10.1103/physrevx.9.031049} {\bibfield  {journal} {\bibinfo  {journal} {Physical Review X}\ }\textbf {\bibinfo {volume} {9}},\ \bibinfo {pages} {031049} (\bibinfo {year} {2019})},\ \Eprint {https://arxiv.org/abs/1811.12428} {1811.12428} \BibitemShut {NoStop}%
\bibitem [{\citenamefont {Sachs}\ and\ \citenamefont {Austern}(1951)}]{Austern.Sachs.1951}%
  \BibitemOpen
  \bibfield  {author} {\bibinfo {author} {\bibfnamefont {R.~G.}\ \bibnamefont {Sachs}}\ and\ \bibinfo {author} {\bibfnamefont {N.}~\bibnamefont {Austern}},\ }\bibfield  {title} {\bibinfo {title} {{Consequences of Guage Invariance for Radiative Transitions}},\ }\href {https://doi.org/10.1103/physrev.81.705} {\bibfield  {journal} {\bibinfo  {journal} {Physical Review}\ }\textbf {\bibinfo {volume} {81}},\ \bibinfo {pages} {705} (\bibinfo {year} {1951})}\BibitemShut {NoStop}%
\bibitem [{\citenamefont {Iskin}(2019)}]{IskinPRA2019}%
  \BibitemOpen
  \bibfield  {author} {\bibinfo {author} {\bibfnamefont {M.}~\bibnamefont {Iskin}},\ }\bibfield  {title} {\bibinfo {title} {Geometric mass acquisition via a quantum metric: An effective-band-mass theorem for the helicity bands},\ }\href {https://doi.org/10.1103/PhysRevA.99.053603} {\bibfield  {journal} {\bibinfo  {journal} {Phys. Rev. A}\ }\textbf {\bibinfo {volume} {99}},\ \bibinfo {pages} {053603} (\bibinfo {year} {2019})}\BibitemShut {NoStop}%
\bibitem [{\citenamefont {Kivelson}(1982)}]{Kivelson1982}%
  \BibitemOpen
  \bibfield  {author} {\bibinfo {author} {\bibfnamefont {S.}~\bibnamefont {Kivelson}},\ }\bibfield  {title} {\bibinfo {title} {Wannier functions in one-dimensional disordered systems: Application to fractionally charged solitons},\ }\href {https://doi.org/10.1103/PhysRevB.26.4269} {\bibfield  {journal} {\bibinfo  {journal} {Phys. Rev. B}\ }\textbf {\bibinfo {volume} {26}},\ \bibinfo {pages} {4269} (\bibinfo {year} {1982})}\BibitemShut {NoStop}%
\bibitem [{\citenamefont {Martin}(2004)}]{Martin.Martin.2004}%
  \BibitemOpen
  \bibfield  {author} {\bibinfo {author} {\bibfnamefont {R.~M.}\ \bibnamefont {Martin}},\ }\bibfield  {title} {\bibinfo {title} {{Electronic Structure}}\ }\href {https://doi.org/10.1017/cbo9780511805769} {10.1017/cbo9780511805769} (\bibinfo {year} {2004})\BibitemShut {NoStop}%
\bibitem [{\citenamefont {Onishi}\ and\ \citenamefont {Fu}(2024{\natexlab{c}})}]{Fu.Onishi.2024.Dielectric}%
  \BibitemOpen
  \bibfield  {author} {\bibinfo {author} {\bibfnamefont {Y.}~\bibnamefont {Onishi}}\ and\ \bibinfo {author} {\bibfnamefont {L.}~\bibnamefont {Fu}},\ }\bibfield  {title} {\bibinfo {title} {{Universal relation between energy gap and dielectric constant}},\ }\bibfield  {journal} {\bibinfo  {journal} {arXiv}\ }\href {https://doi.org/10.48550/arxiv.2401.04180} {10.48550/arxiv.2401.04180} (\bibinfo {year} {2024}{\natexlab{c}}),\ \Eprint {https://arxiv.org/abs/2401.04180} {2401.04180} \BibitemShut {NoStop}%
\bibitem [{\citenamefont {Souza}\ \emph {et~al.}(2024)\citenamefont {Souza}, \citenamefont {Martin},\ and\ \citenamefont {Stengel}}]{Stengel.Souza.2024}%
  \BibitemOpen
  \bibfield  {author} {\bibinfo {author} {\bibfnamefont {I.}~\bibnamefont {Souza}}, \bibinfo {author} {\bibfnamefont {R.~M.}\ \bibnamefont {Martin}},\ and\ \bibinfo {author} {\bibfnamefont {M.}~\bibnamefont {Stengel}},\ }\bibfield  {title} {\bibinfo {title} {{Optical bounds on many-electron localization}},\ }\bibfield  {journal} {\bibinfo  {journal} {arXiv}\ }\href {https://doi.org/10.48550/arxiv.2407.17908} {10.48550/arxiv.2407.17908} (\bibinfo {year} {2024}),\ \Eprint {https://arxiv.org/abs/2407.17908} {2407.17908} \BibitemShut {NoStop}%
\bibitem [{\citenamefont {Traini}(1996)}]{Traini.Traini.1996}%
  \BibitemOpen
  \bibfield  {author} {\bibinfo {author} {\bibfnamefont {M.}~\bibnamefont {Traini}},\ }\bibfield  {title} {\bibinfo {title} {{Electric polarizability of the hydrogen atom: a sum rule approach}},\ }\href {https://doi.org/10.1088/0143-0807/17/1/006} {\bibfield  {journal} {\bibinfo  {journal} {European Journal of Physics}\ }\textbf {\bibinfo {volume} {17}},\ \bibinfo {pages} {30} (\bibinfo {year} {1996})}\BibitemShut {NoStop}%
\bibitem [{\citenamefont {Kohn}(1961)}]{Kohn.Kohn.1961}%
  \BibitemOpen
  \bibfield  {author} {\bibinfo {author} {\bibfnamefont {W.}~\bibnamefont {Kohn}},\ }\bibfield  {title} {\bibinfo {title} {{Cyclotron Resonance and de Haas-van Alphen Oscillations of an Interacting Electron Gas}},\ }\href {https://doi.org/10.1103/physrev.123.1242} {\bibfield  {journal} {\bibinfo  {journal} {Physical Review}\ }\textbf {\bibinfo {volume} {123}},\ \bibinfo {pages} {1242} (\bibinfo {year} {1961})}\BibitemShut {NoStop}%
\bibitem [{\citenamefont {Onishi}\ and\ \citenamefont {Fu}(2024{\natexlab{d}})}]{onishi2024universal}%
  \BibitemOpen
  \bibfield  {author} {\bibinfo {author} {\bibfnamefont {Y.}~\bibnamefont {Onishi}}\ and\ \bibinfo {author} {\bibfnamefont {L.}~\bibnamefont {Fu}},\ }\bibfield  {title} {\bibinfo {title} {Universal relation between energy gap and dielectric constant},\ }\href {https://arxiv.org/abs/2401.04180} {\bibfield  {journal} {\bibinfo  {journal} {arXiv preprint arXiv:2401.04180}\ } (\bibinfo {year} {2024}{\natexlab{d}})}\BibitemShut {NoStop}%
\bibitem [{\citenamefont {Onsager}(1952)}]{Onsager1952interpretation}%
  \BibitemOpen
  \bibfield  {author} {\bibinfo {author} {\bibfnamefont {L.}~\bibnamefont {Onsager}},\ }\bibfield  {title} {\bibinfo {title} {Interpretation of the {de Haas}-{van Alphen} effect},\ }\href {https://doi.org/10.1080/14786440908521019} {\bibfield  {journal} {\bibinfo  {journal} {The London, Edinburgh, and Dublin Philosophical Magazine and Journal of Science}\ }\textbf {\bibinfo {volume} {43}},\ \bibinfo {pages} {1006} (\bibinfo {year} {1952})}\BibitemShut {NoStop}%
\bibitem [{\citenamefont {Roth}(1966)}]{Roth1966semiclassical}%
  \BibitemOpen
  \bibfield  {author} {\bibinfo {author} {\bibfnamefont {L.~M.}\ \bibnamefont {Roth}},\ }\bibfield  {title} {\bibinfo {title} {Semiclassical theory of magnetic energy levels and magnetic susceptibility of {Bloch} electrons},\ }\href {https://doi.org/10.1103/PhysRev.145.434} {\bibfield  {journal} {\bibinfo  {journal} {Physical Review}\ }\textbf {\bibinfo {volume} {145}},\ \bibinfo {pages} {434} (\bibinfo {year} {1966})}\BibitemShut {NoStop}%
\bibitem [{\citenamefont {Mikitik}\ and\ \citenamefont {Sharlai}(1999)}]{Mikitik1999manifestation}%
  \BibitemOpen
  \bibfield  {author} {\bibinfo {author} {\bibfnamefont {G.}~\bibnamefont {Mikitik}}\ and\ \bibinfo {author} {\bibfnamefont {Y.~V.}\ \bibnamefont {Sharlai}},\ }\bibfield  {title} {\bibinfo {title} {Manifestation of {Berry}'s phase in metal physics},\ }\href {https://doi.org/10.1103/PhysRevLett.82.2147} {\bibfield  {journal} {\bibinfo  {journal} {Physical Review Letters}\ }\textbf {\bibinfo {volume} {82}},\ \bibinfo {pages} {2147} (\bibinfo {year} {1999})}\BibitemShut {NoStop}%
\bibitem [{\citenamefont {Gao}\ and\ \citenamefont {Niu}(2017)}]{Gao2017zero}%
  \BibitemOpen
  \bibfield  {author} {\bibinfo {author} {\bibfnamefont {Y.}~\bibnamefont {Gao}}\ and\ \bibinfo {author} {\bibfnamefont {Q.}~\bibnamefont {Niu}},\ }\bibfield  {title} {\bibinfo {title} {Zero-field magnetic response functions in {Landau} levels},\ }\href {https://doi.org/10.1073/pnas.1702595114} {\bibfield  {journal} {\bibinfo  {journal} {Proceedings of the National Academy of Sciences}\ }\textbf {\bibinfo {volume} {114}},\ \bibinfo {pages} {7295} (\bibinfo {year} {2017})}\BibitemShut {NoStop}%
\bibitem [{\citenamefont {Fuchs}\ \emph {et~al.}(2018)\citenamefont {Fuchs}, \citenamefont {Pi{\'e}chon},\ and\ \citenamefont {Montambaux}}]{Fuchs2018landau}%
  \BibitemOpen
  \bibfield  {author} {\bibinfo {author} {\bibfnamefont {J.-N.}\ \bibnamefont {Fuchs}}, \bibinfo {author} {\bibfnamefont {F.}~\bibnamefont {Pi{\'e}chon}},\ and\ \bibinfo {author} {\bibfnamefont {G.}~\bibnamefont {Montambaux}},\ }\bibfield  {title} {\bibinfo {title} {{Landau} levels, response functions and magnetic oscillations from a generalized {Onsager} relation},\ }\href {https://doi.org/10.21468/SciPostPhys.4.5.024} {\bibfield  {journal} {\bibinfo  {journal} {SciPost Physics}\ }\textbf {\bibinfo {volume} {4}},\ \bibinfo {pages} {024} (\bibinfo {year} {2018})}\BibitemShut {NoStop}%
\bibitem [{\citenamefont {Berry}(1984{\natexlab{b}})}]{Berry1984quantal}%
  \BibitemOpen
  \bibfield  {author} {\bibinfo {author} {\bibfnamefont {M.~V.}\ \bibnamefont {Berry}},\ }\bibfield  {title} {\bibinfo {title} {Quantal phase factors accompanying adiabatic changes},\ }\href {https://doi.org/10.1098/rspa.1984.0023} {\bibfield  {journal} {\bibinfo  {journal} {Proceedings of the Royal Society of London. A. Mathematical and Physical Sciences}\ }\textbf {\bibinfo {volume} {392}},\ \bibinfo {pages} {45} (\bibinfo {year} {1984}{\natexlab{b}})}\BibitemShut {NoStop}%
\bibitem [{\citenamefont {Xiao}\ \emph {et~al.}(2010)\citenamefont {Xiao}, \citenamefont {Chang},\ and\ \citenamefont {Niu}}]{Xiao2010rmp}%
  \BibitemOpen
  \bibfield  {author} {\bibinfo {author} {\bibfnamefont {D.}~\bibnamefont {Xiao}}, \bibinfo {author} {\bibfnamefont {M.-C.}\ \bibnamefont {Chang}},\ and\ \bibinfo {author} {\bibfnamefont {Q.}~\bibnamefont {Niu}},\ }\bibfield  {title} {\bibinfo {title} {{Berry} phase effects on electronic properties},\ }\href {https://doi.org/10.1103/RevModPhys.82.1959} {\bibfield  {journal} {\bibinfo  {journal} {Reviews of Modern Physics}\ }\textbf {\bibinfo {volume} {82}},\ \bibinfo {pages} {1959} (\bibinfo {year} {2010})}\BibitemShut {NoStop}%
\bibitem [{\citenamefont {Hwang}\ \emph {et~al.}(2021{\natexlab{a}})\citenamefont {Hwang}, \citenamefont {Rhim},\ and\ \citenamefont {Yang}}]{hwang2021geometric}%
  \BibitemOpen
  \bibfield  {author} {\bibinfo {author} {\bibfnamefont {Y.}~\bibnamefont {Hwang}}, \bibinfo {author} {\bibfnamefont {J.-W.}\ \bibnamefont {Rhim}},\ and\ \bibinfo {author} {\bibfnamefont {B.-J.}\ \bibnamefont {Yang}},\ }\bibfield  {title} {\bibinfo {title} {Geometric characterization of anomalous landau levels of isolated flat bands},\ }\href@noop {} {\bibfield  {journal} {\bibinfo  {journal} {Nature communications}\ }\textbf {\bibinfo {volume} {12}},\ \bibinfo {pages} {6433} (\bibinfo {year} {2021}{\natexlab{a}})}\BibitemShut {NoStop}%
\bibitem [{\citenamefont {Rhim}\ \emph {et~al.}(2020)\citenamefont {Rhim}, \citenamefont {Kim},\ and\ \citenamefont {Yang}}]{rhim2020quantum}%
  \BibitemOpen
  \bibfield  {author} {\bibinfo {author} {\bibfnamefont {J.-W.}\ \bibnamefont {Rhim}}, \bibinfo {author} {\bibfnamefont {K.}~\bibnamefont {Kim}},\ and\ \bibinfo {author} {\bibfnamefont {B.-J.}\ \bibnamefont {Yang}},\ }\bibfield  {title} {\bibinfo {title} {Quantum distance and anomalous landau levels of flat bands},\ }\href@noop {} {\bibfield  {journal} {\bibinfo  {journal} {Nature}\ }\textbf {\bibinfo {volume} {584}},\ \bibinfo {pages} {59} (\bibinfo {year} {2020})}\BibitemShut {NoStop}%
\bibitem [{\citenamefont {Jung}\ \emph {et~al.}(2024)\citenamefont {Jung}, \citenamefont {Lim},\ and\ \citenamefont {Yang}}]{jung2024quantum}%
  \BibitemOpen
  \bibfield  {author} {\bibinfo {author} {\bibfnamefont {J.}~\bibnamefont {Jung}}, \bibinfo {author} {\bibfnamefont {H.}~\bibnamefont {Lim}},\ and\ \bibinfo {author} {\bibfnamefont {B.-J.}\ \bibnamefont {Yang}},\ }\bibfield  {title} {\bibinfo {title} {Quantum geometry and landau levels of quadratic band crossings},\ }\href@noop {} {\bibfield  {journal} {\bibinfo  {journal} {Physical Review B}\ }\textbf {\bibinfo {volume} {109}},\ \bibinfo {pages} {035134} (\bibinfo {year} {2024})}\BibitemShut {NoStop}%
\bibitem [{\citenamefont {Oh}\ \emph {et~al.}(2024)\citenamefont {Oh}, \citenamefont {Rhim},\ and\ \citenamefont {Yang}}]{oh2024revisiting}%
  \BibitemOpen
  \bibfield  {author} {\bibinfo {author} {\bibfnamefont {C.-g.}\ \bibnamefont {Oh}}, \bibinfo {author} {\bibfnamefont {J.-W.}\ \bibnamefont {Rhim}},\ and\ \bibinfo {author} {\bibfnamefont {B.-J.}\ \bibnamefont {Yang}},\ }\bibfield  {title} {\bibinfo {title} {Revisiting the magnetic responses of bilayer graphene from the perspective of the quantum distance},\ }\href@noop {} {\bibfield  {journal} {\bibinfo  {journal} {arXiv preprint arXiv:2406.05939}\ } (\bibinfo {year} {2024})}\BibitemShut {NoStop}%
\bibitem [{\citenamefont {Chang}\ and\ \citenamefont {Niu}(1996)}]{chang1996berry}%
  \BibitemOpen
  \bibfield  {author} {\bibinfo {author} {\bibfnamefont {M.-C.}\ \bibnamefont {Chang}}\ and\ \bibinfo {author} {\bibfnamefont {Q.}~\bibnamefont {Niu}},\ }\bibfield  {title} {\bibinfo {title} {Berry phase, hyperorbits, and the hofstadter spectrum: Semiclassical dynamics in magnetic bloch bands},\ }\href@noop {} {\bibfield  {journal} {\bibinfo  {journal} {Physical Review B}\ }\textbf {\bibinfo {volume} {53}},\ \bibinfo {pages} {7010} (\bibinfo {year} {1996})}\BibitemShut {NoStop}%
\bibitem [{\citenamefont {Rhim}\ and\ \citenamefont {Yang}(2019)}]{rhim2019classification}%
  \BibitemOpen
  \bibfield  {author} {\bibinfo {author} {\bibfnamefont {J.-W.}\ \bibnamefont {Rhim}}\ and\ \bibinfo {author} {\bibfnamefont {B.-J.}\ \bibnamefont {Yang}},\ }\bibfield  {title} {\bibinfo {title} {Classification of flat bands according to the band-crossing singularity of bloch wave functions},\ }\href@noop {} {\bibfield  {journal} {\bibinfo  {journal} {Physical Review B}\ }\textbf {\bibinfo {volume} {99}},\ \bibinfo {pages} {045107} (\bibinfo {year} {2019})}\BibitemShut {NoStop}%
\bibitem [{\citenamefont {Rhim}\ and\ \citenamefont {Yang}(2021)}]{rhim2021singular}%
  \BibitemOpen
  \bibfield  {author} {\bibinfo {author} {\bibfnamefont {J.-W.}\ \bibnamefont {Rhim}}\ and\ \bibinfo {author} {\bibfnamefont {B.-J.}\ \bibnamefont {Yang}},\ }\bibfield  {title} {\bibinfo {title} {Singular flat bands},\ }\href@noop {} {\bibfield  {journal} {\bibinfo  {journal} {Advances in Physics: X}\ }\textbf {\bibinfo {volume} {6}},\ \bibinfo {pages} {1901606} (\bibinfo {year} {2021})}\BibitemShut {NoStop}%
\bibitem [{\citenamefont {Hwang}\ \emph {et~al.}(2021{\natexlab{b}})\citenamefont {Hwang}, \citenamefont {Rhim},\ and\ \citenamefont {Yang}}]{hwang2021general}%
  \BibitemOpen
  \bibfield  {author} {\bibinfo {author} {\bibfnamefont {Y.}~\bibnamefont {Hwang}}, \bibinfo {author} {\bibfnamefont {J.-W.}\ \bibnamefont {Rhim}},\ and\ \bibinfo {author} {\bibfnamefont {B.-J.}\ \bibnamefont {Yang}},\ }\bibfield  {title} {\bibinfo {title} {General construction of flat bands with and without band crossings based on wave function singularity},\ }\href@noop {} {\bibfield  {journal} {\bibinfo  {journal} {Physical Review B}\ }\textbf {\bibinfo {volume} {104}},\ \bibinfo {pages} {085144} (\bibinfo {year} {2021}{\natexlab{b}})}\BibitemShut {NoStop}%
\bibitem [{\citenamefont {Hwang}\ \emph {et~al.}(2021{\natexlab{c}})\citenamefont {Hwang}, \citenamefont {Jung}, \citenamefont {Rhim},\ and\ \citenamefont {Yang}}]{hwang2021wave}%
  \BibitemOpen
  \bibfield  {author} {\bibinfo {author} {\bibfnamefont {Y.}~\bibnamefont {Hwang}}, \bibinfo {author} {\bibfnamefont {J.}~\bibnamefont {Jung}}, \bibinfo {author} {\bibfnamefont {J.-W.}\ \bibnamefont {Rhim}},\ and\ \bibinfo {author} {\bibfnamefont {B.-J.}\ \bibnamefont {Yang}},\ }\bibfield  {title} {\bibinfo {title} {Wave-function geometry of band crossing points in two dimensions},\ }\href@noop {} {\bibfield  {journal} {\bibinfo  {journal} {Physical Review B}\ }\textbf {\bibinfo {volume} {103}},\ \bibinfo {pages} {L241102} (\bibinfo {year} {2021}{\natexlab{c}})}\BibitemShut {NoStop}%
\bibitem [{\citenamefont {McCann}\ and\ \citenamefont {Koshino}(2013)}]{mccann2013electronic}%
  \BibitemOpen
  \bibfield  {author} {\bibinfo {author} {\bibfnamefont {E.}~\bibnamefont {McCann}}\ and\ \bibinfo {author} {\bibfnamefont {M.}~\bibnamefont {Koshino}},\ }\bibfield  {title} {\bibinfo {title} {The electronic properties of bilayer graphene},\ }\href@noop {} {\bibfield  {journal} {\bibinfo  {journal} {Reports on Progress in physics}\ }\textbf {\bibinfo {volume} {76}},\ \bibinfo {pages} {056503} (\bibinfo {year} {2013})}\BibitemShut {NoStop}%
\bibitem [{\citenamefont {Novoselov}\ \emph {et~al.}(2006)\citenamefont {Novoselov}, \citenamefont {McCann}, \citenamefont {Morozov}, \citenamefont {Fal’ko}, \citenamefont {Katsnelson}, \citenamefont {Zeitler}, \citenamefont {Jiang}, \citenamefont {Schedin},\ and\ \citenamefont {Geim}}]{novoselov2006unconventional}%
  \BibitemOpen
  \bibfield  {author} {\bibinfo {author} {\bibfnamefont {K.~S.}\ \bibnamefont {Novoselov}}, \bibinfo {author} {\bibfnamefont {E.}~\bibnamefont {McCann}}, \bibinfo {author} {\bibfnamefont {S.}~\bibnamefont {Morozov}}, \bibinfo {author} {\bibfnamefont {V.~I.}\ \bibnamefont {Fal’ko}}, \bibinfo {author} {\bibfnamefont {M.}~\bibnamefont {Katsnelson}}, \bibinfo {author} {\bibfnamefont {U.}~\bibnamefont {Zeitler}}, \bibinfo {author} {\bibfnamefont {D.}~\bibnamefont {Jiang}}, \bibinfo {author} {\bibfnamefont {F.}~\bibnamefont {Schedin}},\ and\ \bibinfo {author} {\bibfnamefont {A.}~\bibnamefont {Geim}},\ }\bibfield  {title} {\bibinfo {title} {Unconventional quantum hall effect and berry’s phase of 2$\pi$ in bilayer graphene},\ }\href@noop {} {\bibfield  {journal} {\bibinfo  {journal} {Nature physics}\ }\textbf {\bibinfo {volume} {2}},\ \bibinfo {pages} {177} (\bibinfo {year} {2006})}\BibitemShut {NoStop}%
\bibitem [{\citenamefont {Ozawa}\ and\ \citenamefont {Mera}(2021)}]{ozawa2021relations}%
  \BibitemOpen
  \bibfield  {author} {\bibinfo {author} {\bibfnamefont {T.}~\bibnamefont {Ozawa}}\ and\ \bibinfo {author} {\bibfnamefont {B.}~\bibnamefont {Mera}},\ }\bibfield  {title} {\bibinfo {title} {Relations between topology and the quantum metric for chern insulators},\ }\href@noop {} {\bibfield  {journal} {\bibinfo  {journal} {Physical Review B}\ }\textbf {\bibinfo {volume} {104}},\ \bibinfo {pages} {045103} (\bibinfo {year} {2021})}\BibitemShut {NoStop}%
\bibitem [{\citenamefont {Haldane}(1983)}]{haldane1983fractional}%
  \BibitemOpen
  \bibfield  {author} {\bibinfo {author} {\bibfnamefont {F.~D.~M.}\ \bibnamefont {Haldane}},\ }\bibfield  {title} {\bibinfo {title} {Fractional quantization of the hall effect: A hierarchy of incompressible quantum fluid states},\ }\href@noop {} {\bibfield  {journal} {\bibinfo  {journal} {Physical Review Letters}\ }\textbf {\bibinfo {volume} {51}},\ \bibinfo {pages} {605} (\bibinfo {year} {1983})}\BibitemShut {NoStop}%
\bibitem [{\citenamefont {Wang}\ \emph {et~al.}(2021{\natexlab{c}})\citenamefont {Wang}, \citenamefont {Cano}, \citenamefont {Millis}, \citenamefont {Liu},\ and\ \citenamefont {Yang}}]{wang2021exact}%
  \BibitemOpen
  \bibfield  {author} {\bibinfo {author} {\bibfnamefont {J.}~\bibnamefont {Wang}}, \bibinfo {author} {\bibfnamefont {J.}~\bibnamefont {Cano}}, \bibinfo {author} {\bibfnamefont {A.~J.}\ \bibnamefont {Millis}}, \bibinfo {author} {\bibfnamefont {Z.}~\bibnamefont {Liu}},\ and\ \bibinfo {author} {\bibfnamefont {B.}~\bibnamefont {Yang}},\ }\bibfield  {title} {\bibinfo {title} {Exact landau level description of geometry and interaction in a flatband},\ }\href@noop {} {\bibfield  {journal} {\bibinfo  {journal} {Physical review letters}\ }\textbf {\bibinfo {volume} {127}},\ \bibinfo {pages} {246403} (\bibinfo {year} {2021}{\natexlab{c}})}\BibitemShut {NoStop}%
\bibitem [{\citenamefont {Ledwith}\ \emph {et~al.}(2023{\natexlab{b}})\citenamefont {Ledwith}, \citenamefont {Vishwanath},\ and\ \citenamefont {Parker}}]{ledwith2023vortexability}%
  \BibitemOpen
  \bibfield  {author} {\bibinfo {author} {\bibfnamefont {P.~J.}\ \bibnamefont {Ledwith}}, \bibinfo {author} {\bibfnamefont {A.}~\bibnamefont {Vishwanath}},\ and\ \bibinfo {author} {\bibfnamefont {D.~E.}\ \bibnamefont {Parker}},\ }\bibfield  {title} {\bibinfo {title} {Vortexability: A unifying criterion for ideal fractional chern insulators},\ }\href@noop {} {\bibfield  {journal} {\bibinfo  {journal} {Physical Review B}\ }\textbf {\bibinfo {volume} {108}},\ \bibinfo {pages} {205144} (\bibinfo {year} {2023}{\natexlab{b}})}\BibitemShut {NoStop}%
\bibitem [{\citenamefont {Korshunov}\ \emph {et~al.}(2023)\citenamefont {Korshunov}, \citenamefont {Hu}, \citenamefont {Subires}, \citenamefont {Jiang}, \citenamefont {C{\u{a}}lug{\u{a}}ru}, \citenamefont {Feng}, \citenamefont {Rajapitamahuni}, \citenamefont {Yi}, \citenamefont {Roychowdhury}, \citenamefont {Vergniory}, \citenamefont {Strempfer}, \citenamefont {Shekhar}, \citenamefont {Vescovo}, \citenamefont {Chernyshov}, \citenamefont {Said}, \citenamefont {Bosak}, \citenamefont {Felser}, \citenamefont {Bernevig},\ and\ \citenamefont {Blanco-Canosa}}]{santiagoblancocanosa}%
  \BibitemOpen
  \bibfield  {author} {\bibinfo {author} {\bibfnamefont {A.}~\bibnamefont {Korshunov}}, \bibinfo {author} {\bibfnamefont {H.}~\bibnamefont {Hu}}, \bibinfo {author} {\bibfnamefont {D.}~\bibnamefont {Subires}}, \bibinfo {author} {\bibfnamefont {Y.}~\bibnamefont {Jiang}}, \bibinfo {author} {\bibfnamefont {D.}~\bibnamefont {C{\u{a}}lug{\u{a}}ru}}, \bibinfo {author} {\bibfnamefont {X.}~\bibnamefont {Feng}}, \bibinfo {author} {\bibfnamefont {A.}~\bibnamefont {Rajapitamahuni}}, \bibinfo {author} {\bibfnamefont {C.}~\bibnamefont {Yi}}, \bibinfo {author} {\bibfnamefont {S.}~\bibnamefont {Roychowdhury}}, \bibinfo {author} {\bibfnamefont {M.~G.}\ \bibnamefont {Vergniory}}, \bibinfo {author} {\bibfnamefont {J.}~\bibnamefont {Strempfer}}, \bibinfo {author} {\bibfnamefont {C.}~\bibnamefont {Shekhar}}, \bibinfo {author} {\bibfnamefont {E.}~\bibnamefont {Vescovo}}, \bibinfo {author} {\bibfnamefont {D.}~\bibnamefont {Chernyshov}}, \bibinfo {author} {\bibfnamefont {A.~H.}\ \bibnamefont {Said}}, \bibinfo {author} {\bibfnamefont
  {A.}~\bibnamefont {Bosak}}, \bibinfo {author} {\bibfnamefont {C.}~\bibnamefont {Felser}}, \bibinfo {author} {\bibfnamefont {B.~A.}\ \bibnamefont {Bernevig}},\ and\ \bibinfo {author} {\bibfnamefont {S.}~\bibnamefont {Blanco-Canosa}},\ }\bibfield  {title} {\bibinfo {title} {Softening of a flat phonon mode in the kagome scv6sn6},\ }\href {https://doi.org/10.1038/s41467-023-42186-6} {\bibfield  {journal} {\bibinfo  {journal} {Nature Communications}\ }\textbf {\bibinfo {volume} {14}},\ \bibinfo {pages} {6646} (\bibinfo {year} {2023})}\BibitemShut {NoStop}%
\bibitem [{\citenamefont {Chan}\ \emph {et~al.}(2022)\citenamefont {Chan}, \citenamefont {Gr\'emaud},\ and\ \citenamefont {Batrouni}}]{Chan2022BandTouching}%
  \BibitemOpen
  \bibfield  {author} {\bibinfo {author} {\bibfnamefont {S.~M.}\ \bibnamefont {Chan}}, \bibinfo {author} {\bibfnamefont {B.}~\bibnamefont {Gr\'emaud}},\ and\ \bibinfo {author} {\bibfnamefont {G.~G.}\ \bibnamefont {Batrouni}},\ }\bibfield  {title} {\bibinfo {title} {Designer flat bands: Topology and enhancement of superconductivity},\ }\href {https://doi.org/10.1103/PhysRevB.106.104514} {\bibfield  {journal} {\bibinfo  {journal} {Phys. Rev. B}\ }\textbf {\bibinfo {volume} {106}},\ \bibinfo {pages} {104514} (\bibinfo {year} {2022})}\BibitemShut {NoStop}%
\bibitem [{\citenamefont {Li}\ \emph {et~al.}(2024)\citenamefont {Li}, \citenamefont {Deng}, \citenamefont {Chen}, \citenamefont {Efetov},\ and\ \citenamefont {Law}}]{li2024flatbandjosephsonjunctions}%
  \BibitemOpen
  \bibfield  {author} {\bibinfo {author} {\bibfnamefont {Z.~C.~F.}\ \bibnamefont {Li}}, \bibinfo {author} {\bibfnamefont {Y.}~\bibnamefont {Deng}}, \bibinfo {author} {\bibfnamefont {S.~A.}\ \bibnamefont {Chen}}, \bibinfo {author} {\bibfnamefont {D.~K.}\ \bibnamefont {Efetov}},\ and\ \bibinfo {author} {\bibfnamefont {K.~T.}\ \bibnamefont {Law}},\ }\href {https://arxiv.org/abs/2404.09211} {\bibinfo {title} {Flat band josephson junctions with quantum metric}} (\bibinfo {year} {2024}),\ \Eprint {https://arxiv.org/abs/2404.09211} {arXiv:2404.09211 [cond-mat.supr-con]} \BibitemShut {NoStop}%
\bibitem [{\citenamefont {Solnyshkov}\ \emph {et~al.}(2021)\citenamefont {Solnyshkov}, \citenamefont {Leblanc}, \citenamefont {Bessonart}, \citenamefont {Nalitov}, \citenamefont {Ren}, \citenamefont {Liao}, \citenamefont {Li},\ and\ \citenamefont {Malpuech}}]{Solnyshkov2021}%
  \BibitemOpen
  \bibfield  {author} {\bibinfo {author} {\bibfnamefont {D.~D.}\ \bibnamefont {Solnyshkov}}, \bibinfo {author} {\bibfnamefont {C.}~\bibnamefont {Leblanc}}, \bibinfo {author} {\bibfnamefont {L.}~\bibnamefont {Bessonart}}, \bibinfo {author} {\bibfnamefont {A.}~\bibnamefont {Nalitov}}, \bibinfo {author} {\bibfnamefont {J.}~\bibnamefont {Ren}}, \bibinfo {author} {\bibfnamefont {Q.}~\bibnamefont {Liao}}, \bibinfo {author} {\bibfnamefont {F.}~\bibnamefont {Li}},\ and\ \bibinfo {author} {\bibfnamefont {G.}~\bibnamefont {Malpuech}},\ }\bibfield  {title} {\bibinfo {title} {Quantum metric and wave packets at exceptional points in non-hermitian systems},\ }\href {https://doi.org/10.1103/PhysRevB.103.125302} {\bibfield  {journal} {\bibinfo  {journal} {Phys. Rev. B}\ }\textbf {\bibinfo {volume} {103}},\ \bibinfo {pages} {125302} (\bibinfo {year} {2021})}\BibitemShut {NoStop}%
\bibitem [{\citenamefont {Liao}\ \emph {et~al.}(2021)\citenamefont {Liao}, \citenamefont {Leblanc}, \citenamefont {Ren}, \citenamefont {Li}, \citenamefont {Li}, \citenamefont {Solnyshkov}, \citenamefont {Malpuech}, \citenamefont {Yao},\ and\ \citenamefont {Fu}}]{Liao2021ExceptionalPoint}%
  \BibitemOpen
  \bibfield  {author} {\bibinfo {author} {\bibfnamefont {Q.}~\bibnamefont {Liao}}, \bibinfo {author} {\bibfnamefont {C.}~\bibnamefont {Leblanc}}, \bibinfo {author} {\bibfnamefont {J.}~\bibnamefont {Ren}}, \bibinfo {author} {\bibfnamefont {F.}~\bibnamefont {Li}}, \bibinfo {author} {\bibfnamefont {Y.}~\bibnamefont {Li}}, \bibinfo {author} {\bibfnamefont {D.}~\bibnamefont {Solnyshkov}}, \bibinfo {author} {\bibfnamefont {G.}~\bibnamefont {Malpuech}}, \bibinfo {author} {\bibfnamefont {J.}~\bibnamefont {Yao}},\ and\ \bibinfo {author} {\bibfnamefont {H.}~\bibnamefont {Fu}},\ }\bibfield  {title} {\bibinfo {title} {Experimental measurement of the divergent quantum metric of an exceptional point},\ }\href {https://doi.org/10.1103/PhysRevLett.127.107402} {\bibfield  {journal} {\bibinfo  {journal} {Phys. Rev. Lett.}\ }\textbf {\bibinfo {volume} {127}},\ \bibinfo {pages} {107402} (\bibinfo {year} {2021})}\BibitemShut {NoStop}%
\bibitem [{\citenamefont {Amelio}\ and\ \citenamefont {Goldman}(2024)}]{Amelio2024Lasing}%
  \BibitemOpen
  \bibfield  {author} {\bibinfo {author} {\bibfnamefont {I.}~\bibnamefont {Amelio}}\ and\ \bibinfo {author} {\bibfnamefont {N.}~\bibnamefont {Goldman}},\ }\bibfield  {title} {\bibinfo {title} {Lasing in non-hermitian flat bands: Quantum geometry, coherence, and the fate of {Kardar-Parisi-Zhang} physics},\ }\href {https://doi.org/10.1103/PhysRevLett.132.186902} {\bibfield  {journal} {\bibinfo  {journal} {Phys. Rev. Lett.}\ }\textbf {\bibinfo {volume} {132}},\ \bibinfo {pages} {186902} (\bibinfo {year} {2024})}\BibitemShut {NoStop}%
\bibitem [{\citenamefont {Tesfaye}\ and\ \citenamefont {Eckardt}(2024)}]{tesfaye2024}%
  \BibitemOpen
  \bibfield  {author} {\bibinfo {author} {\bibfnamefont {I.}~\bibnamefont {Tesfaye}}\ and\ \bibinfo {author} {\bibfnamefont {A.}~\bibnamefont {Eckardt}},\ }\href {https://arxiv.org/abs/2406.12981} {\bibinfo {title} {Quantum geometry of bosonic bogoliubov quasiparticles}} (\bibinfo {year} {2024}),\ \Eprint {https://arxiv.org/abs/2406.12981} {arXiv:2406.12981 [cond-mat.quant-gas]} \BibitemShut {NoStop}%
\bibitem [{\citenamefont {Yu}\ \emph {et~al.}(2019)\citenamefont {Yu}, \citenamefont {Yang}, \citenamefont {Gong}, \citenamefont {Cao}, \citenamefont {Lu}, \citenamefont {Liu}, \citenamefont {Zhang}, \citenamefont {Plenio}, \citenamefont {Jelezko}, \citenamefont {Ozawa}, \citenamefont {Goldman},\ and\ \citenamefont {Cai}}]{YuNatSciRev2019}%
  \BibitemOpen
  \bibfield  {author} {\bibinfo {author} {\bibfnamefont {M.}~\bibnamefont {Yu}}, \bibinfo {author} {\bibfnamefont {P.}~\bibnamefont {Yang}}, \bibinfo {author} {\bibfnamefont {M.}~\bibnamefont {Gong}}, \bibinfo {author} {\bibfnamefont {Q.}~\bibnamefont {Cao}}, \bibinfo {author} {\bibfnamefont {Q.}~\bibnamefont {Lu}}, \bibinfo {author} {\bibfnamefont {H.}~\bibnamefont {Liu}}, \bibinfo {author} {\bibfnamefont {S.}~\bibnamefont {Zhang}}, \bibinfo {author} {\bibfnamefont {M.~B.}\ \bibnamefont {Plenio}}, \bibinfo {author} {\bibfnamefont {F.}~\bibnamefont {Jelezko}}, \bibinfo {author} {\bibfnamefont {T.}~\bibnamefont {Ozawa}}, \bibinfo {author} {\bibfnamefont {N.}~\bibnamefont {Goldman}},\ and\ \bibinfo {author} {\bibfnamefont {J.}~\bibnamefont {Cai}},\ }\bibfield  {title} {\bibinfo {title} {Experimental measurement of the quantum geometric tensor using coupled qubits in diamond},\ }\href {https://doi.org/10.1093/nsr/nwz193} {\bibfield  {journal} {\bibinfo  {journal} {Nat. Sci. Rev.}\ }\textbf {\bibinfo {volume}
  {7}},\ \bibinfo {pages} {254–260} (\bibinfo {year} {2019})}\BibitemShut {NoStop}%
\bibitem [{\citenamefont {Gianfrate}\ \emph {et~al.}(2020)\citenamefont {Gianfrate}, \citenamefont {Bleu}, \citenamefont {Dominici}, \citenamefont {Ardizzone}, \citenamefont {De~Giorgi}, \citenamefont {Ballarini}, \citenamefont {Lerario}, \citenamefont {West}, \citenamefont {Pfeiffer}, \citenamefont {Solnyshkov}, \citenamefont {Sanvitto},\ and\ \citenamefont {Malpuech}}]{gianfrate2020}%
  \BibitemOpen
  \bibfield  {author} {\bibinfo {author} {\bibfnamefont {A.}~\bibnamefont {Gianfrate}}, \bibinfo {author} {\bibfnamefont {O.}~\bibnamefont {Bleu}}, \bibinfo {author} {\bibfnamefont {L.}~\bibnamefont {Dominici}}, \bibinfo {author} {\bibfnamefont {V.}~\bibnamefont {Ardizzone}}, \bibinfo {author} {\bibfnamefont {M.}~\bibnamefont {De~Giorgi}}, \bibinfo {author} {\bibfnamefont {D.}~\bibnamefont {Ballarini}}, \bibinfo {author} {\bibfnamefont {G.}~\bibnamefont {Lerario}}, \bibinfo {author} {\bibfnamefont {K.}~\bibnamefont {West}}, \bibinfo {author} {\bibfnamefont {L.~N.}\ \bibnamefont {Pfeiffer}}, \bibinfo {author} {\bibfnamefont {D.~D.}\ \bibnamefont {Solnyshkov}}, \bibinfo {author} {\bibfnamefont {D.}~\bibnamefont {Sanvitto}},\ and\ \bibinfo {author} {\bibfnamefont {G.}~\bibnamefont {Malpuech}},\ }\bibfield  {title} {\bibinfo {title} {Measurement of the {Quantum} {Geometric} {Tensor} and of the {Anomalous} {Hall} {Drift}},\ }\href {https://doi.org/https://doi.org/10.1038/s41586-020-1989-2} {\bibfield  {journal}
  {\bibinfo  {journal} {Nature}\ }\textbf {\bibinfo {volume} {578}},\ \bibinfo {pages} {381} (\bibinfo {year} {2020})}\BibitemShut {NoStop}%
\bibitem [{\citenamefont {Zheng}\ \emph {et~al.}(2022)\citenamefont {Zheng}, \citenamefont {Xu}, \citenamefont {Ma}, \citenamefont {Li}, \citenamefont {Dong}, \citenamefont {Zhang}, \citenamefont {Wang}, \citenamefont {Sun}, \citenamefont {Wu}, \citenamefont {Zhao}, \citenamefont {Li}, \citenamefont {Lan}, \citenamefont {Tan},\ and\ \citenamefont {Yu}}]{ZhengChinPhysLett2022}%
  \BibitemOpen
  \bibfield  {author} {\bibinfo {author} {\bibfnamefont {W.}~\bibnamefont {Zheng}}, \bibinfo {author} {\bibfnamefont {J.}~\bibnamefont {Xu}}, \bibinfo {author} {\bibfnamefont {Z.}~\bibnamefont {Ma}}, \bibinfo {author} {\bibfnamefont {Y.}~\bibnamefont {Li}}, \bibinfo {author} {\bibfnamefont {Y.}~\bibnamefont {Dong}}, \bibinfo {author} {\bibfnamefont {Y.}~\bibnamefont {Zhang}}, \bibinfo {author} {\bibfnamefont {X.}~\bibnamefont {Wang}}, \bibinfo {author} {\bibfnamefont {G.}~\bibnamefont {Sun}}, \bibinfo {author} {\bibfnamefont {P.}~\bibnamefont {Wu}}, \bibinfo {author} {\bibfnamefont {J.}~\bibnamefont {Zhao}}, \bibinfo {author} {\bibfnamefont {S.}~\bibnamefont {Li}}, \bibinfo {author} {\bibfnamefont {D.}~\bibnamefont {Lan}}, \bibinfo {author} {\bibfnamefont {X.}~\bibnamefont {Tan}},\ and\ \bibinfo {author} {\bibfnamefont {Y.}~\bibnamefont {Yu}},\ }\bibfield  {title} {\bibinfo {title} {Measuring quantum geometric tensor of non-{A}belian system in superconducting circuits},\ }\href
  {https://doi.org/10.1088/0256-307X/39/10/100202} {\bibfield  {journal} {\bibinfo  {journal} {Chin. Phys. Lett.}\ }\textbf {\bibinfo {volume} {39}},\ \bibinfo {pages} {100202} (\bibinfo {year} {2022})}\BibitemShut {NoStop}%
\bibitem [{\citenamefont {Yi}\ \emph {et~al.}(2023)\citenamefont {Yi}, \citenamefont {Yu}, \citenamefont {Yuan}, \citenamefont {Jiao}, \citenamefont {Yang}, \citenamefont {Jiang}, \citenamefont {Zhang}, \citenamefont {Chen},\ and\ \citenamefont {Pan}}]{Yi2023}%
  \BibitemOpen
  \bibfield  {author} {\bibinfo {author} {\bibfnamefont {C.-R.}\ \bibnamefont {Yi}}, \bibinfo {author} {\bibfnamefont {J.}~\bibnamefont {Yu}}, \bibinfo {author} {\bibfnamefont {H.}~\bibnamefont {Yuan}}, \bibinfo {author} {\bibfnamefont {R.-H.}\ \bibnamefont {Jiao}}, \bibinfo {author} {\bibfnamefont {Y.-M.}\ \bibnamefont {Yang}}, \bibinfo {author} {\bibfnamefont {X.}~\bibnamefont {Jiang}}, \bibinfo {author} {\bibfnamefont {J.-Y.}\ \bibnamefont {Zhang}}, \bibinfo {author} {\bibfnamefont {S.}~\bibnamefont {Chen}},\ and\ \bibinfo {author} {\bibfnamefont {J.-W.}\ \bibnamefont {Pan}},\ }\bibfield  {title} {\bibinfo {title} {Extracting the quantum geometric tensor of an optical {R}aman lattice by {B}loch-state tomography},\ }\href {https://doi.org/10.1103/PhysRevResearch.5.L032016} {\bibfield  {journal} {\bibinfo  {journal} {Phys. Rev. Res.}\ }\textbf {\bibinfo {volume} {5}},\ \bibinfo {pages} {L032016} (\bibinfo {year} {2023})}\BibitemShut {NoStop}%
\bibitem [{\citenamefont {Cuerda}\ \emph {et~al.}(2024)\citenamefont {Cuerda}, \citenamefont {Taskinen}, \citenamefont {K\"allman}, \citenamefont {Grabitz},\ and\ \citenamefont {T\"orm\"a}}]{Cuerda2024}%
  \BibitemOpen
  \bibfield  {author} {\bibinfo {author} {\bibfnamefont {J.}~\bibnamefont {Cuerda}}, \bibinfo {author} {\bibfnamefont {J.~M.}\ \bibnamefont {Taskinen}}, \bibinfo {author} {\bibfnamefont {N.}~\bibnamefont {K\"allman}}, \bibinfo {author} {\bibfnamefont {L.}~\bibnamefont {Grabitz}},\ and\ \bibinfo {author} {\bibfnamefont {P.}~\bibnamefont {T\"orm\"a}},\ }\bibfield  {title} {\bibinfo {title} {Observation of quantum metric and non-hermitian berry curvature in a plasmonic lattice},\ }\href {https://doi.org/10.1103/PhysRevResearch.6.L022020} {\bibfield  {journal} {\bibinfo  {journal} {Phys. Rev. Res.}\ }\textbf {\bibinfo {volume} {6}},\ \bibinfo {pages} {L022020} (\bibinfo {year} {2024})}\BibitemShut {NoStop}%
\bibitem [{\citenamefont {{Tanaka}}\ \emph {et~al.}(2024)\citenamefont {{Tanaka}}, \citenamefont {{{\^I}-j. Wang}}, \citenamefont {{Dinh}}, \citenamefont {{Rodan-Legrain}}, \citenamefont {{Zaman}}, \citenamefont {{Hays}}, \citenamefont {{Kannan}}, \citenamefont {{Almanakly}}, \citenamefont {{Kim}}, \citenamefont {{Niedzielski}}, \citenamefont {{Serniak}}, \citenamefont {{Schwartz}}, \citenamefont {{Watanabe}}, \citenamefont {{Taniguchi}}, \citenamefont {{Grover}}, \citenamefont {{Orlando}}, \citenamefont {{Gustavsson}}, \citenamefont {{Jarillo-Herrero}},\ and\ \citenamefont {{Oliver}}}]{tanaka2024kinetic}%
  \BibitemOpen
  \bibfield  {author} {\bibinfo {author} {\bibfnamefont {M.}~\bibnamefont {{Tanaka}}}, \bibinfo {author} {\bibfnamefont {J.}~\bibnamefont {{{\^I}-j. Wang}}}, \bibinfo {author} {\bibfnamefont {T.~H.}\ \bibnamefont {{Dinh}}}, \bibinfo {author} {\bibfnamefont {D.}~\bibnamefont {{Rodan-Legrain}}}, \bibinfo {author} {\bibfnamefont {S.}~\bibnamefont {{Zaman}}}, \bibinfo {author} {\bibfnamefont {M.}~\bibnamefont {{Hays}}}, \bibinfo {author} {\bibfnamefont {B.}~\bibnamefont {{Kannan}}}, \bibinfo {author} {\bibfnamefont {A.}~\bibnamefont {{Almanakly}}}, \bibinfo {author} {\bibfnamefont {D.~K.}\ \bibnamefont {{Kim}}}, \bibinfo {author} {\bibfnamefont {B.~M.}\ \bibnamefont {{Niedzielski}}}, \bibinfo {author} {\bibfnamefont {K.}~\bibnamefont {{Serniak}}}, \bibinfo {author} {\bibfnamefont {M.~E.}\ \bibnamefont {{Schwartz}}}, \bibinfo {author} {\bibfnamefont {K.}~\bibnamefont {{Watanabe}}}, \bibinfo {author} {\bibfnamefont {T.}~\bibnamefont {{Taniguchi}}}, \bibinfo {author} {\bibfnamefont {J.~A.}\ \bibnamefont {{Grover}}},
  \bibinfo {author} {\bibfnamefont {T.~P.}\ \bibnamefont {{Orlando}}}, \bibinfo {author} {\bibfnamefont {S.}~\bibnamefont {{Gustavsson}}}, \bibinfo {author} {\bibfnamefont {P.}~\bibnamefont {{Jarillo-Herrero}}},\ and\ \bibinfo {author} {\bibfnamefont {W.~D.}\ \bibnamefont {{Oliver}}},\ }\bibfield  {title} {\bibinfo {title} {{Kinetic Inductance, Quantum Geometry, and Superconductivity in Magic-Angle Twisted Bilayer Graphene}},\ }\href {https://doi.org/10.48550/arXiv.2406.13740} {\bibfield  {journal} {\bibinfo  {journal} {arXiv e-prints}\ ,\ \bibinfo {eid} {arXiv:2406.13740}} (\bibinfo {year} {2024})},\ \Eprint {https://arxiv.org/abs/2406.13740} {arXiv:2406.13740 [cond-mat.supr-con]} \BibitemShut {NoStop}%
\bibitem [{\citenamefont {Fang}\ \emph {et~al.}(2024)\citenamefont {Fang}, \citenamefont {Cano},\ and\ \citenamefont {Ghorashi}}]{PhysRevLett.133.106701}%
  \BibitemOpen
  \bibfield  {author} {\bibinfo {author} {\bibfnamefont {Y.}~\bibnamefont {Fang}}, \bibinfo {author} {\bibfnamefont {J.}~\bibnamefont {Cano}},\ and\ \bibinfo {author} {\bibfnamefont {S.~A.~A.}\ \bibnamefont {Ghorashi}},\ }\bibfield  {title} {\bibinfo {title} {Quantum geometry induced nonlinear transport in altermagnets},\ }\href {https://doi.org/10.1103/PhysRevLett.133.106701} {\bibfield  {journal} {\bibinfo  {journal} {Phys. Rev. Lett.}\ }\textbf {\bibinfo {volume} {133}},\ \bibinfo {pages} {106701} (\bibinfo {year} {2024})}\BibitemShut {NoStop}%
\bibitem [{\citenamefont {Bałut}\ \emph {et~al.}(2024)\citenamefont {Bałut}, \citenamefont {Bradlyn},\ and\ \citenamefont {Abbamonte}}]{bałut2024quantumentanglementquantumgeometry}%
  \BibitemOpen
  \bibfield  {author} {\bibinfo {author} {\bibfnamefont {D.}~\bibnamefont {Bałut}}, \bibinfo {author} {\bibfnamefont {B.}~\bibnamefont {Bradlyn}},\ and\ \bibinfo {author} {\bibfnamefont {P.}~\bibnamefont {Abbamonte}},\ }\href {https://arxiv.org/abs/2409.15583} {\bibinfo {title} {Quantum entanglement and quantum geometry measured with inelastic x-ray scattering}} (\bibinfo {year} {2024}),\ \Eprint {https://arxiv.org/abs/2409.15583} {arXiv:2409.15583 [cond-mat.mes-hall]} \BibitemShut {NoStop}%
\bibitem [{\citenamefont {Kang}\ \emph {et~al.}(2024{\natexlab{c}})\citenamefont {Kang}, \citenamefont {Kim}, \citenamefont {Qian}, \citenamefont {Neves}, \citenamefont {Ye}, \citenamefont {Jung}, \citenamefont {Puntel}, \citenamefont {Mazzola}, \citenamefont {Fang}, \citenamefont {Jozwiak}, \citenamefont {Bostwick}, \citenamefont {Rotenberg}, \citenamefont {Fuji}, \citenamefont {Vobornik}, \citenamefont {Park}, \citenamefont {Checkelsky}, \citenamefont {Yang},\ and\ \citenamefont {Comin}}]{Kang2024QGT}%
  \BibitemOpen
  \bibfield  {author} {\bibinfo {author} {\bibfnamefont {M.}~\bibnamefont {Kang}}, \bibinfo {author} {\bibfnamefont {S.}~\bibnamefont {Kim}}, \bibinfo {author} {\bibfnamefont {Y.}~\bibnamefont {Qian}}, \bibinfo {author} {\bibfnamefont {P.~M.}\ \bibnamefont {Neves}}, \bibinfo {author} {\bibfnamefont {L.}~\bibnamefont {Ye}}, \bibinfo {author} {\bibfnamefont {J.}~\bibnamefont {Jung}}, \bibinfo {author} {\bibfnamefont {D.}~\bibnamefont {Puntel}}, \bibinfo {author} {\bibfnamefont {F.}~\bibnamefont {Mazzola}}, \bibinfo {author} {\bibfnamefont {S.}~\bibnamefont {Fang}}, \bibinfo {author} {\bibfnamefont {C.}~\bibnamefont {Jozwiak}}, \bibinfo {author} {\bibfnamefont {A.}~\bibnamefont {Bostwick}}, \bibinfo {author} {\bibfnamefont {E.}~\bibnamefont {Rotenberg}}, \bibinfo {author} {\bibfnamefont {J.}~\bibnamefont {Fuji}}, \bibinfo {author} {\bibfnamefont {I.}~\bibnamefont {Vobornik}}, \bibinfo {author} {\bibfnamefont {J.-H.}\ \bibnamefont {Park}}, \bibinfo {author} {\bibfnamefont {J.~G.}\ \bibnamefont {Checkelsky}},
  \bibinfo {author} {\bibfnamefont {B.-J.}\ \bibnamefont {Yang}},\ and\ \bibinfo {author} {\bibfnamefont {R.}~\bibnamefont {Comin}},\ }\bibfield  {title} {\bibinfo {title} {Measurements of the quantum geometric tensor in solids},\ }\href@noop {} {\bibfield  {journal} {\bibinfo  {journal} {Nature Physics to be published on Nov.25}\ } (\bibinfo {year} {2024}{\natexlab{c}})}\BibitemShut {NoStop}%
\bibitem [{\citenamefont {Kim}\ \emph {et~al.}(2024)\citenamefont {Kim}, \citenamefont {Chung}, \citenamefont {Qian}, \citenamefont {Park}, \citenamefont {Jozwiak}, \citenamefont {Rotenberg}, \citenamefont {Bostwick}, \citenamefont {Kim},\ and\ \citenamefont {Yang}}]{Kim2024QGT}%
  \BibitemOpen
  \bibfield  {author} {\bibinfo {author} {\bibfnamefont {S.}~\bibnamefont {Kim}}, \bibinfo {author} {\bibfnamefont {Y.}~\bibnamefont {Chung}}, \bibinfo {author} {\bibfnamefont {Y.}~\bibnamefont {Qian}}, \bibinfo {author} {\bibfnamefont {S.}~\bibnamefont {Park}}, \bibinfo {author} {\bibfnamefont {C.}~\bibnamefont {Jozwiak}}, \bibinfo {author} {\bibfnamefont {E.}~\bibnamefont {Rotenberg}}, \bibinfo {author} {\bibfnamefont {A.}~\bibnamefont {Bostwick}}, \bibinfo {author} {\bibfnamefont {K.~S.}\ \bibnamefont {Kim}},\ and\ \bibinfo {author} {\bibfnamefont {B.-J.}\ \bibnamefont {Yang}},\ }\bibfield  {title} {\bibinfo {title} {Direct measurement of the quantum metric tensor in solids},\ }\href@noop {} {\bibfield  {journal} {\bibinfo  {journal} {to appear}\ } (\bibinfo {year} {2024})}\BibitemShut {NoStop}%
\bibitem [{\citenamefont {Tovmasyan}\ \emph {et~al.}(2018)\citenamefont {Tovmasyan}, \citenamefont {Peotta}, \citenamefont {Liang}, \citenamefont {T\"orm\"a},\ and\ \citenamefont {Huber}}]{Tovmasyan2018}%
  \BibitemOpen
  \bibfield  {author} {\bibinfo {author} {\bibfnamefont {M.}~\bibnamefont {Tovmasyan}}, \bibinfo {author} {\bibfnamefont {S.}~\bibnamefont {Peotta}}, \bibinfo {author} {\bibfnamefont {L.}~\bibnamefont {Liang}}, \bibinfo {author} {\bibfnamefont {P.}~\bibnamefont {T\"orm\"a}},\ and\ \bibinfo {author} {\bibfnamefont {S.~D.}\ \bibnamefont {Huber}},\ }\bibfield  {title} {\bibinfo {title} {Preformed pairs in flat bloch bands},\ }\href {https://doi.org/10.1103/PhysRevB.98.134513} {\bibfield  {journal} {\bibinfo  {journal} {Phys. Rev. B}\ }\textbf {\bibinfo {volume} {98}},\ \bibinfo {pages} {134513} (\bibinfo {year} {2018})}\BibitemShut {NoStop}%
\bibitem [{\citenamefont {Han}\ \emph {et~al.}(2024)\citenamefont {Han}, \citenamefont {Herzog-Arbeitman}, \citenamefont {Bernevig},\ and\ \citenamefont {Kivelson}}]{han2024quantumgeometricnestingsolvable}%
  \BibitemOpen
  \bibfield  {author} {\bibinfo {author} {\bibfnamefont {Z.}~\bibnamefont {Han}}, \bibinfo {author} {\bibfnamefont {J.}~\bibnamefont {Herzog-Arbeitman}}, \bibinfo {author} {\bibfnamefont {B.~A.}\ \bibnamefont {Bernevig}},\ and\ \bibinfo {author} {\bibfnamefont {S.~A.}\ \bibnamefont {Kivelson}},\ }\href {https://arxiv.org/abs/2401.04163} {\bibinfo {title} {"quantum geometric nesting'' and solvable model flat-band systems}} (\bibinfo {year} {2024}),\ \Eprint {https://arxiv.org/abs/2401.04163} {arXiv:2401.04163 [cond-mat.str-el]} \BibitemShut {NoStop}%
\bibitem [{\citenamefont {Salerno}\ \emph {et~al.}(2023)\citenamefont {Salerno}, \citenamefont {Ozawa},\ and\ \citenamefont {T\"orm\"a}}]{Salerno2023}%
  \BibitemOpen
  \bibfield  {author} {\bibinfo {author} {\bibfnamefont {G.}~\bibnamefont {Salerno}}, \bibinfo {author} {\bibfnamefont {T.}~\bibnamefont {Ozawa}},\ and\ \bibinfo {author} {\bibfnamefont {P.}~\bibnamefont {T\"orm\"a}},\ }\bibfield  {title} {\bibinfo {title} {Drude weight and the many-body quantum metric in one-dimensional bose systems},\ }\href {https://doi.org/10.1103/PhysRevB.108.L140503} {\bibfield  {journal} {\bibinfo  {journal} {Phys. Rev. B}\ }\textbf {\bibinfo {volume} {108}},\ \bibinfo {pages} {L140503} (\bibinfo {year} {2023})}\BibitemShut {NoStop}%
\bibitem [{\citenamefont {Faugno}\ and\ \citenamefont {Ozawa}(2023)}]{faugno2023geometriccharacterizationbodylocalization}%
  \BibitemOpen
  \bibfield  {author} {\bibinfo {author} {\bibfnamefont {W.~N.}\ \bibnamefont {Faugno}}\ and\ \bibinfo {author} {\bibfnamefont {T.}~\bibnamefont {Ozawa}},\ }\href {https://arxiv.org/abs/2311.12280} {\bibinfo {title} {Geometric characterization of many body localization}} (\bibinfo {year} {2023}),\ \Eprint {https://arxiv.org/abs/2311.12280} {arXiv:2311.12280 [cond-mat.dis-nn]} \BibitemShut {NoStop}%
\bibitem [{\citenamefont {Tam}\ \emph {et~al.}(2024{\natexlab{b}})\citenamefont {Tam}, \citenamefont {Herzog-Arbeitman},\ and\ \citenamefont {Yu}}]{tam2024quantum}%
  \BibitemOpen
  \bibfield  {author} {\bibinfo {author} {\bibfnamefont {P.~M.}\ \bibnamefont {Tam}}, \bibinfo {author} {\bibfnamefont {J.}~\bibnamefont {Herzog-Arbeitman}},\ and\ \bibinfo {author} {\bibfnamefont {J.}~\bibnamefont {Yu}},\ }\bibfield  {title} {\bibinfo {title} {Quantum geometry and entanglement in two-dimensional insulators: A view from the corner charge fluctuation},\ }\href@noop {} {\bibfield  {journal} {\bibinfo  {journal} {arXiv preprint arXiv:2406.17023}\ } (\bibinfo {year} {2024}{\natexlab{b}})}\BibitemShut {NoStop}%
\bibitem [{\citenamefont {Wu}\ \emph {et~al.}(2024)\citenamefont {Wu}, \citenamefont {Cai}, \citenamefont {Cheng},\ and\ \citenamefont {Kumar}}]{wu2024corner}%
  \BibitemOpen
  \bibfield  {author} {\bibinfo {author} {\bibfnamefont {X.-C.}\ \bibnamefont {Wu}}, \bibinfo {author} {\bibfnamefont {K.-L.}\ \bibnamefont {Cai}}, \bibinfo {author} {\bibfnamefont {M.}~\bibnamefont {Cheng}},\ and\ \bibinfo {author} {\bibfnamefont {P.}~\bibnamefont {Kumar}},\ }\bibfield  {title} {\bibinfo {title} {Corner charge fluctuations and many-body quantum geometry},\ }\href@noop {} {\bibfield  {journal} {\bibinfo  {journal} {arXiv preprint arXiv:2408.16057}\ } (\bibinfo {year} {2024})}\BibitemShut {NoStop}%
\bibitem [{\citenamefont {Niedermeier}\ \emph {et~al.}(2024)\citenamefont {Niedermeier}, \citenamefont {Nairn}, \citenamefont {Flindt},\ and\ \citenamefont {Lado}}]{Niedermeier2024}%
  \BibitemOpen
  \bibfield  {author} {\bibinfo {author} {\bibfnamefont {M.}~\bibnamefont {Niedermeier}}, \bibinfo {author} {\bibfnamefont {M.}~\bibnamefont {Nairn}}, \bibinfo {author} {\bibfnamefont {C.}~\bibnamefont {Flindt}},\ and\ \bibinfo {author} {\bibfnamefont {J.~L.}\ \bibnamefont {Lado}},\ }\href {https://arxiv.org/abs/2404.06048} {\bibinfo {title} {Quantum computing topological invariants of two-dimensional quantum matter}} (\bibinfo {year} {2024}),\ \Eprint {https://arxiv.org/abs/2404.06048} {arXiv:2404.06048 [quant-ph]} \BibitemShut {NoStop}%
\bibitem [{\citenamefont {Chen}\ \emph {et~al.}(2024)\citenamefont {Chen}, \citenamefont {Ding}, \citenamefont {Shen}, \citenamefont {Zhu},\ and\ \citenamefont {Gong}}]{Chen2024directprobetopologygeometry}%
  \BibitemOpen
  \bibfield  {author} {\bibinfo {author} {\bibfnamefont {T.}~\bibnamefont {Chen}}, \bibinfo {author} {\bibfnamefont {H.-T.}\ \bibnamefont {Ding}}, \bibinfo {author} {\bibfnamefont {R.}~\bibnamefont {Shen}}, \bibinfo {author} {\bibfnamefont {S.-L.}\ \bibnamefont {Zhu}},\ and\ \bibinfo {author} {\bibfnamefont {J.}~\bibnamefont {Gong}},\ }\href {https://arxiv.org/abs/2403.14249} {\bibinfo {title} {Direct probe of topology and geometry of quantum states on ibm q}} (\bibinfo {year} {2024}),\ \Eprint {https://arxiv.org/abs/2403.14249} {arXiv:2403.14249 [quant-ph]} \BibitemShut {NoStop}%
\bibitem [{\citenamefont {Bernevig}\ \emph {et~al.}(2021{\natexlab{b}})\citenamefont {Bernevig}, \citenamefont {Song}, \citenamefont {Regnault},\ and\ \citenamefont {Lian}}]{bernevig2021d}%
  \BibitemOpen
  \bibfield  {author} {\bibinfo {author} {\bibfnamefont {B.~A.}\ \bibnamefont {Bernevig}}, \bibinfo {author} {\bibfnamefont {Z.-D.}\ \bibnamefont {Song}}, \bibinfo {author} {\bibfnamefont {N.}~\bibnamefont {Regnault}},\ and\ \bibinfo {author} {\bibfnamefont {B.}~\bibnamefont {Lian}},\ }\bibfield  {title} {\bibinfo {title} {Twisted bilayer graphene {{III}}. {{Interacting Hamiltonian}} and exact symmetries},\ }\href {https://doi.org/10.1103/PhysRevB.103.205413} {\bibfield  {journal} {\bibinfo  {journal} {Physical Review B}\ }\textbf {\bibinfo {volume} {103}},\ \bibinfo {pages} {205413} (\bibinfo {year} {2021}{\natexlab{b}})},\ \Eprint {https://arxiv.org/abs/2009.12376} {arXiv:2009.12376 [cond-mat]} \BibitemShut {NoStop}%
\bibitem [{\citenamefont {Bernevig}\ \emph {et~al.}(2021{\natexlab{c}})\citenamefont {Bernevig}, \citenamefont {Song}, \citenamefont {Regnault},\ and\ \citenamefont {Lian}}]{bernevig2021e}%
  \BibitemOpen
  \bibfield  {author} {\bibinfo {author} {\bibfnamefont {B.~A.}\ \bibnamefont {Bernevig}}, \bibinfo {author} {\bibfnamefont {Z.-D.}\ \bibnamefont {Song}}, \bibinfo {author} {\bibfnamefont {N.}~\bibnamefont {Regnault}},\ and\ \bibinfo {author} {\bibfnamefont {B.}~\bibnamefont {Lian}},\ }\bibfield  {title} {\bibinfo {title} {Twisted bilayer graphene {{I}}. {{Matrix}} elements, approximations, perturbation theory and a \$k{\textbackslash}cdot p\$ 2-{{Band}} model},\ }\href {https://doi.org/10.1103/PhysRevB.103.205411} {\bibfield  {journal} {\bibinfo  {journal} {Physical Review B}\ }\textbf {\bibinfo {volume} {103}},\ \bibinfo {pages} {205411} (\bibinfo {year} {2021}{\natexlab{c}})},\ \Eprint {https://arxiv.org/abs/2009.11301} {arXiv:2009.11301 [cond-mat]} \BibitemShut {NoStop}%
\bibitem [{\citenamefont {Wu}\ \emph {et~al.}(2021)\citenamefont {Wu}, \citenamefont {Zhang}, \citenamefont {Liu}, \citenamefont {Liu},\ and\ \citenamefont {Zhang}}]{wu2021}%
  \BibitemOpen
  \bibfield  {author} {\bibinfo {author} {\bibfnamefont {Y.-R.}\ \bibnamefont {Wu}}, \bibinfo {author} {\bibfnamefont {X.-F.}\ \bibnamefont {Zhang}}, \bibinfo {author} {\bibfnamefont {C.-F.}\ \bibnamefont {Liu}}, \bibinfo {author} {\bibfnamefont {W.-M.}\ \bibnamefont {Liu}},\ and\ \bibinfo {author} {\bibfnamefont {Y.-C.}\ \bibnamefont {Zhang}},\ }\bibfield  {title} {\bibinfo {title} {Superfluid density and collective modes of fermion superfluid in dice lattice},\ }\href {https://doi.org/10.1038/s41598-021-93007-z} {\bibfield  {journal} {\bibinfo  {journal} {Scientific Reports}\ }\textbf {\bibinfo {volume} {11}},\ \bibinfo {pages} {13572} (\bibinfo {year} {2021})}\BibitemShut {NoStop}%
\bibitem [{\citenamefont {Mitra}(1969)}]{Mitra1969EPC}%
  \BibitemOpen
  \bibfield  {author} {\bibinfo {author} {\bibfnamefont {T.}~\bibnamefont {Mitra}},\ }\bibfield  {title} {\bibinfo {title} {Electron-phonon interaction in the modified tight-binding approximation},\ }\href@noop {} {\bibfield  {journal} {\bibinfo  {journal} {Journal of Physics C: Solid State Physics}\ }\textbf {\bibinfo {volume} {2}},\ \bibinfo {pages} {52} (\bibinfo {year} {1969})}\BibitemShut {NoStop}%
\bibitem [{\citenamefont {McMillan}(1968)}]{McMillan1968SCTc}%
  \BibitemOpen
  \bibfield  {author} {\bibinfo {author} {\bibfnamefont {W.~L.}\ \bibnamefont {McMillan}},\ }\bibfield  {title} {\bibinfo {title} {Transition temperature of strong-coupled superconductors},\ }\href {https://doi.org/10.1103/PhysRev.167.331} {\bibfield  {journal} {\bibinfo  {journal} {Phys. Rev.}\ }\textbf {\bibinfo {volume} {167}},\ \bibinfo {pages} {331} (\bibinfo {year} {1968})}\BibitemShut {NoStop}%
\bibitem [{\citenamefont {Regnault}\ and\ \citenamefont {Bernevig}(2011{\natexlab{b}})}]{regnaultbernevig}%
  \BibitemOpen
  \bibfield  {author} {\bibinfo {author} {\bibfnamefont {N.}~\bibnamefont {Regnault}}\ and\ \bibinfo {author} {\bibfnamefont {B.~A.}\ \bibnamefont {Bernevig}},\ }\bibfield  {title} {\bibinfo {title} {Fractional chern insulator},\ }\href {https://doi.org/10.1103/PhysRevX.1.021014} {\bibfield  {journal} {\bibinfo  {journal} {Phys. Rev. X}\ }\textbf {\bibinfo {volume} {1}},\ \bibinfo {pages} {021014} (\bibinfo {year} {2011}{\natexlab{b}})}\BibitemShut {NoStop}%
  \bibitem [{\citenamefont {Bellissard}\ \emph {et~al.}(1994)\citenamefont {Bellissard}, \citenamefont {van Elst},\ and\ \citenamefont {Schulz‐~Baldes}}]{10.1063/1.530758}%
  \BibitemOpen
  \bibfield  {author} {\bibinfo {author} {\bibfnamefont {J.}~\bibnamefont {Bellissard}}, \bibinfo {author} {\bibfnamefont {A.}~\bibnamefont {van Elst}},\ and\ \bibinfo {author} {\bibfnamefont {H.}~\bibnamefont {Schulz‐~Baldes}},\ }\bibfield  {title} {\bibinfo {title} {The noncommutative geometry of the quantum hall effect},\ }\href {https://doi.org/10.1063/1.530758} {\bibfield  {journal} {\bibinfo  {journal} {Journal of Mathematical Physics}\ }\textbf {\bibinfo {volume} {35}},\ \bibinfo {pages} {5373} (\bibinfo {year} {1994})},\ \Eprint {https://arxiv.org/abs/https://pubs.aip.org/aip/jmp/article-pdf/35/10/5373/19099403/5373\_1\_online.pdf} {https://pubs.aip.org/aip/jmp/article-pdf/35/10/5373/19099403/5373\_1\_online.pdf} \BibitemShut {NoStop}%
\bibitem [{\citenamefont {Tovmasyan}\ \emph {et~al.}(2016)\citenamefont {Tovmasyan}, \citenamefont {Peotta}, \citenamefont {T\"orm\"a},\ and\ \citenamefont {Huber}}]{PhysRevB.94.245149}%
  \BibitemOpen
  \bibfield  {author} {\bibinfo {author} {\bibfnamefont {M.}~\bibnamefont {Tovmasyan}}, \bibinfo {author} {\bibfnamefont {S.}~\bibnamefont {Peotta}}, \bibinfo {author} {\bibfnamefont {P.}~\bibnamefont {T\"orm\"a}},\ and\ \bibinfo {author} {\bibfnamefont {S.~D.}\ \bibnamefont {Huber}},\ }\bibfield  {title} {\bibinfo {title} {Effective theory and emergent $\text{SU}(2)$ symmetry in the flat bands of attractive hubbard models},\ }\href {https://doi.org/10.1103/PhysRevB.94.245149} {\bibfield  {journal} {\bibinfo  {journal} {Phys. Rev. B}\ }\textbf {\bibinfo {volume} {94}},\ \bibinfo {pages} {245149} (\bibinfo {year} {2016})}\BibitemShut {NoStop}%
\bibitem [{\citenamefont {Mao}\ and\ \citenamefont {Chowdhury}(2023)}]{doi:10.1073/pnas.2217816120}%
  \BibitemOpen
  \bibfield  {author} {\bibinfo {author} {\bibfnamefont {D.}~\bibnamefont {Mao}}\ and\ \bibinfo {author} {\bibfnamefont {D.}~\bibnamefont {Chowdhury}},\ }\bibfield  {title} {\bibinfo {title} {Diamagnetic response and phase stiffness for interacting isolated narrow bands},\ }\href {https://doi.org/10.1073/pnas.2217816120} {\bibfield  {journal} {\bibinfo  {journal} {Proceedings of the National Academy of Sciences}\ }\textbf {\bibinfo {volume} {120}},\ \bibinfo {pages} {e2217816120} (\bibinfo {year} {2023})},\ \Eprint {https://arxiv.org/abs/https://www.pnas.org/doi/pdf/10.1073/pnas.2217816120} {https://www.pnas.org/doi/pdf/10.1073/pnas.2217816120} \BibitemShut {NoStop}%
\bibitem [{\citenamefont {Kruthoff}\ \emph {et~al.}(2017)\citenamefont {Kruthoff}, \citenamefont {de~Boer}, \citenamefont {van Wezel}, \citenamefont {Kane},\ and\ \citenamefont {Slager}}]{PhysRevX.7.041069}%
  \BibitemOpen
  \bibfield  {author} {\bibinfo {author} {\bibfnamefont {J.}~\bibnamefont {Kruthoff}}, \bibinfo {author} {\bibfnamefont {J.}~\bibnamefont {de~Boer}}, \bibinfo {author} {\bibfnamefont {J.}~\bibnamefont {van Wezel}}, \bibinfo {author} {\bibfnamefont {C.~L.}\ \bibnamefont {Kane}},\ and\ \bibinfo {author} {\bibfnamefont {R.-J.}\ \bibnamefont {Slager}},\ }\bibfield  {title} {\bibinfo {title} {Topological classification of crystalline insulators through band structure combinatorics},\ }\href {https://doi.org/10.1103/PhysRevX.7.041069} {\bibfield  {journal} {\bibinfo  {journal} {Phys. Rev. X}\ }\textbf {\bibinfo {volume} {7}},\ \bibinfo {pages} {041069} (\bibinfo {year} {2017})}\BibitemShut {NoStop}%
\bibitem [{\citenamefont {Andrews}\ \emph {et~al.}(2024)\citenamefont {Andrews}, \citenamefont {Raja}, \citenamefont {Mishra}, \citenamefont {Zaletel},\ and\ \citenamefont {Roy}}]{PhysRevB.109.245111}%
  \BibitemOpen
  \bibfield  {author} {\bibinfo {author} {\bibfnamefont {B.}~\bibnamefont {Andrews}}, \bibinfo {author} {\bibfnamefont {M.}~\bibnamefont {Raja}}, \bibinfo {author} {\bibfnamefont {N.}~\bibnamefont {Mishra}}, \bibinfo {author} {\bibfnamefont {M.~P.}\ \bibnamefont {Zaletel}},\ and\ \bibinfo {author} {\bibfnamefont {R.}~\bibnamefont {Roy}},\ }\bibfield  {title} {\bibinfo {title} {Stability of fractional chern insulators with a non-landau level continuum limit},\ }\href {https://doi.org/10.1103/PhysRevB.109.245111} {\bibfield  {journal} {\bibinfo  {journal} {Phys. Rev. B}\ }\textbf {\bibinfo {volume} {109}},\ \bibinfo {pages} {245111} (\bibinfo {year} {2024})}\BibitemShut {NoStop}%
\bibitem [{\citenamefont {Li}\ \emph {et~al.}(2021)\citenamefont {Li}, \citenamefont {Kumar}, \citenamefont {Sun},\ and\ \citenamefont {Lin}}]{PhysRevResearch.3.L032070}%
  \BibitemOpen
  \bibfield  {author} {\bibinfo {author} {\bibfnamefont {H.}~\bibnamefont {Li}}, \bibinfo {author} {\bibfnamefont {U.}~\bibnamefont {Kumar}}, \bibinfo {author} {\bibfnamefont {K.}~\bibnamefont {Sun}},\ and\ \bibinfo {author} {\bibfnamefont {S.-Z.}\ \bibnamefont {Lin}},\ }\bibfield  {title} {\bibinfo {title} {Spontaneous fractional chern insulators in transition metal dichalcogenide moir\'e superlattices},\ }\href {https://doi.org/10.1103/PhysRevResearch.3.L032070} {\bibfield  {journal} {\bibinfo  {journal} {Phys. Rev. Res.}\ }\textbf {\bibinfo {volume} {3}},\ \bibinfo {pages} {L032070} (\bibinfo {year} {2021})}\BibitemShut {NoStop}%
\bibitem [{\citenamefont {Wu}\ \emph {et~al.}(2024)\citenamefont {Wu}, \citenamefont {Sarkar}, \citenamefont {Wan}, \citenamefont {Sun},\ and\ \citenamefont {Lin}}]{PhysRevResearch.6.L032063}%
  \BibitemOpen
  \bibfield  {author} {\bibinfo {author} {\bibfnamefont {A.-K.}\ \bibnamefont {Wu}}, \bibinfo {author} {\bibfnamefont {S.}~\bibnamefont {Sarkar}}, \bibinfo {author} {\bibfnamefont {X.}~\bibnamefont {Wan}}, \bibinfo {author} {\bibfnamefont {K.}~\bibnamefont {Sun}},\ and\ \bibinfo {author} {\bibfnamefont {S.-Z.}\ \bibnamefont {Lin}},\ }\bibfield  {title} {\bibinfo {title} {Quantum-metric-induced quantum hall conductance inversion and reentrant transition in fractional chern insulators},\ }\href {https://doi.org/10.1103/PhysRevResearch.6.L032063} {\bibfield  {journal} {\bibinfo  {journal} {Phys. Rev. Res.}\ }\textbf {\bibinfo {volume} {6}},\ \bibinfo {pages} {L032063} (\bibinfo {year} {2024})}\BibitemShut {NoStop}%
\bibitem [{\citenamefont {Li}\ \emph {et~al.}(2024)\citenamefont {Li}, \citenamefont {Su}, \citenamefont {Kim}, \citenamefont {Kee}, \citenamefont {Sun},\ and\ \citenamefont {Lin}}]{PhysRevB.109.245131}%
  \BibitemOpen
  \bibfield  {author} {\bibinfo {author} {\bibfnamefont {H.}~\bibnamefont {Li}}, \bibinfo {author} {\bibfnamefont {Y.}~\bibnamefont {Su}}, \bibinfo {author} {\bibfnamefont {Y.~B.}\ \bibnamefont {Kim}}, \bibinfo {author} {\bibfnamefont {H.-Y.}\ \bibnamefont {Kee}}, \bibinfo {author} {\bibfnamefont {K.}~\bibnamefont {Sun}},\ and\ \bibinfo {author} {\bibfnamefont {S.-Z.}\ \bibnamefont {Lin}},\ }\bibfield  {title} {\bibinfo {title} {Contrasting twisted bilayer graphene and transition metal dichalcogenides for fractional chern insulators: An emergent gauge picture},\ }\href {https://doi.org/10.1103/PhysRevB.109.245131} {\bibfield  {journal} {\bibinfo  {journal} {Phys. Rev. B}\ }\textbf {\bibinfo {volume} {109}},\ \bibinfo {pages} {245131} (\bibinfo {year} {2024})}\BibitemShut {NoStop}%
\bibitem [{\citenamefont {Mendez-Valderrama}\ \emph {et~al.}(2024)\citenamefont {Mendez-Valderrama}, \citenamefont {Mao},\ and\ \citenamefont {Chowdhury}}]{PhysRevLett.133.196501}%
  \BibitemOpen
  \bibfield  {author} {\bibinfo {author} {\bibfnamefont {J.~F.}\ \bibnamefont {Mendez-Valderrama}}, \bibinfo {author} {\bibfnamefont {D.}~\bibnamefont {Mao}},\ and\ \bibinfo {author} {\bibfnamefont {D.}~\bibnamefont {Chowdhury}},\ }\bibfield  {title} {\bibinfo {title} {Low-energy optical sum rule in moir\'e graphene},\ }\href {https://doi.org/10.1103/PhysRevLett.133.196501} {\bibfield  {journal} {\bibinfo  {journal} {Phys. Rev. Lett.}\ }\textbf {\bibinfo {volume} {133}},\ \bibinfo {pages} {196501} (\bibinfo {year} {2024})}\BibitemShut {NoStop}%
\bibitem [{\citenamefont {Mao}\ \emph {et~al.}(2024)\citenamefont {Mao}, \citenamefont {Mendez-Valderrama},\ and\ \citenamefont {Chowdhury}}]{mao2024lowenergyopticalabsorptioncorrelated}%
  \BibitemOpen
  \bibfield  {author} {\bibinfo {author} {\bibfnamefont {D.}~\bibnamefont {Mao}}, \bibinfo {author} {\bibfnamefont {J.~F.}\ \bibnamefont {Mendez-Valderrama}},\ and\ \bibinfo {author} {\bibfnamefont {D.}~\bibnamefont {Chowdhury}},\ }\href {https://arxiv.org/abs/2410.16352} {\bibinfo {title} {Is the low-energy optical absorption in correlated insulators controlled by quantum geometry?}} (\bibinfo {year} {2024}),\ \Eprint {https://arxiv.org/abs/2410.16352} {arXiv:2410.16352 [cond-mat.str-el]} \BibitemShut {NoStop}%
\bibitem [{\citenamefont {Lim}\ \emph {et~al.}(2015)\citenamefont {Lim}, \citenamefont {Fuchs},\ and\ \citenamefont {Montambaux}}]{PhysRevA.92.063627}%
  \BibitemOpen
  \bibfield  {author} {\bibinfo {author} {\bibfnamefont {L.-K.}\ \bibnamefont {Lim}}, \bibinfo {author} {\bibfnamefont {J.-N.}\ \bibnamefont {Fuchs}},\ and\ \bibinfo {author} {\bibfnamefont {G.}~\bibnamefont {Montambaux}},\ }\bibfield  {title} {\bibinfo {title} {Geometry of bloch states probed by st\"uckelberg interferometry},\ }\href {https://doi.org/10.1103/PhysRevA.92.063627} {\bibfield  {journal} {\bibinfo  {journal} {Phys. Rev. A}\ }\textbf {\bibinfo {volume} {92}},\ \bibinfo {pages} {063627} (\bibinfo {year} {2015})}\BibitemShut {NoStop}%
\bibitem [{\citenamefont {Pi\'echon}\ \emph {et~al.}(2016)\citenamefont {Pi\'echon}, \citenamefont {Raoux}, \citenamefont {Fuchs},\ and\ \citenamefont {Montambaux}}]{PhysRevB.94.134423}%
  \BibitemOpen
  \bibfield  {author} {\bibinfo {author} {\bibfnamefont {F.}~\bibnamefont {Pi\'echon}}, \bibinfo {author} {\bibfnamefont {A.}~\bibnamefont {Raoux}}, \bibinfo {author} {\bibfnamefont {J.-N.}\ \bibnamefont {Fuchs}},\ and\ \bibinfo {author} {\bibfnamefont {G.}~\bibnamefont {Montambaux}},\ }\bibfield  {title} {\bibinfo {title} {Geometric orbital susceptibility: Quantum metric without berry curvature},\ }\href {https://doi.org/10.1103/PhysRevB.94.134423} {\bibfield  {journal} {\bibinfo  {journal} {Phys. Rev. B}\ }\textbf {\bibinfo {volume} {94}},\ \bibinfo {pages} {134423} (\bibinfo {year} {2016})}\BibitemShut {NoStop}%
\bibitem [{\citenamefont {Jankowski}\ and\ \citenamefont {Slager}(2024)}]{PhysRevLett.133.186601}%
  \BibitemOpen
  \bibfield  {author} {\bibinfo {author} {\bibfnamefont {W.~J.}\ \bibnamefont {Jankowski}}\ and\ \bibinfo {author} {\bibfnamefont {R.-J.}\ \bibnamefont {Slager}},\ }\bibfield  {title} {\bibinfo {title} {Quantized integrated shift effect in multigap topological phases},\ }\href {https://doi.org/10.1103/PhysRevLett.133.186601} {\bibfield  {journal} {\bibinfo  {journal} {Phys. Rev. Lett.}\ }\textbf {\bibinfo {volume} {133}},\ \bibinfo {pages} {186601} (\bibinfo {year} {2024})}\BibitemShut {NoStop}%
\bibitem [{\citenamefont {Yogendra}\ \emph {et~al.}(2024)\citenamefont {Yogendra}, \citenamefont {Baskaran},\ and\ \citenamefont {Das}}]{yogendra2024fractionalwannierorbitalstightbinding}%
  \BibitemOpen
  \bibfield  {author} {\bibinfo {author} {\bibfnamefont {K.~B.}\ \bibnamefont {Yogendra}}, \bibinfo {author} {\bibfnamefont {G.}~\bibnamefont {Baskaran}},\ and\ \bibinfo {author} {\bibfnamefont {T.}~\bibnamefont {Das}},\ }\href {https://arxiv.org/abs/2407.12559} {\bibinfo {title} {Fractional wannier orbitals and tight-binding gauge fields for kitaev honeycomb superlattices with flat majorana bands}} (\bibinfo {year} {2024}),\ \Eprint {https://arxiv.org/abs/2407.12559} {arXiv:2407.12559 [cond-mat.str-el]} \BibitemShut {NoStop}%
\bibitem [{\citenamefont {Gong}\ \emph {et~al.}(2024)\citenamefont {Gong}, \citenamefont {Du}, \citenamefont {Sun}, \citenamefont {Lu},\ and\ \citenamefont {Xie}}]{gong2024nonlineartransporttheoryorder}%
  \BibitemOpen
  \bibfield  {author} {\bibinfo {author} {\bibfnamefont {Z.-H.}\ \bibnamefont {Gong}}, \bibinfo {author} {\bibfnamefont {Z.~Z.}\ \bibnamefont {Du}}, \bibinfo {author} {\bibfnamefont {H.-P.}\ \bibnamefont {Sun}}, \bibinfo {author} {\bibfnamefont {H.-Z.}\ \bibnamefont {Lu}},\ and\ \bibinfo {author} {\bibfnamefont {X.~C.}\ \bibnamefont {Xie}},\ }\href {https://arxiv.org/abs/2410.04995} {\bibinfo {title} {Nonlinear transport theory at the order of quantum metric}} (\bibinfo {year} {2024}),\ \Eprint {https://arxiv.org/abs/2410.04995} {arXiv:2410.04995 [cond-mat.mes-hall]} \BibitemShut {NoStop}%
  \bibitem [{\citenamefont {Bouhon}\ \emph {et~al.}(2023)\citenamefont {Bouhon}, \citenamefont {Timmel},\ and\ \citenamefont {Slager}}]{bouhon2023quantumgeometryprojectivesingle}%
  \BibitemOpen
  \bibfield  {author} {\bibinfo {author} {\bibfnamefont {A.}~\bibnamefont {Bouhon}}, \bibinfo {author} {\bibfnamefont {A.}~\bibnamefont {Timmel}},\ and\ \bibinfo {author} {\bibfnamefont {R.-J.}\ \bibnamefont {Slager}},\ }\href {https://arxiv.org/abs/2303.02180} {\bibinfo {title} {Quantum geometry beyond projective single bands}} (\bibinfo {year} {2023}),\ \Eprint {https://arxiv.org/abs/2303.02180} {arXiv:2303.02180 [cond-mat.mes-hall]} \BibitemShut {NoStop}%
\end{thebibliography}

%

\appendix

\section{Quantum geometric tensor (Fubini-Study metric)}\label{sec:metric}

\subsection{General Mathematical Formulation}

This section contains the formulation of the QGT in mathematical language. We present it here to link to the available mathematical physics literature.  

Suppose we have $N$ orthonormal row vectors $u_{n \bsl{k}}$, $n=1\ldots N$, in a Hilbert space, where $\bsl{k}$ is a vector of generic parameter. 
The QGT is defined as:
\eqa{
    \left[ Q_{ij}(\bsl{k}) \right]_{mn} =  \partial_{k_i} u^\dagger_{m\bsl{k}}\left(1 - \sum_{l=1}^N u_{l \bsl{k}}u^\dagger_{l\bsl{k}} \right)\partial_{k_j}u_{n\bsl{k}}\ ,
}
where $i,j$ are spatial direction indices. 
For convenience, we denote $\tilde{u}_{\bsl{k}} = (u_{1 \bsl{k}}, u_{2 \bsl{k}},\cdots,  u_{N \bsl{k}})$.
Using $\tilde{u}_{\bsl{k}}$, the QGT can be expressed compactly : 
\begin{equation}
   Q_{ij}(\bsl{k}) = \partial_{k_i}\tilde{u}_{\bsl{k}}^\dagger\left(\mathds{1} - \tilde{u}_{\bsl{k}}\tilde{u}^\dagger_{\bsl{k}}\right)\partial_{k_j}\tilde{u}_{\bsl{k}}\,.
\end{equation}

We use $G_{ij}(\bsl{k})$ to denote $\left[Q_{ij}(\bsl{k})+Q_{ji}(\bsl{k})\right]/2$, and $g_{ij}(\bsl{k}) = {\rm Tr}\left[G_{ij}(\bsl{k}) \right]$ is the quantum metric. As the Berry connection is defined by $\bsl{A}(\bsl{k}) = \ii\tilde{u}^\dagger_{\bsl{k}}\partial_{\bsl{k}}\tilde{u}_{\bsl{k}}$, then the anti-symmetric part of $Q_{ij}$ is proportional to the nonabelian Berry curvature: $\left[ Q_{ij}(\bsl{k})-Q_{ji}(\bsl{k})\right]/2= -\frac{\ii }{2} F_{ij}(\bsl{k}) = -\frac{\ii }{2} \left(\partial_{k_i}A_j(\bsl{k}) -\partial_{k_j}A_i(\bsl{k}) - \ii[A_i(\bsl{k}),A_j(\bsl{k})]\right)$. The QGT is then
\begin{equation}
  Q_{ij}(\bsl{k}) = G_{ij}(\bsl{k}) -\frac{\ii}{2}F_{ij}(\bsl{k})\,,
\end{equation}
 An important property of $G_{ij}$ is its positive definiteness. Suppose we have several complex vectors $c_i \in \mathbb{C}^n$, we obtain:
\eqa{
 & \sum_{ij}c^\dagger_i Q_{ij}(\bsl{k})c_j \\
 & = \sum_{ij}\sum_{l,m=1}^n c^*_{i,l}\left[Q_{ij}(\bsl{k})\right]_{lm} c_{j,m} \\
 & =  \left(\sum_i c^\dagger_i \partial_{k_i}\tilde{u}^\dagger_{\bsl{k}} \right)\left(\mathds{1}-\tilde{u}_{\bsl{k}}\tilde{u}^\dagger_{\bsl{k}}\right) \left(\sum_i \partial_{k_i}\tilde{u}_{\bsl{k}}c_i \right) \\
 & =  \varphi^\dagger \left(\mathds{1}-\tilde{u}_{\bsl{k}}\tilde{u}^\dagger_{\bsl{k}}\right) \varphi \ ,\label{eqn:provepositive}
}
where $ \varphi =  \sum_i \partial_{k_i}\tilde{u}_{\bsl{k}}c_i $.
As $\{u_m(\bsl{k})\}$ are orthonormal vectors, the matrix $\left(\mathds{1}-\tilde{u}_{\bsl{k}}\tilde{u}^\dagger_{\bsl{k}}\right)$ is a projector with eigenvalue $0$ or $1$. Therefore the scalar product $\varphi^\dagger\left(\mathds{1}-\tilde{u}_{\bsl{k}}\tilde{u}^\dagger_{\bsl{k}}\right)\varphi$ is  non-negative. For the proper choice of complex vectors $c_i$ properly, \cref{eqn:provepositive} can be used to prove inequalities between the quantum metric and Berry curvature, \ie, the so-called lower bounds.

\subsection{Geometric Interpretation of Quantum Metric}
The geometric meaning of the quantum metric is the distance between quantum states. The Bloch wave functions of $N$ bands $\tilde{u}_{\bsl{k}}$ define a map from the Brillouin zone torus to $\mathbb{CP}^{N-1}$. The distance between two points $\bsl{k}$ and $\bsl{k} + d\bsl{k}$ is defined as:
\begin{equation}
  d^2(\bsl{k},\bsl{k}+ d\bsl{k}) = \frac{1}{2}{\rm Tr}\left(\tilde{u}_{\bsl{k}}\tilde{u}^\dagger_{\bsl{k}} -\tilde{u}_{\bsl{k} + d\bsl{k}}\tilde{u}^\dagger_{\bsl{k} + d\bsl{k}} \right)^2.
\end{equation} By expanding this equation to the second order we will find $d^2(\bsl{k},\bsl{k} + d\bsl{k}) = \sum_{ij} g_{ij}(\bsl{k})dk_idk_j$ and represents a direct link between the quantum distance and the QGT.

\subsection{Physical Interpretation} 

Now we focus on the quantum geometry in condensed matter system, where $\bsl{k}$ is the Bloch momentum, and $u_{n\bsl{k}}$ should be replaced by $\ket{u_{n\bsl{k}}}$, which is the periodic part of the Bloch state $\ket{\psi_{n\bsl{k}}}$.
The quantum metric is related to the Wannier function localization which is studied in \refcite{Vanderbilt.Marzari.1997}. 
Here the Wannier states is the Fourier transformation of the Bloch state:
\begin{equation}\label{eqn:wannierdef}
  |\bsl{R}n\rangle = \frac{1}{\sqrt{N}}\sum_{\bsl{k}}e^{-i\bsl{k}\cdot\bsl{R}}|\psi_{n\bsl{k}}\rangle\ ,
\end{equation}
where $N$ here is the number of lattice sites.
The Wannier function localization functional can be defined in the Wannier basis:
\begin{equation}
  \Omega = \sum_n\left[\langle 0 n | \hat{{\bsl r}}^2 |0n\rangle - |\langle 0 n|\hat{\bsl{r}}|0n \rangle |^2\right]\,,
\end{equation}
where $\hat{\bsl{r}}$ is the position operator. We now express it in the Bloch basis.
Owing to
$$
|\psi_{n \bsl{k}} \rangle = e^{i\bsl{k}\cdot \hat{\bsl{r}}} | u_{n \bsl{k}}\rangle\ ,
$$
the overlap between the periodic parts of two Bloch functions with different momenta is: 
\begin{equation}\label{eqn:overlap}
    \langle u_{m \bsl{k}}|u_{n \bsl{k}+ \bsl{q}}\rangle = \langle \psi_{m\bsl{k}}| e^{-i\bsl{q}\cdot \hat{\bsl{r}}}|\psi_{n{\bsl{k}+\bsl{q}}}\rangle\,.
  \end{equation} 
The right-hand side of this equation is the Bloch states. We can transform it into Wannier states:
\begin{equation}
  \langle u_{m \bsl{k}}|u_{n \bsl{k}+ \bsl{q}}\rangle = \frac{1}{N} \sum_{\bsl{R}\bsl{R}'}e^{-i\bsl{k}\cdot (\bsl{R}'-\bsl{R})}\langle \bsl{R}' m|e^{-i\bsl{q}\cdot \hat{\bsl{r}}}|\bsl{R}n\rangle e^{i\bsl{q}\cdot\bsl{R}} \ .
\end{equation} 
The Wannier functions have a discrete translation symmetry along the Bravias lattice:
$$
\langle \bsl{R}'m|e^{-i\bsl{q}\cdot \hat{\bsl{r}}}|\bsl{R}n\rangle e^{i\bsl{q}\cdot \bsl{R}} = \langle (\bsl{R}'-\bsl{R})m|  e^{-i\bsl{q}\cdot \hat{\bsl{r}}}| 0n\rangle\,,
$$
 Eq. (\ref{eqn:overlap}) becomes
$$
  \langle u_{m \bsl{k}}|u_{n \bsl{k}+ \bsl{q}}\rangle = \sum_{\bsl{R}}e^{-i\bsl{k}\cdot\bsl{R}}\langle \bsl{R}m| e^{-i\bsl{q}\cdot \hat{\bsl{r}}}|0n\rangle\,.
$$
We take the first and second-order derivatives of $\bsl{q}$ on both sides of this equation, and then evaluate the result at $q=0$:
\begin{align}
  \bra{u_{m \bsl{k}}}\nabla_{\bsl{k}}\ket{u_{n \bsl{k}}} &= -\ii\sum_{\bsl{R}} e^{-i\bsl{k}\cdot \bsl{R}}\langle \bsl{R}m|\hat{\bsl{r}}|0n\rangle \label{eqn:rexpectationk}\\
    \bra{u_{m \bsl{k}}}\nabla^2_{\bsl{k}}\ket{u_{n \bsl{k}}} & = -\sum_{\bsl{R}}e^{-i\bsl{k}\cdot \bsl{R}}\langle \bsl{R}m|\hat{\bsl{r}}^2|0n\rangle \label{eqn:r2expectationk}\,.
\end{align}
Taking the Fourier transformation
\begin{align}
  \langle \bsl{R}m|\hat{\bsl{r}}|0n\rangle &= \ii \frac{1}{N}\sum_{\bsl{k}}\,\bra{u_m(\bsl{k})}\nabla_{\bsl{k}}\ket{u_{n \bsl{k}}} e^{i\bsl{k}\cdot\bsl{R}}\label{eqn:rexpectation}\\
  \langle \bsl{R}m|\hat{\bsl{r}}^2|0n\rangle &= -\frac{1}{N}\sum_{\bsl{k}} \bra{u_m(\bsl{k})}\nabla^2_{\bsl{k}}\ket{u_{n \bsl{k}}} e^{i\bsl{k}\cdot \bsl{R}}\,.\label{eqn:r2expectation}
\end{align}
We divide the localization functional $\Omega$ into the following two parts:
\begin{align}
  \Omega &= \sum_n\left[\langle 0 n | \hat{{\bsl r}}^2 |0n\rangle - |\langle 0 n|\hat{\bsl{r}}|0n \rangle |^2\right]\nonumber\\
  & = \sum_n\left[\langle 0n|\hat{\bsl{r}}^2|0n\rangle - \sum_{\bsl{R}m}|\langle \bsl{R} m| \hat{\bsl{r}} | 0n\rangle |^2\right] + \nonumber\\&
  + \sum_{n}\sum_{\bsl{R}m \neq 0n}|\langle \bsl{R}m|\hat{\bsl{r}}|0n\rangle |^2  \,,\label{eqn:twoparts}
\end{align}
We denote the first and second terms in \cref{eqn:twoparts} by $\Omega_I$ and $\tilde{\Omega}$, respectively. $\tilde{\Omega}$ is obviously always non-negative. Using \cref{eqn:rexpectation} and \cref{eqn:r2expectation},  $\Omega_I$ can be written as the following form: 
\eqa{
  \Omega_I  =&\ -\frac
  {1}{N}\sum_{\bsl{k}} \sum_{n}\bra{u_{n \bsl{k}}}\nabla^2_{\bsl{k}}\ket{u_{n \bsl{k}}} \\ 
  & -  \frac{1}{N}\sum_{\bsl{k}}\sum_{nm} \braket{u_{m  \bsl{k}}}{\nabla_{\bsl{k}} u_{n \bsl{k}}} \cdot \braket{\nabla_{\bsl{k}}  u_{n \bsl{k}}}{ u_{m \bsl{k}}}
}
As $\langle 0 n | \hat{{\bsl r}}^2 |0n\rangle$ is a real number, we can take complex conjugation and obtain
\begin{equation}
 \frac{1}{N}\sum_{\bsl{k}} \braket{u_{n \bsl{k}}}{\nabla^2_{\bsl{k}}u_{n \bsl{k}}} = \frac{1}{N}\sum_{\bsl{k}} \braket{\nabla^2_{\bsl{k}} u_{n \bsl{k}}}{u_{n \bsl{k}}}\,.
\end{equation}
Owign to $\braket{u_{n \bsl{k}}}{u_{n \bsl{k}}} = 1$, the second order derivative gives $\braket{u_{n \bsl{k}}}{\nabla^2_{\bsl{k}}u_{n \bsl{k}}} + \braket{\nabla^2_{\bsl{k}} u_{n \bsl{k}}}{u_{n \bsl{k}}} + 2  \sum_i \braket{\partial_{k_i}u_{n \bsl{k}}}{\partial_{k_i}u_{n \bsl{k}}} = 0$. Consequently, we  obtain
\eqa{
 & -\frac{1}{N}\sum_{\bsl{k}} \braket{u_{n \bsl{k}}}{\nabla^2_{\bsl{k}}u_{n \bsl{k}}}   \\ &= -\frac{1}{2}\frac{1}{N}\sum_{\bsl{k}} \braket{u_{n \bsl{k}}}{\nabla^2_{\bsl{k}}u_{n \bsl{k}}} -\frac{1}{2}\frac{1}{N}\sum_{\bsl{k}} \braket{\nabla^2_{\bsl{k}}u_{n \bsl{k}}}{u_{n \bsl{k}}}  \\
 & = \frac{1}{N}\sum_{\bsl{k}}\sum_i \braket{\partial_{k_i}u_{n \bsl{k}}}{\partial_{k_i}u_{n \bsl{k}}}\ .
}
We use this to replace the first term  in $\Omega_I$:
\eqa{
   \Omega_I  =&\  \frac{1}{N}\sum_{\bsl{k}}\sum_{n,i} \braket{\partial_{k_i}u_{n \bsl{k}}}{\partial_{k_i}u_{n \bsl{k}}}\\ 
  & -  \frac{1}{N}\sum_{\bsl{k}}\sum_{nm,i} \braket{\partial_{k_i}  u_{n \bsl{k}}}{ u_{m \bsl{k}}} \braket{u_{m  \bsl{k}}}{\partial_{k_i} u_{n \bsl{k}}} \\
  & = \frac{1}{N}\sum_{\bsl{k}}{\rm Tr}\left[g(\bsl{k})\right]
}
Since the quantum metric is invariant under gauge transformation, $\Omega_I$ is gauge invariant. If the integral of ${\rm Tr}\left[g(\bsl{k})\right]$ has a nonzero lower bound, the gauge-invariant part of the Wannier function localization functional will also be bounded. Because $\tilde{\Omega}$ is always positive by definition, the lower bound of the gauge invariant part $\Omega_I$ is also the lower bound of the functional $\Omega$ itself.

\section{Examples of Quantum Geometric Contribution to Spin Stiffness}
\label{app:spin_stiffness}

In this appendix, we provide examples for the quantum geometric contribution to the spin stiffness.
Let $H_0$ be the single-particle Hamiltonian for TBG projected on
the active 8 lowest-energy bands, 2 per spin and valley degree of freedom.
The Hamiltonian, $H_I$ describing the Coulomb interaction when projected
on the active bands of $H_0$, takes the form of a positive semidefinite Hamiltonian
that can be written in the form~\cite{bernevig2021d}
\eq{
 H_I = \frac{1}{A} \sum_{\GG,\qq}O_{\qq,\GG}O_{-\qq,-\GG}
 \label{eq.HI}
}
where $A$ is the sample area,
$\{\GG\}$ are the reciprocal lattice vectors of the moir\'e lattice,
$\{\qq\}$ are the wave vectors within the moir\'e BZ and
\begin{align}
 O_{\qq,\GG} =&\!\!\!\! \sum_{\substack{\kk,m,n,\\ \eta,s}} [V(\GG+\qq)]^{1/2}M_{m,n}^{\eta}(k,\qq+\GG) \nonumber \\
  &\times (\rho_{ \kk,\qq, m,n,s}^{(\eta)}-\delta_{\qq,0}\delta_{m,n}/2).
  \label{eq.OOp}
\end{align}
In Eq.~(\ref{eq.OOp}) 
$m (n)\pm 1$, $\eta$, and $s$ are the band, valley, and spin indices, respectively,
$V(\qq)$ is the Coulomb interaction,  
$\rho_{\kk,\qq,m,n,s}^{(\eta)}$ is the density operator in the band basis
and 
$M_{m,n}^{\eta}(k,\qq+\GG) = \braket{u_{\eta,m (\kk+\qq)}}{u_{\eta,n \kk}}$
are the overlap matrices, {\em form factors}, between the bands' $\ket{u_{\eta, m\bsl{k}}}$
for valley $\eta$.
The presence of the overlap matrices in the expression of the operators
$O_{\qq,\GG}$ is responsible for quantum geometry terms in the expressions
for the dispersion of the 
excitations.

Starting from $H_I$ we can write the energy $E_{G}(\qq)$ of the neutral, long wavelength ($q\to 0$), 
``Goldstone'' excitations in terms of the interaction potential and form factors. 
Up to second order in $\qq$, $E_G(\qq)$ can be written as:
\eq{
 E_G({\qq}) = \frac{1}{2}\sum_{ij}[D_s^{(s)}]_{ij} q_i q_j
 \label{eq.energy.sw}
}
The stiffness $[D_s^{(s)}]_{ij}$ can be expressed analytically in some limits. 

In \refcite{bernevig2021f}, it was shown that in the first chiral limit, the limit in which 
the interlayer coupling between A sublattices is zero,
the form factors take a simple, diagonal, form~\cite{bernevig2021f}:
\eq{
 M_{e_Y}^{(\eta)}(\kk,\qq+\GG) = \alpha_0(\kk,\qq+\GG)\xi^0\tau^0 + i\xi^y\tau^0\alpha_2(\kk,\qq+\GG)
 }
where $e_Y$ is the Euler's class ($C=\pm e_Y=\pm 1$), $\xi^i$, $\tau^i$ are Pauli-matrices acting on the band and valley indices, respectively,
and $\alpha_i$ are real functions.
In this limit, when $M_{m,n}^{\eta}(\kk,\qq+\GG)$ for $\qq=0$ do not depend on $\kk$, flat metric condition that is valid 
at, or close to, the first magic angle~\cite{bernevig2021e}, we have~\cite{bernevig2021f}:
\begin{align}
 [D_s^{(s)}]_{ij} = &\frac{1}{2A}\sum_{\kk,\qq,\GG}V(\GG+\qq)\nonumber \\
 \times [&\alpha_0(\kk,\qq+\GG)\partial_{k_i}\partial_{k_j}\alpha_0(\kk,\qq+\GG) \nonumber \\
 & + \alpha_2(\kk,\qq+\GG)\partial_{k_i}\partial_{k_j}\alpha_2(\kk,\qq+\GG)  \nonumber \\
 & + 2\partial_{k_i}\alpha_0(\kk,\qq+\GG)\partial_{k_j}\alpha_0(\kk,\qq+\GG) \nonumber \\
 &  + 2\partial_{k_i}\alpha_2(\kk,\qq+\GG)\partial_{k_j}\alpha_2(\kk,\qq+\GG)]
   \label{eq.rhos.01}
\end{align}
from which the explicit relation between the spin stiffness and the metric of the bands can be extracted.
Within mean-field theory, and a single-mode approximation, Ref.~\cite{wu2021} obtained an explicit, but approximate,
relation between the spin-wave stiffness of the spin-and-valley maximally polarized
sate at filling $\nu=3$ and the quantum geometric properties of the highest-energy filled band:
\eq{
 [D_s^{(s)}]_{ij}  = \frac{1}{A N_{MC}}\delta_{ij}\sum_{\kk\qq,l,m}\Omega_{\kk}^2\exp(-q_l g_{lm} q_m),
 \label{eq.rhos.02}
}
where $N_{MC}$ is the number of moir\'e cells within $A$.

For the case of saturated ferromagnetism, using the variational approach, the authors of Ref.~\cite{kang2024} found that in the strongly correlated limit,
\eq{
  [D_s^{(s)}]_{ij}  = \frac{1}{N_c N_{\rm occ}}\sum_{\kk}^{N_c}\sum_n^{N_{\rm occ}}[\partial_{k_i}\partial_{k_j}\epsilon_{n \kk} + 2\Delta(\kk)g_{ij}(\kk)],
 \label{eq.rhos.03}
}
where $\epsilon_{n \kk}$ are the eigenvalues of the single-particle Hamiltonian, $N_c$ is the number of unit cells in the sample,
and $N_{\rm occ}$ is the number of occupied states per unit cell. Notice that in this
approximation, the occupied (and unoccupied) bands are degenerate, so $g_{ij}(\kk)$
in Eq.~(\ref{eq.rhos.02}) is the Abelian (minimal) quantum metric.

Equations~\ceq{eq.rhos.01},\ceq{eq.rhos.02},\ceq{eq.rhos.03} exemplify, for specific cases,
the role of the bands' quantum geometry in determining the spin, or pseudospin, stiffness
and therefore the dispersion of the associated Goldstone modes.

\section{Quantum Geometry and Electron-Phonon Coupling}

\label{app:EPC}

In this appendix, we provide additional details on the Gaussian approximation (GA) based on \refcite{Yu05032023GeometryEPC} and use it to illustrate how quantum geometry contributes to the EPC.

Under the tight-binding approximation and Frohlich two-center approximation~\cite{Mitra1969EPC}, the non-interacting electron Hamiltonian and the EPC Hamiltonian are directly given by the smooth hopping function \( t(\bsl{r}) \), which satisfies \( [t(\bsl{r})]^* = t(-\bsl{r}) \) to ensure Hermiticity.

Here, \( t(\bsl{r}) \) does not carry any orbital, sublattice, or spin indices, as we focus on one type of atom and one spinless \( s \) orbital per atom, though more than one atom per unit cell may be present.

Using the hopping function, the electron Hamiltonian (without the Coulomb interaction) that accounts for atomic motions reads 
\eqa{
\label{eq:H_el+elph_Frohlich}
& H_{el+ion-motions} =\sum_{\bsl{R}\bsl{\tau}, \bsl{R}'\bsl{\tau}'}  
c^\dagger_{\bsl{R}+\bsl{\tau}} c_{\bsl{R}'+\bsl{\tau}'}\\
& \qquad \times t(\bsl{R}+\bsl{\tau}+\bsl{u}_{\bsl{R}+\bsl{\tau}}-\bsl{R}'-\bsl{\tau}'-\bsl{u}_{\bsl{R}'+\bsl{\tau}'})\ ,
}
where \( \bsl{R} \) labels the lattice point, \( \bsl{\tau} \) labels the positions of the sublattices in the \( \bsl{R}=0 \) unit cell, \( c^\dagger_{\bsl{R}+\bsl{\tau}} \) creates an electron in the spinless \( s \) orbital at \( \bsl{R}+\bsl{\tau} \), and \( \bsl{u}_{\bsl{R}+\bsl{\tau}} \) denotes the displacement of the atom at \( \bsl{R}+\bsl{\tau} \).

Since the hopping function generally decays exponentially as \( |\bsl{r}| \) increases, \( t(\bsl{R}+\bsl{\tau}+\bsl{u}_{\bsl{R}+\bsl{\tau}}-\bsl{R}'-\bsl{\tau}'-\bsl{u}_{\bsl{R}'+\bsl{\tau}'}) \) can be expanded in a series of \( (\bsl{u}_{\bsl{R}+\bsl{\tau}}-\bsl{u}_{\bsl{R}'+\bsl{\tau}'}) \).

The zeroth-order term yields the non-interacting electron Hamiltonian under the tight-binding approximation:
\eq{
\label{eq:H_el_s}
H_{el}=\sum_{\bsl{R}\bsl{\tau}, \bsl{R}'\bsl{\tau}'}  
t(\bsl{R}+\bsl{\tau}-\bsl{R}'-\bsl{\tau}')
c^\dagger_{\bsl{R}+\bsl{\tau}} 
c_{\bsl{R}'+\bsl{\tau}'}\ ,
}
and the first-order term is the leading order term for EPC, given by
\eqa{
\label{eq:H_elph_Frohlich}
& H_{el-ph} = \sum_{\bsl{R}\bsl{\tau}, \bsl{R}'\bsl{\tau}'}  \sum_{ \alpha_{\bsl{\tau}} \alpha'_{\bsl{\tau}'}} c^\dagger_{ \bsl{R}+\bsl{\tau},\alpha_{\bsl{\tau}}} 
c_{\bsl{R}'+\bsl{\tau}',\alpha_{\bsl{\tau}'}'} \\
& \qquad \times
(\bsl{u}_{\bsl{R}+\bsl{\tau}}-\bsl{u}_{\bsl{R}'+\bsl{\tau}'})\cdot \left.\nabla_{\bsl{r}} t(\bsl{r})\right|_{\bsl{r} = \bsl{R}+\bsl{\tau}-\bsl{R}'-\bsl{\tau}'}
\ .
}
Higher-order terms are typically neglected.

The GA assumes that the hopping function has a Gaussian form:
\eq{
t(\bsl{r}) = t_0 \exp\left(\gamma \frac{r^2}{2}\right)\ ,
}
where \( r=|\bsl{r}| \), and \( \gamma<0 \) is determined by the standard deviation of the Gaussian function.
As a result, we have
\eq{
\label{eq:t_gradient_Gaussian_s}
\nabla_{\bsl{r}} t(\bsl{r}) = \gamma \bsl{r} t(\bsl{r})\ .
}

\eqnref{eq:t_gradient_Gaussian_s} converts the spatial derivative to the position vector in the EPC Hamiltonian (\eqnref{eq:H_elph_Frohlich}), along with an additional factor of \( \gamma \).
To better understand how this conversion connects the EPC Hamiltonian to the electron Hamiltonian, we transform the Hamiltonian to momentum space.
Specifically, the Fourier transformation rule for the basis states reads
\eqa{
\label{eq:FT_Gaussian}
& c_{\bsl{k},\bsl{\tau}}^\dagger = \frac{1}{\sqrt{N}} \sum_{\bsl{R}} e^{\ii \bsl{k}\cdot (\bsl{R}+\bsl{\tau})} c_{\bsl{R}+\bsl{\tau}}^\dagger\\
& u_{\bsl{q}\bsl{\tau} i} = \frac{1}{\sqrt{N}} \sum_{\bsl{R}} e^{-\ii \bsl{q}\cdot (\bsl{R}+\bsl{\tau})} u_{\bsl{R}+\bsl{\tau}, i}\ ,
}
from which we obtain 
\eq{
u_{\bsl{q}\bsl{\tau} i}^\dagger = u_{-\bsl{q}\bsl{\tau} i}\ .
}
For the electron Hamiltonian,
\eqa{
\label{eq:H_el_gen_k_gaussian_s}
H_{el} & = \sum_{\bsl{k}}^{\text{\BZ}} c^\dagger_{\bsl{k}} h(\bsl{k}) c_{\bsl{k}} = \sum_{\bsl{k}}^{\text{\BZ}}\sum_{n} \epsilon_{n \bsl{k}} \gamma^\dagger_{\bsl{k},n} \gamma_{\bsl{k},n}\ ,
}
where 
\( c^\dagger_{\bsl{k}}= (..., c^\dagger_{\bsl{k},\bsl{\tau}} , ...) \),
\eq{
\label{eq:h_k_gaussian_s}
\left[ h(\bsl{k}) \right]_{\bsl{\tau}\bsl{\tau}'} = \sum_{\bsl{R}}t(\bsl{R}+\bsl{\tau}-\bsl{\tau}')  e^{-\ii \bsl{k}\cdot (\bsl{R} + \bsl{\tau} - \bsl{\tau}')}\ ,
}
\eq{
 h(\bsl{k}) U_{n \bsl{k}} = \epsilon_{n \bsl{k}} U_{n \bsl{k}}\ ,
}
and \( \gamma^\dagger_{\bsl{k},n} = c^\dagger_{\bsl{k}} U_{n \bsl{k}} \).
Under the tight-binding approximation, the term ``quantum geometry" (or band geometry) typically refers to the momentum dependence of 
\eq{
\label{eq:P_n_Gaussian}
P_{n \bsl{k}} = U_{n \bsl{k}} U_{n\bsl{k}}^\dagger\ ,
}
and we will particularly use the quantum metric
\eqa{
\label{eq:FSM_g_two_expressions_Gaussian}
g_{n,ij}(\bsl{k}) & = \frac{1}{2}\Tr\left[ \partial_{k_i} P_{n \bsl{k}}  \partial_{k_j} P_{n \bsl{k}} \right] \\
& = \frac{1}{2}\Tr\left[ \partial_{k_i} P_{n \bsl{k}} P_{n \bsl{k}} \partial_{k_j} P_{n \bsl{k}} \right] + (i\leftrightarrow j)\ .
}

For the EPC in momentum space, we have
\eqa{
\label{eq:H_el-ph_Frolich_k}
 & H_{el-ph}    \\
 & = \frac{1}{\sqrt{N}}\sum_{\bsl{k}_1}^{\BZ} \sum_{\bsl{k}_2}^{\BZ}  \sum_{ \bsl{\tau} ,  i}     c^\dagger_{\bsl{k}_1}\left[ \chi_{\bsl{\tau}} f_{i}(\bsl{k}_2) - f_{i}(\bsl{k}_1)\chi_{\bsl{\tau}} \right] c_{\bsl{k}_2}   u^\dagger_{\bsl{k}_2 - \bsl{k}_1,\bsl{\tau},i} \ ,
}
where 
\eq{
\left[ \chi_{\bsl{\tau}} \right]_{\bsl{\tau}_1 \bsl{\tau}_2} = \delta_{ \bsl{\tau}, \bsl{\tau}_{1}} \delta_{\bsl{\tau}_1 \bsl{\tau}_2 }
}
is the projection matrix onto the \(\bsl{\tau}\) sublattice,
\eq{
\label{eq:g_k_Frolich}
\left[f_{i}(\bsl{k})\right]_{\bsl{\tau}_1 \bsl{\tau}_2 } = \sum_{ \bsl{R} } e^{- \ii \bsl{k} \cdot ( \bsl{R}+ \bsl{\tau}_1 - \bsl{\tau}_2)}   \left. \partial_{r_i} t(\bsl{r})\right|_{\bsl{r} = \bsl{R}+\bsl{\tau}_1-\bsl{\tau}_2}  \ ,
}
and \( i=x,y,z \) labels the spatial direction.
It is clear from \eqnref{eq:H_el-ph_Frolich_k} that the form of the EPC Hamiltonian is governed by \( f_{i}(\bsl{k}) \), which we focus on below.

Using \eqnref{eq:t_gradient_Gaussian_s}, we find that the EPC term \( f_{i}(\bsl{k}) \) relates to the electron matrix Hamiltonian \( h(\bsl{k}) \) as
\eq{
\left[f_{i}(\bsl{k})\right]_{\bsl{\tau}_1 \bsl{\tau}_2 }  = \ii \gamma \partial_{k_i} \left[h(\bsl{k})\right]_{\bsl{\tau}_1 \bsl{\tau}_2 }\ ,
}
which means that 
\eq{
\label{eq:g_k_Frolich_Gaussian_s}
f_{i}(\bsl{k}) = \ii \gamma \partial_{k_i}  h(\bsl{k})\ .
}
We refer to \eqnref{eq:g_k_Frolich_Gaussian_s} as the Gaussian form of the EPC.

The Gaussian form of the EPC allows us to define the energetic and geometric components of the EPC.
Note that the electron matrix Hamiltonian contains information about both the bands and the projection matrix, i.e.,
\eq{
h(\bsl{k}) = \sum_{n}\epsilon_{n \bsl{k}} P_{n \bsl{k}}\ ,
}
where the projection matrix \( P_{n \bsl{k}} \) is defined in \eqnref{eq:P_n_Gaussian}.
Then, we have
\eq{
\label{eq:g_g_E_g_geo_Gaussian_s}
f_{i}(\bsl{k}) = \ii \gamma \partial_{k_i}  h(\bsl{k})  = f_{i}^E(\bsl{k}) + f^{geo}(\bsl{k})\ ,
}
where 
\eq{
\label{eq:g_E_Gaussian_s}
f_{i}^E(\bsl{k}) = \ii \gamma   \sum_{n} \partial_{k_i} \epsilon_{n \bsl{k}} P_{n \bsl{k}} 
}
is the energetic component of the EPC, which vanishes if all electron bands are exactly flat, and 
\eq{
\label{eq:g_geo_Gaussian_s}
f_{i}^{geo}(\bsl{k}) =  \ii \gamma  \sum_{n}\epsilon_{n \bsl{k}} \partial_{k_i} P_{n \bsl{k}}
}
is the geometric component of the EPC, as \( f_{i}^{geo}(\bsl{k}) \) depends on the geometric properties of the Bloch eigenvector \( U_{n \bsl{k}} \) (i.e., the momentum dependence of \( P_{n \bsl{k}} \)).
In the one-band case (i.e., with only one atom per unit cell), \( f_{i}^{geo}(\bsl{k}) \) must vanish since \( P_{n \bsl{k}}=1 \) is independent of momentum (\( n \) can only take one value in the one-band case), while \( f_{i}^E(\bsl{k}) \) can still be nonzero since the energy band can still vary with momentum.


The key quantity we study is the dimensionless EPC constant \( \lambda \)~\cite{McMillan1968SCTc}, which is given by
\eqa{
\label{eq:lambda_omegabar_Gaussian_s}
\lambda  = \frac{2}{ N} D(\mu)  \frac{1}{ \mcomega}  \left\langle \Gamma \right\rangle \ ,
}
where \( \mu \) is the chemical potential, \( D(\mu) \) is the electron density of states at \( \mu \), and \( \mcomega \) is the mean-squared phonon frequency as defined in \refcite{McMillan1968SCTc}.
The EPC matrix element \( \Gamma_{n n'}(\bsl{k},\bsl{k}') \) is defined by
\eqa{
\label{eq:Gamma_nm_Gaussian_s}
  \Gamma_{n n'}(\bsl{k},\bsl{k}') & = \frac{1}{2 m} \sum_{\bsl{\tau},i}  \Tr\left\{ P_{n\bsl{k}}  \left[ \chi_{\bsl{\tau}} f_{i}(\bsl{k}') - f_{i}(\bsl{k})\chi_{\bsl{\tau}} \right] \right.\\
  & \left. \times P_{n'\bsl{k}'}   \left[ \chi_{\bsl{\tau}} f_{i}(\bsl{k}) - f_{i}(\bsl{k}')\chi_{\bsl{\tau}} \right] \right\} \ ,
 }
where \( m \) is the atomic mass, and 
 \eq{
\left\langle \Gamma \right\rangle =\frac{ \sum_{\bsl{k},\bsl{k}'}^{\BZ}\sum_{n,n'} \delta\left(\mu - \epsilon_{n \bsl{k}} \right) \delta\left(\mu - \epsilon_{n' \bsl{k}'} \right) \Gamma_{n n'}(\bsl{k},\bsl{k}') }{\sum_{\bsl{k},\bsl{k}'}^{\BZ}\sum_{n,n'} \delta\left(\mu - \epsilon_{n \bsl{k}} \right) \delta\left(\mu - \epsilon_{n' \bsl{k}'} \right) }\ .
}
As discussed in \refcite{Yu05032023GeometryEPC}, we will focus on \( \left\langle \Gamma \right\rangle \) and treat \( \mcomega \) as a parameter determined by first-principles calculations, primarily because \( \mcomega \) can be well-approximated by the frequency of specific phonon modes in graphene and \(\mathrm{MgB}_2\)~\cite{Yu05032023GeometryEPC}.
By combining \eqnref{eq:Gamma_nm_Gaussian_s} with \eqnref{eq:g_g_E_g_geo_Gaussian_s}, we can express 
\eq{
  \Gamma_{n n'}(\bsl{k},\bsl{k}') = \Gamma_{n n'}^{E-E}(\bsl{k},\bsl{k}')  + \Gamma_{n n'}^{geo-geo}(\bsl{k},\bsl{k}') + \Gamma_{n n'}^{E-geo}(\bsl{k},\bsl{k}') \ ,
 }
where
\eqa{
\label{eq:Gamma_X}
& \Gamma_{n n'}^{E-E}(\bsl{k},\bsl{k}') = \frac{1}{2 m} \sum_{\bsl{\tau},i}  \Tr\left\{ P_{n\bsl{k}}  \left[ \chi_{\bsl{\tau}} f_{i}^E(\bsl{k}') - f_{i}^E(\bsl{k})\chi_{\bsl{\tau}} \right] \right. \\
& \qquad \times \left. P_{n'\bsl{k}'}   \left[ \chi_{\bsl{\tau}} f_{i}^E(\bsl{k}) - f_{i}^E(\bsl{k}')\chi_{\bsl{\tau}} \right] \right\} \\
& \Gamma_{n n'}^{geo-geo}(\bsl{k},\bsl{k}')  = \frac{1}{2 m} \sum_{\bsl{\tau},i}  \Tr\left\{ P_{n\bsl{k}}  \left[ \chi_{\bsl{\tau}} f_{i}^{geo}(\bsl{k}') - f_{i}^{geo}(\bsl{k})\chi_{\bsl{\tau}} \right] \right. \\
& \qquad \times \left. P_{n'\bsl{k}'}   \left[ \chi_{\bsl{\tau}} f_{i}^{geo}(\bsl{k}) - f_{i}^{geo}(\bsl{k}')\chi_{\bsl{\tau}} \right] \right\} \\
& \Gamma_{n n'}^{E-geo}(\bsl{k},\bsl{k}') = \frac{1}{2 m} \sum_{\bsl{\tau},i}  \Tr\left\{ P_{n\bsl{k}}  \left[ \chi_{\bsl{\tau}} f_{i}^{E}(\bsl{k}') - f_{i}^{E}(\bsl{k})\chi_{\bsl{\tau}} \right] \right. \\
& \qquad \times \left. P_{n'\bsl{k}'}   \left[ \chi_{\bsl{\tau}} f_{i}^{geo}(\bsl{k}) - f_{i}^{geo}(\bsl{k}')\chi_{\bsl{\tau}} \right] \right\} + c.c.\ .
}
By defining
\eq{
\label{eq:Gamma_mu_X}
\left\langle \Gamma \right\rangle^{X} =\frac{ \sum_{\bsl{k},\bsl{k}'}^{\BZ}\sum_{n,n'} \delta\left(\mu - \epsilon_{n \bsl{k}} \right) \delta\left(\mu - \epsilon_{n' \bsl{k}'} \right) \Gamma_{n n'}^X(\bsl{k},\bsl{k}') }{\sum_{\bsl{k},\bsl{k}'}^{\BZ}\sum_{n,n'} \delta\left(\mu - \epsilon_{n \bsl{k}} \right) \delta\left(\mu - \epsilon_{n' \bsl{k}'} \right) }
}
for \( X= E-E,\ geo-geo,\ E-geo \),
we arrive at 
\eq{
\lambda = \lambda_E + \lambda_{geo} + \lambda_{E-geo}\ ,
}
where
\eqa{
\label{eq:lambda_E}
\lambda_E = \frac{2}{ N} D(\mu)  \frac{1}{ \mcomega} \left\langle \Gamma \right\rangle^{E-E}
}
is the energetic contribution to \( \lambda \) as it depends only on \( f^E_i \) and not on \( f^{geo}_i \), 
\eqa{
\label{eq:lambda_geo}
\lambda_{geo} = \frac{2}{ N} D(\mu)  \frac{1}{ \mcomega} \left\langle \Gamma \right\rangle^{geo-geo}
}
is the geometric contribution to \( \lambda \) as it depends only on \( f^{geo}_i \) and not on \( f^E_i \),
and 
\eqa{
\label{eq:lambda_E_geo}
\lambda_{E-geo} = \frac{2}{ N} D(\mu)  \frac{1}{ \mcomega} \left\langle \Gamma \right\rangle^{E-geo}
}
is the cross-term contribution to \( \lambda \) as it depends on both \( f^{geo}_i \) and \( f^E_i \).

We now further examine the expressions of \( \left\langle \Gamma \right\rangle^{geo-geo} \) and \( \lambda_{geo} \).
We can split \( \left\langle \Gamma \right\rangle^{geo-geo} \) into two parts:
\eqa{
  \left\langle \Gamma \right\rangle^{geo-geo}  = \left\langle \Gamma \right\rangle^{geo-geo,2} +  \left\langle \Gamma \right\rangle^{geo-geo,1}\ ,
}
where 
\eqa{
  & \left\langle \Gamma \right\rangle^{geo-geo,1}   =  - \frac{1}{D^2(\mu)} \sum_{\bsl{\tau},i}  \frac{1}{m}  \sum_{\bsl{k}_1,\bsl{k}_2}^{\BZ}\sum_{n,m} \delta\left(\mu - \epsilon_{n\bsl{k}_1} \right) \\
  & \times \delta\left(\mu - \epsilon_{m\bsl{k}_2} \right)  \Tr\left[ f_{i}^{geo}(\bsl{k}_1) P_{n\bsl{k}_1}  f_{i}^{geo}(\bsl{k}_1)\chi_{\bsl{\tau}} P_{m \bsl{k}_2}   \chi_{\bsl{\tau}}  \right]\\
  &  \left\langle \Gamma \right\rangle^{geo-geo,2}   = \frac{1}{D^2(\mu)}  \frac{1}{2} \frac{1}{m}  \sum_{\bsl{\tau},i}  \\
  & \times \Tr\left[ \left(\sum_{\bsl{k}_1}^{\BZ}\sum_{n}  \delta\left(\mu - \epsilon_{n\bsl{k}_1} \right) \chi_{\bsl{\tau}} f_{i}^{geo}(\bsl{k}_1)  P_{n \bsl{k}_1} \right)^2    \right] +c.c.\ .
}
Similarly, \( \lambda_{geo} \) can be decomposed into two parts:
\eq{
\lambda_{geo} = \lambda_{geo,1}+ \lambda_{geo,2}\ ,
}
where 
\eq{
\lambda_{geo,1/2} = \frac{2}{ N} D(\mu)  \frac{1}{ \mcomega} \left\langle \Gamma \right\rangle^{geo-geo,1/2}\ .
}

For convenience in illustrating the explicit geometric dependence, we re-write \( \left\langle \Gamma \right\rangle^{geo-geo,1/2} \).
\( \left\langle \Gamma \right\rangle^{geo-geo,1} \) can be re-written as
\eqa{
\label{eq:Gamma_ave_geo-geo,1_GA}
  &  \left\langle \Gamma \right\rangle^{geo-geo,1}  =   \frac{ \gamma^2}{D(\mu)} \sum_{i}  \frac{1}{m}  \sum_{\bsl{k}}^{\BZ}\sum_{n,n_1,n_2} \delta\left(\mu - \epsilon_{n \bsl{k}} \right) \\
  & \qquad \times \epsilon_{n_1 \bsl{k}} \epsilon_{n_2 \bsl{k}} \Tr\left[ \partial_{k_i} P_{n_1 \bsl{k}}  P_{n\bsl{k}}  \partial_{k_i} P_{n_2 \bsl{k}} M    \right]\ ,
}
where
\eqa{
\label{eq:M_expression_GA}
M  & = \frac{1}{D(\mu) }\sum_{\bsl{\tau}} \sum_m \sum_{\bsl{k}_2}^{\BZ}\delta\left(\mu - \epsilon_{m\bsl{k}_2} \right) \chi_{\bsl{\tau}} P_{m \bsl{k}_2}   \chi_{\bsl{\tau}} \\
& = \sum_{\bsl{\tau}} a_{\bsl{\tau}} \chi_{\bsl{\tau}}\ ,
}
and
\eq{
\label{eq:a_tau_expression_GA}
a_{\bsl{\tau}} = \frac{1}{D(\mu) } \sum_m \sum_{\bsl{k}_2}^{\BZ}\delta\left(\mu - \epsilon_{m\bsl{k}_2} \right) \left[ P_{m \bsl{k}_2} \right]_{\bsl{\tau}\bsl{\tau}} 
}

In the following, we consider the two-band case (i.e., a system with only two electron bands) for clearer illustration of the explicit geometric dependence.

For \( \lambda_{geo,1} \), we have 
\eqa{
\label{eq:lambda_geo_1}
\lambda_{geo,1} 
& = \frac{2\Omega \gamma^2}{ (2\pi)^3 m \mcomega} \sum_{n,i,\bsl{\tau}} \int_{FS_n} d\sigma_{\bsl{k}}\frac{\Delta E^2(\bsl{k})}{|\nabla_{\bsl{k}} \epsilon_{n \bsl{k}}|}  a_{\bsl{\tau}} \left[g_{n,\bsl{\tau},ii}(\bsl{k}) \right]_{ii}\ ,
}
where \( \Delta E(\bsl{k}) = | E_2(\bsl{k}) - E_1(\bsl{k})| \),
\eq{
\left[ g_{n,\bsl{\tau}}(\bsl{k}) \right]_{ij}=  \frac{1}{2}\Tr\left[ \partial_{k_i} P_{n\bsl{k}}  P_{n\bsl{k}}  \partial_{k_j} P_{n\bsl{k}} \chi_{\bsl{\tau}} \right] +(i\leftrightarrow j)\ ,
}
and \( a_{\bsl{\tau}} \) is defined in \eqnref{eq:a_tau_expression_GA}.
As discussed in \refcite{Yu05032023GeometryEPC}, \( g_{n,\bsl{\tau}}(\bsl{k}) \) is an orbital-selective quantum metric, as it is defined by inserting the projection matrix \( \chi_{\bsl{\tau}} \) into the original definition of the quantum metric.
Therefore, in the two-band case, \( \lambda_{geo,1} \) directly depends on the linear combination of the orbital-selective quantum metric.

We can further simplify \( \lambda_{geo,2} \) to 
\eqa{
\label{eq:lambda_geo,2_GA_twoband}
& \lambda_{geo,2} = \frac{N \gamma^2}{ m D(\mu) \mcomega}  \sum_{\bsl{\tau},i} \\
& \times \left( \frac{\Omega}{(2\pi)^3} \sum_{n}\int_{FS_n} d\sigma_{\bsl{k}}\frac{\Delta E(\bsl{k})}{|\nabla_{\bsl{k}} \epsilon_{n \bsl{k}}|}  \mathcal{A}_{i,n,\bsl{\tau}}(\bsl{k})\right)^2  + c.c.  \ ,
}
where 
\eq{
\label{eq:A_n_tau}
\bsl{\mathcal{A}}_{n,\bsl{\tau}}(\bsl{k}) =  \Tr\left[ \chi_{\bsl{\tau}} \nabla_{\bsl{k}} P_{n \bsl{k}} P_{n \bsl{k}} \right] 
}
is an orbital-selective complex vector field.
Thus, the explicit geometric dependence in \( \lambda_{geo,2} \) arises from the orbital-selective complex vector field.

Finally, in the two-band case, we can also explicitly derive the geometric dependence of \( \lambda_{E-geo} \), which is given by
\eqa{
& \lambda_{E-geo} =  \frac{\Omega}{(2\pi)^3} \frac{4  \gamma^2}{m \mcomega}  \sum_{n} \int_{FS_n} d\sigma_{\bsl{k}}\frac{\Delta E(\bsl{k})}{|\nabla_{\bsl{k}} \epsilon_{n \bsl{k}}|}  (-1)^n \\
& \qquad \times \sum_{\bsl{\tau},i}        \Re[\mathcal{A}_{\bsl{\tau},n,i} (\bsl{k})] \left(  \partial_{k_i}\epsilon_{n \bsl{k}} a_{\bsl{\tau}} - b_{\bsl{\tau},i}  \right)\ , 
}
where \( a_{\bsl{\tau}} \) is defined in \eqnref{eq:a_tau_expression_GA}, \( \mathcal{A}_{\bsl{\tau},n,i} (\bsl{k}) \) is the orbital-selective complex vector field in \eqnref{eq:A_n_tau}, and
\eq{
b_{\bsl{\tau},i} =\frac{1}{D(\mu)}  \sum_{\bsl{k}'}^{\BZ}\sum_{n'}   \delta\left(\mu - \epsilon_{n' \bsl{k}'} \right)  \Tr\left[  \partial_{k'_i}\epsilon_{n' \bsl{k}'} P_{n'\bsl{k}'} \chi_{\bsl{\tau}} \right]\ .
}
Thus, we see that the explicit geometric quantity in \( \lambda_{E-geo} \) is again the orbital-selective complex vector field, similar to \( \lambda_{geo,2} \).

\section{Quantum Geometry of Nearly Flat Chern Bands}
\label{app:FCI}

In this appendix, we will compute the quantum metric of the nearly flat Chern bands of the two-band model in \refcite{Sun2011}, which was used in \refcite{regnaultbernevig} to demonstrate FCIs.
The model is defined on the square lattice with lattice constant set to 1; the matrix Hamiltonian in the momentum space reads
\eq{
\label{eq:two-band}
h(\bsl{k}) = \mat{ m(\bsl{k}) & r(\bsl{k}) \\  r^*(\bsl{k}) & - m(\bsl{k}) }\ ,
}
where
\eq{
m(\bsl{k}) = 2 t_2 [\cos(k_x) - \cos(k_y) ] + M\ ,
}
and
\eq{
r(\bsl{k})  = t_1 e^{\ii \phi} \left[ 1 + e^{\ii (k_y - k_x)} \right] + t_1 e^{-\ii \phi} \left[ e^{\ii k_y} + e^{- \ii  k_x} \right]  \ .
}

\begin{figure}
		\begin{center}
			\includegraphics[width=\columnwidth]{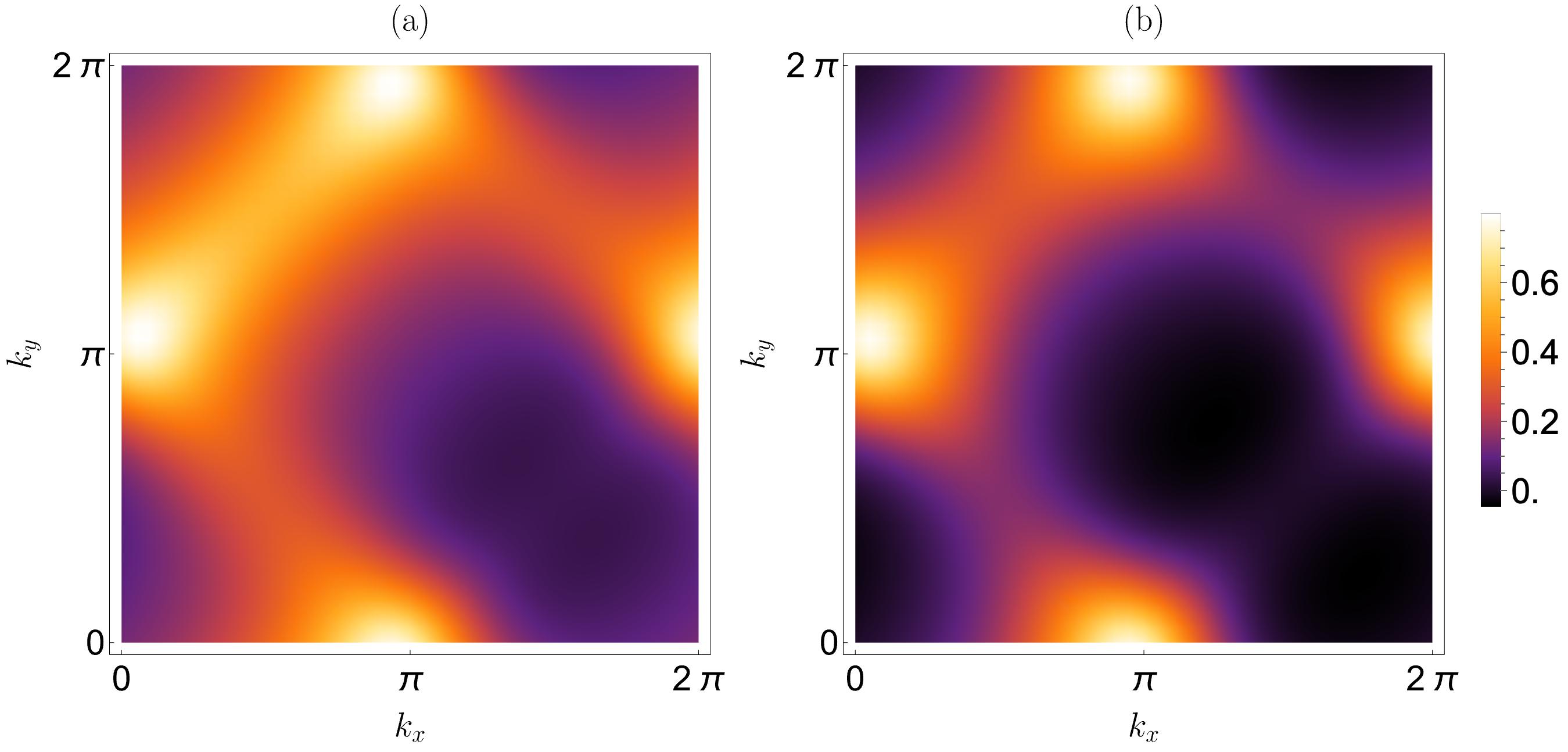}
		\end{center}
		\caption{
			The plots of the trace of quantum metric ($\Tr[g(\bsl{k})]$) and the Berry curvature of the lower band of the model in \eqnref{eq:two-band} in (a) and (b), respectively.
		}
		\label{Trg_BC_plot}
	\end{figure}

As discussed in \refcite{regnaultbernevig}, clear numerical evidence of the FCIs can be found for $t_2 = (2-\sqrt{2})/2 t_1 > 0 $, $\phi=\pi/4$, and $M=0$, when (i) we consider the fractional fillings of the lower band and (ii) we artificially flatten the dispersion of the lower band.
Clearly, flattening the dispersion is to minic the exact flatness of the Landau level. 
One the other hand, as shown in \cref{Trg_BC_plot}, the quantum metric and Berry curvature of the lower band have considerable fluctuations, which are not that close to the uniform ones in Landau levels.
In particular, the integration over $\Tr[g(\bsl{k})]$, \ie, $\frac{1}{2\pi}\int d^2 k \Tr[g(\bsl{k})]$, is 1.71 which is not that close to the Chern number 1 of that band.
Nevertheless, FCIs can still exist despite the fluctuations of the quantum geometry demonstrates.

\end{document}